\input harvmac
\input amssym.def
\input amssym.tex

\def\^{{\wedge}}
\def\*{{\star}}
\def\bar{\overline}
\def\hat{\widehat}

\def\ad#1{{\rm ad}#1}
\def\e#1{{\rm e}^{\, #1}}
\def\wt#1{\widetilde#1}

\def\Td{\mathop{\rm Td}}

\def\Vol{\mathop{\rm Vol}}
\def\Map{\mathop{\rm Map}}
\def\Sym{\mathop{\rm Sym}}
\def\Im{\mathop{\rm Im}}
\def\mod{\mathop{\rm mod}}
\def\Res{\mathop{\rm Res}}
\def\sgn{\mathop{\rm sign}}
\def\rk{\mathop{\rm rk}}

\def\BC{{\Bbb C}}
\def\BP{{\Bbb P}}

\def\BR{{\Bbb R}}
\def\BZ{{\Bbb Z}}

\def\CA{{\cal A}}
\def\CC{{\cal C}}
\def\CD{{\cal D}}
\def\CE{{\cal E}}

\def\CG{{\cal G}}
\def\CH{{\cal H}}

\def\CL{{\cal L}}
\def\CM{{\cal M}}

\def\CQ{{\cal Q}}
\def\CR{{\cal R}}
\def\CS{{\cal S}}
\def\CX{{\cal X}}
\def\CV{{\cal V}}

\def\Fg{{\frak g}}
\def\Fh{{\frak h}}

\def\Ft{{\frak t}}

\noblackbox

\def\urlfont{\hyphenpenalty=10000 \hyphenchar\tentt='057 \tt}

\newbox\tmpbox\setbox\tmpbox\hbox{\abstractfont PUPT-2150}
\Title{\vbox{\baselineskip12pt\hbox{\hss hep-th/0503126}
\hbox{PUPT-2150}}}
{\vbox{
\centerline{Non-Abelian Localization For Chern-Simons Theory}}}
\smallskip
\centerline{Chris Beasley}
\smallskip
\centerline{\it{Joseph Henry Laboratories, Princeton University}}
\centerline{\it{Princeton, New Jersey 08544}}
\medskip
\centerline{and}
\medskip
\centerline{Edward Witten}
\smallskip
\centerline{\it{School of Natural Sciences, Institute for Advanced
Studies}}
\centerline{\it{Princeton, New Jersey 08540}}
\bigskip\bigskip

We reconsider Chern-Simons gauge theory on a Seifert manifold $M$ (the
total space of a nontrivial circle bundle over a Riemann surface
$\Sigma$).  When $M$ is a Seifert manifold, Lawrence and Rozansky have
shown from the exact solution of Chern-Simons theory that the
partition function has a remarkably simple structure and can be
rewritten entirely as a sum of local contributions from the flat
connections on $M$.  We explain how this empirical fact follows from
the technique of non-abelian localization as applied to the
Chern-Simons path integral.  In the process, we show that the
partition function of Chern-Simons theory on $M$ admits a topological
interpretation in terms of the equivariant cohomology of the moduli
space of flat connections on $M$.

\Date{March 2005}

\lref\AchucarroVZ{
A.~Achucarro and P.~K.~Townsend,
``A Chern-Simons Action For Three-Dimensional Anti-De Sitter Supergravity
Theories,''
Phys.\ Lett.\ B {\bf 180} (1986) 89--92.}

\lref\AganagicJS{
M.~Aganagic, H.~Ooguri, N.~Saulina and C.~Vafa,
``Black Holes, q-Deformed 2d Yang-Mills, and Non-Perturbative
Topological Strings,'' {\urlfont hep-th/0411280}.}

\lref\AganagicQG{
M.~Aganagic, M.~Marino and C.~Vafa,
``All Loop Topological String Amplitudes From Chern-Simons Theory,''
Commun.\ Math.\ Phys.\  {\bf 247} (2004) 467--512,
{\urlfont hep-th/0206164}.}

\lref\AtiyahCS{
M.~F. Atiyah, ``Circular Symmetry and Stationary-Phase
Approximation,'' in {\it Colloquium in Honor of Laurent Schwartz},
Vol. 1, Ast\'erisque {\bf 131} (1985) 43--59.}

\lref\AtiyahKN{
M.~Atiyah, {\it The Geometry and Physics of Knots}, Cambridge
University Press, Cambridge, 1990.}

\lref\AtiyahRB{
M.~Atiyah and R.~Bott,
``The Moment Map and Equivariant Cohomology,''
Topology {\bf 23} (1984) 1--28.}

\lref\AtiyahYM{
M.~Atiyah and R.~Bott,
``Yang-Mills Equations Over Riemann Surfaces,''
Phil. Trans. R. Soc. Lond. {\bf A308} (1982) 523--615.}

\lref\AtiyahFM{
M.~Atiyah, ``On Framings of Three-Manifolds'', Topology {\bf 29}
(1990) 1--7.}

\lref\APS{
M.~F.~Atiyah, V.~Patodi, and I.~Singer, ``Spectral Asymmetry and
Riemannian Geometry, I, II, III,''  Math. Proc. Camb. Phil. Soc. {\bf
77} (1975) 43--69;  {\bf 78} (1975) 405--432;  {\bf 79} (1976)
71--99.}

\lref\Blair{
D.~E.~Blair, {\it Riemannian Geometry of Contact and Symplectic
Manifolds}, Birkh\"auser, Boston, 2002.}

\lref\BlauRS{
M.~Blau and G.~Thompson,
``Localization and Diagonalization: A Review of Functional Integral
Techniques For Low Dimensional Gauge Theories and Topological Field
Theories,'' J.\ Math.\ Phys.\  {\bf 36} (1995) 2192--2236, {\urlfont
hep-th/9501075}.}

\lref\BottT{
R.~Bott and L.~Tu, {\it Differential Forms in Algebraic Topology},
Springer-Verlag, New York, 1982.}

\lref\Bourbaki{
N.~Bourbaki, {\it Lie Groups and Lie Algebras}, Vol.2, Springer-Verlag,
Berlin, 1989.}

\lref\BuscherT{
T.~Buscher, ``Path Integral Derivation of Quantum Duality in Nonlinear
Sigma Models,''  Phys. Lett. {\bf B201} (1988) 466--472.}

\lref\CordesFC{
S.~Cordes, G.~W.~Moore and S.~Ramgoolam,
``Lectures on 2-d Yang-Mills Theory, Equivariant Cohomology and
Topological Field Theories,''
Nucl.\ Phys.\ Proc.\ Suppl.\  {\bf 41} (1995) 184--244,
{\urlfont hep-th/9411210}.}

\lref\CoxKZ{
D.~Cox and S.~Katz, {\it Mirror Symmetry and Algebraic Geometry},
American Mathematical Society, Providence, Rhode Island, 1999.}

\lref\deHaroRZ{
S.~de Haro and M.~Tierz, ``Discrete and Oscillatory
Matrix Models in Chern-Simons Theory,'' {\urlfont hep-th/0501123}.}

\lref\deHaroWN{
S.~de Haro, ``Chern-Simons Theory, 2d Yang-Mills, and
Lie Algebra Wanderers,'' {\urlfont hep-th/0412110}.}

\lref\deHaroUZ{
S.~de Haro, ``Chern-Simons Theory in Lens Spaces From
2d Yang-Mills on the Cylinder,'' JHEP {\bf 0408} (2004) 041, {\urlfont
hep-th/0407139}.}

\lref\deHaroID{
S.~de Haro and M.~Tierz, ``Brownian Motion, Chern-Simons Theory, and
2d Yang-Mills,'' Phys.\ Lett.\ {\bf B601} (2004) 201--208, {\urlfont
hep-th/0406093}.}

\lref\DeserWH{
S.~Deser, R.~Jackiw, and S.~Templeton, ``Topologically
Massive Gauge Theories,'' Annals Phys.\  {\bf 140} (1982) 372--411.}

\lref\DiaconescuQF{
D.~E.~Diaconescu, B.~Florea and A.~Grassi,
``Geometric Transitions, del Pezzo Surfaces, and Open String Instantons,''
Adv.\ Theor.\ Math.\ Phys.\  {\bf 6} (2003) 643--702,
{\urlfont hep-th/0206163}.}

\lref\DiaconescuSF{
D.~E.~Diaconescu, B.~Florea and A.~Grassi,
``Geometric Transitions and Open String Instantons,''
Adv.\ Theor.\ Math.\ Phys.\  {\bf 6} (2003) 619--642,
{\urlfont hep-th/0205234}.}

\lref\Donaldson{
S.~K.~Donaldson, ``Moment Maps and Diffeomorphisms,'' in {\it Sir
Michael Atiyah: A Great Mathematician of the Twentieth Century}, Asian
J.\ Math.\ {\bf 3} (1999) 1--15.}

\lref\Drezet{
J.-M. Drezet and M.~S.~Narasimhan, ``Groupe de Picard des vari\'et\'es
de modules de fibr\'es semi-stables sur les courbes alg\'ebriques,''
Invent.\ Math.\ {\bf 97} (1989) 53--94.}

\lref\Duistermaat{
J.~J.~Duistermaat and G.~J.~Heckman, ``On the Variation in the
Cohomology of the Symplectic Form of the Reduced Phase Space,''
Invent.\ Math.\ {\bf 69} (1982) 259--268; Addendum, Invent.\ Math.\
{\bf 72} (1983) 153--158.}

\lref\Etnyre{
J.~B.~Etnyre,
``Introductory Lectures on Contact Geometry,'' in {\it Topology and
Geometry of Manifolds (Athens, GA 2001)}, Proc.\ Sympos.\ Pure Math.\
{\bf 71}, Amer.\ Math.\ Soc., Providence, RI, 2003,
{\urlfont math.SG/0111118}.}

\lref\FreedRG{
D.~Freed and R.~Gompf, ``Computer Calculation of Witten's 3-Manifold
Invariant,'' Commun.\ Math.\ Phys.\  {\bf 141} (1991) 79--117.}

\lref\Furuta{
M.~Furuta and B. Steer, ``Seifert Fibred Homology 3-Spheres and the
Yang-Mills Equations on Riemann Surfaces with Marked Points,'' Adv. in
Math. {\bf 96} (1992) 38--102.}

\lref\GeigesHJ{
H.~Geiges, ``Contact Geometry,'' {\urlfont math.SG/0307242}.}

\lref\GuilleminS{
V.~Guillemin and S.~Sternberg, {\it Symplectic Techniques in Physics},
Cambridge University Press, Cambridge, 1984.}

\lref\GuilleminSII{
V.~Guillemin and S.~Sternberg, {\it Supersymmetry and Equivariant de
Rham Theory}, Springer, Berlin, 1999.}

\lref\JeffreyLC{
L.~Jeffrey, ``Chern-Simons-Witten Invariants of Lens Spaces and Torus
Bundles, and the Semiclassical Approximation,'' Commun.\ Math.\ Phys.\
{\bf 147} (1992) 563--604.}

\lref\JeffreyTH{
L.~Jeffrey, {\it On Some Aspects of Chern-Simons Gauge Theory},
D.Phil. thesis, University of Oxford, 1991.}

\lref\Garoufalidis{
S.~Garoufalidis, {\it Relations Among $3$-Manifold Invariants},
Ph.D. thesis, University of Chicago, 1992.}

\lref\GopakumarII{
R.~Gopakumar and C.~Vafa,
``M-theory and Topological Strings. I,''
{\urlfont hep-th/9809187}.}

\lref\GopakumarKI{
R.~Gopakumar and C.~Vafa,
``On the Gauge Theory/Geometry Correspondence,''
Adv.\ Theor.\ Math.\ Phys.\  {\bf 3} (1999) 1415--1443,
{\urlfont hep-th/9811131}.}

\lref\GukovNA{
S.~Gukov,
``Three-Dimensional Quantum Gravity, Chern-Simons Theory, and the
A-Polynomial,'' {\urlfont hep-th/0306165}.}

\lref\HansenSK{
S.~K.~Hansen,  ``Reshetikhin-Turaev Invariants of Seifert $3$-Manifolds
and a Rational Surgery Formula,'' Algebr. Geom. Topol. {\bf 1} (2001)
627--686, {\urlfont math.GT/0111057}.}

\lref\HansenTT{
S.~K.~Hansen and T.~Takata, ``Reshetikhin-Turaev Invariants of Seifert
$3$-Manifolds for Classical Simple Lie Algebras,'' J. Knot Theory
Ramifications {\bf 13} (2004) 617--668, {\urlfont math.GT/0209403}.}

\lref\LawrenceRZ{
R.~Lawrence and L.~Rozansky,
``Witten-Reshetikhin-Turaev Invariants of Seifert Manifolds,''
Commun.\ Math.\ Phys.\  {\bf 205} (1999) 287--314.}

\lref\Kawasaki{
T.~Kawasaki, ``The Riemann-Roch Theorem for Complex $V$-manifolds,''
Osaka J. Math. {\bf 16} (1979) 151--159.}

\lref\MarinoFK{
M.~Marino,
``Chern-Simons Theory, Matrix Integrals, and Perturbative Three-Manifold
Invariants,''
Commun.\ Math.\ Phys.\ {\bf 253} (2004) 25--49,
{\urlfont hep-th/0207096}.}

\lref\Martinet{
J.~Martinet, ``Formes de contact sur les variet\'et\'es de dimension
3,'' Springer Lecture Notes in Math {\bf 209} (1971) 142--163.}

\lref\MigdalZG{
A.~A.~Migdal,
``Recursion Equations In Gauge Field Theories,''
Zh.\ Eksp.\ Teor.\ Fiz.\  {\bf 69} (1975) 810--822
[Sov.\ Phys.\ JETP {\bf 42} (1975) 413--418].}

\lref\Moore{
G.~Moore and N.~Seiberg,
``Lectures on RCFT,'' in {\it Superstrings '89: Proceedings of the
Trieste Spring School}, pp. 1--129, Ed. by M.~Green et al, World
Scientific, Singapore, 1990.}

\lref\NeilJR{
J.R. Neil, ``Combinatorial Calculation of the Various Normalizations of
the Witten Invariants for $3$-Manifolds,'' J. Knot Theory Ramifications
{\bf 1} (1992) 407--449.}

\lref\OrlikPK{
P.~Orlik, {\it Seifert Manifolds},
Lecture Notes in Mathematics {\bf 291}, Ed. by A. Dold and B. Eckmann,
Springer-Verlag, Berlin, 1972.}

\lref\PressleySG{
A.~Pressley and G.~Segal, {\it Loop Groups}, Clarendon Press, Oxford,
1986.}

\lref\RocekPS{
M.~Rocek and E.~Verlinde, ``Duality, Quotients, and Currents,''
Nucl.\ Phys.\ B {\bf 373} (1992) 630--646,
{\urlfont hep-th/9110053}.}

\lref\RozanskyL{
L. Rozansky, ``A Large $k$ Asymptotics of Witten's Invariant of Seifert
Manifolds,'' Comm.\ Math.\ Phys.\ {\bf 171} (1995) 279--322, {\urlfont
hep-th/9303099}.}

\lref\RozanskyWV{
L.~Rozansky,
``Residue Formulas for the Large $k$ Asymptotics of Witten's Invariants of
Seifert Manifolds: The Case of SU(2),''
Commun.\ Math.\ Phys.\  {\bf 178} (1996) 27--60, {\urlfont
hep-th/9412075}.}

\lref\SatakeI{
I.~Satake, ``On a Generalization of the Notion of Manifold,''
Proc. Nat. Acad. Sci. USA {\bf 42} (1956) 359--363.}

\lref\SatakeII{
I.~Satake, ``The Gauss-Bonnet Theorem for V-manifolds,''
J. Math. Soc. Japan {\bf 9} (1957) 464--492.}

\lref\TelemanCW{
C.~Teleman and C.~T.~Woodward, ``The Index Formula on the Moduli of
$G$-bundles,'' {\urlfont math.AG/0312154}.}

\lref\WittenFB{
E.~Witten,
``Chern-Simons Gauge Theory as a String Theory,''
Prog.\ Math.\  {\bf 133} (1995) 637--678,
{\urlfont hep-th/9207094}.}

\lref\WittenGF{
E.~Witten, ``On S-Duality in Abelian Gauge Theory,''
Selecta Math.\  {\bf 1} (1995) 383--410,
{\urlfont hep-th/9505186}.}

\lref\WittenHC{
E.~Witten,
``(2+1)-Dimensional Gravity As An Exactly Soluble System,''
Nucl.\ Phys.\ B {\bf 311} (1988) 46--78.}

\lref\WittenHF{
E.~Witten,
``Quantum Field Theory and the Jones Polynomial,''
Commun.\ Math.\ Phys.\  {\bf 121} (1989) 351--399.}

\lref\WittenWE{
E.~Witten,
``On Quantum Gauge Theories in Two-Dimensions,''
Commun.\ Math.\ Phys.\  {\bf 141} (1991) 153--209.
}

\lref\WittenXU{
E.~Witten,
``Two-dimensional Gauge Theories Revisited,''
J.\ Geom.\ Phys.\  {\bf 9} (1992) 303--368,
{\urlfont hep-th/9204083}.}

\lref\WoodwardCT{
C.~T.~Woodward,
``Localization for the Norm-Square of the Moment Map and the
Two-Dimensional Yang-Mills Integral,'' {\urlfont math.SG/0404413}.}

\lref\Zelobenko{
D.~P.~Zelobenko, {\it Compact Lie Groups and Their Representations},
Translations of Mathematical Monographs, Vol. 40, American
Mathematical Society, Providence, Rhode Island, 1973.}

\newsec{Introduction}

Chern-Simons gauge theory is remarkable for the deep connections it
bears to an array of otherwise disparate topics in mathematics and
physics.  For instance, Chern-Simons theory is intimately related to
the theory of knot invariants and the topology of three-manifolds
\refs{\WittenHF,\AtiyahKN}, to two-dimensional rational conformal
field theory \Moore\ via a holographic correspondence, to
three-dimensional quantum gravity
\refs{\DeserWH\AchucarroVZ\WittenHC{--}\GukovNA}, to the open  string
field theory of the topological A-model \WittenFB, and via a large $N$
duality to the Gromov-Witten theory of non-compact Calabi-Yau threefolds
\refs{\GopakumarII\GopakumarKI\DiaconescuSF\DiaconescuQF{--}\AganagicQG}.

Of course, Chern-Simons theory is also a topological gauge theory,
though of a very exotic sort.  In the case of a more conventional
topological gauge theory such as topological Yang-Mills theory on a
Riemann surface or on a four-manifold (for a review of both topics,
see \CordesFC), the theory can be fundamentally interpreted in terms
of the cohomology ring of some classical moduli space of connections.
In this sense, such gauge theories are themselves essentially
classical.  In contrast, Chern-Simons theory is intrinsically a
quantum theory, and it is exotic precisely because it does not admit a
general mathematical interpretation in terms of the cohomology of some
classical moduli space of connections.

Yet if we consider Chern-Simons theory not on a general three-manifold
$M$ but only on three-manifolds which are of a simple sort and which
perhaps carry additional geometric structure, then we might expect
Chern-Simons theory itself to simplify.  In particular, we might hope
that the theory in this case admits a more conventional mathematical
interpretation in terms of the cohomology of some classical moduli
space of connections.

For instance, in the very special case that $M$ is just the product of
$S^1$ and a Riemann surface $\Sigma$, so that $M = S^1 \times \Sigma$,
then the partition function $Z$ of Chern-Simons theory on $M$ does
have a well-known topological interpretation.  In this case, $Z$ is
the dimension of the Chern-Simons Hilbert space, obtained from
canonical quantization on $\BR \times \Sigma$.  In turn, this Hilbert
space can be interpreted geometrically as the space of global
holomorphic sections of a certain line bundle over the moduli space
$\CM_0$ of flat connections on $\Sigma$.

If we consider for simplicity Chern-Simons theory with gauge group $G
= SU(r+1)$ at level $k$, then the relevant line bundle over $\CM_0$ is the
$k$-th power of a universal determinant line $\CL$ on $\CM_0$.  Of
course, the moduli space $\CM_0$ is singular at the points corresponding
to the reducible flat connections on $\Sigma$.  However, suitably
interpreted, the index theorem in combination with the Kodaira
vanishing theorem for the higher cohomology of $\CL^k$ still yields a
topological expression for $Z$,
\eqn\INDX{ Z(k) \,=\, \dim{H^0(\CM_0,\CL^k)} \,=\, \chi(\CM_0,\CL^k)
\,=\, \int_{\CM_0} \exp{\left(k \, \Omega'\right)} \Td(\CM_0)\,,}
where $\Omega' = c_1(\CL)$ is the first Chern class of $\CL$ and
$\Td(\CM_0)$ is the Todd class of $\CM_0$.

In this paper, we show that the Chern-Simons partition function
has an analogous topological interpretation on a related but much
broader class of three-manifolds.  Specifically, we consider the case
that $M$ is a Seifert manifold, so that $M$ can be succinctly
described as the total space of a {\it nontrivial} circle bundle over
a Riemann surface $\Sigma$, \eqn\FB{ S^1 \longrightarrow M
\buildrel\pi\over\longrightarrow\Sigma\,,}
where, as we later explain, $\Sigma$ is generally allowed to have
orbifold points and the circle bundle is allowed to be a corresponding
orbifold bundle.

In this case, our fundamental result is to reinterpret the Chern-Simons
partition function as a topological quantity determined entirely by a
suitable equivariant cohomology ring on the moduli space
of flat connections on $M$.  Because the moduli space of flat
connections on $M$ is directly related to the moduli space of
solutions of the Yang-Mills equation on $\Sigma$, our result
implies that Chern-Simons theory on $M$ can be also be interpreted as a
two-dimensional topological theory on $\Sigma$ akin, in a way which we
make precise, to two-dimensional Yang-Mills theory.  This
two-dimensional interpretation of Chern-Simons theory on $M$ has also
been noted recently by Aganagic and collaborators in \AganagicJS,
where the theory is identified with a $q$-deformed version of
two-dimensional Yang-Mills theory.  For other work on relations
between Chern-Simons theory and two-dimensional Yang-Mills theory, see
\refs{\deHaroID\deHaroUZ\deHaroWN{--}\deHaroRZ}.

Of course, physical Yang-Mills theory on a Riemann surface $\Sigma$
also has a  well-known topological interpretation in terms of
intersection theory on the moduli space $\CM_0$ of flat connections on
$\Sigma$.  This interpretation follows from the technique of
non-abelian localization, as applied to the Yang-Mills path integral
\WittenXU.  In an analogous fashion, we arrrive at our new
interpretation of Chern-Simons theory by applying non-abelian
localization to the Chern-Simons path integral,
\eqn\CSZ{ Z(k) \,=\, \int \! \CD \! A \; \exp{\left[i {k \over {4
\pi}} \int_M \! \Tr\left( A \^ d A + {2 \over 3} A \^ A \^ A
\right)\right]}\,.}

As we recall in Section 4, non-abelian localization provides a method for
computing symplectic integrals of the canonical form
\eqn\PZSM{ Z(\epsilon) \,=\, \int_X \!
\exp{\left[\Omega - {1 \over {2 \epsilon}} \left(\mu,\mu\right)
\right]}\,.}
Here $X$ is an arbitrary symplectic manifold with symplectic
form $\Omega$.  We assume that a Lie group $H$ acts on $X$ in a
Hamiltonian fashion, with moment map $\mu: X \to \Fh^*$, where $\Fh^*$
is the dual of the Lie algebra $\Fh$ of $H$.  Finally,
$(\cdot,\cdot)$ is an invariant quadratic form on $\Fh$
and dually on $\Fh^*$ which we use to define the action $S = \ha
(\mu,\mu)$, and $\epsilon$ is a coupling parameter.

As we briefly review in Section 2, the path integral of Yang-Mills
theory on a Riemann surface immediately takes the canonical form in
\PZSM, where the affine space of all connections on a fixed principal
bundle plays the role of $X$ and where the group of gauge
transformations plays the role of $H$.  In contrast, the path integral
\CSZ\ of Chern-Simons theory on a Seifert manifold is not manifestly
of this required form.  Nonetheless, in Section 3 we show that this path
integral can be cast into the form \PZSM\ for which non-abelian
localization applies.  More abstractly, we show that Chern-Simons
theory on a Seifert manifold has a symplectic interpretation
generalizing the classic interpretation due to Atiyah and Bott
\AtiyahYM\ of two-dimensional Yang-Mills theory.

Because the path integral of Chern-Simons theory on a Seifert manifold
$M$ assumes the canonical form \PZSM, we deduce as an immediate corollary
that the path integral localizes on critical points of the
Chern-Simons action, which are the flat connections on $M$.  In fact,
this observation has been made previously by Lawrence and Rozansky
\refs{\LawrenceRZ,\RozanskyWV} (and later generalized by Mari\~no in
\MarinoFK) as an entirely empirical statement deduced from the known
formula for the exact partition function.  For a selection of explicit
computations of the Chern-Simons partition function, see for instance
\refs{\FreedRG\JeffreyTH\JeffreyLC\Garoufalidis\NeilJR\RozanskyL\HansenSK
{--}\HansenTT}.

Considering $SU(2)$ Chern-Simons theory on a Seifert homology sphere
$M$, Lawrence and Rozansky managed to recast the known formula for
$Z(k)$, which initially involves an unwieldy sum over the integrable
representations of an $SU(2)$ WZW model at level $k$, into a simple
sum of contour integrals and residues which can be formally identified
with the contributions from the flat connections on $M$ in the
stationary phase approximation to the path integral.

A very simple example of a Seifert manifold is $S^3$, by virtue of the
Hopf fibration over $\BC\BP^1$.  The result of Lawrence and Rozansky
in the case of $SU(2)$ Chern-Simons theory on $S^3$ then amounts to
rewriting the well-known expression for $Z(k)$ as below,
\eqn\LOCLRM{
Z(k) \,=\, \sqrt{2 \over {k+2}} \sin{\left({\pi \over
{k+2}}\right)}\,=\, {1 \over {2 \pi i}} \, \e{-{{i \pi} \over
{k+2}}} \int_{-\infty}^{+\infty} \!\! dx \; \sinh^2{\left(\ha \e{{i
\pi} \over 4} \, x\right)} \> \exp{\left(-{{(k+2)} \over {8 \pi}}
x^2\right)}\,.}
We note that, when the hyperbolic sine is expressed as a sum of
exponentials, the integral in \LOCLRM\ becomes a sum of elementary
Gaussian integrals which conspire to produce the standard expression
for $Z(k)$.  Because the only flat connection on $S^3$ is the trivial
connection, the integral over $x$ in \LOCLRM\ is to be identified with the
stationary phase contribution from the trivial connection to the
path integral.

So one immediate application of our work here is to provide an
underlying mathematical explanation for the phenomenological results in
\refs{\LawrenceRZ\RozanskyWV{--}\MarinoFK}.  In fact, we will apply
localization to the Chern-Simons path integral to derive directly the
expression of Lawrence and Rozansky in \LOCLRM\ for the partition
function on $S^3$.  One amusing aspect of this computation is that we
will see the famous shift in the level $k \to k + 2$.

In order to perform concrete computations in Chern-Simons theory
using localization, we must have a thorough understanding of the local
symplectic geometry near each flat connection.  As we will see, this
local geometry shares important features with the local geometry near
the higher, unstable critical points of Yang-Mills theory on a
Riemann surface.

Thus, as a warmup for our computations in Chern-Simons theory, we
begin in Section 4 by discussing localization for Yang-Mills theory.
We first review the computation in \WittenXU\ of the contribution to
the path integral from flat Yang-Mills connections, corresponding to
the stable minima of the Yang-Mills action, and then we extend this
result to compute precisely the contributions from the higher,
unstable critical points as well.  Localization at the unstable
critical points of Yang-Mills theory has been studied previously in
the physics literature by Blau and Thompson \BlauRS\ and (most
recently) in the mathematics literature by Woodward and Teleman
\refs{\TelemanCW,\WoodwardCT}, but we find it useful to supplement
these references with another discussion more along the lines of
\WittenXU.  Of course, the roots of our work on localization trace
back to the beautiful equivariant interpretation by Atiyah and Bott
\AtiyahRB\ of the Duistermaat-Heckman formula \Duistermaat.

In Section 5 we then apply localization to perform path integral
computations in Chern-Simons theory on a Seifert manifold.  As
mentioned above, these computations depend on the nature of the local
symplectic geometry near each critical point, and for illustration
we consider two extreme cases.

First, we consider localization at the trivial connection on a Seifert
homology sphere.  In this case, the first homology group of $M$ is
zero, $H_1(M, \BZ) = 0$, and the trivial connection is an isolated
flat connection.  On the other hand, all constant gauge
transformations on $M$ fix the trivial connection, and this large
isotropy group, isomorphic to the gauge group $G$ itself, plays an
important role in the localization.  Here we directly derive a formula
found by Lawrence and Rozansky in \LawrenceRZ\ and generalized by
Mari\~no in \MarinoFK.

Second, we consider localization on a smooth component of the moduli
space of flat connections.  Such a component consists of irreducible
connections, for which the isotropy group arises solely from the
center of $G$.  In this case, we derive a formula originally obtained
by Rozansky in \RozanskyWV\ by again working empirically from the
known formula for the partition function.

Finally, although we will not elaborate on this perspective here, one
of the original motivations for our study of localization in
Chern-Simons theory was to place computations in this theory into a
theoretical framework analogous to the framework of abelian
localization in the topological $A$-model of open and closed strings
(see Chapter 9 of \CoxKZ\ for a nice mathematical review of abelian
localization in the closed string $A$-model).

\bigskip\noindent{\it Special Note}

We would like to thank Raoul Bott for his inspiration. Many of us
learned much of our differential topology from the book by Bott
and Tu \BottT.  The second author first learned of equivariant
cohomology from Bott, in 1983. This was in the context of Bott
explaining the mathematical context for certain results that had
been suggested in \ref\witten{E. Witten, ``Supersymmetry and Morse
Theory,'' J. Diff. Geom. {\bf 17} (1982) 661.}, following an
earlier lecture given by Bott at a physics conference
\ref\bottmorse{R. Bott, ``Morse Theoretic Aspects Of Yang-Mills
Theory,'' in {\it Recent Developments In Gauge Theories}, ed. G.
't Hooft, et. al., Plenum Press, New York, 1980.} where the second
author and many other physicists had heard of Morse theory for the
first time.

\newsec{The Symplectic Geometry of Yang-Mills Theory on a Riemann Surface}

A central theme of this paper is the close relationship between
Chern-Simons theory on a Seifert manifold $M$ and Yang-Mills theory on
the associated Riemann surface $\Sigma$.  Thus, as a prelude to our
discussion of the path integral of Chern-Simons theory on $M$, we
begin by recalling how the path integral of Yang-Mills theory on
$\Sigma$ can be understood as a symplectic integral of the canonical
form \PZSM.

In fact, we start by considering the path integral of Yang-Mills
theory on a compact Riemannian manifold $\Sigma$ of {\it arbitrary}
dimension, so that
\eqn\YMD{\eqalign{
&Z(\epsilon) \,=\, {1 \over {\Vol(\CG(P))}} \,
\left({1 \over {2 \pi \epsilon}}\right)^{\Delta_{\CG(P)}/2} \,
\int_{\CA(P)} \! \CD \! A \; \exp{\left[ {1 \over {2 \epsilon}}
\int_\Sigma \! \Tr\left(F_A \^ \* F_A\right)\right]}\,,\cr
&\Delta_{\CG(P)} \,=\, \dim \CG(P)\,.\cr}}

Here $F_A = dA + A\^A$ is the curvature of the connection $A$.  We
assume that the Yang-Mills gauge group $G$ is compact, connected, and
simple.  If $G = SU(r+1)$, then ``$\Tr$'' in \YMD\ denotes the trace in
the fundamental representation.  With our conventions, $A$ is an
anti-hermitian element of the Lie algebra of $SU(r+1)$, so that the
trace determines a negative-definite quadratic form.  For more general
$G$, ``$\Tr$'' denotes the unique invariant, negative-definite quadratic
form on the Lie algebra $\Fg$ of $G$ which is normalized so that, for
simply-connected $G$, the Chern-Simons level $k$ in \CSZ\ obeys the
conventional integral quantization.  Of course, the parameter
$\epsilon$ is related to the Yang-Mills coupling $g$ via $\epsilon =
g^2$.

In order to define $Z$ formally, we fix a principal $G$-bundle $P$
over $\Sigma$.  Then the space $\CA(P)$ over which we integrate is the
space of connections on $P$.  The group $\CG(P)$ of gauge
transformations acts on $\CA(P)$, and we have normalized $Z$ in \YMD\
by dividing by the volume of $\CG(P)$ and a formal power of
$\epsilon$.  As we review in Section 4, this normalization of $Z$ is
the natural normalization when $\Sigma$ is a Riemann surface and we apply
non-abelian localization to compute $Z$.

The space $\CA(P)$ is an affine space, which means that, if we choose
a particular basepoint $A_0$ in $\CA(P)$, then we can identify
$\CA(P)$ with its tangent space at $A_0$.  This tangent space is the
vector space of sections of the bundle $\Omega^1_\Sigma \otimes
\ad(P)$ of one-forms on $\Sigma$ taking values in the adjoint bundle
associated to $P$.  In other words, an arbitrary connection $A$ on $P$
can be written as $A = A_0 + \eta$ for some section $\eta$ of
$\Omega^1_\Sigma \otimes \ad(P)$.

Of course, to discuss an integral over $\CA(P)$ even formally, we must
also discuss the measure $\CD \! A$ that appears in \YMD.  Because the
space $\CA(P)$ is affine, we can define $\CD \! A$ up to an overall
multiplicative constant by taking any translation-invariant measure on
$\CA(P)$.

In general, the Yang-Mills action is only defined once we choose a
metric on $\Sigma$, which induces a corresponding duality
operator $\*$, as appears in \YMD.  This duality operator $\*$ induces
a metric on $\CA(P)$ such that if $\eta$ is any tangent vector to
$\CA(P)$, then the norm of $\eta$ is defined by
\eqn\YMDA{ (\eta,\eta) \,=\, - \int_\Sigma\!\Tr\left(\eta \^ \*
\eta\right)\,.}
Thus, a convenient way to represent the path integral measure and to
fix its normalization is to take $\CD \! A$ to be the Riemannian
measure induced by the metric \YMDA\ on $\CA(P)$.  We also use the
operator $\*$ to define a similar invariant metric on $\CG(P)$, which
formally determines the volume of $\CG(P)$.

Although we generally require a metric on $\Sigma$ to define physical
Yang-Mills theory, when $\Sigma$ is a Riemann surface we actually need
much less geometric structure to define the theory.  In this case, to
define the Yang-Mills action in \YMD\ we only require a duality
operator $\*$ which relates the zero-forms and the two-forms on
$\Sigma$.  In turn, to define such an operator we require only a
symplectic structure with associated symplectic form $\omega$ on
$\Sigma$, so that $\*$ is defined by $\* 1 = \omega$.

The symplectic form $\omega$ is invariant under all area-preserving
diffeomorphisms of $\Sigma$, and this large group acts as a symmetry
of two-dimensional Yang-Mills theory.  More precisely, this symmetry
group is ``large'' in the sense that its complexification is the full
group of orientation-preserving diffeomorphisms of $\Sigma$
\Donaldson.  This fact is fundamentally responsible for the
topological nature of two-dimensional Yang-Mills theory.

Furthermore, when $\Sigma$ is a Riemann surface, the affine space
$\CA(P)$ acquires additional geometric structure.  First, $\CA(P)$ has
a natural symplectic form $\Omega$.  If $\eta$ and $\xi$ are any two
tangent vectors to $\CA(P)$, then $\Omega$ is defined by
\eqn\OMYM{ \Omega(\eta, \xi) = - \int_\Sigma \Tr
\left( \eta \^ \xi \right)\,.}
Clearly $\Omega$ is closed and non-degenerate.  Second, $\CA(P)$ has a
natural complex structure.  This complex structure is associated to
the duality operator $\*$ itself, since $\*^2 = -1$ when acting on the
tangent space of $\CA(P)$.  Finally, the metric on $\CA(P)$ is
manifestly Kahler with respect to this symplectic form and complex
structure, since we see that the metric defined by \YMDA\ can be
rewritten as $\Omega(\,\cdot\,,\*\,\cdot\,)$.

An important consequence of the fact that the metric on $\CA(P)$ is
Kahler when $\Sigma$ is a Riemann surface is that the Riemannian
measure $\CD A$ on $\CA(P)$ is actually the same as the symplectic
measure defined by $\Omega$.  If $X$ is a symplectic manifold of
dimension $2n$ with symplectic form $\Omega$, then the symplectic
measure on $X$ is given by the top-form $\Omega^n / n!$.  This measure
can be represented uniformly for $X$ of arbitrary dimension by the
expression $\exp{(\Omega)}$, where we implicitly pick out from the
series expansion of the exponential the term which is of top degree on
$X$.  Consequently, because the Riemannian and the symplectic measures
on $\CA(P)$ agree, we can formally replace $\CD A$ in the Yang-Mills
path integral \YMD\ by the expression $\exp{(\Omega)}$, as in the
canonical symplectic integal \PZSM.  This natural symplectic measure
on $\CA(P)$ makes no reference to the metric on $\Sigma$.

\medskip\noindent{\it The Yang-Mills Action as the Square of the Moment
Map}\smallskip

Of course, as an affine space, $\CA(P)$ is pretty boring.  What makes
Yang-Mills theory interesting is the fact that $\CA(P)$ is acted on by
the group $\CG(P)$ of gauge transformations.  In fact, another special
consequence of considering Yang-Mills theory on a Riemann surface is that
the action of $\CG(P)$ on $\CA(P)$ is Hamiltonian with respect to the
symplectic form $\Omega$.

To recall what the Hamiltonian condition implies, we consider the
general situation that a connected Lie group $H$ with Lie algebra
$\Fh$ acts on a symplectic manifold $X$ preserving the symplectic form
$\Omega$.   The action of $H$ on $X$ is then Hamiltonian when there
exists an algebra homomorphism from $\Fh$ to the algebra of functions
on $X$ under the Poisson bracket.  The Poisson bracket of functions
$f$ and $g$ on $X$ is given by $\{f,g\} = -V_f(g)$, where $V_f$ is the
Hamiltonian vector field associated to $f$.  This vector field is
determined by the relation ${d f = \iota_{V_f} \Omega}$, where
$\iota_{V_f}$ is the interior product with $V_f$.  More explicitly, in
local canonical coordinates on $X$, the components of $V_f$ are
determined by $f$ as ${V_f^m = - (\Omega^{-1})^{m n} \, \partial_n
f}$, where $\Omega^{-1}$ is an ``inverse'' to $\Omega$ that arises by
considering the symplectic form as an isomorphism $\Omega: TM
\rightarrow T^*M$ with inverse $\Omega^{-1}:T^*M \rightarrow TM$.  In
coordinates, $\Omega^{-1}$ is defined by $(\Omega^{-1})^{l m} \,
\Omega_{m n} \,=\, \delta^l_n$, and  ${\{f, g\} = \Omega_{m n} V^m_f
V^n_g}$.  The algebra homomorphism from the Lie algebra $\Fh$ to the
algebra of functions on $X$ under the Poisson bracket is then
specified by a moment map $\mu: X \longrightarrow \Fh^*$, under which
an element $\phi$ of $\Fh$ is sent to the function
$\langle\mu,\phi\rangle$ on $X$, where $\langle\cdot,\cdot\rangle$ is
the dual pairing between $\Fh$ and $\Fh^*$.

The moment map by definition satisfies the relation
\eqn\MOMMAPEQ{d\langle\mu,\phi\rangle = \iota_{V(\phi)} \Omega\,,}
where $V(\phi)$ is the vector field on $X$ which is generated by the
infinitesimal action of $\phi$.  In terms of $\mu$, the Hamiltonian
condition then becomes the condition that $\mu$ also satisfy
\eqn\HOMMUII{ \left\{ \langle\mu,\phi\rangle , \langle\mu,\psi\rangle
\right\} = \langle\mu,[\phi,\psi]\rangle\,.}
Geometrically, the equation \HOMMUII\ is an infinitesimal
expression of the condition that the moment map $\mu$ commute with the
action of $H$ on $X$ and the coadjoint action of $H$ on $\Fh^*$.

Returning from this abstract discussion to the case of Yang-Mills
theory on $\Sigma$, we first consider the moment map for the action of
$\CG(P)$ on $\CA(P)$, as originally discussed in \AtiyahYM.  Elements
of the Lie algebra of $\CG(P)$ are represented by sections of the
adjoint bundle $\ad(P)$ on $\Sigma$, so if $\phi$ is such a section
then the corresponding vector field $V(\phi)$ on $\CA(P)$ is given as
usual by
\eqn\DGA{ V(\phi) \,=\, d_A \phi \,=\, d\phi + [A,\phi]\,.}
We then compute directly using \OMYM,
\eqn\DOMII{ \iota_{V(\phi)} \Omega \, = \,- \int_\Sigma \Tr \left( d_A
\phi \^ \delta A \right) \, = \, \int_\Sigma \Tr \left(\phi \> d_A \delta
A\right) \, = \, \delta \! \int_\Sigma \Tr \left(F_A \phi\right)\,.}
Here we write $\delta$ for the exterior derivative acting on $\CA(P)$,
so that, for instance, $\delta A$ is regarded as a one form on
$\CA(P)$.  Thus, the relation \MOMMAPEQ\ determines, up to an additive
constant, that the moment map $\mu$ for the action of $\CG(P)$ on
$\CA(P)$ is
\eqn\YMMO{ \mu = F_A\,.}
Here we regard $F_A$, being a section of $\Omega^2_\Sigma \otimes
\ad(P)$, as an element of the dual of the Lie algebra of $\CG(P)$.

One can then check directly that $\mu$ in \YMMO\ satisfies the
condition \HOMMUII\ that it arise from a Lie algebra homomorphism, and
this condition fixes the arbitrary additive constant that could
otherwise appear in $\mu$ to be zero.  Thus, $\CG(P)$ acts in a
Hamiltonian fashion on $\CA(P)$ with moment map given by $\mu = F_A$.
In particular, if we introduce the obvious positive-definite,
invariant quadratic form on the Lie algebra of $\CG(P)$, defined by
\eqn\QFG{ (\phi,\phi) \,=\, - \int_\Sigma \! \Tr\left(\phi \^ \*
\phi\right)\,,}
then the Yang-Mills action $S$ is proportional to the square of the
moment map,
\eqn\YMS{ S \,=\, - \ha \int_\Sigma \! \Tr\left(F_A \^ \* F_A\right)
\,=\, \ha \left(\mu,\mu\right)\,.}
As a result, the path integral of Yang-Mills theory on $\Sigma$ can be
recast completely in terms of the symplectic data associated to the
Hamiltonian action of $\CG(P)$ on $\CA(P)$,
\eqn\PZYMII{ Z(\epsilon) \,=\,  {1 \over {\Vol(\CG(P))}} \,
\left({1 \over {2 \pi \epsilon}}\right)^{\Delta_{\CG(P)}/2} \,
\int_{\CA(P)} \! \exp{\left[ \Omega - {1 \over {2 \epsilon}}
\left(\mu,\mu\right) \right]}\,,}
precisely as in \PZSM.

\newsec{The Symplectic Geometry of Chern-Simons Theory on a Seifert
Manifold}

In this section, we explain how the path integral of Chern-Simons
theory on a Seifert manifold can be recast as a symplectic integral of
the canonical form \PZSM\ which is suitable for non-abelian
localization.  More generally, we explain some beautiful facts about
the symplectic geometry of Chern-Simons theory on a Seifert manifold.

To set up notation, we consider Chern-Simons theory on a
three-manifold $M$ with compact, connected, simply-connected, and
simple gauge group $G$.  With these assumptions, any principal
$G$-bundle $P$ on $M$ is necessarily trivial, and we denote by
$\CA$ the affine space of connections on the trivial bundle.  We
denote by $\CG$ the group of gauge transformations acting on $\CA$.

We begin with the Chern-Simons path integral,
\eqn\PZCS{\eqalign{
Z(\epsilon) &= {1 \over {\Vol(\CG)}} \, \left({1 \over {2 \pi
\epsilon}}\right)^{\Delta_{\CG}} \, \int_\CA \! \CD \! A \;
\exp{\left[{i \over {2 \epsilon}} \, \int_M \! \Tr\left( A \^ d A + {2
\over 3} A \^ A \^ A \right)\right]}\,,\cr
\epsilon &= {{2 \pi} \over k}\,,\qquad \Delta_{\CG} = \dim \CG\,.}}

We have introduced a coupling parameter $\epsilon$ by analogy to the
canonical integral in \PZSM, and we have included a number of formal
factors in $Z$.  First, we have the measure $\CD A$ on $\CA$, which we
define up to norm as a translation-invariant measure on $\CA$.  As is
standard, we have also divided the path integral by the volume of the
gauge group $\CG$.  Finally, to be fastidious, we have normalized $Z$ by a
formal power of $\epsilon$ which, as in \YMD, will be natural in
defining $Z$ by localization.

\subsec{A New Formulation of Chern-Simons Theory}

At the moment, we make no assumption about the three-manifold $M$.
However, if $M$ is an $S^1$ bundle over a Riemann surface $\Sigma$, or
an orbifold thereof, then to reduce Chern-Simons theory on $M$ to a
topological theory on $\Sigma$ we must eventually decouple one of the
three components of the gauge field $A$.  This observation motivates the
following general reformulation of Chern-Simons theory, which proves
to be key to the rest of the paper.

In order to decouple one of the components of $A$, we begin by
choosing a one-dimensional subbundle of the cotangent bundle $T^* M$
of $M$.  Locally on $M$, this choice can be represented by the choice
of an everywhere non-zero one-form $\kappa$, so that the subbundle of
$T^* M$ consists of all one-forms proportional to $\kappa$.  However,
if $t$ is any non-zero function, then clearly $\kappa$ and $t \,
\kappa$ generate the same subbundle in $T^* M$.  Thus, our choice of a
one-dimensional subbundle of $T^* M$ corresponds locally to the choice
of an equivalence class of one-forms under the relation
\eqn\EQVA{ \kappa \sim t \, \kappa\,.}
We note that the representative one-form $\kappa$ which generates the
subbundle need only be defined locally on $M$.  Globally, the
subbundle might or might not be generated by a non-zero one-form which is
defined everywhere on $M$; this condition depends upon whether the
sign of $\kappa$ can be consistently defined under \EQVA\ and thus
whether the subbundle is orientable or not.

We now attempt to decouple one of the three components of $A$.
Specifically, our goal is to reformulate Chern-Simons theory on $M$ as
a theory which respects a new local symmetry under which $A$ varies as
\eqn\SHFTA{ \delta A \,=\, \sigma \kappa\,.}
Here $\sigma$ is an {\it arbitrary} section of the bundle $\Omega^0_M
\otimes \Fg$ of Lie algebra-valued functions on $M$.

The Chern-Simons action certainly does not respect the local
``shift'' symmetry in \SHFTA.  However, we can trivially introduce this
shift symmetry into Chern-Simons theory if we simultaneously introduce
a new scalar field $\Phi$ on $M$ which transforms like $A$ in the adjoint
representation of the gauge group.  Under the shift symmetry, $\Phi$
transforms as
\eqn\SHFTP{ \delta \Phi \,=\, \sigma\,.}

Now, if $\kappa$ in \SHFTA\ is scaled by a non-zero function $t$ so
that $\kappa \to t \, \kappa$, then this rescaling can be absorbed into
the arbitrary section $\sigma$ which also appears in \SHFTA\ so that
the transformation law for $A$ is well-defined.  However, from the
transformation \SHFTP\ of $\Phi$ under the same symmetry, we see that
because we absorb $t$ into $\sigma$ we must postulate an inverse
scaling of $\Phi$, so that $\Phi \to t^{-1} \Phi$.  As a result,
although $\kappa$ is only locally defined up to scale, the product
$\kappa \, \Phi$ is well-defined on $M$.

The only extension of the Chern-Simons action which now incorporates
both $\Phi$ and the shift symmetry is the Chern-Simons functional
$CS(\,\cdot\,)$ of the shift invariant combination $A - \kappa \,
\Phi$.  Thus, we consider the theory with action
\eqn\SAPI{ S(A,\Phi) \,=\, CS(A - \kappa \, \Phi)\,,}
or more explicitly,
\eqn\SAPII{ S(A, \Phi) \,=\, CS(A) - \int_M \Big[ 2 \kappa \^ \Tr(\Phi
F_A) - \kappa \^ d \kappa \, \Tr(\Phi^2)\Big]\,.}

To proceed, we play the usual game used to derive field theory
dualities by path integral manipulations, as for $T$-duality in two
dimensions \refs{\BuscherT,\RocekPS} or abelian $S$-duality in four
dimensions \WittenGF.  We have introduced a new degree of freedom,
namely $\Phi$, into Chern-Simons theory, and we have simultaneously
enlarged the symmetry group of the theory so that this degree of
freedom is completely gauge trivial.  As a result, we can either use
the shift symmetry \SHFTP\ to gauge $\Phi$ away, in which case we
recover the usual description of Chern-Simons theory, or we can
integrate $\Phi$ out, in which case we obtain a new description of
Chern-Simons theory which respects the action of the shift symmetry
\SHFTA\ on $A$.

\medskip\noindent{\it A Contact Structure on $M$}\smallskip

Hitherto, we have supposed that the one-dimensional subbundle of $T^*
M$ represented by $\kappa$ is arbitrary, but at this point we must
impose an important geometric condition on this subbundle.  From the
action $S(A,\Phi)$ in \SAPII, we see that the term quadratic in $\Phi$ is
multiplied by the local three-form $\kappa \^ d\kappa$.  In order for
this quadratic term to be everywhere non-degenerate on $M$, so that we can
easily perform the path integral over $\Phi$, we require that $\kappa
\^ d\kappa$ is also everywhere non-zero on $M$.

Although $\kappa$ itself is only defined locally and up to rescaling
by a non-zero function $t$, the condition that $\kappa \^ d\kappa \neq
0$ pointwise on $M$ is a globally well-defined condition on the
subbundle generated by $\kappa$.  For when $\kappa$ scales as $\kappa
\to t\,\kappa$ for any non-zero function $t$, we easily see that
$\kappa \^ d\kappa$ also scales as $\kappa \^ d\kappa \to t^2 \,
\kappa \^ d\kappa$.  Thus, the condition that $\kappa \^ d\kappa \neq
0$ is preserved under arbitrary rescalings of $\kappa$.

The structure which we thus introduce on $M$ is the choice of a
one-dimensional subbundle of $T^* M$ for which any local generator
$\kappa$ satisfies $\kappa \^ d\kappa \neq 0$ at each point of $M$.
This geometric structure, which appears so naturally here, is known as a
contact structure \refs{\Etnyre\GeigesHJ{--}\Blair}.  More generally, on
an arbitrary manifold $M$ of odd dimension $2n+1$, a contact
structure on $M$ is defined as a one-dimensional subbundle of $T^* M$
for which the local generator $\kappa$ satisfies $\kappa \^
(d\kappa)^n \neq 0$ everywhere on $M$.

In many ways, a contact structure is the analogue of a symplectic
structure for manifolds of odd dimension.  The fact that we must
choose a contact structure on $M$ for our reformulation of
Chern-Simons theory is thus closely related to the fact, mentioned
previously, that we must choose a symplectic structure on the Riemann
surface $\Sigma$ in order to define Yang-Mills theory on $\Sigma$.

We will say a bit more about contact structures on Seifert manifolds
later, but for now, we just observe that, by a classic theorem of Martinet
\Martinet, any compact, orientable\foot{We note that, because $\kappa
\^ d\kappa \to t^2 \, \kappa \^ d\kappa$ under a local rescaling of
$\kappa$ and because $t^2$ is always positive, the sign of the local
three-form $\kappa \^ d\kappa$ is well-defined.  So any three-manifold
with a contact structure is necessarily orientable.} three-manifold
possesses a contact structure.

\medskip\noindent{\it Path Integral Manipulations}\smallskip

Thus, we choose a contact structure on the three-manifold $M$, and we
consider the theory defined by the path integral
\eqn\PZCSII{\eqalign{
Z(\epsilon) \,&=\, {1 \over {\Vol(\CG)}} \, {1 \over {\Vol(\CS)}}
\, \left({1 \over {2 \pi \epsilon}}\right)^{\Delta_{\CG}} \, \times
\cr
&\times \, \int \CD \! A \, \CD \! \Phi \; \exp{\left[ {i \over {2
\epsilon}} \left( CS\left(A\right) - \int_M \! 2 \kappa \^
\Tr\left(\Phi F_A\right) \,+\, \int_M \! \kappa \^ d \kappa \,
\Tr\left(\Phi^2\right)\right)\right]}\,.\cr}}
Here the measure $\CD \! \Phi$ is defined independently of any metric on
$M$ by the invariant, positive-definite quadratic form
\eqn\DPHI{ \left(\Phi,\Phi\right) = - \int_M \kappa \^
d\kappa \, \Tr\left(\Phi^2\right)\,,}
which is invariant under the scaling ${\kappa \to t\, \kappa}$,
${\Phi \to t^{-1} \, \Phi}$.  We similarly use this quadratic form
to define formally the volume of the group $\CS$ of shift symmetries,
as appears in the normalization of \PZCSII.

Using the shift symmetry \SHFTP, we can fix $\Phi = 0$ trivially, with
unit Jacobian, and the resulting group integral over $\CS$ produces a
factor of $\Vol(\CS)$ to cancel the corresponding factor in the
normalization of $Z(\epsilon)$.  Hence, the new theory defined by
\PZCSII\ is fully equivalent to Chern-Simons theory.

On the other hand, because the field $\Phi$ appears only quadratically
in the action \SAPII, we can also perform the path integral over
$\Phi$ directly.  Upon integrating out $\Phi$, the new action
$S(A)$ for the gauge field becomes
\eqn\SAA{ S(A) \,=\, \int_M \! \Tr \left( A \^ d A + {2 \over 3} A \^ A
\^ A \right) \,-\, \int_M {1 \over {\kappa \^ d \kappa}}
\Tr\Big[ (\kappa \^ F_A)^2 \Big]\,.}

We find it convenient to abuse notation slightly by writing ``$1 /
\kappa \^ d\kappa$'' in \SAA.  To explain this notation precisely, we
observe that, as $\kappa \^ d\kappa$ is nonvanishing, we can always
write $\kappa \^ F_A = \varphi \, \kappa \^ d\kappa$ for some function
$\varphi$ on $M$ taking values in the Lie algebra $\Fg$.  Thus, we set
$\kappa \^ F_A / \kappa \^ d\kappa = \varphi$, and the second term in
$S(A)$ becomes $\int_M \kappa \^ \Tr\left(F_A \varphi\right)$.  As our
notation in \SAA\ suggests, this term is invariant under the
transformation $\kappa \to t\,\kappa$, since $\varphi$ transforms as
$\varphi \to t^{-1} \, \varphi$.

By construction, the new action $S(A)$ in \SAA\ is invariant under the
action of the shift symmetry \SHFTA\ on $A$.  We can directly check this
invariance once we note that, under the shift symmetry, the expression
$\kappa \^ F_A$ transforms as
\eqn\SHFTF{ \kappa \^ F_A \,\longrightarrow\, \kappa \^ F_A + \sigma \,
\kappa\^d\kappa\,.}

The partition function $Z(\epsilon)$ now takes the form
\eqn\PZCSIII{\eqalign{Z(\epsilon) \,&=\, {1 \over {\Vol(\CG)}} \, {1
\over {\Vol(\CS)}} \, \left({{-i} \over {2 \pi
\epsilon}}\right)^{\Delta_{\CG}/2} \, \times\cr
&\times \, \int_\CA \CD \! A \; \exp{\left[{i \over {2
\epsilon}}\left( \int_M \! \Tr \left( A \^ d A + {2 \over 3} A \^ A
\^ A \right) \,-\, \int_M {1 \over {\kappa \^ d \kappa}} \Tr\left[
\left(\kappa \^ F_A\right)^2 \right]\right)\right] }\,,\cr}}
where the Gaussian integral over $\Phi$ cancels some factors of
$2 \pi \epsilon$ in the normalization of $Z$.  As is standard, in
integrating over $\Phi$ we assume that the integration contour has
been slightly rotated off the real axis, effectively giving $\epsilon$
a small imaginary part, to regulate the oscillatory Gaussian integral.
Thus, the theory described by the path integral \PZCSIII\ is fully
equivalent to Chern-Simons theory, but now one component of $A$
manifestly decouples.

\subsec{Contact Structures on Seifert Manifolds}

Our reformulation of Chern-Simons theory in \PZCSIII\ applies to any
three-manifold $M$ with a specified contact structure.  However, in
order to apply non-abelian localization to Chern-Simons theory on $M$,
we require that $M$ has additional symmetry.

Specifically, we require that $M$ admits a locally-free $U(1)$ action,
which means that the generating vector field on $M$ associated to the
infinitesimal action of $U(1)$ is nowhere vanishing.  A free $U(1)$
action on $M$ clearly satisfies this condition, but more generally it
is satisfied by any $U(1)$ action such that no point on $M$ is fixed
by all of $U(1)$ (at such a point the generating vector field would
vanish).  Such an action need not be free, since some points on $M$
could be fixed by a cyclic subgroup of $U(1)$.  The class of
three-manifolds which admit a $U(1)$ action of this sort are precisely
the Seifert manifolds \OrlikPK.

To proceed further to a symplectic description of Chern-Simons theory, we
now restrict attention to the case that $M$ is a Seifert manifold.  We
first review a few basic facts about such manifolds, for which a
complete reference is \OrlikPK.

\medskip\noindent{\it $M$ Admits a Free $U(1)$ Action}\smallskip

For simplicity, we begin by assuming that the three-manifold $M$
admits a free $U(1)$ action.  In this case, $M$ is the total
space of a circle bundle over a Riemann surface $\Sigma$,
\eqn\FBII{ S^1 \longrightarrow M
\buildrel\pi\over\longrightarrow\Sigma\,,}
and the free $U(1)$ action simply arises from rotations in the
fiber of \FBII.  The topology of $M$ is completely determined
by the genus $g$ of $\Sigma$ and the degree $n$ of the bundle.
Assuming that the bundle is nontrivial, we can always arrange by a
suitable choice of orientation for $M$ that $n \ge 1$.

At this point, one might wonder why we restrict attention to the case
of nontrivial bundles over $\Sigma$.  As we now explain, in this case
$M$ admits a natural contact structure which is invariant under the
action of $U(1)$.  As a result, our reformulation of Chern-Simons
theory in \PZCSIII\ still respects this crucial symmetry of $M$.

To describe this $U(1)$ invariant contact structure on $M$, we
simply exhibit an invariant one-form $\kappa$, defined globally on
$M$, which satisfies the contact condition that $\kappa \^
d\kappa$ is nowhere vanishing.  To describe $\kappa$, we begin by
choosing a symplectic form $\omega$ on $\Sigma$ which is
normalized so that \eqn\INTKII{ \int_\Sigma \omega \,=\, 1\,.}
Regarding $M$ as the total space of a principal $U(1)$-bundle, we
take $\kappa$ to be a connection on this bundle (and hence a
real-valued one-form on $M$) whose curvature satisfies \eqn\CONII{
d\kappa = n \, \pi^* \omega\,,} where we recall that $n \ge 1$ is
the degree of the bundle.  For a nice, explicit description of
$\kappa$ in this situation, see the description of the angular
form in \refs{\BottT, \S 6}.

We let $R$ (for ``rotation'') be the non-vanishing vector field on $M$
which generates the $U(1)$ action and which is normalized so that its
orbits have unit period.  By the fundamental properties of a
connection, $\kappa$ is invariant under the $U(1)$ action and
satisfies $\langle\kappa, R\rangle = 1$.  Here we use
$\langle\,\cdot\,,\,\cdot\,\rangle$ generally to denote the canonical
dual pairing.  Thus, $\kappa$ pulls back to a non-zero one-form which
generates the integral cohomology of each $S^1$ fiber of $M$, and we
immediately see from \CONII\ that $\kappa \^ d\kappa$ is everywhere
non-vanishing on $M$ so long as the bundle is nontrivial.

For future reference, we note that the integral of $\kappa \^ d\kappa$
over $M$ is determined as follows.  Because $\kappa$ satisfies
$\langle\kappa, R\rangle = 1$, where $R$ is the generator of the
$U(1)$ action whose orbits correspond to the $S^1$ fibers over
$\Sigma$ in \FBII, the integral of $\kappa$ over any such fiber is
given by
\eqn\INTKI{ \int_{S^1} \kappa \,=\, 1\,.}
Upon integrating over the $S^1$ fiber of $M$, we see from \INTKII,
\CONII, and \INTKI\ that
\eqn\VOLM{ \int_M \kappa \^ d\kappa \,=\, n \, \int_M \kappa \^ \pi^*
\omega \,=\, n \, \int_\Sigma \omega \,=\, n\,.}

\medskip\noindent{\it Orbifold Generalization}\smallskip

Of course, in the above construction we have assumed that $M$ admits a
free $U(1)$ action, which is a more stringent requirement than the
condition that no point of $M$ is completely fixed by the $U(1)$
action.  However, an arbitrary Seifert manifold does admit an orbifold
description precisely analogous to the description of $M$ as a principal
$U(1)$-bundle over a Riemann surface.  This point of view is taken in
a nice paper by Furuta and Steer \Furuta\ for an application somewhat
related to ours, and we follow their basic exposition below.

To generalize our previous discussion to the case of an arbitrary
Seifert manifold, we simply replace the Riemann surface $\Sigma$ with
an orbifold, and we replace the principal $U(1)$-bundle over $\Sigma$
with its orbifold counterpart.  Concretely, the orbifold base
$\hat\Sigma$ of $M$ is now described by a Riemann surface of genus
$g$ with $N$ marked points $p_j$, $j=1,\ldots,N$, at which the
coordinate neighborhoods are modeled not on $\BC$ but on $\BC /
\BZ_{\alpha_j}$ for some cyclic group $\BZ_{\alpha_j}$, which acts on
the local coordinate $z$ at $p_j$ as \eqn\ZAORB{ z \mapsto \zeta \cdot
z\,,\qquad \zeta = \e{2 \pi i / \alpha_j}\,.}
The choice of the particular orbifold points $p_j$ is topologically
irrelevant, and the orbifold base $\hat\Sigma$ can be completely specified
by the genus $g$ and the set of integers $\{\alpha_1,\ldots,\alpha_N\}$.

We now consider a line $V$-bundle over $\hat\Sigma$.  Such an object
is precisely analogous to a complex line bundle, except that the local
trivialization over each orbifold point $p_j$ of $\hat\Sigma$ is
now modeled on $\BC \times \BC/\BZ_{\alpha_j}$, where $\BZ_{\alpha_j}$
acts on the local coordinates $(z,s)$ of the base and fiber as
\eqn\ZAORBF{  z \mapsto \zeta \cdot z\,,\qquad s \mapsto
\zeta^{\beta_j} \cdot s\,,\qquad \zeta = \e{2 \pi i / \alpha_j}\,,}
for some integers $0 \le \beta_j < \alpha_j$.

Given such a line $V$-bundle over $\hat\Sigma$, an arbitrary Seifert
manifold $M$ can be described as the total space of the associated
$S^1$ fibration.  Of course, we require that $M$ itself is smooth.
This condition implies that each pair of integers $(\alpha_j,\beta_j)$
above must be relatively prime so that the local action \ZAORBF\ of the
orbifold group $\BZ_{\alpha_j}$ on $\BC \times S^1$ is free (in
particular, we require $\beta_j \neq 0$ above).

The $U(1)$ action on $M$ again arises from rotations in the fibers
over $\hat\Sigma$, but this action is no longer free.  Rather, the points
in the $S^1$ fiber over each ramification point $p_j$ of $\hat\Sigma$
are fixed by the cyclic subgroup $\BZ_{\alpha_j}$ of $U(1)$, due to
the orbifold identification in \ZAORBF.

Once the integers $\{\beta_1,\ldots,\beta_N\}$ are fixed, the
topological isomorphism class of a line $V$-bundle on $\hat\Sigma$
is specified by a single integer $n$, the degree.  Thus, in total, the
description of an arbitrary Seifert manifold $M$ is given by the
Seifert invariants \eqn\SFRT{\Big[g;n;(\alpha_1,\beta_1), \ldots,
(\alpha_N,\beta_N)\Big]\,,\quad \gcd(\alpha_j,\beta_j) = 1\,.}

Because the basic notions of bundles, connections, curvatures, and
(rational) characteristic classes generalize immediately from smooth
manifolds to orbifolds \refs{\SatakeI, \SatakeII}, our previous
construction of an invariant contact form $\kappa$ as a connection on
a principal $U(1)$-bundle immediately generalizes to the orbifold
situation here.   In this case, if $\hat\CL$ denotes the line
$V$-bundle over $\hat\Sigma$ which describes $M$, with Seifert
invariants \SFRT, then $\hat\CL$ is nontrivial so long as its Chern
class is non-zero (and positive by convention),
\eqn\CHRNCL{ c_1(\hat\CL) \,=\, n + \sum_{j=1}^N {{\beta_j} \over
{\alpha_j}}\,>\,0\,,}
which generalizes our previous condition that $n \ge 1$.  In
particular, $n$ can now be any integer such that the expression in
\CHRNCL\ is positive.

In the Chern-Weil description of the Chern class, $c_1(\hat\CL)$
is represented by smooth curvature in the bulk of the orbifold
$\hat\Sigma$.  In contrast, the degree $n$ receives contributions
from both the bulk curvature in $\hat\Sigma$ and from local,
delta-function curvatures at the orbifold points of $\hat\Sigma$.
That is why $n$ is an integer but the orbifold first Chern class
$c_1(\hat\CL)$ is not.  The delta-function contributions to $n$
are cancelled by the rational numbers $\beta_j/\alpha_j$ appearing
explicitly in the formula \CHRNCL\ for $c_1(\hat\CL)$.

From \CHRNCL, to define a contact structure on $M$ we choose the
connection $\kappa$ so that its curvature is given by
\eqn\CHRNCLII{ d\kappa \,=\, \left(n + \sum_{j=1}^N {{\beta_j} \over
{\alpha_j}}\right) \pi^* \hat\omega\,,}
where $\hat\omega$ is a symplectic form on $\hat\Sigma$ of unit volume, as
in \INTKII.  Then, exactly as in \VOLM, the integral of $\kappa \^
d\kappa$ over $M$ is determined by the Chern class of $\hat\CL$,
\eqn\VOLMII{ \int_M \kappa \^ d\kappa \,=\, n + \sum_{j=1}^N
{{\beta_j} \over {\alpha_j}}\,.}

For future reference, we also note that the Riemann-Roch formula for a
line bundle on a Riemann surface has a direct generalization to the
case of a line $V$-bundle on an orbifold \Kawasaki, so that
\eqn\RR{ \chi(\hat\CL) = \dim_{\BC} H^0(\hat\Sigma,\hat\CL)
- \dim_{\BC} H^1(\hat\Sigma,\hat\CL) \,=\, n + 1 - g\,,}
which justifies calling $n$ the degree of $\hat\CL$.

In this discussion, we have used the notation $\hat\Sigma$ and
$\hat\CL$ to distinguish these orbifold quantities from their smooth
counterparts $\Sigma$ and $\CL$.  In the future, we will not make this
artificial distinction, and in our discussion of Chern-Simons
theory we will use $\Sigma$ and $\CL$ to denote general orbifold
quantities.

\subsec{A Symplectic Structure For Chern-Simons Theory}

We now specialize to the case of Chern-Simons theory on a Seifert
manifold $M$, which carries a distinguished $U(1)$ action and an invariant
contact form $\kappa$.  Initially, the path integral of Chern-Simons
theory on $M$ is an integral over the affine space $\CA$ of all
connections on $M$.  Unlike the case of two-dimensional Yang-Mills
theory, $\CA$ is not naturally symplectic and cannot play the role of
the symplectic manifold $X$ that appears in the canonical symplectic
integral \PZSM.

However, we now reap the reward of our reformulation of Chern-Simons
theory to decouple one component of $A$.  Specifically, we consider
the following two-form $\Omega$ on $\CA$.  If $\eta$ and $\xi$ are any
two tangent vectors to $\CA$, and hence are represented by sections of
the bundle $\Omega^1_M \otimes \Fg$ on $M$, then we define $\Omega$ by
\eqn\BO{\Omega(\eta,\xi) = - \int_M \kappa\^\Tr\left( \eta \^ \xi
\right)\,.}

Because $\kappa$ is a globally-defined one-form on $M$, this
expression is well-defined.  Further, $\Omega$ is closed and
invariant under all the symmetries.  In particular, $\Omega$ is
invariant under the group $\CS$ of shift symmetries, and by virture of
this shift invariance $\Omega$ is degenerate along tangent vectors to
$\CA$ of the form $\sigma \kappa$, where $\sigma$ is an arbitrary
section of $\Omega^0_M \otimes \Fg$.  However, unlike the gauge
symmetry $\CG$, which acts nonlinearly on $\CA$, the shift symmetry
$\CS$ acts in a simple, linear fashion on $\CA$.  Thus, we can
trivially take the quotient of $\CA$ by the action of $\CS$, which we
denote as $\bar\CA$, \eqn\CMI{ \bar\CA \,=\, \CA / \CS\,.}
Under this quotient, the presymplectic form $\Omega$ on $\CA$
descends immediately to a symplectic form on $\bar\CA$, which becomes
a symplectic space naturally associated to Chern-Simons theory on
$M$.  In the following, $\bar\CA$ plays the role of the abstract
symplectic manifold $X$ in \PZSM.

\medskip\noindent{\it More About the Path Integral Measure}\smallskip

Our reformulation of the Chern-Simons action $S(A)$ in \SAA\ is invariant
under the shift symmetry $\CS$, so $S(A)$ descends to the quotient
$\bar\CA$ of $\CA$ by $\CS$.  But we should also think (at least
formally) about the path integral measure $\CD \! A$.  As in
Yang-Mills theory, we define $\CD \! A$ up to norm as a
translation-invariant measure on $\CA$, and a convenient way both to
describe $\CD \! A$ and to fix its normalization is to consider this
measure as induced from a Riemannian metric on $\CA$.  In turn, we
describe this metric on $\CA$ as induced from a corresponding metric
on $M$, so that a tangent vector $\eta$ to $\CA$ has norm
\eqn\GAII{ (\eta,\eta) \,=\, - \int_M \! \Tr{\left(\eta \^ \*
\eta\right)}\,.}
We normalize the volume of $\CG$ in \PZCS\ using the similarly
induced, invariant metric on $\CG$.

We assume that $U(1)$ acts on $M$ by isometries, so that the
metric on $M$ associated to the operator $\*$ in \GAII\ takes the form
\eqn\GM{ ds^2_M \,=\, \pi^* ds^2_\Sigma \,+\, \kappa \otimes \kappa\,.}
Here $ds^2_\Sigma$ represents any Kahler metric on $\Sigma$ which is
normalized so that the corresponding Kahler form pulls back to
$d\kappa$.  As a result of this normalization convention, the duality
operator $\*$ defined by the metric \GM\ satisfies $\* 1 \,=\,
\kappa \^ d\kappa$.

Tangent vectors to the orbits of the shift symmetry $\CS$ are
described by sections of $\Omega^1_M \otimes \Fg$ which take the form
$\sigma \kappa$, where $\sigma$ is any function taking values in $\Fg$
on $M$.  Similarly, tangent vectors to the quotient $\bar\CA$ are
naturally represented by sections of $\Omega^1_M \otimes \Fg$
which are annihilated by the interior product $\iota_R$ with the
vector field $R$, the generator of the $U(1)$ action on $M$.  When the
metric on $M$ takes the form in \GM, the one-forms annihilated by
$\iota_R$ are orthogonal to the one-forms proportional to $\kappa$.
Thus, the tangent space to $\CS$ is orthogonal to the tangent space to
$\bar\CA$ in the corresponding metric \GAII\ on $\CA$.

We can exhibit the orthogonal decomposition of the metric in \GAII\
explicitly
as \eqn\GMPV{ (\eta,\eta) = - \int_M \kappa \^ d\kappa \,
\Tr\left[\left(\iota_R \eta\right)^2\right] - \int_M \kappa \^
\Tr\Big[\Pi(\eta) \^ \*_2 \Pi(\eta)\Big]\,.}
The first term in \GMPV\ describes the metric on $\CS$ which we have
already introduced in \DPHI, and the second term describes the induced
metric on $\bar\CA$.  The form of the first term follows immediately
from the fact that $\* \kappa = d\kappa$.

In the second term of \GMPV, we have introduced two natural operators.
First, we introduce the the operator $\Pi$ which projects from the
tangent space of $\CA$ to the tangent space of $\bar\CA$, so that
$\Pi$ is given by
\eqn\PROJA{ \Pi(\eta) = \eta - \left( \iota_R \eta \right) \kappa\,.}
Trivially, $\iota_R \circ \Pi = 0$.

Second, we introduce an effective ``two-dimensional'' duality operator
$\*_2$ on $M$ which induces a corresponding complex structure on
$\bar\CA$.  This operator is defined globally on $M$ by \eqn\DLM{\*_2
\,=\, - \iota_R \circ \*\,.}  Using that $\* \kappa = d\kappa$ and $\*
1 = \kappa \^ d\kappa$, we see immediately that $\*_2 \, \kappa = \*_2
\, (\kappa \^ d\kappa) = 0$ and that $\*_2 \, 1 = - d\kappa$.  Also,
one can easily check (for instance by considering local coordinates)
that $\*_2$ satisfies $(\*_2)^2 = -1$ when acting on one-forms in the
image of $\Pi$, representing tangent vectors to $\bar\CA$.  This
latter property is important, since it implies that $\*_2$ defines a
complex structure on $\bar\CA$ exactly as in two-dimensional
Yang-Mills theory.

With this notation in place, the form of the second term in \GMPV\
follows immediately from the simple computation below,
\eqn\IXBIV{\eqalign{
\Pi(\eta) \^ \* \Pi(\eta) \,&=\, \iota_R \Big(\kappa \^
\Pi(\eta)\Big) \^ \* \Pi(\eta)\,,\cr
&=\, - \kappa \^ \Pi(\eta) \^ \, \iota_R\!\Big( \*
\Pi(\eta)\Big)\,,\cr
&=\, \kappa \^ \Pi(\eta) \^ \*_2 \Pi(\eta)\,.\cr}}
In passing from the first to the second line of \IXBIV, we have
``integrated by parts'' with respect to the operator $\iota_R$, as
$\iota_R\left(\kappa \^ \Pi(\eta) \^ \* \Pi(\eta)\right)$ is trivially
zero on the three-manifold $M$ by dimensional reasons.

We thus see from the second term in \GMPV\ that the induced metric on
$\bar\CA$ is Kahler with respect to the symplectic form $\Omega$ in
\BO\ and the complex structure $\*_2$.  Hence the Riemannian measure
induced on $\bar\CA$ from \GMPV\ is identical to the symplectic
measure induced by $\Omega$.

Finally, because the measure along the orbits of $\CS$ in $\CA$ is the
same as the invariant measure \DPHI\ which we defined on $\CS$ itself,
we can trivially integrate over these orbits, which simply contribute
a factor of the volume $\Vol(\CS)$ to the path integral.
Consequently, the Chern-Simons path integral in \PZCSIII\ reduces to
an integral over $\bar\CA$ with its symplectic measure,
\eqn\PZCSIV{\eqalign{
Z(\epsilon) &= {1 \over {\Vol(\CG)}} \, \left({{-i} \over
{2 \pi \epsilon}}\right)^{\Delta_{\CG}/2} \,
\int_{\bar\CA} \exp{\left[\Omega + {i \over {2 \epsilon}} \,
S\left(A\right)\right]}\,,\cr
S(A) &= \int_M \! \Tr{\left( A \^ d A + {2 \over 3} A \^ A \^ A
\right)} - \int_M {1 \over {\kappa \^ d \kappa}}
\Tr\Big[ (\kappa \^ F_A)^2 \Big]\,.\cr}}

\subsec{Hamiltonian Symmetries of Chern-Simons Theory}

To complete our symplectic description of the Chern-Simons path
integral on $M$, we must show that the action $S(A)$ in \PZCSIV\ is
the square of a moment map $\mu$ for the Hamiltonian action of some
symmetry group $\CH$ on the symplectic space $\bar\CA$.

By analogy to the case of Yang-Mills theory on $\Sigma$, one might
naively guess that the relevant symmetry group for Chern-Simons theory
would also be the group $\CG$ of gauge transformations.  One can
easily check that the action of $\CG$ on $\CA$ descends under the
quotient to a well-defined action on $\bar\CA$, and clearly the
symplectic form $\Omega$ on $\bar\CA$ is invariant under $\CG$.
However, one interesting aspect of non-abelian localization for
Chern-Simons theory is the fact that the group $\CH$ which we use for
localization must be somewhat more complicated than $\CG$ itself.

A trivial objection to using $\CG$ for localization is that, by
construction, the square of the moment map $\mu$ for any Hamiltonian
action on $\bar\CA$ defines an invariant function on $\bar\CA$, but
the action $S(A)$ is not invariant under the group $\CG$.  Instead,
the action $S(A)$ is the sum of a manifestly gauge invariant term and
the usual Chern-Simons action, and the Chern-Simons action shifts by
integer multiples of $2 \pi$ under ``large'' gauge transformations,
those not continuously connected to the identity in $\CG$.

This trivial objection is easily overcome.  We consider not the
disconnected group $\CG$ of all gauge transformations but only the
identity component $\CG_0$ of this group, under which $S(A)$ is
invariant.

We now consider the action of $\CG_0$ on $\bar\CA$, and our first task
is to determine the corresponding moment map $\mu$.  If $\phi$ is an
element of the Lie algebra of $\CG_0$, described by a section of the
bundle $\Omega^0_M \otimes \Fg$ on $M$, then the corresponding vector
field $V(\phi)$ generated by $\phi$ on $\CA$ is given by $V(\phi) = d_A
\phi$.  Thus, from our expression for the symplectic form $\Omega$ in
\BO\ we see that \eqn\BOII{ \iota_{V(\phi)} \Omega \,=\, - \int_M
\kappa \^ \Tr\left(d_A \phi \^ \delta A\right)\,.}
Integrating by parts with respect to $d_A$, we can rewrite \BOII\ in
the form $\delta\langle\mu,\phi\rangle$, where
\eqn\MU{\langle\mu,\phi\rangle = \int_M \kappa\^\Tr\Big(\phi
F_A\Big) - \int_M d\kappa\^\Tr\Big(\phi (A - A_0)\Big)\,.}
Here $A_0$ is an arbitrary connection, corresponding to a basepoint in
$\CA$, which we must choose so that the second term in \MU\ can be
honestly interpreted as the integral of a differential form on $M$.
In the case that the gauge group $G$ is simply-connected, so that the 
principal $G$-bundle over $M$ is necessarily trivial, the choice of a
basepoint connection $A_0$ corresponds geometrically to the choice of a
trivialization for the bundle on $M$.  We will say more about this
choice momentarily, but we first observe that the expression for $\mu$
in \MU\ is invariant under the shift symmetry and immediately descends
to a moment map for the action of $\CG$ on $\bar\CA$.

The fact that we must choose a basepoint $A_0$ in $\CA$ to define the
moment map is very important in the following, and it is fundamentally
a reflection of the affine structure of $\CA$.  In general, an affine
space is a space which can be identified with a vector space only
after some basepoint is chosen to represent the origin.  In the case
at hand, once $A_0$ is chosen, we can identify $\CA$ with the vector
space of sections $\eta$ of the bundle $\Omega^1_M \otimes \Fg$ on
$M$, via $A = A_0 + \eta$, as we used in \MU.  However, $\CA$ is not
naturally itself a vector space, since $\CA$ does not intrinsically
possess a distinguished origin.  This statement corresponds to the
geometric statement that, though our principal $G$-bundle on $M$ is
trivial, it does not possess a canonical trivialization.

In terms of the moment map $\mu$, the choice of $A_0$ simply
represents the possibility of adding an arbitrary constant to $\mu$.
In general, our ability to add a constant to $\mu$ means that
$\mu$ need {\it not} determine a Hamiltonian action of $\CG_0$ on
$\bar\CA$.  Indeed, as we show below, the action of $\CG_0$ on
$\bar\CA$ is not Hamiltonian and we cannot simply use $\CG_0$ to
perform localization.

In order not to clutter the expressions below, we assume henceforth
that we have fixed a trivialization of the $G$-bundle on $M$ and we
simply set $A_0 = 0$.

To determine whether the action of $\CG_0$ on $\bar\CA$ is
Hamiltonian, we must check the condition \HOMMUII\ that $\mu$
determine a homomorphism from the Lie algebra of $\CG_0$ to the
algebra of functions on $\bar\CA$ under the Poisson bracket.  So we
directly compute
\eqn\HAM{\eqalign{
\Big\{ \langle\mu,\phi\rangle, \langle\mu,\psi\rangle \Big\} \,&=\,
\Omega\Big(d_A \phi, d_A \psi\Big)\,=\, - \int_M \kappa \^
\Tr\left(d_A \phi \^ d_A \psi\right)\,,\cr
\,&=\, \int_M \kappa\^\Tr\Big([\phi,\psi] F_A\Big) -
\int_M d \kappa\^\Tr\Big(\phi \, d_A \psi\Big)\,,\cr
\,&=\, \langle\mu,[\phi,\psi]\rangle -
\int_M d\kappa \, \Tr\Big(\phi \, d \psi\Big)\,.\cr}}

Thus, the failure of $\mu$ to determine an algebra homomorphism is
measured by the cohomology class of the Lie algebra cocycle
\eqn\COC{\eqalign{ c(\phi,\psi) \,&=\, \Big\{ \langle\mu,\phi\rangle,
\langle\mu,\psi\rangle \Big\} - \langle\mu,[\phi,\psi]\rangle\,,\cr
&\,=\, - \int_M d\kappa \^ \Tr\Big(\phi \, d \psi\Big) \,=\, - \int_M
\kappa \^ d\kappa \Tr\Big(\phi \lie_R \psi\Big)\,.\cr}}
In the second line of \COC, we have rewritten the cocycle more
suggestively by using the Lie derivative $\lie_R$ along the vector
field $R$ on $M$ which generates the $U(1)$ action.  The class of this
cocycle is not zero, and no Hamiltonian action on $\bar\CA$ exists for the
group $\CG_0$.

\medskip\noindent{\it Some Facts About Loop Groups}\smallskip

The cocycle appearing in \COC\ has a very close relationship to a
similar cocycle that arises in the theory of loop groups, and some
well-known loop group constructions feature heavily in our study of
Chern-Simons theory.  We briefly review these ideas, for which a
general reference is \PressleySG.

When $G$ is a finite-dimensional Lie group, we recall that the loop
group $LG$ is defined as the group of smooth maps $\Map(S^1,G)$ from
$S^1$ to $G$.  Similarly, the Lie algebra $L\Fg$ of $LG$ is
the algebra $\Map(S^1,\Fg)$ of smooth maps from $S^1$ to $\Fg$.
When $\Fg$ is simple, then the Lie algebra $L\Fg$ admits a unique,
$G$-invariant cocycle up to scale, and this cocycle is directly
analogous to the cocycle we discovered in \COC.  If $\phi$ and $\psi$
are elements in the Lie algebra $L\Fg$, then this cocycle is defined by
\eqn\LGCOC{ c(\phi,\psi) = -\int_{S^1} \Tr\Big(\phi \, d \psi\Big) \,=\,
-\int_{S^1} \, dt \, \Tr\Big(\phi \lie_R \psi\Big)\,.}
In passing to the last expression, we have by analogy to \COC\
introduced a unit-length parameter $t$ on $S^1$, so that $\int_{S^1}
dt = 1$, and we have introduced the dual vector field $R =
\partial/\partial t$ which generates rotations of $S^1$.

In general, if $\Fg$ is any Lie algebra and $c$ is a nontrivial cocycle,
then $c$ determines a corresponding central extension $\wt\Fg$ of $\Fg$,
\eqn\ZNTRL{ \BR \longrightarrow \wt\Fg \longrightarrow \Fg\,.}
As a vector space, $\wt\Fg = \Fg \oplus \BR$, and the Lie algebra of
$\wt\Fg$ is given by the bracket
\eqn\BRAK{  \Big[ (\phi, a), (\psi, b) \Big] = \Big(
[\phi,\psi], c(\phi,\psi) \Big)\,,}
where $\phi$ and $\psi$ are elements of $\Fg$, and $a$ and $b$ are
elements of $\BR$.

In the case of the Lie algebra $L\Fg$, the cocycle $c$ appearing in
\LGCOC\ consequently determines a central extension $\wt{{L\Fg}}$ of
$L\Fg$.  When $G$ is simply connected, the extension determined
by $c$ or any integral multiple of $c$ lifts to a corresponding
extension of $LG$ by $U(1)$,
\eqn\LGE{ U(1) \longrightarrow \wt{{LG}} \longrightarrow LG\,.}
Topologically, the extension $\wt{{LG}}$ is the total space of the $S^1$
bundle over $LG$ whose Euler class is represented by the cocyle of the
extension, interpreted as an invariant two-form on $LG$.  The fact
that the Euler class must be integral is responsible for the
corresponding quantization condition on the cocycle of the extension.

When $\Fg$ is simple, the algebra $L\Fg$ has a non-degenerate,
invariant inner product which is unique up to scale and is given by
\eqn\INVLG{ \left(\phi,\psi\right) \,=\, -\int_{S^1} dt \,
\Tr\left(\phi\psi\right)\,.}
On the other hand, the corresponding extension $\wt{{L\Fg}}$ does {\it
not} possess a non-degenerate, invariant inner product, since any
element of $\wt{{L\Fg}}$ can be expressed as a commutator, so that
$[\wt{{L\Fg}},\wt{{L\Fg}}] = \wt{{L\Fg}}$, and the center of
$\wt{{L\Fg}}$ is necessarily orthogonal to every commutator under an
invariant inner product.

However, we can also consider the semidirect product $U(1) \ltimes
\wt{{LG}}$.  Here, the rigid $U(1)$ action on $S^1$ induces a natural
$U(1)$ action on $\wt{{LG}}$ by which we define the product, and the
important observation about this group $U(1) \ltimes \wt{{LG}}$ is
that it does admit an invariant, non-degenerate inner product on its
Lie algebra.

Explicitly, the Lie algebra of $S^1 \ltimes \wt{{LG}}$ is identified
with $\BR \oplus \wt{{L\Fg}} = \BR \oplus L\Fg \oplus \BR$ as a
vector space, and the Lie algebra is given by the bracket
\eqn\BRAKII{ \Big[(p,\phi,a),(q,\psi,b)\Big] \,=\,
\Big(0,[\phi,\psi] + p \lie_R \psi - q \lie_R
\phi,c(\phi,\psi)\Big)\,,}
where $\lie_R$ is the Lie derivative with respect to the vector field $R$
generating rotations of $S^1$.  We then consider the manifestly
non-degenerate inner product on $\BR \oplus \wt{{L\Fg}}$ which is
given by \eqn\FRMLG{ \Big((p,\phi,a),(q,\psi,b)\Big) \,=\, - \int_M dt
\, \Tr(\phi \psi) - p b - q a\,.}  One can directly check that this
inner product is invariant under the adjoint action determined by
\BRAKII.  We note that although this inner product is non-degenerate,
it is not positive-definite because of the last two terms in \FRMLG.

\medskip\noindent{\it Extension To Chern-Simons Theory}\smallskip

We now return to our original problem, which is to find a Hamiltonian
action of a group $\CH$ on $\bar\CA$ to use for localization.  The
natural guess to consider the identity component $\CG_0$ of the gauge
group does not work, because the cocycle $c$ in \COC\ obstructs the
action of $\CG_0$ on $\bar\CA$ from being Hamiltonian.

However, motivated by the loop group constructions, we consider now
the central extension $\wt{{\CG_0}}$ of $\CG_0$ by $U(1)$ which is
determined by the cocycle $c$ in \COC,
\eqn\ZNTRLG{ U(1) \longrightarrow \wt{{\CG_0}} \longrightarrow
\CG_0\,.}
We assume that the central $U(1)$ subgroup of $\wt{{\CG_0}}$ acts
trivially on $\bar\CA$, so that the moment map for the central
generator $(0,a)$ of the Lie algebra is constant.  Then, by
construction, we see from \COC\ and \BRAK\ that the new moment map for
the action of $\wt{{\CG_0}}$ on $\bar\CA$, which is given by
\eqn\MUII{ \left\langle\mu,\left(\phi,a\right)\right\rangle \,=\,
\int_M \kappa\^\Tr\left(\phi F_A\right) - \int_M d\kappa\^\Tr\left(\phi A
\right) + a\,,} satisfies the Hamiltonian condition
\eqn\HAMII{ \Big\{ \big\langle\mu,(\phi,a)\big\rangle,
\big\langle\mu,(\psi,b)\big\rangle \Big\} =
\Big\langle\mu,\big[(\phi,a),(\psi,b)\big]\Big\rangle\,.}
The action of the extended group $\wt{\CG_0}$ on $\bar\CA$ is thus
Hamiltonian with moment map in \MUII.

But $\wt{{\CG_0}}$ is still not the group $\CH$ which we must use to
perform non-abelian localization in Chern-Simons theory!  In order to
realize the action $S(A)$ as the square of the moment map $\mu$ for some
Hamiltonian group action on $\bar\CA$, the Lie algebra of the group
must first possess a non-degenerate, invariant inner product.  Just
as for the loop group extension $\wt{{LG}}$, the group $\wt{{\CG_0}}$
does not possess such an inner product.

However, we can elegantly remedy this problem, just as it was remedied
for the loop group, by also considering the action of $U(1)$ on $M$.  The
$U(1)$ action on $M$ induces an action of $U(1)$ on $\wt{{\CG_0}}$, so
we consider the associated semidirect product $U(1) \ltimes
\wt{\CG_0}$.  Then a non-degenerate, invariant inner product on the
Lie algebra of $U(1) \ltimes \wt{\CG_0}$ is given by
\eqn\FRM{ \Big((p,\phi,a),(q,\psi,b)\Big) =
- \int_M \kappa\^d\kappa \, \Tr(\phi \psi) - p b - q a\,,}
in direct correspondence with \FRMLG.  As for the loop group, this
quadratic form is of indefinite signature, due to the hyperbolic form
of the last two terms in \FRM.

Finally, the $U(1)$ action on $M$ immediately induces a corresponding
$U(1)$ action on $\CA$.  Since the contact form $\kappa$ is invariant
under this action, the induced $U(1)$ action on $\CA$ descends to a
corresponding action on the quotient $\bar\CA$.  In general, the
vector field upstairs on $\CA$ which is generated by an arbitrary
element $(p,\phi,a)$ of the Lie algebra of $U(1) \ltimes \wt{\CG_0}$ is
then given by \eqn\DA{ \delta A = d_A \phi + p \, \lie_R A\,,}
where $R$ is the vector field on $M$ generating the action of $U(1)$.
Clearly the moment for the new generator $(p,0,0)$ is given by
\eqn\MUIII{
\Big\langle\mu,(p,0,0)\Big\rangle \,=\, - \ha \, p \int_M
\kappa\^\Tr\left(\lie_R A\^A\right)\,.}
This moment is manifestly invariant under the shift symmetry and
descends to $\bar\CA$.

In fact, the action of $U(1) \ltimes \wt{\CG_0}$ on $\bar\CA$ is
Hamiltonian, with moment map
\eqn\MUIV{\Big\langle\mu,(p,\phi,a)\Big\rangle \,=\,
-\ha p \int_M \kappa\^\Tr\left(\lie_RA\^A\right) \,+\, \int_M
\kappa\^\Tr\left(\phi F_A\right) \,-\, \int_M d\kappa\^\Tr\left(\phi
A\right) \,+\, a\,.}
To check this statement, it suffices to compute
$\Big\{\langle\mu,(p,0,0)\rangle,\langle\mu,(0,\psi,0)\rangle\Big\}$,
which is the only nontrivial Poisson bracket that we have not already
computed.  Thus,
\eqn\HAMIII{\eqalign{
\Big\{\big\langle\mu,(p,0,0)\big\rangle,\big\langle\mu,(0,\psi,0)
\big\rangle\Big\}
\,&=\, \Omega\Big(p \, \lie_RA,d_A\psi\Big)\,=\, -p \int_M
\kappa\^\Tr\left(\lie_RA \^ d_A \psi\right)\,,\cr
\,&=\, p \int_M \kappa\^\Tr\left(\lie_R\psi \, F_A\right) - p \int_M
d\kappa\^\Tr\left(\lie_R\psi \, A\right)\,,\cr
\,&=\, \Big\langle\mu,(0,p\,\lie_R\psi,0)\Big\rangle\,,}}
as required by the Lie bracket \BRAKII.

Thus, we identify $\CH = U(1) \ltimes \wt{\CG_0}$ as the relevant group
of Hamiltonian symmetries which we use for localization in
Chern-Simons theory.

\subsec{The Action $S(A)$ as the Square of the Moment Map}

By construction, the square $(\mu,\mu)$ of the moment map $\mu$ in
\MUIV\ for the Hamiltonian action of $\CH$ on $\bar\CA$ is a function on
$\bar\CA$ invariant under $\CH$.  The new Chern-Simons action
$S(A)$ in \SAA\ is also a function on $\bar\CA$ invariant under
$\CH$.  Given the high degree of symmetry, we certainly expect
that $(\mu,\mu)$ and $S(A)$ agree up to normalization.  We now
check this fact directly and fix the relative normalization.

We first observe that, in terms of the invariant form $(\cdot,\cdot)$
in \FRM\ on the Lie algebra of $\CH$, we can express the moment map
dually as determined by the inner product with the vector
${\left(-1,\, -\left({\kappa\^F_A - d\kappa\^A}\right) /
\kappa\^d\kappa,\, \ha\int_M\kappa\^\Tr(\lie_R A\^A)\right)}$ in the
Lie algebra of $\CH$, so that
\eqn\DL{ \Big\langle\mu,(p,\phi,a)\Big\rangle = \left(\left(-1, -\left(
{{\kappa\^F_A - d\kappa\^A} \over {\kappa\^d\kappa}}\right),
\ha\int_M\kappa\^\Tr(\lie_R A\^A)\right),(p,\phi,a)\right)\,.}
Thus, by duality, the square of $\mu$ is determined to be
\eqn\MUSQR{\eqalign{
(\mu,\mu)&=\left\langle\mu,\left(-1,-\left({{\kappa\^F_A - d\kappa\^A}
\over {\kappa\^d\kappa}}\right),\ha\int_M\kappa\^\Tr(\lie_RA\^A)\right)
\right\rangle\,,\cr
&=\int_M\kappa\^\Tr\Big(\lie_RA\^A\Big) -
\int_M\kappa\^d\kappa \, \Tr\Big(\big({{\kappa\^F_A - d\kappa\^A} \over
{\kappa\^d\kappa}}\big)^2\Big)\,.\cr}}

To simplify the first term of \MUSQR, we use the fact that the Lie
derivative $\lie_R$ can be expressed as an anti-commutator $\lie_R =
\{ \iota_R, d\}$, so that
\eqn\LXA{ \int_M\kappa\^\Tr\Big(\lie_RA\^A\Big)\,=\,\int_M\kappa\^\Tr\Big(
\{\iota_R,d\}A\^A\Big)\,.}
We now observe that $\iota_R A$ can be expressed as
\eqn\IXA{ \iota_R A = {{A \^ d\kappa} \over {\kappa\^d\kappa}}\,.}
Using this fact and integrating by parts\foot{We observe that
trivially ${\iota_R \left(\kappa \^ \Tr\left(dA \^ A\right)\right) =
0}$.} with respect to the outermost operator $d$ or $\iota_R$ in both
of the two terms from the anti-commutator \LXA, we find that
\eqn\LXAII{\eqalign{
\int_M\kappa\^\Tr\Big(\lie_RA\^A\Big)\,&=\,\int_M \Big[\iota_R \kappa \^
\Tr\left(dA \^ A\right) \,-\, \kappa \^ \Tr\left(dA \; \iota_R
A\right)\, +\cr
&\qquad +\, d\kappa \^ \Tr\left(\iota_R A \; A\right) - \kappa \^
\Tr\left(\iota_R A \; d A\right)\Big]\,,\cr
&=\,\int_M \Biggl[\Tr\left(A\^dA\right) \,-\, 2
\kappa\^\Tr\left({{d\kappa\^A} \over {\kappa\^d\kappa}}\,dA\right)\,+\cr
&\qquad +\, d\kappa\^\Tr\left({{d\kappa\^A} \over
{\kappa\^d\kappa}}\,A\right)\Biggr]\,.\cr}}

Consequently, after some algebra, we find that \MUSQR\ becomes
\eqn\MUSQRII{ (\mu,\mu)= - \int_M {1 \over {\kappa\^d\kappa}}
\Tr\Big(\big(\kappa\^F_A\big)^2\Big)  + \int_M \Tr\Big(A\^dA\Big)
+ 2 \int_M \kappa\^\Tr\Big((\iota_RA) A\^A\Big)\,.} In arriving at
\MUSQRII, we have observed that the terms involving $\kappa$ in
\LXAII\ are cancelled by corresponding terms from the second term
in \MUSQR, arising from the perfect square
${\left(\left({\kappa\^F_A - d\kappa\^A}\right) /
\kappa\^d\kappa\right)^2}$, after expanding $F_A=dA+A\wedge A$.
The last term in \MUSQRII, cubic in $A$, arises from the
cross-term in this perfect square when we express  $F_A = dA +
A\^A$ and we apply the identity \IXA.

To simplify the last term of \MUSQRII, we observe that
\eqn\AAAO{ 0 = \iota_R \Big(\kappa\^\Tr(A\^A\^A)\Big) = -3
\kappa\^\Tr\Big((\iota_RA) A\^A\Big) + \Tr\Big(A\^A\^A\Big)\,,}
so that
\eqn\MUSQRIII{
(\mu,\mu)= - \int_M {1 \over {\kappa\^d\kappa}}
\Tr\Big(\big(\kappa\^F_A\big)^2\Big)  + \int_M \Tr\Big(A\^dA + {2
\over 3} A\^A\^A\Big)\,.}

We thus find the beautiful result,
\eqn\CSMU{ S(A) = (\mu,\mu)\,.}
We finally write the Chern-Simons path integral as a symplectic integral
over $\bar\CA$ of the canonical form,
\eqn\PZCSV{ Z(\epsilon) \,=\, {1 \over {\Vol(\CG)}} \, \left({{-i} \over
{2 \pi \epsilon}}\right)^{\Delta_{\CG}/2} \, \int_{\bar\CA}
\exp{\left[\Omega + {i \over {2 \epsilon}}
\left(\mu,\mu\right)\right]}\,.}

\newsec{Non-Abelian Localization and Two-Dimensional Yang-Mills
Theory}

In this section, we recall following \WittenXU\ how the technique of
non-abelian localization can be generally applied to study a
symplectic integral of the canonical form
\eqn\ZE{ Z(\epsilon) \,=\, {1 \over {\Vol(H)}} \,
\left({1 \over {2 \pi \epsilon}}\right)^{\Delta_H/2} \,
\int_X \! \exp{\left[ \Omega - {1 \over {2 \epsilon}}
\left(\mu,\mu\right) \right]}\,,\qquad \Delta_H = \dim H\,.}
Here $X$ is a symplectic manifold with symplectic form
$\Omega$, and $H$ is a Lie group which acts on $X$ in a Hamiltonian
fashion with moment map $\mu$.  Finally, $(\,\cdot\,,\,\cdot\,)$ is an
invariant, positive-definite\foot{In the case of Chern-Simons theory,
the corresponding quadratic form \FRM\ on $\Fh$ has indefinite
signature, due to the hyperbolic summand associated to the two extra
$U(1)$ generators of $\CH$ relative to the group of gauge
transformations $\CG_0$.  Also, invariance under large gauge
transformations requires the Chern-Simons symplectic integral \PZCSV\
to be oscillatory, instead of exponentially damped.  These features do
not essentially change our discussion of localization below, and we
reserve further comment until Section 5.} quadratic form on the Lie
algebra $\Fh$ of $H$ and dually on $\Fh^*$ which we use to define the
``action'' $S = \ha (\mu,\mu)$ and the volume $\Vol(H)$ of $H$ that
appear in \ZE.

Later in this section, we also review and extend the ideas of
\WittenXU\ to apply non-abelian localization to Yang-Mills theory on a
Riemann surface.

\subsec{General Aspects of Non-Abelian Localization}

To apply non-abelian localization to an integral of the form \ZE,
we first observe that $Z(\epsilon)$ can be rewritten as \eqn\ZEII{
Z(\epsilon) \,=\, {1 \over {\Vol(H)}} \, \int_{\Fh \times X}
\left[{{d\phi} \over {2\pi}}\right] \exp{\left[\Omega - i \,
\langle\mu,\phi\rangle - {\epsilon \over 2}
(\phi,\phi)\right]}\,.} Here $\phi$ is an element of the Lie
algebra $\Fh$ of $H$, and $\left[d\phi\right]$ is the Euclidean
measure on $\Fh$ that is determined by the same invariant form
$\left(\,\cdot,\cdot\right)$ which we use to define the volume
$\Vol(H)$ of $H$.  The Gaussian integral over $\phi$ in \ZEII\
leads immediately to the expression in \ZE. The measure
$[d\phi/2\pi]$ includes a factor of $1/2\pi$ for each real
component of $\phi$.

\medskip\noindent{\it A BRST Symmetry}\smallskip

The advantage of writing $Z$ in the form \ZEII\ is that, once we
introduce $\phi$, then $Z$ becomes invariant under a BRST symmetry,
and this BRST symmetry leads directly to a localization formula for
\ZEII.

To describe this BRST symmetry, we recall that the moment map satisfies
\eqn\MOMMAPEQII{d\langle\mu,\phi\rangle \,=\, \iota_{V(\phi)}
\Omega\,,}
where $V(\phi)$ is the vector field on $X$ associated to
the infinitesimal action of $\phi$.  Because of the relation
\MOMMAPEQII, the argument of the exponential in \ZEII\ is immediately
annihilated by the BRST operator $D$ defined by
\eqn\CRTD{ D \,=\, d + i \, \iota_{V(\phi)}\,.}

To exhibit the action of $D$ locally, we choose a basis $\phi^a$ for
$\Fh$, and we introduce local coordinates $x^m$ on $X$.  We also
introduce the notation $\chi^m \equiv dx^m$ for the
corresponding basis of local one-forms on $X$, and we expand the
vector field $V(\phi)$ into components as ${V(\phi) = \phi^a \, V_a^m
\, \partial / \partial x^m}$.  Then the action of $D$ in \CRTD\ is
described in terms of these local coordinates by
\eqn\LCRTD{\eqalign{
D x^m \,&=\, \chi^m\,,\cr
D \chi^m \,&=\, i \, \phi^a \, V^m_a\,,\cr
D \phi^a \,&=\, 0\,.\cr}}

From this local description \LCRTD, we see that the action of $D$
preserves a ghost number, or grading, under which $x$ carries charge
$0$, $\chi$ carries charge $+1$, $\phi$ carries charge $+2$, and $D$
itself carries charge $+1$.

The most important property of a BRST operator is that it squares to
zero.  In this case, either from \CRTD\ or from \LCRTD, we see that
$D$ squares to the Lie derivative along the vector field $V(\phi)$,
\eqn\DSQR{ D^2 = i \, \{d, \iota_{V(\phi)}\} = i \, \lie_{V(\phi)}\,.}
Thus, $D^2 = 0$ exactly when $D$ acts on the subspace of $H$-invariant
functions $\CO(x, \chi, \phi)$ of $x$, $\chi$, and $\phi$.

For simplicity, we restrict attention to functions $\CO(x,\chi,\phi)$
which are polynomial in $\phi$.  Then an arbitrary function of this form
can be expanded as a sum of terms
\eqn\PWRO{ \CO(x)_{m_1 \ldots m_p \, a_1 \ldots a_q} \; \chi^{m_1} \cdots
\chi^{m_p} \, \phi^{a_1} \cdots \phi^{a_q}\,,}
for some $0 \le p \le \dim X$ and $q \ge 0$.  (The restriction on $p$
arises from the fact that $\chi$ satisfies Fermi statistics, whereas
$\phi$ satisfies Bose statistics.)

Globally, each term of the form \PWRO\ is specified by a section of
the bundle ${\Omega^p_X \otimes \Sym^q(\Fh^*)}$ of $p$-forms on $X$
which take values in the $q$-th symmetric tensor product of the dual
$\Fh^*$ of the Lie algebra of $H$.  Thus, if we consider the complex
${\left(\Omega_X^* \otimes \Sym^*\left(\Fh^*\right)\right)^H}$ of all
$H$-invariant differential forms on $X$ which take values in the ring
of polynomial functions on $\Fh$, then we see that $D$ defines a
cohomology theory associated to the action of $H$ on $X$.  This
cohomology theory is known as the Cartan model of the $H$-equivariant
cohomology of $X$.  With the exception of the last computation in
Section 5.3, our applications will not require a greater familiarity
with equivariant cohomology than what we have described here.
However, in Section 5.3 we will need to use a few additional
properties of equivariant cohomology that we discuss in Appendix C,
and we recommend \refs{\AtiyahRB, \GuilleminSII} as basic references.

\medskip\noindent{\it Localization for $Z$}\smallskip

Because the argument of the exponential in \ZEII\ is annihilated by
$D$ and because this argument is manifestly invariant under $H$, the
integrand of the symplectic integral $Z$ determines an equivariant
cohomology class on $X$.  Furthermore, by the usual arguments, $Z$ is
formally unchanged  by the addition of any $D$-exact invariant form to its
integrand.  This formal statement can fail if $X$ is not compact and
$Z$ suffers from divergences, as we analyze in great detail in
Appendix A, but for the moment we ignore this issue and assume $X$ is
compact.  Thus, $Z$ depends only on the equivariant cohomology class
of its integrand.

We now explain how this fact leads immediately to a localization
formula for $Z$.  We first observe that we can add to the argument of
the exponential in \ZEII\ an arbitrary term of the form $t \, D
\lambda$, where $\lambda$ is any $H$-invariant one-form on $X$ and $t$
is a real parameter, so that
\eqn\ZEIII{ Z(\epsilon) = {1 \over {\Vol(H)}} \, \int_{\Fh \times X}
\left[{{d\phi} \over {2\pi}}\right] \exp{\left[\Omega - i \,
\langle\mu,\phi\rangle - {\epsilon \over 2} (\phi,\phi) + t \, D
\lambda\right]}\,.}
This deformation of the integrand of \ZEII\ is $D$-exact and does not
change $Z$.  In particular, $Z$ does not depend on $t$.

By definition, $D \lambda$ is given explicitly by
\eqn\DLAM{ D \lambda \,=\, d \lambda + i \, \langle\lambda,
V(\phi)\rangle\,.}
As before, $\langle\,\cdot\,,\,\cdot\,\rangle$ denotes the canonical
dual pairing, so that in components the last term of \DLAM\ is given
by $\lambda_m V^m_a \phi^a$.

Thus, apart from a polynomial in $t$ that arises from expanding the
term $\exp{(t \, d\lambda)}$, all of the dependence on $t$ in the
integrand of $Z$ arises from the factor ${\exp{\left[i \, t
\, \langle\lambda, V(\phi)\rangle\right]}}$ that now appears in \ZEIII.
So if we consider the limit $t\rightarrow \infty$, then the stationary
phase approximation to the integral is valid, and all contributions
to $Z$ localize around the critical points of the function
$\langle\lambda, V(\phi)\rangle$.

We expand this function in the basis $\phi^a$ for $\Fh$ which we
introduced previously,
\eqn\DLAMII{ \langle\lambda, V(\phi)\rangle \,=\, \phi^a \,
\langle\lambda, V_a\rangle\,.}
Thus, the critical points of $\langle\lambda, V(\phi)\rangle$ arise
from the simultaneous solutions in $\Fh \times X$ of the equations
\eqn\DLAMIII{\eqalign{
\langle\lambda, V_a\rangle \,&=\, 0\,,\cr
\phi^a \, d \langle\lambda, V_a\rangle \,&=\, 0\,.\cr}}
The first equation in \DLAMIII\ implies that $Z$ necessarily localizes
on points in $\Fh \times X$ for which $\langle\lambda, V_a\rangle$
vanishes.  As for the second equation in \DLAMIII, we see that it is
invariant under an overall scaling of $\phi$ in the vector space $\Fh$.
Consequently, upon integrating over $\phi$ in \ZEIII, we see that the
critical locus of the function $\langle\lambda, V(\phi)\rangle$ in
$\Fh \times X$ projects onto the vanishing locus of $\langle\lambda,
V_a\rangle$ in $X$.  So $Z$ localizes on the subset of $X$ where
${\langle\lambda, V_a\rangle = 0}$.

By making a specific choice of the one-form $\lambda$, we can describe
the localization of $Z$ more precisely.  In particular, we now show
that $Z$ localizes on the set of critical points of the function $S =
\ha (\mu,\mu)$ on $X$.

We begin by choosing an almost complex structure $J$ on $X$.  That is,
$J: TX \rightarrow TX$ is a linear map from $TX$ to itself such that
$J^2 = -1$.  We assume that $J$ is compatible with the symplectic form
$\Omega$ in the sense that $\Omega$ is of type $(1,1)$ with respect to
$J$ and the associated metric $G(\cdot,\cdot) = \Omega(\cdot, J
\cdot)$ on $X$ is positive-definite.  Such an almost complex structure
always exists.

Using $J$ and $S$, we now introduce the invariant one-form
\eqn\LCLM{ \lambda \,=\, J \, dS \,=\, (\mu, J \, d\mu)\,.}
In components, ${\lambda \,=\, dx^m J_m^n \partial_n S \,=\, dx^m
\mu^a J^n_m \partial_n \mu_a}$.

The integral $Z$ now localizes on the subset of $X$ where
${\langle\lambda, V_a\rangle= 0}$.  Comparing to \LCLM, we see that
this subset certainly includes all critical points of $S$, since by
definition $dS = 0$ at these points.

Conversely, we now show that if ${\langle\lambda, V_a\rangle = 0}$ at
some point on $X$, then this point is a critical point of $S$.  To
prove this assertion, we use the inverse $\Omega^{-1}$ to $\Omega$,
which arises by considering the symplectic form as
an isomorphism $\Omega: TM \rightarrow T^*M$ with inverse
$\Omega^{-1}:T^*M \rightarrow TM$.  In components, $\Omega^{-1}$ is
defined by $(\Omega^{-1})^{l m} \, \Omega_{m n} \,=\, \delta^l_n$.

In terms of $\Omega^{-1}$, the moment map equation \MOMMAPEQII\ is
equivalent to the relation
\eqn\INVMOMEQ{ V \,=\, \Omega^{-1} \, d\mu\,,}
or ${V_a^m \,=\, (\Omega^{-1})^{m n} \, \partial_n \mu_a}$.  Thus,
\eqn\SII{ \Omega^{-1} \, dS \,=\, \left(\mu, \, \Omega^{-1} d\mu\right)
\,=\, \left( \mu, V\right)\,,}
or ${(\Omega^{-1})^{m n} \, \partial_n S \,=\, \mu^a V^m_a}$.

In particular, the condition that ${\langle\lambda, V_a\rangle = 0}$
implies that
\eqn\SIII{ 0 \,=\, \left(\mu,\,\langle\lambda, V\rangle\right) \,=\,
\langle\lambda,\,\Omega^{-1} dS\rangle \,=\, \langle J\, dS,\,
\Omega^{-1} dS\rangle\,,}
or more explicitly, ${0 \,=\, \mu^a \lambda_m V^m_a \,=\, \lambda_m \,
(\Omega^{-1})^{m n} \, \partial_n S \,=\, (\Omega^{-1})^{m n} J^l_m \,
\partial_l S \,\partial_n S}$.  We recognize the last expression in
\SIII\ as the norm of the one-form $dS$ with respect to the metric $G$
on $X$.  As $G$ is positive-definite, we conclude that the condition
${\langle\lambda, V_a\rangle = 0}$ implies the vanishing of $dS$.
Thus, the symplectic integral $Z$ localizes precisely on the critical
set of $S$.

\subsec{Non-Abelian Localization For Yang-Mills Theory, Part I}

In the rest of this section, we apply non-abelian localization to
perform path integral computations in two-dimensional Yang-Mills
theory on a smooth Riemann surface $\Sigma$.  These computations are an
essential warmup for our later computations in Chern-Simons theory.

As we discussed in Section 2, the Yang-Mills path integral is
naturally a symplectic integral of the canonical form $\ZE$, where the
abstract symplectic manifold $X$ is now the affine space $\CA(P)$ of
connections on a fixed principal $G$-bundle $P$ over $\Sigma$, and
where the abstract group $H$ is now the group $\CG(P)$ of gauge
transformations.  Also, the moment map for the action of $\CG(P)$ on
$\CA(P)$ is simply the curvature of the connection, $\mu = F_A$.

As a result of our general discussion above, the Yang-Mills path
integral localizes on critical points of the Yang-Mills action.
These critical points fall into two qualitatively different sorts.
Because the action $S = \ha (\mu,\mu)$ is quadratic in the moment
map $\mu$, so that $dS = (\mu, d\mu)$, we see that the critical
locus of $S$ includes all points where $\mu$ vanishes, as well as
other points where $\mu$ is generally non-zero.  The points at
which $\mu = 0$ are clearly stable minima of $S$, and any other
critical points at which $\mu \neq 0$ are higher extrema of $S$,
which in our applications are unstable. In the case of Yang-Mills
theory, the stable minima of the action are the flat connections
on $\Sigma$, and the higher extrema are connections with non-zero
curvature which represent classical solutions of Yang-Mills
theory, so that $d_A \* F_A = 0$ with $F_A \neq 0$.

For our application to Chern-Simons theory, we must understand
localization at both the flat and the non-flat solutions of classical
Yang-Mills theory.  So in the rest of Section 4.2, we review following
\WittenXU\ how non-abelian localization works for flat connections,
and then in Section 4.3 we discuss the generalization for solutions of
Yang-Mills theory with curvature.

\medskip\noindent{\it Localization on a Smooth Component of the Moduli
Space of Flat Connections}\smallskip

We assume that $\CM_0$ is a smooth component of the moduli space of
flat connections on $\Sigma$.  For ease of future notation, we make the
identifications
\eqn\IDS{\eqalign{
X \,&=\, \CA(P)\,,\cr
H \,&=\, \CG(P)\,,\cr
\mu \,&=\, F_A\,.\cr}}
We now identify $\CM_0$ abstractly as a symplectic quotient of the zero
locus $\mu^{-1}(0) \subset X$ by the free action of the group $H$, so
that $\CM_0 = \mu^{-1}(0) / H$.

The fundamental result of \WittenXU, whose derivation we now recall,
is that the local contribution $Z(\epsilon)|_{\CM_0}$ to the path integral
from $\CM_0$ is given by the topological expression
\eqn\ZV{ Z(\epsilon)|_{\CM_0} \,=\, \int_{\CM_0} \exp{\left(\Omega +
\epsilon \, \Theta\right)}\,.}
Here $\Omega$ is the symplectic form on $\CM_0$ induced from the
corresponding symplectic form on $X$ (also denoted previously by
$\Omega$), and $\Theta$ is a characteristic class of degree four on
$\CM_0$ which appears explicitly as part of the derivation of \ZV.  In
particular, when the coupling $\epsilon$ is zero, then $Z(0)|_{\CM_0}$ is
the symplectic volume of $\CM_0$.

To derive \ZV\ by localization, we start by considering the local
geometry of the zero set $\mu^{-1}(0)$ in $X$.  Thus, we let
$N$ be a small open neighborhood of $\mu^{-1}(0)$ in $X$, so
that $\mu^{-1}(0) \subset N \subset X$.  We assume that this
neighborhood is chosen so that $N$ is preserved by the action of
$H$ and so that $N$ retracts equivariantly onto $\mu^{-1}(0)$.
By composing this retraction with the quotient by the action of
$H$,  we define a projection $pr: N \rightarrow \CM_0$.  Thus,
denoting the fiber of $pr$ by $F$, we have the following equivariant
bundle
\eqn\PRES{F \longrightarrow N \buildrel pr\over\longrightarrow
\CM_0\,.}

The symplectic integral which describes the local contribution of
$\CM_0$ to $Z$ is now given by
\eqn\ZEIIII{ Z(\epsilon)|_{\CM_0} \,=\, {1 \over {\Vol(H)}} \,
\int_{\Fh \times N} \left[{{d\phi} \over {2\pi}}\right]
\exp{\left[\Omega - i \, \langle\mu,\phi\rangle - {\epsilon \over 2}
(\phi,\phi) + t \, D \lambda\right]}\,,}
where $\lambda$ is the invariant one-form that we introduced in \LCLM\
to localize $Z$.  Because $N$ is noncompact, this integral in \ZEIIII\
is only defined by localization, so that we require $t \neq 0$.

As explained in detail in \WittenXU, because $N$ retracts
equivariantly onto $\CM_0$ and because the action of $H$ is free near
$\mu^{-1}(0)$, the equivariant cohomology class of degree two\foot{We
recall that $\phi$ carries degree $+2$ with respect to equivariant
cohomology.} represented by the expression $\Omega - i \,
\langle\mu,\phi\rangle$ in \ZEIIII\ is simply the pullback by $pr$ of
the induced symplectic form on $\CM_0$.  Similarly, the equivariant
cohomology class of degree four represented by $-\ha(\phi,\phi)$ in
\ZEIIII\ is the pullback by $pr$ of an ordinary cohomology class
$\Theta$ of degree four on $\CM_0$.  Since $H$ acts freely on
$\mu^{-1}(0)$, $\Theta$ represents a degree four characteristic class
of $\mu^{-1}(0)$ regarded as a principal $H$-bundle over $\CM_0$.

Thus, as the only term appearing in the argument of the exponential
in \ZEIIII\ which does not pull back from $\CM_0$ is $t \, D\lambda$
itself, to derive \ZV\ from \ZEIIII\ we must only show that the
integral of $\exp{(t D \lambda)}$ over the fiber $F$ of \PRES\
produces a trivial factor of $1$,
\eqn\COTB{ {1 \over {\Vol(H)}} \, \int_{\Fh \times F}
\left[{{d\phi} \over {2\pi}}\right] \exp{\left[t \, D
\lambda\right]} \,=\, 1\,.}
This computation is what we must essentially generalize to discuss
localization at non-flat Yang-mills solutions, so we review it in
detail.

\medskip\noindent{\it A Local Model For $F$ From Hodge Theory}\smallskip

In order to perform the direct computation of the integral in \COTB,
we first identify the correct local model for the geometry of
$F$.  By assumption, the group $H$ acts freely on $F$, so $F$ must
contain a copy of $H$.  Since $F$ must also be symplectic, the
simplest local model for $F$ is just the cotangent bundle $T^* H$ of
$H$, with its canonical symplectic structure.

In fact, the simple guess that $F = T^* H$ is precisely correct, and
it has an important infinite-dimensional interpretation in the
context of Yang-Mills theory.  To explain this interpretation, we
consider the tangent space to $\CA(P)$ at a point corresponding to a
flat connection $A$.  As we have discussed, the tangent space to
$\CA(P)$ at $A$ can be identified with the space of smooth sections
$\Gamma(\Sigma, \Omega^1_\Sigma \otimes \ad(P))$ of the bundle of
one-forms on $\Sigma$ taking values in the adjoint bundle $\ad(P)$.

By definition, the flatness of $A$ implies that the covariant
derivative $d^{}_A$ satisfies $d_A^2 = 0$.  Because of this fact,
$d^{}_A$ has many of the same properties as the de Rham exterior
derivative $d$, and the usual Hodge decomposition for $d$ has an
immediate analogue for $d^{}_A$.

In the case of the covariant derivative $d^{}_A$, the Hodge decomposition
implies that the vector space ${\Gamma(\Sigma, \Omega^1_\Sigma \otimes
\ad(P))}$ decomposes into three subspaces, orthogonal with respect to
the metric induced by $\*$ on $\CA(P)$, of the form
\eqn\HDG{ \Gamma(\Sigma, \Omega^1_\Sigma \otimes \ad(P)) \,=\, \CH_1
\oplus \Im(d_A^{}) \oplus \Im(d_A^\dagger)\,.}
Here $d_A^\dagger = - \* \,d^{}_A\, \*$ is the standard adjoint
to $d^{}_A$ with respect to the metric on $\CA(P)$.  Also, $\CH_1$
denotes the finite-dimensional subspace of harmonic one-forms taking
values in $\ad(P)$, so that elements of $\CH_1$ are annihilated by the
Laplacian ${\Delta_A = d^{}_A d_A^\dagger + d_A^\dagger d^{}_A}$.
Finally, $\Im(d^{}_A)$ and $\Im(d_A^\dagger)$ denote the images of
$d^{}_A$ and $d_A^\dagger$ when these operators act
respectively on sections of the bundles $\ad(P)$ and $\Omega^2_\Sigma
\otimes \ad(P)$ on $\Sigma$.

Concretely, the Hodge decomposition implies that, if $\eta$ is any
section of $\Omega^1_\Sigma \otimes \ad(P)$, then $\eta$ can be
uniquely written as a sum of three terms, all orthogonal,
\eqn\HDGII{ \eta \,=\, \xi \,+\, d^{}_A \phi \,+\, d_A^\dagger
\Psi\,,}
where $\xi$ satisfies $\Delta_A \xi = 0$ and where $\phi$ and $\Psi$ are
respectively sections of the bundles $\ad(P)$ and $\Omega^2_\Sigma
\otimes \ad(P)$.

To interpret the Hodge decomposition \HDG\ as a geometric
statement, we note that the finite-dimensional vector space
$\CH_1$ of harmonic one-forms can be identified with the tangent
space to the moduli space $\CM_0$ of flat connections at $A$.  For
instance, since $d_A^2 = 0$, we can consider the cohomology of
$d^{}_A$.  As usual, we identify the harmonic forms in $\CH_1$ as
representatives of cohomology classes in
$H^1\!\left(\Sigma,\ad\left(P\right)\right)$.  These cohomology
classes describe infinitesimal deformations of the flat connection
$A$.

On the other hand, since we assume that the gauge group $\CG(P)$ acts
freely at $A$, $d^{}_A$ has no kernel when acting on sections of
$\ad(P)$.  Otherwise, if a section $\phi$ of $\ad(P)$ did satisfy $d^{}_A
\phi = 0$, then the gauge transformation generated by $\phi$ would fix
$A$.  Equivalently, we have that
$H^0\!\left(\Sigma,\ad\left(P\right)\right) = 0$.

Because $d^{}_A$ has no kernel when acting on sections of
$\ad(P)$, $d^{}_A$ can be formally inverted and the image of $d_A$
in ${\Gamma(\Sigma,\Omega^1_\Sigma \otimes \ad(P))}$ identified with
the space of sections of $\ad(P)$ itself.  Of course, a section $\phi$
of $\ad(P)$, as appears in \HDGII, is interpreted geometrically as a
tangent vector to the gauge group $\CG(P)$.

Similarly, we can also identify the image of the adjoint
$d_A^\dagger$ with the space of sections of the bundle
$\Omega^2_\Sigma \otimes \ad(P)$.  Such a section $\Psi$, as in
\HDGII, is interpreted geometrically as a cotangent vector to the
gauge group $\CG(P)$.

Furthermore, if we recall the natural symplectic form $\Omega$ on
$\CA(P)$ in \OMYM, we see that $\Im(d_A)$ is isotropic with
respect to $\Omega$.  For if $\phi$ and $\psi$ are any two sections of
the bundle $\ad(P)$ on $\Sigma$, then
\eqn\ISCG{ \Omega(d_A \phi, d_A \psi) \,=\, - \int_\Sigma \Tr(d_A
\phi \^ d_A \psi) \,=\, \int_\Sigma \Tr(\phi \, d_A^2 \psi) \,=\,
0\,.}
This fact crucially relies on the flatness of $A$, since we use that
$d_A^2 = 0$ in deducing the last equality of \ISCG.  Of course,
the fact that $\Im(d^{}_A)$ is isotropic with respect to $\Omega$
is mirrored by the fact that $H$ is a Lagrangian submanifold of $T^* H$.

Thus, the Hodge decomposition \HDG\ applied to $\Gamma(\Sigma,
\Omega^1_\Sigma \otimes \ad(P))$ locally reflects the geometric
statement that $F$ is modeled on the cotangent bundle $T^* H$.  In
this example, it may seem perverse to translate the simple statement
that $F = T^* H$ into the infinite-dimensional statement of the Hodge
decomposition.  However, when we consider the corresponding local
geometry for higher critical points, this infinite-dimensional
perspective allows us to deduce directly how the simple symplectic
model based on $T^* H$ must be modified to describe higher critical
points of Yang-Mills theory.

\medskip\noindent{\it Computing a Symplectic Integral on $T^*
H$}\smallskip

Having identified the symplectic model for $F$ as the cotangent bundle
$T^* H$, we compute in the remainder of this subsection the symplectic
integral
\eqn\COTBII{ {1 \over {\Vol(H)}} \, \int_{\Fh \times T^* H}
\left[{{d\phi} \over {2\pi}}\right] \exp{\left[t \, D
\lambda\right]}\,.}
We review this short computation from \WittenXU\ simply because we
must generalize it to discuss localization at non-flat Yang-Mills
connections.

Thus, we consider the symplectic manifold $T^* H$ with its canonical
symplectic structure.  By convention, the action of $H$ on $T^* H$ is
induced from the right action of $H$ on itself.  By passing to a basis
of right-invariant one-forms and using the invariant metric
$(\cdot,\cdot)$ on $H$, we identify $T^* H \cong H \times \Fh$.  Under
this identification, we introduce coordinates $(g,\gamma)$ on $H
\times \Fh$.

In these coordinates, the canonical right-invariant one-form on
$H$ which takes values in $\Fh$ is given by \eqn\CARTAN{ \theta
\,=\, dg g^{-1}\,.} In terms of $\theta$, the canonical symplectic
structure on $T^* H$ is given by the invariant two-form
\eqn\SYMH{\eqalign{ \Omega \,&=\, d(\gamma,\theta)\,=\,
(d\gamma,\theta) + (\gamma,d\theta)\,,\cr &= \left(d\gamma + \ha
[\gamma, \theta], \theta\right)\,,\cr}} where in passing to the
second line of \SYMH\ we recall that ${d\theta = \theta \^ \theta
= \ha [\theta, \theta]}$. Also, if $\phi$ is an element of $\Fh$,
then the corresponding vector field $V(\phi)$ on $T^* H$ which is
generated by the infinitesimal right-action of $\phi$ is given by
\eqn\VECTH{ \delta g \,=\, - g \phi\,,\qquad \delta \gamma \,=\,
0\,.}

To proceed, we require an explicit formula for the invariant
one-form $\lambda$ appearing in \COTBII.  Abstractly, $\lambda =
(\mu, J \, d\mu)$ is determined by the moment map $\mu$ for the
$H$-action on $T^* H$ and an almost complex structure $J$
compatible with $\Omega$ in \SYMH, both of which are easy to
determine.  A convenient formula for $\lambda$ was obtained in
\WittenXU.  In brief,  one has
${\langle\mu,\phi\rangle=-(\gamma,g\phi g^{-1})}$, and one defines
a $G$-invariant almost complex structure compatible with $\Omega$
by \eqn\SYMJ{
J(\theta)=-\left(d\gamma+\ha[\gamma,\theta]\right)\,,\qquad
J\left(d\gamma + \ha[\gamma,\theta]\right)=\theta\,.} One then
finds that ${(\mu, J\, d\mu) = (\gamma, \theta)}$ after using the
fact that $[\gamma,\gamma]=0$.  So finally \eqn\LOCLAMTG{ \lambda
\,=\, (\mu, J\, d\mu) \,=\, (\gamma,\theta)\,.}

Thus, from \VECTH, \LOCLAMTG, and the definition of $D$ in \CRTD, we see
that
\eqn\LOCLAMII{ D \lambda \,=\, \Omega - i \left(\gamma,g \phi
g^{-1}\right)\,.}
Without loss, we set $t = 1$ in \COTBII\ and we change variables from
$\phi$ to $g \phi g^{-1}$, under which the measure $[d \phi]$ on $\Fh$
is invariant.  Then the symplectic integral takes the simple form
\eqn\COTBIII{ {1 \over {\Vol(H)}} \, \int_{\Fh \times T^* H}
\left[{{d\phi} \over {2\pi}}\right] \exp{\Big[ \Omega  - i\,
\left(\gamma,\phi\right)\Big]}\,.}

The integral over $\gamma$ can be done using the fact that
\eqn\DLTAF{ \int_{-\infty}^{+\infty} dy \, \exp{(-i x y)} \,=\, 2
\pi \, \delta(x)\,,} and the resulting multi-dimensional delta
function can be used to perform the integral over $\phi$.  We note
that the factors of $2\pi$ from \DLTAF\ nicely cancel the factors
of $2 \pi$ in the measure for $\phi$.  Finally, the remaining
integral over $g$ in $H$ produces a factor of the volume $\Vol(H)$
which cancels the prefactor in \COTBIII.  Thus, assuming $T^* H$
is suitably oriented, the symplectic integral over $T^* H$ is
indeed 1, as claimed in \COTB.

\subsec{Non-Abelian Localization For Yang-Mills Theory, Part II}

We now study localization at the higher, unstable critical points of the
Yang-Mills action, which correspond to non-flat connections which
solve the Yang-Mills equation on $\Sigma$.  Localization at the higher
critical points of two-dimensional Yang-Mills theory has recently been
discussed from a mathematical perspective by Woodward and Teleman
\refs{\TelemanCW,\WoodwardCT}, but we find it useful to proceed with a
more naive discussion along the lines of \WittenXU.  We begin with
some generalities about non-flat connections which solve the
Yang-Mills equation on $\Sigma$.

We first introduce the notation $f$ for the section of $\ad(P)$ dual
to the curvature $F_A$,
\eqn\DEFF{ f \,=\, \* F_A\,.}
Then, by definition, any Yang-Mills solution on $\Sigma$ satisfies the
classical equation of motion
\eqn\CLASF{ d_A f \,=\, 0\,.}
This equation simply expresses the geometric condition that $f$ be a
covariantly constant section of $\ad(P)$, and we can consequently
regard $f$ as an element of the Lie algebra $\Fg$ of $G$.

Because $f$ is constant, $f$ yields a reduction of the structure group
$G$ of the bundle to the subgroup $G_f \subset G$ which commutes with
$f$.  In physical terms, the background curvature breaks the gauge
group from $G$ to $G_f$.

As a result of the reduction from $G$ to $G_f$,  any non-flat
Yang-Mills solution for gauge group $G$ can be succinctly described
as a flat connection for gauge group $G_f$ which is twisted by a
constant curvature line bundle associated to the $U(1)$ subgroup of
$G$ generated by $f$.

In general, we denote by $\CM_f$ the moduli space of Yang-Mills
connections whose curvature lies in the conjugacy class of $f$.  We
have already discussed localization on the moduli space $\CM_0$ of
flat connections, for which $G_0 = G$.  At the opposite extreme, $f$
breaks $G$ to a maximal torus $G_f$ commuting with $f$.  We refer to
such a Yang-Mills solution as ``maximally reducible,'' and one basic goal
in this section is to obtain an explicit formula, as in \ZV, for the
contribution to the path integral from the corresponding moduli space
$\CM_f$ of maximally reducible Yang-Mills solutions.  Of course, we
could also consider the local contributions from Yang-Mills solutions
between the extremes of the flat and maximally reducible connections,
but this further generalization is not necessary for our discussion of
Chern-Simons theory.

Because $f$ is constant, the adjoint action of $f$ determines a bundle
map from $\ad(P)$ to itself, and a good idea is to decompose
$\ad(P)$ under this action.  With our conventions, $f$ is
anti-hermitian, so following \AtiyahYM\ we introduce a hermitian
operator $\Lambda$,
\eqn\HLAM{ \Lambda \,=\, i \left[ f , \,\cdot\, \right]\,,}
which acts on a section $\phi$ of $\ad(P)$ as ${\Lambda \, \phi \,=\, i
\left[f, \phi\right]}$.

When we consider the action of $\Lambda$, it is natural to work with
complex, as opposed to real, quantities.  So we now consider in place
of the real bundle $\ad(P)$ the complex bundle $\ad_\BC(P) = \ad(P)
\otimes \BC$.  When we complexify $\ad(P)$, the $(1,0)$ and $(0,1)$
parts of an $\ad(P)$-valued connection become independent complex
variables.  After picking a local complex coordinate $z$ on $\Sigma$,
these can be written locally as $A_z$ and $A_{\bar z}$.

Under the action of $\Lambda$, the bundle $\ad_\BC(P)$ decomposes into
a direct sum of subbundles, each associated to a distinct eigenvalue of
$\Lambda$.  For our purposes, we need only consider the decomposition
of $\ad_\BC(P)$ into the positive, zero, and negative eigenspaces of
$\Lambda$,
\eqn\BIGAD{ \ad_\BC(P) \,=\, \ad_{+}(P) \oplus \ad_{0}(P) \oplus
\ad_{-}(P)\,,}
where $\ad_{\pm}(P)$ and $\ad_{0}(P)$ denote respectively the
subbundles of $\ad_\BC(P)$ associated to these eigenspaces.  The
eigenspace decomposition of $\ad_\BC(P)$ in \BIGAD\ will play an
important role shortly.

\medskip\noindent{\it Example: $G=SU(2)$}\smallskip

As a simple example of these ideas, we consider the higher Yang-Mills
critical points when the gauge group $G$ is $SU(2)$.   In this case,
all non-flat Yang-Mills solutions are maximally reducible, since any
$f \neq 0$ reduces the structure group to a maximal torus $U(1)
\subset SU(2)$.

The rank-one case $G = SU(2)$ of Yang-Mills theory is also the
essential case to understand for our application to Chern-Simons gauge
theory, with gauge group of arbitrary rank.  As we explain in
Section 5, near a flat Chern-Simons connection on the three-manifold
$M$, the local geometry in the symplectic manifold $\bar\CA$ of \CMI\
can be modeled on the geometry of infinitely-many copies of the
geometry near a higher $SU(2)$ Yang-Mills critical point.  This
correspondence arises heuristically by identifying the background Yang-Mills
curvature $f$, which generates the torus $U(1) \subset SU(2)$, with
the geometric curvature of $M$ regarded as a principal $U(1)$-bundle
over the surface $\Sigma$.

In the case of Yang-Mills theory, since $f$ reduces the structure
group of the $SU(2)$ bundle to $U(1)$, the $SU(2)$ bundle on
$\Sigma$ splits as a direct sum of line bundles. As $f$ itself is
associated to a constant curvature line bundle on $\Sigma$, up to
conjugacy $f$ takes the form
\eqn\CFII{ f = 2 \pi i \pmatrix{n&0\cr0&-n\cr}\,,}
for some integer $n \neq 0$.  Because the Weyl group of $SU(2)$ acts
on $f$ by sending $n \to -n$, without loss we can assume that $n > 0$.

Introducing the standard generators of $\frak{su}(2)$ regarded as a
complex Lie algebra,
\eqn\BS{ \sigma_z \,=\, \pmatrix{i&0\cr0&-i\cr}\,,\quad
\sigma_+ \,=\, \pmatrix{0&1\cr0&0\cr}\,,\quad \sigma_- \,=\,
\pmatrix{0&0\cr1&0\cr}\,,}
we see that $\Lambda$ acts on $\frak{su}(2)$, and hence on $\ad_\BC(P)$,
with eigenvalues $0$ and $\pm 4 \pi n$.  Thus, in this case the
general decomposition of $\ad_\BC(P)$ in \BIGAD\ takes the simple form
\eqn\DCADP{ \ad_\BC(P) \,=\, \CL^{-1}(-2n) \oplus \CO \oplus \CL(2n)\,.}
Here $\CO$ is the trivial line bundle on $\Sigma$, $\CL$ is an
arbitrary flat line bundle on $\Sigma$, and we use
the standard notation ${\CL(2n) = \CL \otimes \CO(2n)}$, where
$\CO(2n)$ is the $2n$-th tensor power of a fixed line bundle $\CO(1)$
of degree one on $\Sigma$.

Thus, for each $n > 0$, the choice of a non-flat connection
solving the Yang-Mills equation on $\Sigma$ is determined by the
choice of the flat line bundle $\CL$ on $\Sigma$.  Such a line
bundle is specified by the $U(1)$ holonomy of its connection, and
hence the moduli space of flat line bundles on $\Sigma$ is
parametrized by a complex torus, the Jacobian of $\Sigma$.  If
$\Sigma$ has genus $g$, with $2g$ periods, then the Jacobian has
complex dimension $g$.  Thus, for fixed $f \neq 0$, the moduli
space $\CM_f$ of higher critical points of $SU(2)$ Yang-Mills
theory on $\Sigma$ is simply a complex torus of dimension $g$.

More generally, if we consider an arbitrary gauge group $G$ of rank
$r$ such that $f$ breaks $G$ to a maximal torus, then the
corresponding moduli space $\CM_f$ is again a complex torus of
dimension $g \, r$ which describes the holonomy in $U(1)^r$.

\medskip\noindent{\it The Partition Function of $SU(2)$ Yang-Mills
Theory}\smallskip

One of our basic goals in the rest of this section is to compute
directly the contributions from higher critical points to the
partition function $Z$ of $SU(2)$ Yang-Mills theory.  Of course, $Z$
can be computed exactly \MigdalZG, and we can readily extract
from the known expression for $Z$ a formula for the local
contributions from the higher critical points.  So before we delve
into our path integral computation, we present now the answer which
we expect to reproduce and we preview its most interesting features.

In general, if the gauge group $G$ is simply-connected, then the
partition function of Yang-Mills theory on a unit area Riemann surface of
genus $g$ is given by a sum over representations $\CR$ of $G$ of the form
\eqn\TDYMZ{ Z(\epsilon) \,=\, \left(\Vol(G)\right)^{2g
- 2} \, \sum_\CR \, {1 \over {\dim (\CR)^{2g-2}}} \, \exp{\left(-\ha
\epsilon \, \wt C_2\left(\CR\right)\right)}\,.}
Here $\wt C_2(\CR)$ is a renormalized\foot{The renormalized quadratic
Casimir $\wt C_2(\CR)$ differs from the usual quadratic Casimir solely
by an additive constant.} version of the quadratic Casimir
associated to the representation $\CR$, and the volume $\Vol(G)$ of
$G$ is determined in our conventions by the invariant form $-\Tr$ on
the Lie algebra $\Fg$.  We recall that for $G=SU(r+1)$, ``$\Tr$''
denotes the trace in the fundamental representation.

Finally, because of the possibility of weighting the Yang-Mills path
integral on $\Sigma$ by a purely topological factor $\exp{\left(c
\left(2g-2\right)\right)}$ for an arbitrary constant $c$, we have
fixed the prefactor in \TDYMZ\ so that $Z(0)$ agrees, at least
up to a sign which we will not try to fix, with the symplectic volume
of the moduli space $\CM_0$ of flat connections on $\Sigma$ as
computed in \WittenWE\ from the theory of Reidemeister-Ray-Singer
torsion.  Our choice of $c$ differs from the choice in \WittenWE\
simply because the symplectic form $\Omega$ in \OMYM\ which we use
here is related to the integral symplectic form $\Omega'$ used in
\WittenWE\ by $\Omega = 4 \pi^2 \, \Omega'$.

We now evaluate \TDYMZ\ in the case $G = SU(2)$.  In this case, each
representation is labelled by its dimension, so we denote by $\CR_n$ the
$SU(2)$ representation of dimension $n$. The renormalized quadratic
Casimir of $\CR_n$, which is just the usual quadratic Casimir with an
additive constant, is then
\eqn\CTWOR{ \wt C_2(\CR_n) \,=\, \ha \, n^2\,.}
Finally, using the metric on $SU(2)$ determined by $-\Tr$, the volume
of $SU(2)$ is given\foot{This fact follows immediately if we recall
that the volume of an $S^3$ of unit radius is $2 \pi^2$.  However, in
our metric on $SU(2)$, the $U(1)$ subgroup associated to the
normalized generator $T_z = {1 \over \sqrt{2}} \, \sigma_z$, as in
\BS, has length $2 \pi \sqrt{2}$, so $SU(2)$ has radius $r = \sqrt{2}$ in
our metric.} by ${\Vol(SU(2)) = 2^{5/2} \pi^2}$.  Thus, the partition
function \TDYMZ\ of $SU(2)$ Yang-Mills theory on $\Sigma$ becomes
\eqn\TDYMZII{ Z(\epsilon) \,=\, \left(32 \pi^4 \right)^{g-1}
\, \sum_{n=1}^\infty \, {1 \over {n^{2g-2}}} \, \exp{\left(-{{\epsilon \,
n^2} \over 4}\right)}\,.}

In order to extract the contributions of the higher critical points
from \TDYMZII, we first differentiate $Z(\epsilon)$ with respect to
$\epsilon$ to cancel the prefactor $n^{-2(g-1)}$ in the summand of
\TDYMZII,
\eqn\DTDYMZ{ {{\partial^{g-1} Z(\epsilon)} \over {\partial
\epsilon^{g-1}}} \,=\, \left(-8 \pi^4\right)^{g-1} \, \sum_{n=1}^\infty
\, \exp{\left(-{{\epsilon \, n^2} \over 4}\right)} \,=\, \ha \,
\left(-8 \pi^4\right)^{g-1} \, \left(-1 \,+\, \sum_{n \in \BZ} \,
\exp{\left(-{{\epsilon \, n^2} \over 4}\right)}\right)\,.}
To obtain a manifestly convergent expression in the weak coupling
regime of small $\epsilon$, we apply Poisson summation to the last
term of \DTDYMZ\ to obtain
\eqn\DTDYMZII{ {{\partial^{g-1} Z(\epsilon)} \over {\partial
\epsilon^{g-1}}} \,=\, \ha \, \left(-8 \pi^4\right)^{g-1} \, \left(-1
\,+\, \sqrt{{{4 \pi} \over \epsilon}} \, \sum_{n \in \BZ} \,
\exp{\left(-{{\left(2 \pi n\right)^2} \over \epsilon}\right)}\right)\,.}

Finally, to identify the contribution in \DTDYMZII\ from higher
Yang-Mills critical points, we observe that at a higher critical point
of degree $n$, the classical Yang-Mills action $S_n$ determined by $f$ in
\CFII\ is given by $S_n = (2 \pi n)^2 / \epsilon$ (assuming $\Sigma$
has unit area).  The semiclassical contribution to $Z$ from such a
critical point is weighted by the usual exponential factor
$\exp{(-S_n)}$, which we see directly in the last term of \DTDYMZII.
Thus, the locus $\CM_n$ of higher critical points of degree $n$
contributes to the sum in \DTDYMZII\ as
\eqn\LOCN{ {{\partial^{g-1} Z(\epsilon)} \over {\partial
\epsilon^{g-1}}}\Bigg|_{\CM_n} \,=\, \left(-8 \pi^4\right)^{g-1} \,
\sqrt{{{4 \pi} \over \epsilon}} \, \exp{\left(-{{\left(2 \pi
n\right)^2} \over \epsilon}\right)}\,.}
We note that a trivial factor of two in \LOCN\ arises from the action
of the Weyl group, since the two terms in \DTDYMZII\ for both $\pm n$
arise from the higher critical points of degree $n$.

This expression \LOCN\ is what we compute using localization, and it
has a number of interesting features.  Most fundamentally, we see
that the natural quantity to consider is not $Z$ but its derivative
$\partial^{g-1} Z(\epsilon) / \partial \epsilon^{g-1}$.  In discussing
the higher critical points, we lose nothing by considering this
derivative, since any terms in $Z$ that are polynomial in $\epsilon$,
and hence are annihilated by the derivative, arise as contributions
from the moduli space $\CM_0$ of flat connections.  Moreover, although
the formula in \LOCN\ is expressed in terms of elementary functions,
its integral with respect to $\epsilon$ cannot be expressed so
simply.

We also see from \LOCN\ that the local contributions from the higher
critical points to $\partial^{g-1} Z(\epsilon) / \partial
\epsilon^{g-1}$ are essentially independent of $g$ and $n$, apart from
a numerical prefactor and the usual exponential dependence on the
classical action $S_n$.

Finally, we see that the only dependence on $\epsilon$ in \LOCN\
besides the classical dependence on $S_n$ is through the prefactor
proportional to $\epsilon^{-1/2}$.  As we will see, this prefactor
reflects the geometric fact that the gauge group does not act freely
on the locus of non-flat Yang-Mills solutions.  To explain this fact,
we note that for any Yang-Mills solution the section $f$ of $\ad(P)$
satisfies $d_A f = 0$, so that $f \neq 0$ generates a $U(1)$ subgroup
of the gauge group $\CG(P)$ that fixes the corresponding point of
$\CA(P)$.

This geometric observation about higher critical points of Yang-Mills
theory is actually a general property of any higher critical points of
the abstract symplectic model with quadratic action $S = \ha (\mu,\mu)$.
Namely, the abstract Hamiltonian group $H$ can never act freely at a
higher critical point of $S$.

By definition, such a higher critical point $x_0$ in the symplectic
manifold $X$ is described by the conditions ${dS = (\mu, d\mu) = 0}$
with ${\mu \neq 0}$ at $x_0$.  To show that $H$ does not act freely at
$x_0$, we now exhibit a Hamiltonian vector field which vanishes at
$x_0$.  We first recall the quantity $V = \Omega^{-1} d\mu$ which we
introduced in Section 4.1.  Geometrically $V$, or ${V^m_a =
(\Omega^{-1})^{m n} \partial_n \mu_a}$ in components, is a linear map
from the Lie algebra $\Fh$ of $H$ to the space of Hamiltonian vector
fields on $X$.  From \INVMOMEQ\ and \SII, we see that $V$ trivially
satisfies ${(\mu, V) = \mu^a V^m_a = 0}$ at $x_0$.  But since
$\mu(x_0)$ is non-zero, we can consider on $X$ the Hamiltonian
vector field generated by $\mu(x_0)$ itself.  This vector field is given
by ${(\mu(x_0)\,, V) = \mu(x_0)^a \, V^m_a}$, and by our observations
above it vanishes at $x_0$.

\medskip\noindent{\it The Hodge Decomposition at a Higher Yang-Mills
Critical Point}\smallskip

In many respects, localization at an irreducible, flat Yang-Mills
solution is precisely opposite to localization at a maximally
reducible, non-flat Yang-Mills solution.  In both cases, the local
geometry in $\CA(P)$ near these critical points can be described as
the total space $N$ of an equivariant bundle with infinite-dimensional
fiber $F$ over a finite-dimensional moduli space $\CM_f$,
\eqn\PRESII{F \longrightarrow N \buildrel pr\over\longrightarrow
\CM_f\,.}
However, in the case of a flat connection the interesting
contributions to the integral over $N$ arise from the moduli space
$\CM_0$ itself, and the integral over the infinite-dimensional fiber
$F = T^* H$ contributes a trivial factor of $1$.  In contrast, for a
maximally reducible Yang-Mills solution, the integral over $\CM_f$
is essentially trivial, and the interesting contributions arise from
the fiber $F$.  Therefore, the most important aspect of our discussion
of non-abelian localization at higher critical points in Yang-Mills
theory is to identify the correct symplectic model for $F$, analogous
to the identification $F = T^* H$ used previously.

At this point, we can immediately see that a local symplectic model
for $F$ based on $T^* H$ does not correctly describe the geometry near
$\CM_f$ if $f \neq 0$.  First, as we have already observed, the gauge
group does not act freely at points on $\CM_f$, as we used in
identifying $F$ with $T^* H$ when we considered the geometry near
$\CM_0$.  Second, if $\phi$ and $\psi$ are any two sections of
$\ad(P)$ representing tangent vectors to $\CG(P)$, then the
computation in \ISCG\ shows that the symplectic form $\Omega$ at a
point on $\CM_f$ satisfies
\eqn\ISCGII{ \Omega(d_A \phi, d_A \psi) \,=\, - \int_\Sigma \Tr(d_A
\phi \^ d_A \psi) \,=\, \int_\Sigma \Tr(\phi \, d_A^2 \psi) \,=\,
\int_\Sigma \Tr\left(\phi \, \left[F_A, \psi\right]\right)\,.}
Here we just use the fact that $d_A^2 = F_A$ is nonzero, and we observe
that the last expression in \ISCGII\ need not vanish for suitable
$\phi$ and $\psi$.  Thus, the orbit of $\CG(P)$ through any point on
$\CM_f$ is no longer an isotropic submanifold of $\CA(P)$, as would be
required to model this orbit on $H$ embedded in the cotangent bundle
$T^* H$ with its canonical symplectic structure.

Now, the fact that $F$ is not modelled on $T^* H$ at a higher
critical point of Yang-Mills theory must be reflected in a breakdown
of the naive Hodge decomposition for the corresponding covariant
derivative $d_A$, so that
\eqn\NOHDG{ \Gamma(\Sigma, \Omega^1_\Sigma \otimes \ad(P)) \,\neq\, \CH_1
\oplus \Im(d_A^{}) \oplus \Im(d_A^\dagger)\,.}
Thus, a natural strategy to determine the correct symplectic model for
$F$ is just to consider how the Hodge decomposition is modified
when $A$ is a non-flat solution of the Yang-Mills equation.

In expanding around a flat connection, the tangent space to the
moduli space $\CM_0$ of flat connections is given by
$H_{d_A}^1(\Sigma, {\rm ad}(P))$.  For a non-flat Yang-Mills
connection, $d_A$ only squares to zero when restricted to
$\ad_0(P)$, the subspace of $\ad(P)$ that commutes with $f$.   On
the other hand, deformations of a Yang-Mills solution
automatically preserve $f$ up to gauge transformation, simply
because $f$ automatically has integral eigenvalues.  So tangent
vectors to ${\cal M}_f$ can always be represented by
$\ad_0(P)$-valued one-forms, which represent deformations of the
Yang-Mills solution by flat connections valued in the subgroup of
$G$ that commutes with $f$.  So the tangent space to ${\cal M}_f$
is ${\cal H}_1=H_{d_A}^1(\Sigma,\ad_0(P))$.  By standard Hodge
theory, this can also be defined as \eqn\jiko{{\cal
H}_1=H^1_{\bar\partial}(\Sigma,{\rm ad}_0(P)).} Similarly, the Lie
algebra of the unbroken subgroup $G_f$, which leaves fixed the
given Yang-Mills connection, is \eqn\piko{{\cal
H}_0=H^0_{d_A}(\Sigma,{\rm
ad}_0(P))=H^0_{\bar\partial}(\Sigma,\ad_0(P)).}

What we have said so far is a fairly direct generalization of the
usual statements in the flat case. However, if $A$ is a non-flat
Yang-Mills solution, then the usual Hodge theory needs to be
modified from the flat case in two essential ways. First, once we
get out of $\ad_0(P)$, the image of $d_A$ and the image of
$d_A^\dagger$ are no longer transverse. They have a nonzero,
finite-dimensional intersection that we will call ${\cal E}_0$:
\eqn\SETM{\Im(d_A^{})\cap \Im(d_A^\dagger)={\cal E}_0.}
 Second, the image of $d_A$ plus
the image of $d_A^\dagger$ plus the tangent space ${\cal H}_1$ to
the moduli space no longer generates
$T_P=\Gamma(\Sigma,\Omega^1_\Sigma\otimes\ad(P))$. The quotient
$T_P/({\rm Im}(d_A^{})\oplus \Im(d_A^\dagger))$ is another
finite-dimensional vector space ${\cal E}_1$.  The bundles ${\cal
E}_0$ and ${\cal E}_1$ both have natural complex structures. They
will turn out to be
 \eqn\AUXHDG{\eqalign{ \CE_0 \,&=\,
H^0_{\bar\partial}(\Sigma, \ad_+(P))\,,\cr \CE_1 \,&=\,
H^1_{\bar\partial}(\Sigma, \ad_+(P)) \oplus H^1_{\bar
\partial}(\Sigma,\ad_-(P)).\cr}}
We will often regard  these complex vector spaces as  real vector
spaces of twice the dimension.

Thus, the correct generalization of \NOHDG\ is informally
\eqn\HDGII{ \Gamma(\Sigma, \Omega^1_\Sigma \otimes \ad(P)) \,=\,
\CH_1 \oplus \Im(d_A^{}) \oplus \Im(d_A^\dagger) \ominus \CE_0
\oplus \CE_1\,.} As indicated by our use of ``$\ominus$'', the
expression in \HDGII\ is to be interpreted somewhat in the sense
of $K$-theory. Since $\Im(d_A)$ and $\Im(d_A^\dagger)$ have a
non-trivial intersection ${\cal E}_0$, this extra copy of ${\cal
E}_0$ must be removed to get the right description of $
\Gamma(\Sigma, \Omega^1_\Sigma \otimes \ad(P))$.

The definition of the Dolbeault cohomology groups in \AUXHDG\
requires a complex structure on $\Sigma$.  Abstractly, this
complex structure is induced from the duality operator $\*$ on
$\Sigma$. Because $\*^2 = -1$ when $\*$ acts on any one-form on
$\Sigma$, we can define the bundles $\Omega^{0,1}$ and
$\Omega^{1,0}$ of complex one-forms of either type on $\Sigma$ by
the respective $+i$ and $-i$ eigenspaces of $\*$.  This
decomposition by type determines the complex structure and hence
the Dolbeault $\bar\partial$ operator appearing in \AUXHDG.

However, for the following we find it useful to give an explicit
formula for the operator $\bar\partial$, acting on the bundle
$\ad_\BC(P)$, in terms of $\*$ and the covariant derivative $d_A$.  We
define the operators $\bar\partial{}^{(p)}$ acting on complex
$p$-forms on $\Sigma$ taking values in $\ad_\BC(P)$ by
\eqn\DBARP{\eqalign{
&\bar\partial{}^{(0)} \,=\, d_A \,-\, i \, \* d_A\,,\cr
&\bar\partial{}^{(1)} \,=\, -i \, d_A \,+\, d_A \*\,,\cr
&\bar\partial{}^{(2)} \,=\, 0\,.\cr}}
Again because $\*^2 = -1$ when acting on one-forms on $\Sigma$, one can
easily check the essential requirement that $\bar\partial{}^{(1)}
\circ \bar\partial{}^{(0)} = 0$.  From the expression for
$\bar\partial{}^{(1)}$ in \DBARP, we also see that $\bar\partial{}^{(1)}$
annihilates all one-forms in the $+i$ eigenspace of $\*$, which we have
identified with the space of one-forms of type $(0,1)$.

The subbundle $\ad_0(P)$ has a de Rham cohomology (with respect to
$d_A$) that we have already encountered.  The subbundles
$\ad_+(P)$ and $\ad_-(P)$ do not have de Rham cohomology, but they
have   Dolbeault cohomology groups  \eqn\DLBHC{
H^0_{\bar\partial}(\Sigma, \ad_+(P))\,,\quad
H^0_{\bar\partial}(\Sigma, \ad_-(P))\,,\quad
H^1_{\bar\partial}(\Sigma, \ad_+(P))\,,\quad H^1_{\bar
\partial}(\Sigma, \ad_-(P))}
that we should expect will enter somehow. Of these cohomology
groups, $H^0_{\bar\partial}(\Sigma, \ad_{-}(P))$ is zero by the
Kodaira vanishing theorem \AtiyahYM, which is the reason that
$\CE_0$ in \AUXHDG\ only involves $\ad_+(P)$.  (We also note
parenthetically that $H^1_{\bar\partial}(\Sigma,\ad_+(P))$ is
similarly zero for critical points associated to line bundles of
sufficiently high degree.)  So we are left to show that $\CE_0$
corresponds to the finite-dimensional intersection of
$\Im(d_A^{})$ and $\Im(d_A^\dagger)$ and  $\CE_1$ describes the
tangent vectors to ${\cal A}(P)$ not contained in
$\Im(d_A^{})\oplus \Im(d_A^\dagger)\oplus {\cal H}_1$.

We identify $\CE_0$ as described in \SETM\ immediately from our
formula for $\bar\partial{}^{(0)}$ in \DBARP. It is convenient to
write $\ad(P)=\ad_0(P)\oplus \ad_\perp(P)$, with $\ad_\perp(P)$
(whose complexification is $\ad_+(P)\oplus \ad_-(P)$) the
orthocomplement of $\ad_0(P)$.  By standard Hodge theory, if we
restrict to $\ad_0(P)$, $\Im(d_A^{})\cap \Im(d_A^\dagger)=0$.  So
the nontrivial intersection of $\Im(d_A^{})$ and
$\Im(d_A^\dagger)$ occurs in $\ad_\perp(P)$. Such an intersection
arises if there is $\phi\in \Gamma(\Sigma,\ad_\perp(P))$ and
$\Psi\in \Omega^2(\Sigma,\ad_\perp(P))$ such that
$d_A^{}\phi=d_A^\dagger\Psi$.  If so, let $\psi=\*\Psi$,
whereupon, since $d_A^\dagger = -\*d_A\*$ and $\*^2=-1$, we have
$d_A\phi=-\*d_A\psi$.  So if $\varphi=\phi+i\psi$, we have
$\bar\partial{}^{(0)}\varphi=(d_A-i\*d_A)\varphi=0$.  Hence $
\varphi\in H^0_{\bar\partial}(\Sigma,\ad_+(P)\oplus \ad_-(P))$.
But by Kodaira vanishing, $\ad_-(P)$ does not contribute, and
$\varphi\in H^0_{\bar\partial}(\Sigma,\ad_+(P))$.   This argument
can also be run backwards, to map
$H^0_{\bar\partial}(\Sigma,\ad_+(P))$     to $\CE_0$.  This
explains the claim that ${\cal
E}_0=H^0_{\bar\partial}(\Sigma,\ad_+(P))$.

Finally, we can identify $\CE_1$, the subspace of
$\Gamma(\Sigma,\ad_\perp(P))$ that is orthogonal to the image of
$d_A$ and the image of $d_A^\dagger$. We begin with the
tautological observation that the orthocomplement of the image of
$d_A^{}$ is precisely the  kernel of $d_A^\dagger$, and similarly
the orthocomplement of the image of $d_A^\dagger$ is precisely the
 the kernel of $d_A^{}$. Thus, $\CE_1$, the orthocomplement to the
image of $d_A^{}$ and $d_A^\dagger$, consists of forms annihilated by both
$d_A^{\dagger}$ and $d_A$.\foot{Notice that although $d_A^2$ and
$d_A^\dagger{}^2$ are nonzero in general, they annihilate
$\Omega^1(\Sigma,\ad_\perp(P))$ for dimensional reasons, as a
result of which $d_A$ and $d_A^\dagger $ can have a kernel!} Given
the formula $\bar\partial{}^{(1)}=-id_A+d_A\star$, it follows that
$\bar\partial{}^{(1)}$ annihilates $\CE_1$.  Moreover,
$\bar\partial^\dagger{}^{(1)}$, the $\bar\partial^\dagger$
operator acting on one-forms, is
$\bar\partial^\dagger{}^{(1)}=d_A^\dagger - i d_A^\dagger\*$, and so
annihilates $\CE_1$.  This reasoning can also be read backwards to
show that a form annihilated by $\bar\partial{}^{(1)}$ and its adjoint
$\bar\partial^\dagger{}^{(1)}$ is annihilated by $d_A$ and
$d_A^\dagger$ and hence is contained in $\CE_1$. By Hodge theory,
the joint kernel of $\bar\partial$ and  $\bar\partial^\dagger$ is
the same as the cohomology of $\bar\partial$. So finally,
$\CE_1=H^1_{\bar\partial}(\Sigma,\ad_+(P)\oplus \ad_-(P))$, as we
have claimed.

\medskip\noindent{\it A New Symplectic Model For Localization}\smallskip

The Hodge decomposition \HDGII\ implicitly describes the local
symplectic model to use at a higher Yang-Mills critical point.
We now present this model and compute via localization the canonical
symplectic integral in this case.

Abstractly, our local model for $F$ now differs in two ways from the
model based on the cotangent bundle $T^* H$.  First, $H$ no longer
acts freely at the given critical point.  We let $H_0 \subset H$ denote
the subgroup of $H$ which fixes the critical point.  Thus, the orbit
of $H$ through the critical point can be identified with $H / H_0$.
In the case of Yang-Mills theory, the vector space $\CH_0$ of harmonic
sections of $\ad_0(P)$ is abstractly identified with the Lie algebra
$\Fh_0$ of $H_0$.

Second, because of the appearance of $\CE_0$ and $\CE_1$ in the Hodge
decomposition in \HDGII, the naive model based on the cotangent bundle
of the orbit $H / H_0$ must be modified in the following way.  If we
simply wanted to discuss the cotangent bundle of the orbit $H / H_0$,
then we could again pass to a basis of right-invariant forms and use the
invariant metric $(\cdot,\cdot)$ on $\Fh$ to present $T^* (H / H_0)$ as
a homogeneous bundle
\eqn\COTFB{ T^* (H / H_0) \,\cong\, H \times_{H_0} \left(\Fh \ominus
\Fh_0\right)\,.}
Here $\Fh \ominus \Fh_0$ denotes the orthogonal complement to $\Fh_0$
in $\Fh$, and ``$\times_{H_0}$'' indicates that we identify points
$(g,\gamma)$ in the product $H \times (\Fh \ominus \Fh_0)$ under the
following action of $H_0$,
\eqn\KACT{ h \cdot (g,\gamma) \,=\, \left(h g\,, h \gamma
h^{-1}\right)\,,\qquad h \in H_0\,.}

To incorporate the appearance of $\CE_0$ and $\CE_1$ in \HDGII, we
now introduce abstractly a subspace $E_0$ of the Lie algebra $\Fh$
which has a trivial intersection with $\Fh_0$ and is preserved
under the adjoint action of $H_0$, so that infinitesimally
$[\Fh_0, E_0] \subseteq E_0$. This condition certainly holds in
Yang-Mills theory for the vector space $\CE_0$. Similarly, we
introduce another vector space $E_1$ on which $H_0$ acts in some
representation.    We assume that, like the subspace $E_0$, the
representation $E_1$ admits a metric invariant under the action of
$H_0$.

We now describe our model for $F$ as a homogeneous bundle over the
orbit $H/H_0$ which generalizes \COTFB.  To describe this bundle, we
need only specify the fiber of $F$ over the identity coset of $H/H_0$
and the action of $H_0$ on the fiber.  Thus, as in the modified Hodge
decomposition \HDGII, we subtract $E_0$ from the cotangent fiber of
$H/H_0$ in \COTFB, meaning that we take the orthogonal complement to
$E_0$ in $\Fh \ominus \Fh_0$, and we also add $E_1$ to the cotangent
fiber of $H/H_0$.  So the resulting fiber of $F$ over the
identity is given by ${\Fh \ominus \Fh_0 \ominus E_0 \oplus E_1}$.  By our
assumptions on $E_0$ and $E_1$, this vector space transforms as a
representation of $H_0$.

In summary, the local model for $F$ is given abstractly by the following
homogeneous bundle over $H/H_0$,
\eqn\COTFBII{ F \,=\, H \times_{H_0} \left(\Fh \ominus \Fh_0 \ominus E_0
\oplus E_1\right)\,.}
We now use $\gamma$ to denote an element of the orthogonal complement
$\Fh^\perp$ to $\Fh_0 \oplus E_0$ in $\Fh$,
\eqn\FHPERP{ \gamma \,\in\, \Fh^\perp \,=\, \Fh \ominus \Fh_0 \ominus
E_0\,,}
and we use $v$ to denote a vector in $E_1$.  So in \COTFBII, we
identify points $(g,\gamma,v)$ in the product ${H \times (\Fh^\perp
\oplus E_1)}$ under the following action of $H_0$,
\eqn\KACTII{ h \cdot (g,\gamma,v) \,=\, \left(h g\,, h \gamma
h^{-1}\,, h \cdot v\right)\,,\qquad h \in H_0\,.}

To specify completely our local model, we must also discuss the
symplectic structure and the Hamiltonian $H$-action on $F$.  We will
be somewhat brief, since we are just applying standard techniques to
construct symplectic bundles, as explained for instance in Ch. 35--41
of \GuilleminS.

In order to construct a symplectic structure on $F$, we must make some
additional assumptions about the representations $E_0$ and $E_1$ of
$H_0$.  We first introduce an element $\gamma_0$ of $\Fh_0$.  Abstractly,
$\gamma_0$ corresponds to the value of the moment map
at the given critical point, and in the Yang-Mills context $\gamma_0$
is identified with $f$.

As in Yang-Mills theory, we assume that the hermitian operator $\Lambda$,
\eqn\HLAMII{ \Lambda \,=\, i \left[\gamma_0,\,\cdot\,\right]\,,}
annihilates $\Fh_0$ and acts on the vector spaces $E_0$ and $E_1$
with strictly non-zero eigenvalues.  The first assumption implies
that $\gamma_0$ is central in $\Fh_0$ and is invariant under the adjoint
action of $H_0$,
\eqn\INVGAM{ H_0 \; \gamma_0 \; H_0^{-1} \,=\, \gamma_0\,.}

Because the action of $\gamma_0$ preserves the invariant metrics
on $E_0$ and $E_1$, the action of $\gamma_0$ is represented by a
real, anti-symmetric matrix.  By our second assumption above, this
matrix is non-degenerate.  Consequently, the decomposition of
$E_0$, and similarly $E_1$, into the positive and negative
eigenspaces of $\Lambda$ defines a complex structure which is
invariant under the action of $H_0$ and for which the invariant
metric $(\cdot,\cdot)$ is hermitian.

Having introduced $\gamma_0$, we now describe the symplectic structure on
$F$.  As in Section 4.2, we let $\theta$ be the canonical
right-invariant one-form on $H$ taking values in $\Fh$,
\eqn\CRNTH{ \theta \,=\, dg g^{-1}\,.}
We recall that in the case of the cotangent bundle $T^* H$ or $T^*
(H/H_0)$, we can immediately describe the sympletic structure with
the manifestly closed and non-degenerate two-form $\Omega_0$,
\eqn\SYMHII{ \Omega_0 \,=\, d(\gamma,\theta)\,,}
which reduces on the orbit $H/H_0$, where $\gamma = 0$, to the
canonical form $(d\gamma,\theta)$.

Similarly, when we consider the homogeneous bundle $F$ in \COTFBII,
$\Omega_0$ in \SYMHII\ still descends to a closed two-form on $F$.
However, because $\gamma$ now takes values in $\Fh^\perp$ as in
\FHPERP, the restriction of $\Omega_0$ to the orbit $H/H_0$ is
degenerate on the subspace $E_0$ of the tangent space to the orbit.
Thus, if we ignore the vector space $E_1$ for the moment, then to
construct a symplectic structure on the homogeneous bundle with fiber
$\Fh^\perp$ over $H/H_0$ we must supplement the canonical two-form
$\Omega_0$ with an additional two-form which is non-degenerate on $E_0$.

What other two-form should we consider?  For motivation, while
keeping $E_1=0$, let us consider the opposite case from the
cotangent bundle. As the cotangent bundle has $E_0=0$,  the other
extreme is for $E_0$ to be all of $\Fh\ominus \Fh_0$, so that
$\Fh\ominus\Fh_0\ominus E_0=0$ and $F=H/H_0$.  Since we have
postulated that $\gamma_0$ acts non-degenerately on $E_0$, while
commuting with $\Fh_0$, it follows in this case that $\Fh_0$ is
precisely the subalgebra of $\Fh$ that commutes with $\gamma_0$.
Therefore, $H/H_0$ is precisely the orbit of $\gamma_0$ in the Lie
algebra of $H$. Such an orbit is called a coadjoint orbit (for
compact Lie groups the difference between the adjoint
representation and its dual is not important here) and has a
natural symplectic structure, namely \eqn\SYMCOAD{ \Omega_1 \,=\,
d(\gamma_0,\theta) \,=\, \ha \left(\theta, \left[\gamma_0,
\theta\right]\right)\,,} where we observe that $d \theta = \theta
\^ \theta = \ha [\theta,\theta]$ in deducing the second equality
of \SYMCOAD. Because $\gamma_0$ is invariant under the adjoint
action of $H_0$ in \INVGAM, $\Omega_1$ is also invariant under the
action of $H_0$ in \KACTII\ and descends to a manifestly closed
and nondegenerate  two-form on $H/H_0$.  Indeed, coadjoint orbits
are the basic examples of homogeneous symplectic manifolds.

In fact, we have already seen the coadjoint form $\Omega_1$ arise in
the context of Yang-Mills theory.  We recall from \ISCGII\ that the
restriction of the Yang-Mills symplectic form $\Omega$ on the affine
space $\CA(P)$ to the orbit of $\CG(P)$ through a non-flat Yang-Mills
solution is given by
\eqn\OMVWHKIII{ \Omega(d_A \phi, d_A \psi) \,=\, \int_\Sigma
\Tr\left(\phi \, \left[F_A, \psi\right]\right)\,.}
Upon identifying the abstract element $\gamma_0$ with $f$, we see that
$\Omega_1$ in \SYMCOAD\ precisely represents \OMVWHKIII.

The general case, still with $E_1=0$, is a mixture of the
cotangent bundle and the coadjoint orbit.  We thus naturally add
the two two-forms that arise in those two cases and consider the
sum \eqn\SUMOM{ \Omega_0 + \Omega_1 \,=\, d(\gamma +
\gamma_0,\theta)\,,} which restricts on the orbit $H/H_0$, where
$\gamma = 0$, to the simple expression \eqn\SUMOMII{
\left(\Omega_0 + \Omega_1\right)|_{H/H_0} \,=\, (d\gamma, \theta)
+ \ha \left(\theta, \left[\gamma_0, \theta\right]\right)\,.} We
see immediately from \SUMOMII\ that $\Omega_0 + \Omega_1$ defines
a symplectic form on a neighborhood of $H/H_0$ in the homogeneous
bundle with fiber $\Fh^\perp$.  For instance, since the expression
in \SUMOM\ is manifestly invariant under the right action of $H$
on $H/H_0$, we need only consider \SUMOMII\ as restricted to the
tangent space $(\Fh \ominus \Fh_0) \oplus \Fh^\perp$ of the bundle
at the identity coset on $H/H_0$. The top power of \SUMOMII\ on
this tangent space is then manifestly non-zero, since all tangent
vectors in $\Fh^\perp$ are paired by $\Omega_0$ and the remaining
tangent vectors to the orbit in $E_0$ are paired by $\Omega_1$.

Finally, we need to include $E_1$. By assumption, $E_1$ has a
metric and a complex structure invariant under the action of
$H_0$, so that $E_1$ has an associated symplectic form $\wt\Omega$
invariant under $H_0$.

In order to pass from the symplectic form $\wt\Omega$ on $E_1$ to
a closed two-form on $F$ which is non-degenerate on the $E_1$
fiber at the identity coset of $H/H_0$ and compatible with the
bundle structure of $F$, we must further suppose that $H_0$ acts
on $E_1$ in a Hamiltonian fashion with moment map $\wt\mu$. We can
always choose $\wt\mu$ to vanish at the origin of $E_1$. We also
observe that since the action of $H_0$ on $E_1$ is linear, of the
form $\delta v = \psi \cdot v$ for $v$ in $E_1$ and $\psi$ in
$\Fh_0$, the moment map $\wt\mu$ depends quadratically on $v$ and
satisfies $d\wt\mu = 0$ at the origin of $E_1$.

With these observations in hand, we consider the two-form
$\Omega_2$ defined below, \eqn\SYMV{ \Omega_2 \,=\, \wt\Omega +
d\langle\wt\mu\,,\theta\rangle\,.} This two-form is manifestly
closed, as $\tilde\Omega$ is closed.  It also is clearly invariant
under the action of $H_0$ in \KACTII.

Finally, to explain the appearance of the second term in \SYMV, we
note that the action of $\Fh_0$ on $F$ can be described as follows.
For $\psi\in \Fh_0$, the corresponding vector field $V(\psi)$ on $F$
acts by
\eqn\VECTH{ \delta g \,=\, \psi g\,,\qquad \delta \gamma \,=\,
\left[\psi,\gamma\right]\,,\qquad \delta v \,=\, \psi \cdot v.}
In order that $\Omega_2$ descend under the quotient by $H_0$ which
defines the bundle, we require that $\Omega_2$ be invariant under
$H_0$ (as we have already seen) and that $\Omega_2$ be annihilated by
contraction with $V(\psi)$.  By the defining moment map relation, the
contraction of $V(\psi)$ with $\wt\Omega$ is ${\iota_{V(\psi)}
\wt\Omega = d\langle\wt\mu, \psi\rangle}$.  As for the second term in
\SYMV, the one-form $\langle\wt\mu, \theta\rangle$ is invariant under
the action of $H_0$ and hence annihilated by the Lie derivative
${\lie_{V(\psi)} = \{d, \iota_{V(\psi)}\}}$.  Thus we see that
${\iota_{V(\psi)} \, d\langle\wt\mu,\theta\rangle =
-d \, \iota_{V(\psi)}\langle\wt\mu,\theta\rangle =
-d\langle\wt\mu,\psi\rangle}$, which cancels the contraction of
$\iota_{V(\psi)}$ with $\wt\Omega$.

Because $\wt\mu = d\wt\mu = 0$ at the origin of $E_1$, the
restriction of $\Omega_2$ to the orbit $H/H_0$ in $F$ is simply the
symplectic form $\wt\Omega$ on $E_1$.  Thus, the sum of $\Omega_0$,
$\Omega_1$, and $\Omega_2$ defines a symplectic form $\Omega$ on a
neighborhood of the orbit $H/H_0$ in $F$,
\eqn\OMVW{\eqalign{
\Omega \,&=\, \Omega_0 + \Omega_1 + \Omega_2\,,\cr
&=\, d\left(\gamma + \gamma_0\,,\theta\right) \,+\,
d\langle\wt\mu\,,\theta\rangle + \wt\Omega\,.\cr}}

Having placed a symplectic structure on $F$, we are left to consider
the action of $H$ on $F$.  As in the model based on the cotangent
bundle, we assume that $H$ acts from the right on the orbit $H/H_0$ in
$F$, so that
\eqn\GACT{ h \cdot (g,\gamma,v) \,=\, (g h^{-1}, \gamma, v)\,,\qquad h \in
H\,.}
The corresponding element $\phi$ in $\Fh$ generates the vector field
\eqn\VPACT{ \delta g \,=\, - g \phi\,,\qquad \delta \gamma
\,=\,0\,,\qquad \delta v\,=\, 0\,.}
Since the one-form $\theta$ appearing in $\Omega$ is right-invariant, the
symplectic form $\Omega$ is manifestly invariant under $H$.

Finally, using \OMVW\ and \VPACT, one can easily check that the action
of $H$ on $F$ is Hamiltonian with moment map $\mu$ given by
\eqn\MOMTWGV{ \langle\mu,\phi\rangle \,=\, \left(\gamma + \gamma_0, g \phi
g^{-1}\right) + \left\langle\wt\mu,g \phi g^{-1}\right\rangle\,.}  In
particular, we see that the value of $\mu$ at the point corresponding
to the identity coset on the orbit $H/H_0$ is just the dual of
$\gamma_0$ in $\Fh^*$, as we have claimed.

\medskip\noindent{\it Computing the Symplectic Integral over
$F$}\smallskip

For our applications to both Yang-Mills theory and Chern-Simons
theory, we now compute the canonical symplectic integral over $F$,
\eqn\ZVWHK{ Z(\epsilon) \,=\, {1 \over {\Vol(H)}} \, \int_{\Fh \times F}
\left[{{d\phi} \over {2\pi}}\right] \exp{\left[\Omega - i \,
\langle\mu,\phi\rangle - {\epsilon \over 2} (\phi,\phi) + t \,
D\lambda \right]}\,.}
In this expression, $\lambda$ is the canonical one-form defined as in
\LCLM\ by ${\lambda = J \, dS}$, where ${S = \ha (\mu,\mu)}$ and $J$ is a
compatible almost-complex structure, and $t$ is a non-zero parameter.

Before we delve into computations, let us make a few remarks about how
this symplectic integral over $F$ is to be interpreted.  We start by
considering the canonical symplectic integral \ZEIII\ of the same form
as \ZVWHK\ but defined as an integral over a compact symplectic
manifold $X$ instead of $F$.  Because $X$ is compact, this integral is
convergent for arbitrary $t$, including $t = 0$, and does not depend
on either $t$ or $\lambda$.

By our general analysis of Section 4.1, in the limit $t \rightarrow
\infty$ and for $\lambda$ of the canonical form, the integral over $X$
localizes on the critical set of $S$ and reduces to a finite sum of
contributions from the components of this set.  Although the global
integral over $X$ is perfectly defined, independent of $t$ and
$\lambda$, the contributions from the critical locus of $S$ are only
defined via localization, with $t \neq 0$ and $\lambda$ of the
canonical form.  For instance, at a higher critical point of $S$, for
which we model the normal symplectic geometry on $F$, the unstable
modes of $S$ make the integral over the non-compact fibers of $F$
ill-defined when $t=0$.  Thus, the symplectic integral $Z(\epsilon)$
over $F$ as in \ZVWHK\ represents a {\it definition} of the local
contribution from an unstable critical point of $S$ in $X$.

Although we use the canonical one-form ${\lambda = J \, dS}$ to define via
localization the integral over $F$ in \ZVWHK, we are free to
compute $Z(\epsilon)$ using any other invariant form $\lambda'$ which
is homotopic to $\lambda$ on $F$.  In particular, though $\lambda$ is
defined globally on $X$, $\lambda'$ need only be defined locally on
$F$.

The reason that we might want to compute $Z(\epsilon)$ using some
alternative form $\lambda'$ instead of the canonical one-form $\lambda$ is
just that generically the integral over $F$ defined by $\lambda$ is not
Gaussian even in the limit $t \rightarrow \infty$ and cannot
be easily evaluated in closed form.  See the appendix of \WittenXU\
for a simple example of this behavior.  However, by making a
convenient choice for $\lambda'$, we can greatly simplify our
computation and essentially reduce it to the evaluation of Gaussian
integrals.

So in order to compute $Z(\epsilon)$ in \ZVWHK, we first make a
convenient choice for $\lambda'$.  Since the motivation for our choice is
fundamentally to simplify the evaluation of $Z(\epsilon)$, we next
evaluate \ZVWHK\ using $\lambda'$ in place of $\lambda$.  Finally, in
Appendix A, we perform the analysis required to show that
$Z(\epsilon)$ as defined using the canonical one-form $\lambda$ can be
equivalently evaluated using $\lambda'$.

To describe our choice for $\lambda'$, we introduce a projection
$\Pi_{\Fh_0}$ onto $\Fh_0$ and a projection $\Pi_{E_0}$ onto $E_0$ in
the Lie algebra $\Fh$ of $H$.  We define these projections using the
invariant metric on $\Fh$, so that they are invariant under the
adjoint action of $H_0$ on $\Fh$.  We then introduce the quantities
\eqn\QTS{\eqalign{
&\theta_{\Fh_0} \,=\, \Pi_{\Fh_0}(\theta)\,,\qquad (g \phi
g^{-1})_{\Fh_0} \,=\, \Pi_{\Fh_0}(g \phi g^{-1})\,,\cr
&\theta_{E_0} \,=\, \Pi_{E_0}(\theta)\,,\qquad (g \phi g^{-1})_{E_0}
\,=\, \Pi_{E_0}(g \phi g^{-1})\,.\cr}}

We now define $\lambda'$ as \eqn\LOCLAM{ \lambda' \,=\,
\left(\gamma\,,\theta\right) \,-\, i \left(\theta_{E_0}, g \phi
g^{-1}\right) + i \left(\left(g \phi g^{-1}\right)_{\Fh_0} \cdot
v\,, dv\right) - i \left(\left(g \phi g^{-1}\right)_{\Fh_0}\cdot
v\,, \theta_{\Fh_0} \cdot v\right)\,.} The first term in \LOCLAM\
has the same form as the canonical one-form which we used for
localization on $T^* H$.  However, we recall that now $\gamma$
takes values not in $\Fh$ but in ${\Fh^\perp = \Fh \ominus \Fh_0
\ominus E_0}$.  As before, this first term has degree one under
the grading on equivariant cohomology.  The other three terms are
associated to the new vector spaces $E_0$ and $E_1$ that appear at
a higher critical point.  Since $\phi$ carries charge $+2$ under
the grading on equivariant cohomology, these terms are all of
degree three.

The most basic requirement that $\lambda'$ must satisfy is that it
descends to an invariant form on $F$ under the quotient by $H_0$
which defines the homogeneous bundle.  So we first observe that
$\lambda'$ is manifestly invariant under the action of $H_0$ in
\KACTII. Furthermore, if $V(\psi)$ denotes the vector field on the
product ${H \times (\Fh^\perp \oplus E_1)}$ generated by $\psi$ in
$\Fh_0$ as in \VECTH, then the first two terms in $\lambda'$ are
trivially annihilated upon contraction with $V(\psi)$ since both
$\gamma$ and $\theta_{E_0}$ take values in the orthocomplement to
$\Fh_0$.  Because of the identity \eqn\IOTLAM{ \iota_{V(\psi)} \,
dv \,=\, \psi \cdot v\,=\, \left(\iota_{V(\psi)}
\theta_{\Fh_0}\right) \cdot v\,,} the last two terms in $\lambda'$
are also annihilated upon contraction with $V(\psi)$.  So
$\lambda'$ descends to a well-defined form on $F$.

Finally, to check that $\lambda'$ is invariant under the action of $H$
on $F$ in \GACT, we simply note that $\phi$ transforms under the
adjoint action of $H$ so that the quantity $g \phi g^{-1}$ is
invariant.  Since $\theta$ is also invariant under the action of $H$,
$\lambda'$ is manifestly invariant.

To motivate our definition \LOCLAM, we now use $\lambda'$ to
compute the symplectic integral over $F$.  We first compute $D
\lambda'$.  As we saw when we considered localization on $T^* H$,
the final expression for $D\lambda'$ will only involve $\phi$ in
the invariant combination $g \phi g^{-1}$.  Thus, even before
presenting our formula for $D\lambda'$, we make the change of
variables from $\phi$ to $g \phi g^{-1}$ in the symplectic
integral in order to simplify slightly our result.  If we recall
that $D = d + i \, \iota_{V(\phi)}$ and we use the formula in
\VPACT\ for $V(\phi)$, we find by a straightforward computation
that \eqn\DLOCLAM{\eqalign{ D \lambda' \,&=\, \left(d\gamma,
\theta\right) \,-\, i\, \left(\gamma, \phi\right) \,-\, i
\left(\theta_{E_0},\left[\phi_{\Fh_0}, \theta_{E_0}\right]\right)
\,-\, \left(\phi_{E_0}, \phi_{E_0}\right) \,+\,\cr &+\, i
\left(\phi_{\Fh_0} \cdot dv, dv\right) \,-\, \left(\phi_{\Fh_0}
\cdot v, \phi_{\Fh_0} \cdot v\right) \,+\, {{\cal X}}\,.\cr}}

Here ${\cal X}$ consists of extra terms in $D\lambda'$ that will
not actually contribute to the symplectic integral in the limit
$t\to\infty$.  Explicitly, \eqn\DLOCLAMX{\eqalign{ {\cal X} \,&=\,
\left(\gamma, \ha[\theta,\theta]\right) \,-\,
i\left(\ha\left[\theta^\perp, \theta^\perp\right],
\phi_{E_0}\right) \,-\, i\left(\left[\theta^\perp,
\theta_{E_0}\right], \phi^\perp\right) \,-\,
i\left(\ha\left[\theta_{E_0}, \theta_{E_0}\right],
\phi^\perp\right)\,-\,\cr &-\, i \left(\phi_{\Fh_0}\cdot v,
\ha\left[\theta, \theta\right]_{\Fh_0}\cdot v\right)\quad \mod \,
\theta_{\Fh_0}\,.\cr}} (Terms involving $\theta_{\Fh_0}$ in
$D\lambda'$, some of which are omitted here, actually cancel since
$D\lambda'$ is a pullback from $F$.)  We use the fact that
${d\theta = \ha [\theta, \theta]}$ to simplify somewhat the form
of ${\cal X}$, and we use the natural notation $\theta^\perp$ and
$\phi^\perp$ to denote the projections of $\theta$ and $\phi$ onto
$\Fh^\perp$.

In \DLOCLAM, the first two terms arise from the action of $D$ on
the first term in $\lambda'$, the next two arise from the action
of $D$ on the second term in $\lambda'$, and the final two terms
arise from the action of $D$ on the last two terms in $\lambda'$.
We remark that our choice of the $i$'s that appear in the
definition \LOCLAM\ of $\lambda'$ was made to ensure that the
quadratic terms in \DLOCLAM\ involving $\phi_{E_0}$ and
$\phi_{\Fh_0} \cdot v$ are both negative-definite.

We now consider the canonical symplectic integral in \ZVWHK\ with
$\lambda'$ in place of $\lambda$ and in the limit $t \rightarrow
\infty$. This symplectic integral is an integral over the product
$\Fh \times F$. We can perform this integral over $\Fh \times F$
in two steps. First, we hold the projection $\phi_{\Fh_0}$ of the
variable $\phi$ in $\Fh_0 \subset \Fh$ fixed, and we perform the
integral over the remaining variables in ${\wt F = (\Fh \ominus
\Fh_0) \times F}$.  This integral produces a measure on $\Fh_0$,
which we then use to perform the remaining integral over $\Fh_0$.
The utility of this way of performing the symplectic integral is
that, with our ansatz for $\lambda'$, we will see that the first
integral over $(\Fh \ominus \Fh_0) \times F$ can be performed
directly as a Gaussian integral in the limit $t \rightarrow
\infty$ and under the assumption that $\phi_{\Fh_0}$ acts in a
non-degenerate fashion on $E_0$ and $E_1$.

To prove this fact, we first consider the symplectic integral over
${\wt F = (\Fh \ominus \Fh_0) \times F}$ which arises if ${\cal
X}$ is omitted from $D\lambda'$.  So we consider the integral
\eqn\ZVWHKII{\eqalign{ I(\phi_{\Fh_0}) \,=\, {1 \over {\Vol(H)}}
\, \int_{\wt F} \, \left[{{d\phi} \over {2\pi}}\right] \,
&\exp{\bigl[t \left(d\gamma, \theta\right) - i t \left(\gamma,
\phi\right) - i t \left(\theta_{E_0}, \left[\phi_{\Fh_0},
\theta_{E_0}\right]\right) - t \left(\phi_{E_0},
\phi_{E_0}\right)\bigr]} \,\times\,\cr \times \, &\exp{\bigl[i t
\left(\phi_{\Fh_0} \cdot dv, dv\right) - t \left(\phi_{\Fh_0}
\cdot v, \phi_{\Fh_0} \cdot v\right)\bigr]}\,.\cr}} For fixed
$\phi_{\Fh_0}$ acting non-degenerately on $E_0$ and $E_1$, this
integral \ZVWHKII\ is a Gaussian integral, which we now evaluate.
In performing this integral, we recall that the vector spaces
$E_0$ and $E_1$ carry a complex structure, invariant under the
action of $\phi_{\Fh_0}$, for which the metric $(\cdot,\cdot)$ is
hermitian.

Assuming $E_1$ is suitably oriented, the Gaussian integral over $v$
in $E_1$ first produces a factor
\eqn\GVINT{ \det\left({{\phi_{\Fh_0}} \over {2
\pi}}\Big|_{E_1}\right)^{-1}\,.}
This expression does not depend on $t$, due to a cancellation between the
factors of $t$ that arise from the Gaussian integral over $v$ and the
factors of $t$ that appear in the measure on $E_1$.

The remainder of the integration is similar, but is actually
perhaps more easily explained if we adopt a physicist's notation
rather than the mathematical notation in which \ZVWHKII\ has been
written.  In mathematical notation, $\theta=dg \,g^{-1}$ is a
one-form; we are supposed to expand the exponential to produce a
top-form which is then integrated.  In physics notation, $\theta$
is understood as a fermionic variable, and \ZVWHKII\ must be
reexpressed to contain an extra factor $dg\,d\theta$ in the
measure.

In the physics notation, we now perform the Gaussian integrals
over $\phi_{E_0}$ and $\theta_{E_0}$.  The powers of $t$ cancel,
just as in the integration over $v$ (which in physics notation
would have been an integral over $v$ and an independent fermionic
variable $\hat v=dv$), and we are left with a determinantal factor
\eqn\kipo{\det\left({{\phi_{\Fh_0}} \over {2 \pi}}\Big|_{E_0}\right),}
which now appears in the numerator as it comes from a fermionic
integration.  The factors of $2\pi$ come from the Gaussian
integral together with the measure $[d\phi/2\pi]$.

Similarly, in physics notation, $\gamma$ and $\hat\gamma=d\gamma$
are treated as independent bosonic and fermion variables and the
measure contains an extra factor $d\gamma\,d\hat\gamma$. Likewise,
we integrate separately over $H/H_0$ and over fermionic variables
$\theta$.  In fact, we have already performed the integration over
$\theta_{E_0}$, so we are only left with the component of $\theta
$ in $\Fh^\perp$.  The integral over $\gamma$ gives a delta
function setting to zero the projection of $\phi$ to $\Fh^\perp$.
The integral over $\hat\gamma$ gives a delta function setting to
zero the component of $\theta$ in $\Fh^\perp$, and canceling the
power of $t$ generated by the $\gamma$ integral.  Finally, the
integration over $H/H_0$ produces a factor of ${\rm Vol}(H)/{\rm
Vol}(H_0)$.

So finally, simplifying the notation by setting
$\psi=\phi_{\Fh_0}$, the result arising from the Gaussian
integration is
 \eqn\ZVWHKIII{ I(\psi) \,=\, {1 \over {\Vol(H_0)}}\,
\det\left({{\psi} \over {2 \pi}}\Big|_{E_0}\right) \,
\det\left({{\psi} \over {2 \pi}}\Big|_{E_1}\right)^{-1}\,,\qquad
\psi \in \Fh_0\,.}   Of course, a conventional mathematical
exposition of the calculation would arrive at the same result
after grouping the factors a little differently.

The result \ZVWHKIII\ for the integral \ZVWHKII\ is independent of
$t$.  We now observe that the terms in ${\cal X}$ which we omitted
from $D\lambda'$ when computing \ZVWHKIII\ are all of at least
third order in the integration variables on ${\wt F = (\Fh \ominus
\Fh_0) \times F}$ (which do {\it not} include the constant
$\phi_{\Fh_0}$).  Thus, upon rescaling all the integration
variables by $t^{-\ha}$ so that the quadratic terms in \ZVWHKII\
become independent of $t$, we see that any contributions from
terms in ${\cal X}$ to the symplectic integral fall off at least
as fast as $t^{-\ha}$ for large $t$.  Thus, our Gaussian
evaluation of the symplectic integral over $\wt F$ is exact as $t
\to \infty$.

So we are left to consider the remaining integral over $\Fh_0$, which
is now given formally by
\eqn\ZVWHKIV{ Z'(\epsilon) \,=\, {1 \over {\Vol(H_0)}} \, \int_{\Fh_0}
\left[{{d\psi} \over {2\pi}}\right] \, \det\left({{\psi} \over {2
\pi}}\Big|_{E_0}\right) \, \det\left({{\psi} \over {2
\pi}}\Big|_{E_1}\right)^{-1}\, \, \exp{\left[-i \left(\gamma_0,
\psi\right) - {\epsilon \over 2} \left(\psi, \psi\right)\right]}\,.}
In obtaining this expression, we recall from \MOMTWGV\ that the
value of the moment map $\mu$ at the identity coset on the orbit $H/H_0$
is $\gamma_0$.  Also, we denote this quantity as $Z'(\epsilon)$,
instead of $Z(\epsilon)$, to emphasize that we compute it with
$\lambda'$ instead of the canonical form $\lambda$ that defines the
local contributions to $Z(\epsilon)$.

Now, this formal integral over $\Fh_0$ in \ZVWHKIV\ might or might not
actually be defined.  Due to the exponential factor in the integrand
of \ZVWHKIV, the integral is certainly convergent at large $\psi$.
However, on the locus in $\Fh_0$ where the determinant of $\psi$
acting on $E_1$ vanishes (for instance at the origin of $\Fh_0$), the
measure $I(\psi)$ in \ZVWHKIII\ might be singular if there is no
compensating zero from the determinant of $\psi$ acting on
$E_0$.  If $I(\psi)$ is singular, then the integral in \ZVWHKIV\ could
fail to be convergent at the singularity.  Since $Z(\epsilon)$ as
defined using the canonical one-form $\lambda$ is always finite, our
computation using $\lambda'$ cannot generally be valid.

On the other hand, because $E_0$ and $E_1$ are both finite-dimensional
vector spaces, with
\eqn\DIME{ \dim_\BC E_0 \,=\, d_0\,, \qquad \dim_\BC E_1 \,=\, d_1\,,}
the determinants appearing in $I(\psi)$ in \ZVWHKIII\ are just invariant
polynomials, homogeneous of degrees $d_0$ and $d_1$, of $\psi$ in
$\Fh_0$.  For our application to $SU(2)$ Yang-Mills theory, for which
$H_0 = U(1)$, we need only consider the simplest case that $\Fh_0 =
\BR$ is one-dimensional.  In this case, the invariant polynomials are
just monomials
\eqn\DETPHI{ \det\left({{\psi} \over {2 \pi}}\Big|_{E_0}\right)\,=\,
c_0 \, \psi^{d_0}\,,\qquad \det\left({{\psi} \over {2
\pi}}\Big|_{E_1}\right) \,=\, c_1 \, \psi^{d_1}\,,}
for some constants $c_0$ and $c_1$.

Assuming \DETPHI, we see that \ZVWHKIV\ becomes
\eqn\ZVWHKV{ Z'(\epsilon) \,=\, {1 \over {\Vol(H_0)}} \, \int_{\Fh_0}
\left[{{d\psi} \over {2\pi}}\right]
\, \left({{c_0} \over {c_1}}\right) \, \psi^{d_0 - d_1}\, \exp{\left[-i
\left(\gamma_0,\psi\right) - {\epsilon \over 2} \left(\psi,
\psi\right)\right]}\,.}
Although this expression in \ZVWHKV\ is ill-defined if $d_1 > d_0$, we
can still apply our previous work to compute using $\lambda'$ a
completely well-defined integral.  Namely, instead of considering the
symplectic integral $Z'(\epsilon)$, we introduce the differential
operator $Q$,
\eqn\BIGQD{ Q \,=\, \left(-2 {\partial \over {\partial
\epsilon}}\right)^{\ha (d_1 - d_0)}\,,}
and we consider instead the quantity
\eqn\NEWZINT{ Q \cdot Z'(\epsilon) \,=\, {1 \over {\Vol(H)}} \,
\int_{\Fh \times F} \left[{{d\phi} \over {2\pi}}\right] \,
(\phi,\phi)^{\ha (d_1 - d_0)} \, \exp{\left[\Omega - i \,
\langle\mu,\phi\rangle - {\epsilon \over 2}
(\phi, \phi) + t \, D\lambda' \right]}\,.}
Using the same definition for $\lambda'$ and proceeding exactly as
before, we compute
\eqn\NEWZINTII{\eqalign{
Q \cdot Z'(\epsilon) \,&=\, {1 \over {\Vol(H_0)}} \, \int_{\Fh_0}
\left[{{d\psi} \over {2\pi}}\right] \, \left({{c_0} \over {c_1}}\right)
\, \exp{\left[-i \left(\gamma_0,\psi\right) - {\epsilon \over 2}
\left(\psi, \psi\right)\right]}\,,\cr
&= {1 \over {\Vol(H_0)}} \, \left({{c_0} \over {c_1}}\right) \,
{1 \over {\sqrt{2 \pi \epsilon}}} \,
\exp{\left[-{{\left(\gamma_0,\gamma_0\right)} \over {2
\epsilon}}\right]}\,.\cr}}

The fact that the differential operator $Q$ in \BIGQD\ can be used to
cancel the determinants of $\psi$ in \DETPHI\ that arise from
localization is a special consequence of our assumption that $\dim
\Fh_0 = 1$.  For an arbitrary Lie algebra $\Fh_0$, we cannot generally
express these determinants as functions of only the quadratic
invariant $(\psi,\psi)$ that appears in the canonical symplectic
integral.  As a result, in the general case we cannot cancel such
determinants simply by differentiating $Z(\epsilon)$ with respect to
the coupling $\epsilon$.  Though we will not require the
generalization for this paper, we explain in Appendix B how to extend
the discussion above to the case of general $\Fh_0$.

We see from \NEWZINTII\ that, although our computation using
$\lambda'$ does not always give a sensible answer for $Z'(\epsilon)$,
it does give a sensible answer for the derivative $Q \cdot
Z'(\epsilon)$.  Knowledge of this derivative implicitly determines the
contribution of a higher critical point to $Z'(\epsilon)$, as the only
ambiguity in integrating \NEWZINTII\ is a polynomial in $\epsilon$
which cannot arise from a higher critical point.  Finally, as we show
in Appendix A, the quantity $Q \cdot Z'(\epsilon)$ in \NEWZINTII\
defined using $\lambda'$ agrees with the corresponding quantity $Q
\cdot Z(\epsilon)$ defined using the canonical one-form $\lambda$.
Hence, provided we take derivatives when necessary, we can use
$\lambda'$ for localization computations on $F$.

Our computation also shows that it may be easier to consider the
contributions of higher critical points not to $Z(\epsilon)$ but to
the derivative $Q \cdot Z(\epsilon)$.  We have already seen an example
of this phenomenon in our discussion of $SU(2)$ Yang-Mills theory.  In
that case, we found it more natural to compute the contributions of
higher Yang-Mills critical points to the derivative $\partial^{g-1}
Z(\epsilon) / \partial \epsilon^{g-1}$ in \LOCN\ as opposed to
$Z(\epsilon)$ itself.

\medskip\noindent{\it Application to Higher Critical Points of
Yang-Mills Theory}\smallskip

To finish this section, we apply our abstract study of localization on
$F$ to compute the path integral contributions from maximally
reducible Yang-Mills solutions.  We focus on the specific case of
$SU(2)$ Yang-Mills theory, for which we reproduce the explicit
expression in \LOCN\ for the contributions from the locus $\CM_n$ of
degree $n$ critical points.

As we have discussed, if $f = \* F_A$ is the curvature of a maximally
reducible Yang-Mills solution for gauge group $G$ of rank $r$, then
$f$ breaks the gauge group to a maximal torus $G_f = U(1)^r$.  In
terms of our abstract model, we thus identify the stabilizer group
$H_0$ with the subgroup $U(1)^r \subset \CG(P)$ of constant gauge
transformations in this maximal torus.  As we have also discussed,
this fact implies that  the corresponding moduli space $\CM_f$ of
maximally reducible Yang-Mills solutions is just a complex torus of
dimension $g r$.

Now, our description of the local symplectic model $F$ for the normal
geometry over a higher Yang-Mills critical point is completely general,
since in deriving the model for $F$ we did not make any assumptions
about the reducibility of the connection.  However, if we wish to use
this local model to compute contributions from arbitrary higher
Yang-Mills critical points, we will generally find that both the
integral over $F$ and the integral over the associated moduli space
$\CM_f$ make nontrivial contributions to $Z(\epsilon)$ which depend on
$\epsilon$.

In contrast, if we restrict to the special case that $\CM_f$ describes
maximally reducible Yang-Mills solutions, then only the integral over
$F$ is nontrivial, and the integral over the torus $\CM_f$ contributes a
multiplicative factor $\Vol(\CM_f)$ independent of $\epsilon$, where
\eqn\VOLMF{ \Vol\left(\CM_f\right) \,=\, \int_{\CM_f}
\exp{\left(\Omega\right)}\,.}
From a physical perspective, the contribution from $\CM_f$ to
$Z(\epsilon)$ does not involve the coupling $\epsilon$ because abelian
gauge theory is free.  From a mathematical perspective, the Donaldson
theory of $U(1)$ bundles is simple, as the corresponding universal
bundle is a line bundle having only a first Chern class, which is
proportional to $\Omega$.

In the case of $SU(2)$ Yang-Mills theory, the stabilizer group $H_0$
is just $U(1)$, and $\Fh_0$ has dimension one.  Thus, we can apply our
computation of the integral over $F$ in \NEWZINTII\ to conclude that
the local contribution from the moduli space $\CM_n$ of higher
critical points of degree $n$ is described by
\eqn\ZMF{ \left(-2 {\partial \over {\partial \epsilon}}\right)^{\ha
(d_1 - d_0)} \cdot Z(\epsilon)\Big|_{\CM_n} \,=\, {{\Vol(\CM_n)} \over
{\Vol(H_0)}} \, \left({{c_0} \over {c_1}}\right) \,
{1 \over {\sqrt{2 \pi \epsilon}}} \,
\exp{\left[-{{\left(2 \pi n\right)^2} \over \epsilon}\right]}\,.}
We immediately see that this expression has the same form as the
expression that appeared earlier in \LOCN.

To make a precise comparison of our formula \ZMF\ to \LOCN, we must
compute the various constants appearing in \ZMF.  To start, we
introduce the normalized generator $T_0$ of $H_0$,
\eqn\GENTZ{ T_0 \,=\, {1 \over \sqrt{2}} \, \sigma_z \,=\, {1 \over
\sqrt{2}} \, \pmatrix{i&0\cr0&-i\cr}\,,}
which satisfies $\Tr(T_0^2) = -1$.  From \GENTZ, we immediately see
that the volume of $H_0$ in our metric on $\Fh_0$ is
\eqn\VOLK{ \Vol(H_0) \,=\, 2 \pi \sqrt{2}\,.}

In the case of $SU(2)$ Yang-Mills theory, we have already
identified in \DCADP\ the bundles $\ad_\pm(P)$ with the line
bundles $\CL(+2n)$ and $\CL^{-1}(-2n)$.  Thus, from \AUXHDG, the
complex vector spaces $\CE_0$ and $\CE_1$, abstractly identified with
$E_0$ and $E_1$, are now given by the following Dolbeault cohomology
groups, 
\eqn\AUXHDGII{\eqalign{ E_0 \,&=\, H^0_{\bar\partial}(\Sigma,
\CL(2n))\,,\cr E_1 \,&=\, H^1_{\bar\partial}(\Sigma, \CL(2n))
\oplus H^1_{\bar
\partial}(\Sigma,\CL^{-1}(-2n))\,.\cr}} The index theorem, in
combinating with the vanishing of
$H^0_{\bar\partial}(\Sigma,\CL^{-1}(-2n))$, implies that
\eqn\INDXES{\eqalign{ &\chi(\CL(2n)) \,=\, \dim_\BC
H^0_{\bar\partial}(\Sigma, \CL(2n)) - \dim_\BC
H^1_{\bar\partial}(\Sigma, \CL(2n))\,=\, 2n + 1 - g\,,\cr
&\chi(\CL^{-1}(-2n)) \,=\, \dim_\BC H^1_{\bar
\partial}(\Sigma,\CL^{-1}(-2n)) \,=\, 2n - 1 + g\,.\cr}}
Thus, from \INDXES\ we determine the exponent $\ha (d_1 - d_0)$
appearing in \ZMF\ to be
\eqn\EXPIND{ \ha (d_1 - d_0) \,=\, \ha
\Big[\chi\left(\CL^{-1}(-2n)\right) -
\chi\left(\CL(2n)\right)\Big] \,=\, g - 1\,.}

To fix the ratio $c_0 / c_1$ appearing in \ZMF, which is determined by
the determinant of $\psi / 2 \pi$ acting on $E_0$ and $E_1$ as in
\DETPHI, we recall that $\CL(2n)$ and $\CL^{-1}(-2n)$ arise from the
standard generators $\sigma_\pm$ of the complex Lie algebra of
$SU(2)$, as in \BS.  Since $\sigma_z$ in \GENTZ\ acts with eigenvalues
$\pm 2 i$ on $\sigma_\pm$, we see that $\psi \equiv \psi \cdot T_0$
acts on sections of $\CL(2n)$ and $\CL^{-1}(-2n)$ with eigenvalues $\pm i
\sqrt{2} \, \psi$.  Thus, in this case,
\eqn\DETPHII{\eqalign{
\det\left({{\psi} \over {2 \pi}}\Big|_{E_0}\right) \,
\det\left({{\psi} \over {2 \pi}}\Big|_{E_1}\right)^{-1} \,&=\,
\left({{i \sqrt{2} \, \psi} \over {2 \pi}}\right)^{2n + 1 - g} \,
\left({{-i \sqrt{2}\, \psi} \over {2 \pi}}\right)^{-2n + 1 - g}\,,\cr
&=\, \left({{\psi^2} \over {2 \pi^2}}\right)^{1-g}\,.\cr}}
So
\eqn\DETPHIII{ \left({{c_0} \over {c_1}}\right) \,=\, (2 \pi^2)^{g-1}\,.}

Finally, we must compute the symplectic volume $\Vol(\CM_n)$. This
is equivalent to the moduli space of flat connections for the
group $U(1)$, and appears with the same symplectic structure as if
we were doing $U(1)$ gauge theory. The symplectic form is hence
equivalent to $\Omega=\sum_{i=1}^g dx_i\wedge dy_i$, where our
normalization is such that each of $dx_i$ and $dy_i$ have period
$2 \pi \sqrt{2}$ on the appropriate one-cycle.  (This is the same
factor that appeared in \VOLK.)   Thus, \eqn\VOLMN{ \Vol(\CM_n)
\,=\, \left(8 \pi^2\right)^g\,.}

So from \VOLK, \EXPIND, \DETPHIII, and \VOLMN, we evaluate \ZMF\ as
\eqn\ZMFIII{
{{\partial^{g-1} Z(\epsilon)} \over {\partial
\epsilon^{g-1}}}\Big|_{\CM_n} \,=\, \left(-8 \pi^4\right)^{g-1} \,
\sqrt{{{4 \pi} \over \epsilon}} \, \exp{\left(-{{\left(2 \pi
n\right)^2} \over \epsilon}\right)}\,,}
which agrees with \LOCN.

\newsec{Non-Abelian Localization For Chern-Simons Theory}

We now discuss non-abelian localization for Chern-Simons theory on a
Seifert manifold $M$.  As we recall from Section 3, the Chern-Simons
path integral then takes the symplectic form
\eqn\PZCSVII{ Z(\epsilon) \,=\, {1 \over {\Vol(\CG)}} \, \left({1 \over
{2 \pi i \epsilon}}\right)^{\Delta_{\CG}/2} \, \int_{\bar\CA}
\exp{\left[\Omega - {1 \over {2 i \epsilon}}
\left(\mu,\mu\right)\right]}\,.}
Our general discussion in Section 4 implies that $Z(\epsilon)$
localizes on critical points of the action $S = \ha (\mu, \mu)$.
Explicitly,
\eqn\SAACS{ S \,=\, \int_M \! \Tr \left( A \^ d A + {2 \over 3} A \^ A
\^ A \right) \,-\, \int_M {1 \over {\kappa \^ d \kappa}}
\Tr\Big[ (\kappa \^ F_A)^2 \Big]\,.}

Our first task is thus to classify the critical points of $S$.  We
claim that, up to the action of the shift symmetry, the critical
points of $S$ correspond precisely to the flat connections on $M$.  To
prove this statement, we simply observe that the critical points of
$S$ satisfy the equation of motion
\eqn\EOMSA{ F_A - \left({{\kappa \^ F_A} \over {\kappa \^
d\kappa}} \right) d\kappa - \kappa \^ d_A \left( {{\kappa \^ F_A} \over
{\kappa \^ d\kappa}} \right) \,=\, 0\,,}
where the first term of \EOMSA\ arises from the variation of the
Chern-Simons functional and the last two terms arise from the
variation of the last term in \SAACS.  To classify solutions of
\EOMSA, we recall that $S$ is invariant under the shift symmetry
$\delta A \,=\, \sigma \kappa$, where $\sigma$ is an arbitrary
function on $M$ taking values in the Lie algebra $\Fg$ of the gauge
group $G$.  Under the shift symmetry, the quantity $\kappa \^ F_A$
transforms as
\eqn\SHFTFII{ \kappa \^ F_A \,\longrightarrow\, \kappa \^ F_A + \sigma \,
\kappa\^d\kappa\,.}
Thus, since $\kappa \^ d\kappa$ is everywhere non-zero on $M$, we can
unambiguously fix a gauge for the shift symmetry by the condition
\eqn\GFSHFT{ \kappa \^ F_A \,=\, 0\,.}
In this gauge, any solution of the equation of motion \EOMSA\ is
precisely a flat connection on $M$.  So, as we certainly expect, the
Chern-Simons path integral localizes around points of $\bar\CA$ which
represent flat connections on $M$.

It is interesting to contrast this situation to the case of Yang-Mills
theory on a Riemann surface $\Sigma$.  In that case, the path integral
receives contributions from two qualitatively different kinds of
critical points, for which the moment map ${\mu = F_A}$ satisfies either
${\mu = 0}$ or ${\mu \neq 0}$, and the critical point is respectively
stable or unstable.  Since the critical points of Chern-Simons theory are
described by flat connections on $M$, one might naively suppose that
these critical points are analogous to the stable critical points of
Yang-Mills theory, which are also described by flat connections.
However, let us recall our expression from Section 3 for the
Chern-Simons moment map,
\eqn\MUIVV{\Big\langle\mu,(p,\phi,a)\Big\rangle \,=\,
-\ha p \int_M \kappa\^\Tr\left(\lie_RA\^A\right) \,+\, \int_M
\kappa\^\Tr\left(\phi F_A\right) \,-\, \int_M d\kappa\^\Tr\left(\phi
A\right) \,+\, a\,.}
The last term of \MUIVV\ is simply a constant piece of $\mu$ dual to
the generator $a$ of the central extension of the group $\CG_0$, and
this generator acts trivially on $\bar\CA$.  As a result of this term,
the Chern-Simons moment map is everywhere non-zero, and the critical
points of Chern-Simons theory are actually of the same kind as the
higher, unstable critical points of Yang-Mills theory.

Our goal in the rest of the paper is now to compute the local
contributions to  $Z(\epsilon)$ from two especially simple sorts
of flat connections on $M$.  First, we compute the contribution to
$Z(\epsilon)$ from the trivial connection when $M$ is a Seifert
homology sphere.  Second, we compute the contribution to
$Z(\epsilon)$ from a smooth component in the moduli space of
irreducible flat connections when $M$ is a principal $U(1)$-bundle
over a Riemann surface.  As we will see, these local computations
in Chern-Simons theory are direct generalizations of the local
computation at a higher critical point of two-dimensional
Yang-Mills theory.  The two cases we consider are the extreme
cases in which the connection is either trivial or irreducible.
Other cases are intermediate between these.

\medskip\noindent{\it The Normalization of $Z(\epsilon)$}\smallskip

Before we perform any detailed computations, we must make a few
general remarks about the normalization of $Z(\epsilon)$.  As we see
from \PZCSVII, we have normalized the Chern-Simons path integral with
the formal prefactor
\eqn\PREZG{ {1 \over {\Vol(\CG)}} \, \left({1 \over
{2 \pi i \epsilon}}\right)^{\Delta_{\CG}/2}\,, \qquad \Delta_{\CG}
\,=\, \dim \CG\,,}
which is defined in terms of the group $\CG$ of gauge transformations.

On the other hand, as we discussed in Section 3, the Hamiltonian group
which we use for localization in Chern-Simons theory is {\it not}
$\CG$ but rather the group ${\CH = U(1) \ltimes \wt{\CG_0}}$, where
$\wt{\CG_0}$ is a central extension by $U(1)$ of the identity
component $\CG_0$ of $\CG$.  We also introduce the group ${\CH' = U(1)
\ltimes \wt\CG}$, which arises from the corresponding central extension
$\wt\CG$ of the full group $\CG$ of all gauge transformations.

When we apply non-abelian localization to Chern-Simons theory, the
path integral which we compute most directly is not given by \PZCSVII\
but by the canonically normalized symplectic integral
\eqn\AWWC{ Z_0(\epsilon) \,=\, {1 \over {\Vol(\CH')}} \, \,
\int_{\Fh \times \bar\CA} \left[{{d\phi} \over {2\pi}}\right]
\exp{\left[\Omega - i \, \langle\mu,\phi\rangle - {{i \epsilon} \over 2}
(\phi,\phi)\right]}\,,}
as we computed abstractly in Section 4.  The appearance of the volume
of the disconnected group $\CH'$ in \AWWC, as opposed to the connected
group $\CH$, accounts for the action of gauge transformations in the
disconnected components of $\CG$ on critical points in $\bar\CA$.
Also, because the Chern-Simons path integral is oscillatory, an
imaginary coupling $i\epsilon$ now appears in \AWWC.

If we perform the Gaussian integral over $\phi$ in \AWWC, then
$Z_0(\epsilon)$ becomes
\eqn\AWWCII{ Z_0(\epsilon) \,=\, {i \over {\Vol(\CH')}} \, \left({1
\over {2 \pi i \epsilon}}\right)^{\Delta_\CH/2} \, \int_{\bar \CA}
\exp{\left[\Omega - {1 \over {2 i \epsilon}}
\left(\mu,\mu\right)\right]}\,,\qquad \Delta_\CH \,=\, \dim \CH\,.}
In computing this integral over $\phi$, we must be careful to remember
that the quadratic form $(\,\cdot\,,\,\cdot\,)$ on the Lie algebra
$\Fh$ of $\CH$ is the direct sum of a positive-definite form on the
Lie algebra of the gauge group  $\CG$ and a hyperbolic form (with
signature $(+,-)$) on the two additional generators in $\CH$ relative
to $\CG$.  Had the form on $\Fh$ been positive-definite, the Gaussian
integral over each generator in $\Fh$ would have contributed an
identical factor $(2 \pi i \epsilon)^{-\ha}$ to the prefactor in front of
\AWWCII.  However, due to the hyperbolic summand in
$(\,\cdot\,,\,\cdot\,)$,  the phases that result from the Gaussian
integral over the two generators in the hyperbolic subspace
of $\Fh$ actually cancel.  To account for this cancellation, we
include the extra factor of `$i$' appearing in \AWWCII.

Although $Z_0(\epsilon)$ in \AWWCII\ takes the same form as the physical
Chern-Simons path integral $Z(\epsilon)$ in \PZCSVII, evidently the
prefactor \PREZG\ which fixes the normalization of $Z(\epsilon)$
differs from the corresponding prefactor in $Z_0(\epsilon)$ by the ratio
\eqn\PREZCH{ {{\Vol(\CH')} \over {i \Vol(\CG)}} \cdot
\left({1 \over{2 \pi i \epsilon}}\right)^{\ha (\Delta_{\CG} -
\Delta_{\CH})} \,=\, \Vol(U(1)^2) \cdot 2 \pi \epsilon\,.}
The finite factors $\Vol(U(1)^2)$ and $2 \pi \epsilon$ arise in the
obvious way from the two extra generators in $\CH$ relative to $\CG$.

When we perform localization computations in Chern-Simons theory, we
apply our abstract localization computations in Section 4 to compute
$Z_0(\epsilon)$.  By our observation above, for the purpose of
computing the physical Chern-Simons path integral $Z(\epsilon)$, we must
multiply the results from our abstract local computations by the
finite factor in \PREZCH.  As we will see, this expression turns out to
cancel nicely against corresponding factors from the local computation.

\subsec{A Two-Dimensional Interpretation of Chern-Simons Theory on $M$}

Our symplectic interpretation of Chern-Simons theory on $M$
fundamentally relies on the fact that the shift symmetry decouples one
component of the gauge field $A$.  As a result, we can essentially
perform Kaluza-Klein reduction over the $S^1$ fiber of $M$ to the base
$\Sigma$ to express Chern-Simons theory as a two-dimensional
topological theory on $\Sigma$.  From this two-dimensional
perspective, we can immediately apply our localization computations in
Section 4 to Chern-Simons theory.

In fact, the two-dimensional topological theory on $\Sigma$ arising
from Chern-Simons theory on $M$ is closely related to Yang-Mills
theory on $\Sigma$, a point also recently emphasized in \AganagicJS.
At the level of the classical moduli spaces, the relationship
between Chern-Simons theory on $M$ and Yang-Mills theory on $\Sigma$
was noted long ago by Furuta and Steer in \Furuta.  These authors
identify a correspondence between the moduli space of flat connections
on $M$ and certain components of the moduli space of Yang-Mills
solutions on $\Sigma$.  Since the relationship between flat
connections on $M$ and Yang-Mills solutions on $\Sigma$ underlies
our study of Chern-Simons theory, we now explain the fundamental
aspects of this correspondence.

\medskip\noindent{\it Flat Connections on $M$ From Yang-Mills
Solutions on $\Sigma$}\smallskip

We start by considering the moduli space of flat connections on $M$.  As
before, we suppose that the gauge group $G$ is compact, connected,
simply-connected, and simple.

A flat connection on $M$ is determined by its holonomies, and the
moduli space of flat connections on $M$, up to gauge equivalence, can
be concretely described as the space of group homomorphisms from the
fundamental group $\pi_1(M)$ to $G$, up to conjugacy.  Hence the
structure of the moduli space of flat connections on $M$ is determined
by $\pi_1(M)$.

On the other hand, because $M$ is a Seifert manifold, and hence
generally a $U(1)$ $V$-bundle over an orbifold $\Sigma$, the structure
of $\pi_1(M)$ is closely tied to the structure of the orbifold
fundamental group $\pi_1(\Sigma)$.  This topological fact underlies
the close relationship between flat connections on $M$ and Yang-Mills
solutions on $\Sigma$, and to explain it we now present the group
$\pi_1(M)$.

As in Section 3, we describe $M$ using the Seifert invariants
\eqn\SFRTII{\Big[g;n;(\alpha_1,\beta_1), \ldots,
(\alpha_N,\beta_N)\Big]\,,\quad \gcd(\alpha_j,\beta_j) = 1\,.}
We recall that $g$ is the genus of $\Sigma$, $n$ is the degree of the
$U(1)$ $V$-bundle over $\Sigma$, and the relatively prime integers
$(\alpha_j,\beta_j)$ for $j=1,\ldots,N$ specify the local geometry of
$M$ near the $N$ orbifold points on $\Sigma$.

To present $\pi_1(M)$, we introduce elements
\eqn\GENPM{\eqalign{
&a_p\,, b_p\,,\quad p=1\,,\ldots\,,g\,,\cr
&c_j\,,\quad j=1\,,\ldots\,,N\,,\cr
&h\,.\cr}}
Then $\pi_1(M)$ is generated by these elements in \GENPM\ subject to the
following relations,
\eqn\PIM{\eqalign{
&\left[a_p, h\right] \,=\, \left[b_p, h\right] \,=\,
\left[c_j, h\right] \,=\,1\,,\cr
&c_j^{\alpha_j} h^{\beta_j} \,=\, 1\,,\cr
& \prod_{p=1}^g \, \left[ a_p , b_p \right] \, \prod_{j=1}^N c_j \,=\,
h^n\,.\cr}}

We will not give a formal proof of this presentation of
$\pi_1(M)$, which follows from the standard surgery construction
of $M$ and which can be found in \OrlikPK, but we will describe
the geometric interpretation of the generators in \GENPM.  The
generator $h$, which is a central element of $\pi_1(M)$ by the
first line of \PIM, arises from the generic $S^1$ fiber over
$\Sigma$.  Since $\Sigma$ has genus $g$, the generators $a_p$ and
$b_p$ for $p=1,\ldots,g$ arise from the $2g$ non-contractible
cycles on $\Sigma$.  Finally, the generators $c_p$ for
$p=1,\ldots,N$ arise from small one-cycles in $\Sigma$ about each
of the orbifold points.  We note that from the presentation of
$\pi_1(M)$ in \GENPM\ and \PIM\ one can immediately compute the
corresponding homology group $H_1(M, \BZ)$ as the abelianization
of $\pi_1(M)$.

For example, with a view to our application below, let us
determine the condition to have $H_1(M)=0$.  This requires $g=0$
(or the homology of $\Sigma$ will appear in $H_1(M)$).  So
$\pi_1(M)$ has generators $c_j$, $j=1,\dots,N$, and $c_{0}=h$.
There are $N+1$ relations, namely
$c_j^{\alpha_j}c_{0}{}^{\beta_j}=1, \,j=1\dots,N$, and
$\prod_{j=1}^Nc_j\cdot c_{0}{}^{-n}=1$.  So we can write the
relations in the general form $\prod_{j=0}^{N}c_j^{K_{j,l}}=1$ in
terms of an $N+1\times N+1$ matrix $K$.  A general element of
$H_1(M)$ of the form $\prod_{j=0}^{N}c_j^{v_j}$ is trivial if and
only if one can write $v_j=\sum_{j'}K_{jj'}w_{j'}$ for some
integer-valued vector $w$.  So $H_1(M)$ is trivial if and only if
$\det(K)=\pm 1$.  With the actual form of $K$, one can work out
this determinant and find that the condition is that \eqn\CHRNCL{
n + \sum_{j=1}^N {{\beta_j} \over {\alpha_j}} =\pm\prod_{j=1}^n{1\over
\alpha_j}.}  The left hand side is also equal to the orbifold
first Chern class $c_1({\cal L})$ of the line $V$-bundle ${\cal
L}$ discussed in Section 3.2.

With the presentation of $\pi_1(M)$ in \GENPM\ and \PIM, we can
immediately present $\pi_1(\Sigma)$ as well.  Thus,
$\pi_1(\Sigma)$ is generated by the elements $a_p$, $b_p$, and
$c_j$ in \GENPM, omitting the generator $h$ which arises from the
$S^1$ fiber, and the relations in $\pi_1(\Sigma)$ are given by the
relations in \PIM\ upon setting $h=1$.  A very succinct description of
this relation between $\pi_1(M)$ and $\pi_1(\Sigma)$ is to recognize
$\pi_1(M)$ as a central extension of $\pi_1(\Sigma)$,
\eqn\ZPIONE{ 1 \longrightarrow \BZ \longrightarrow \pi_1(M)
\longrightarrow \pi_1(\Sigma) \longrightarrow 1\,,}
where $h$ is the generator of $\BZ$ above.

Given the close relationship between the groups $\pi_1(M)$ and
$\pi_1(\Sigma)$ expressed in \ZPIONE, we can immediately deduce a
relationship between flat connections on $M$ and Yang-Mills solutions
on $\Sigma$.  To describe this relationship, we consider a
homomorphism $\rho$, \eqn\GPRHO{ \rho: \pi_1(M) \longrightarrow G\,,}
which describes the holonomies of a given flat connection on $M$.

Because $h$ is central in $\pi_1(M)$, the image of $\rho$ must lie in
the centralizer $G_{\rho(h)}$ of the element $\rho(h)$ in $G$.  To
simplify the following discussion, we suppose that $\rho(h)$ actually
lies in the  center $\Gamma$ of $G$, implying that $G_{\rho(h)} = G$.
This condition is necessary whenever the connection described by $\rho$ is
irreducible, and it certainly holds also when the connection is
trivial, which are the two main cases we consider when we perform
computations in Chern-Simons theory.  We refer to \Furuta\ for a
discussion of the general case.

Clearly if $\rho(h) = 1$, so that the corresponding flat connection on
$M$ has trivial holonomy around the $S^1$ fiber over $\Sigma$,
then $\rho$ factors through the extension \ZPIONE\ to induce a
homomorphism from $\pi_1(\Sigma)$ to $G$.  Hence $\rho$ describes
a flat connection on $M$ that pulls back from a flat Yang-Mills
connection on $\Sigma$.

More generally, when $\rho(h)$ is non-trivial in $\Gamma$, then the
corresponding flat connection on $M$ has non-trivial holonomy around
the $S^1$ fiber of $M$ and is not the pull back of a flat
$G$-connection on $\Sigma$.  However, if we pass from $G$ to the
quotient group ${\bar G = G / \Gamma}$, so that we consider the
connection on $M$ as a flat connection on the trivial $\bar G$-bundle,
then the holonomy of this connection around the $S^1$ fiber of $M$
becomes trivial.

As a result, the homomorphism $\rho$ can be interpreted as describing
a flat connection on $M$ which arises from the pull back of a flat
Yang-Mills connection on a generally non-trivial $V$-bundle over
$\Sigma$ whose structure group is now $\bar G$, as opposed to $G$.
In general, a flat connection on a non-trivial $\bar G$-bundle over
$\Sigma$ can be described as a flat connection on the trivial
$G$-bundle over $\Sigma$ such that the connection has non-trivial
monodromies in $\Gamma$ around the orbifold points as well as around
one additional, arbitrarily chosen smooth point of $\Sigma$.  These
monodromies represent the obstruction to smoothly extending the given flat
connection to the trivial $G$-bundle over all of $\Sigma$, and hence
they describe the non-trivial $\bar G$-structure on the bundle.

In the case at hand, we see from the relations \PIM\ which describe
$\pi_1(M)$ as an extension of $\pi_1(\Sigma)$ that the relevant
monodromies are determined by the holonomies of the connection on $M$
associated to the elements $h^{\beta_j}$ and $h^n$, so that these
holonomies determine the topology of the corresponding $\bar G$-bundle
on $\Sigma$.  For instance, if we consider the simplest case that the
gauge group $G$ is $SU(2)$ and $M$ arises from a principal
$U(1)$-bundle over a smooth Riemann surface $\Sigma$ such that the
degree $n$ is odd, then flat connections on $M$ whose holonomies
satisfy ${\rho(h) = \rho(h)^n = -1}$ correspond bijectively to flat
$SU(2)$ connections on $\Sigma$ which have monodromy $-1$ around a
specified puncture.  Such flat $SU(2)$ connections can then be
identified with flat connections on the topologically non-trivial
principal $SO(3)$-bundle over $\Sigma$.

On the other hand, if the degree $n$ of the principal $U(1)$-bundle is
even, then $\rho(h)^n = 1$ for both $\rho(h) = \pm 1$, so points in
both of these components of the moduli space of flat connections on
$M$ are identified with flat $SU(2)$ connections on $\Sigma$.

\medskip\noindent{\it The Local Symplectic Geometry Near a Critical
Point of Chern-Simons Theory}\smallskip

The discussion above shows that irreducible flat connections on $M$
can be identified with corresponding flat Yang-Mills connections on
$\Sigma$.  We now extend this observation to give a ``two-dimensional''
description of the local symplectic geometry in $\bar\CA$ around such
a critical point of Chern-Simons theory.

Because $\bar\CA$ is the quotient of the affine space $\CA$ by the
shift symmetry $\CS$, we are free to work in any convenient gauge for
$\CS$.  For instance, in order to identify the critical points of the
new Chern-Simons action $S$ in \SAACS, we found it convenient to
impose the gauge condition \GFSHFT.

However, in order to describe the local geometry in $\bar\CA$ in terms
of geometric quantities on $\Sigma$, we make a new gauge choice for $\CS$,
corresponding to the gauge condition
\eqn\GFSHFTII{ \iota_R A \,=\, 0\,.}
Because $A$ transforms under the shift symmetry as $\delta A = \sigma
\, \kappa$, the quantity $\iota_R A$ transforms as $\iota_R A \to
\iota_R A + \sigma$, and the gauge condition in \GFSHFTII\ is
unambiguous.

To describe a critical point of the action $S$ in the gauge \GFSHFTII,
we consider as above a flat Yang-Mills connection $B_0$ on a generally
non-trivial $V$-bundle with structure group $\bar G$ over $\Sigma$.
Then, in the gauge \GFSHFTII, the full tangent space to the symplectic
manifold $\bar\CA$ at $B_0$ is described by the space of sections
$\xi$ of the bundle $\Omega^1_M \otimes \Fg$ which satisfy the gauge
condition
\eqn\GETA{ \iota_R \xi \,=\, 0\,.}

Because our symplectic description of Chern-Simons theory respects the
geometric $U(1)$ action on $M$, we naturally consider the
decomposition of the tangent space to $\bar\CA$ under the action of
this $U(1)$.  In terms of the section $\xi$, this statement simply
means that we consider the Fourier decomposition of $\xi$ into
eigenmodes of the operator $\lie_R$.  Thus we write
\eqn\FTETA{ \xi \,=\, \sum_{t = - \infty}^{+ \infty} \xi_t\,,}
where, in addition to the gauge condition \GETA, each eigenmode
$\xi_t$ satisfies
\eqn\FTETAII{ \lie_R \xi_t \,=\, - 2 \pi i t \cdot \xi_t\,.}

We can similarly perform this Fourier decomposition on the tangent
space to the group of gauge transformations $\CG$.  Thus, if $\phi$ is a
section of $\Omega^0_M \otimes \Fg$, we write
\eqn\FTPHI{ \phi \,=\, \sum_{t = - \infty}^{+ \infty} \phi_t\,,}
where
\eqn\FTPHIII{ \lie_R \phi_t \,=\, - 2 \pi i t \cdot \phi_t\,.}

To describe these eigenmodes $\xi_t$ and $\phi_t$ geometrically on
$\Sigma$, we recall that $\CL$ denotes the line $V$-bundle
over $\Sigma$ associated to the Seifert manifold $M$.  Since non-trivial
representations of the $U(1)$ action on $M$ are associated to non-zero
powers of $\CL$ on $\Sigma$, we can describe the modes $\xi_t$
and $\phi_t$ geometrically on $\Sigma$ as being respectively
sections of the bundles $\Omega^1_{\Sigma} \otimes \ad(P) \otimes
{\CL}^t$ and $\Omega^0_{\Sigma} \otimes \ad(P) \otimes
{\CL}^t$.  Here we have also replaced the trivial bundle $\Fg$ on $M$
by the possibly nontrivial $\bar G$-bundle $\ad(P)$ on $\Sigma$.

So, at least formally, the tangent space to $\bar\CA$ at $B_0$
decomposes into the following sum of spaces of sections on
$\Sigma$, \eqn\TANGA{ T\bar \CA= \bigoplus_{t=-\infty}^{+\infty}
\Gamma\left(\Sigma, \Omega^1_\Sigma \otimes \ad(P) \otimes
\CL^t\right)\,,} and similarly for the Lie algebra of $\CG$,
\eqn\TANGG{T\CG = \bigoplus_{t=-\infty}^{+\infty}
\Gamma\left(\Sigma, \Omega^0_\Sigma \otimes \ad(P) \otimes
\CL^t\right)\,.} By assumption, the covariant derivative $d_{B_0}$
commutes with the Lie derivative $\lie_R$, $[d_{B_0}\,, \lie_R] =
0$, so these decompositions are compatible with the action of
$d_{B_0}$.

As in Section 4.2, the local structure of the space of fields over
which we integrate near a given component ${\cal M}$ of the moduli
space of critical points is a fibration \eqn\kurlo{F\longrightarrow
N\buildrel pr\over\longrightarrow {\cal M}.} As before, $F$ is given by
a symplectic bundle \eqn\COTFBIII{ F \,=\, \CH \times_{H_0} \left(\Fh
\ominus \Fh_0 \ominus \CE_0 \oplus \CE_1\right), } where the
invariance group $H_0$ and the exceptional bundles $\CE_0$ and $\CE_1$
must be identified. As we observed at the start of this section,
because the Chern-Simons moment map is non-vanishing, the local model
is analogous to the geometry near a higher critical point of
Yang-Mills theory, with some $\CE_0$ and $\CE_1$.

In the model  \COTFBIII\ for $F$, ${\CH = U(1) \ltimes
\wt{\CG_0}}$ is the Hamiltonian group which we use for
localization, and $H_0$ is the subgroup of $\CH$ which fixes
$B_0$.  In general, $H_0$ is a finite-dimensional group of the
form \eqn\NEWHO{ H_0 \,=\, U(1)^2 \times K_0\,.} One $U(1)$ factor
in $H_0$ arises from the action of $\lie_R$ on $\bar\CA$, which
fixes $B_0$ by assumption, and the other $U(1)$ factor arises from
the central $U(1)$ in $\wt{\CG_0}$.  This $U(1)$ acts trivially on
all of $\bar\CA$. Finally, $K_0$ denotes the group of
 gauge transformations acting on $\ad(P)$ which
fix $B_0$.  These gauge transformations are generated by
covariantly constant sections $\phi$ of $\ad(P) \otimes \CL^0$, so
that $\phi$ is annihilated by $\lie_R$, and consequently $K_0$
commutes with both $U(1)$ factors in $H_0$.

To identify $\CE_0$ and $\CE_1$, we must look at the images of
$d_{B_0}$ and of $\star_2 d_{B_0}$ mapping $T\CG$ to $T\bar \CA$.
The bundle $\ad(P)\otimes {\cal L}^t$ has connection
$C=B_0+t\kappa$ ($\kappa$ is the constant curvature connection on
$\CL$ introduced in Section 3.2).  For fixed $t$, the
three-dimensional operators $d_{B_0}$ and $\star_2 d_{B_0}$ reduce
to two-dimensional operators $d_C$ and $\star d_C$. As $B_0$ is
flat, the connection $C$ has curvature equal to $t$ times a
positive two-form. So the analysis of the intersection and unions
of the images of $d_C$ and $\star d_C$ precisely follows Section
4.3, with the following dictionary between quantities in the
two-dimensional analysis of that section and quantities in the
present three-dimensional problem: \eqn\ico{\eqalign{\ad_0(P)
&\longleftrightarrow \ad(P) \cr
 \ad_+(P) & \longleftrightarrow\bigoplus_{t>0} \, \ad(P)\otimes {\cal
 L}^t\cr
 \ad_-(P) & \longleftrightarrow\bigoplus_{t<0} \, \ad(P)\otimes {\cal
 L}^t.\cr}}
In two dimensions, we decomposed $\ad(P)$ into $\ad_0(P)$,
$\ad_+(P)$, and $\ad_-(P)$ according to the sign of the curvature.
Here, curvature comes only from $\CL$. So finally, we get
\eqn\AUXHDGII{\eqalign{ \CE_0 \,&=\, \bigoplus_{t \neq 0}
H^0_{\bar\partial}(\Sigma, \ad(P) \otimes \CL^t)\,=\, \bigoplus_{t
\ge 1} H^0_{\bar\partial}(\Sigma, \ad(P) \otimes (\CL^t \oplus
\CL^{-t}))\,,\cr \CE_1 \,&=\, \bigoplus_{t \neq 0}
H^1_{\bar\partial}(\Sigma, \ad(P) \otimes \CL^t)\,=\, \bigoplus_{t
\ge 1} H^1_{\bar\partial}(\Sigma, \ad(P) \otimes (\CL^t \oplus
\CL^{-t}))\,.\cr}} Unlike in the case of Yang-Mills theory, these
exceptional bundles $\CE_0$ and $\CE_1$ now have infinite
dimension, since the cohomology groups in \AUXHDGII\ are non-zero
for infinitely many $t$'s.

\subsec{Localization at the Trivial Connection on a Seifert Homology
Sphere}

We are finally prepared to carry out a computation in Chern-Simons
theory using non-abelian localization.  We consider localization
at the trivial connection when $M$ is a Seifert manifold that also
is a homology sphere, that is, it has $H_1=0$. We start by stating
some necessary facts about the topology of $M$ in this case.

\medskip\noindent{\it Seifert Homology Spheres and a Slight
Generalization}\smallskip

We recall that we generally characterize $M$ with the Seifert
invariants \eqn\SFRTII{\Big[g;n;(\alpha_1,\beta_1), \ldots,
(\alpha_N,\beta_N)\Big]\,,\quad \gcd(\alpha_i,\beta_i) = 1\,.} As
we have explained above, $M$ is a homology sphere, with
$H_1(M,\BZ) = 0$, if and only if the invariants in \SFRTII\
satisfy \eqn\HMSPH{ g \,=\, 0\,,\qquad\qquad c_1(\CL_0) \,=\, n +
\sum_{j=1}^N {{\beta_j} \over {\alpha_j}} \,=\, \pm \prod_{j=1}^N
{1 \over {\alpha_j}}\,.} Here $\CL_0$ denotes the line $V$-bundle
over the orbifold $\Sigma$ which describes $M$.

To interpret geometrically the condition on $\CL_0$ in \HMSPH, we
note that this condition implies the arithmetic condition that the
numbers $\alpha_j$ be pairwise relatively prime, so that
\eqn\PRPR{ \gcd(\alpha_j, \alpha_{j'}) = 1\,,\qquad j \neq j'\,.}
In turn, as explained in Section 1 of \Furuta, this arithmetic
condition on the orders of the orbifold points of $\Sigma$ implies
that the Picard group of line $V$-bundles on $\Sigma$ is isomorphic to
$\BZ$, just as for $\BC\BP^1$.  In analogy to the case of $S^3$, which
arises from a generator of the Picard group of $\BC\BP^1$, the
condition on $c_1(\CL_0)$ in \HMSPH\ is then precisely the condition that
$\CL_0$ generate the Picard group of $\Sigma$.

As previously, we orient $M$ so that $c_1(\CL_0)$ is positive, and we
introduce the notation $\beta_j^0$ to distinguish the orbifold
invariants of this fundamental line $V$-bundle $\CL_0$ on $\Sigma$,
\eqn\CLO{ c_1(\CL_0) \,=\, n + \sum_{j=1} \, {{\beta_j^0} \over
{\alpha_j}} \,=\, \prod_{j=1}^N {1 \over {\alpha_j}}\,.}

The reason that we distinguish the invariants $\beta_j^0$ of $\CL_0$
is that, more generally, we will also consider the case that $M$ arises
not from the fundamental line $V$-bundle $\CL_0$ on $\Sigma$ but from
some multiple $\CL_0^d$ for $d \ge 1$.  In this case, we simply
require that $g=0$ in \HMSPH\ and that the invariants $\alpha_j$ be
relatively prime to each $\beta_j$ and also pairwise relatively prime,
as in \PRPR.  The Seifert manifold arising from $\CL_0^d$ is a quotient by
the cyclic group $\BZ_d$ of the Seifert manifold associated to
$\CL_0$, and in this case $H_1(M, \BZ) = \BZ_d$.  So the integer $d$
can be characterized topologically as the order of $H_1(M, \BZ)$,
\eqn\ORDH{ d \,=\, |H_1(M, \BZ)|\,.}
These Seifert manifolds are still rational homology spheres, with
$H_1(M,\BR) = 0$, and the trivial connection on $M$ is an isolated
flat connection.

We note that when the Seifert manifold $M$ is described by a smooth,
degree $n$ line-bundle over $\BC\BP^1$, then $M$ is a lens space, and
the Seifert invariant $n$ coincides with $d$ in \ORDH.

\medskip\noindent{\it The Result of Lawrence and Rozansky}\smallskip

Our basic results on localization for Chern-Simons theory imply that
the Chern-Simons partition function $Z$ can be expressed as a sum of
local contributions from the flat connections on $M$.  In the case $G
= SU(2)$ and with $M$ as above, Lawrence and Rozansky \LawrenceRZ\
have already made this simple structure of $Z$ explicit by working
backwards from the previously known formula for $Z$.  Our goal here is to
compute directly one term in their formula, the local contribution
from the trivial connection.  However, because the general result in
\LawrenceRZ\ is both very elegant and very suggestive, we now pause to
present it.

To express\foot{Our notation differs somewhat from \LawrenceRZ, and we
have normalized $Z(\epsilon)$ so that the partition function on $S^2
\times S^1$ is $1$, whereas the authors of \LawrenceRZ\ normalize the
partition function on $S^3$ to be $1$.} $Z$ as in \LawrenceRZ, we find
it useful to introduce the numerical quantities
\eqn\QRZE{\eqalign{
&\epsilon_r \,=\, {{2 \pi} \over {k + 2}}\,,\cr
&P \,=\, \prod_{j=1}^N \, \alpha_j \quad\hbox{if } N \ge 1\,,\qquad P
= 1 \quad\hbox{otherwise}\,,\cr
&\theta_0 \,=\, 3 - {d \over P} + 12 \, \sum_{j=1}^N \, s(\beta_j,
\alpha_j)\,.\cr}}
Here $\epsilon_r$ is the renormalized coupling incorporating the
famous shift $k \to k + 2$ in the level in the case $G=SU(2)$, and
$s(\beta,\alpha)$ is the Dedekind sum,
\eqn\DEDEK{ s(\beta, \alpha) \,=\, {1 \over {4 \alpha}} \, \sum_{l =
1}^{\alpha - 1} \, \cot\left({{\pi l} \over \alpha}\right)
\cot\left({{\pi l \beta} \over \alpha}\right)\,.}
For brevity, we also introduce the analytic functions
\eqn\FFSS{\eqalign{
F(z) \,&=\, \left(2 \sinh{\left({z \over 2}\right)}\right)^{2 - N}
\cdot \prod_{j=1}^N\, \left(2 \sinh{\left({z \over {2
\alpha_j}}\right)}\right)\,,\cr
G^{(l)}(z) \,&=\, {i \over {4 \epsilon_r}} \left({d \over P}\right) z^2
- {{2 \pi \, l} \over {\epsilon_r}} \, z\,.\cr}}

Then, from the results of \LawrenceRZ, the partition function
$Z(\epsilon)$ of Chern-Simons theory on $M$ can be written as
\eqn\ZLWRZ{\eqalign{ &Z(\epsilon) \,=\, (-1) \, {{\exp{\left({{3
\pi i} \over 4} - {i \over 4} \theta_0 \epsilon_r\right)}} \over
{4 \sqrt{P}}}  \; \Bigg\{ \sum_{l=0}^{d-1} \, {1 \over {2 \pi i}}
\, \int_{\CC^{(l)}} \! dz \; F(z) \, \exp{\left[G^{(l)}(z)\right]}
\,-\,\cr &- \sum_{m=1}^{2 P - 1} \Res\left({{F(z)
\exp{\left[G^{(0)}(z)\right]}} \over {1 - \exp{\left(- {{2 \pi}
\over {\epsilon_r}} \, z\right)}}}\right)\Bigg|_{z= 2 \pi i \, m}
- \sum_{l=1}^{d-1} \sum_{m=1}^{[{{2 P l} \over d}]} \Res\left(F(z)
\, \exp{\left[G^{(l)}(z)\right]}\right)\Bigg|_{z=- 2 \pi i \, m}
\,\Bigg\}\,.\cr}} Here $\CC^{(l)}$ for $l=0,\ldots,d-1$ denote a
set of contours in the complex plane over which we evalute the
integrals in the first line of \ZLWRZ. In particular, $\CC^{(0)}$
is the diagonal line contour through the origin, \eqn\CCRC{
\CC^{(0)} \,=\, \e{{i \pi} \over 4} \times \BR\,,} and the other
contours $\CC^{(l)}$ for $l > 0$ are diagonal line contours
parallel to $\CC^{(0)}$ running through the stationary phase point
of the integrand, given by $z = - 4 \pi i \, l \, (P / d)$. Also,
 ``$\Res$'' denotes the residue of the given analytic function
evaluated at the given point.

We now wish to point out a few general features of this result
\ZLWRZ\ from the perspective of non-abelian localization.

First, the $d$ contour integrals in the first term of \ZLWRZ\ are
identified in \LawrenceRZ\ with the local contributions from the
$d$ reducible flat connections on $M$.  In particular, the
integral arising from $l=0$ above is the local contribution from
the trivial connection, which takes the form
\eqn\ZLWRZII{\eqalign{ Z(\epsilon)\Big|_{\{0\}} \,&=\, (-1) \,
{{\exp{\left({{3 \pi i} \over 4} - {i \over 4} \theta_0
\epsilon_r\right)}} \over {4 \sqrt{P}}} \,\times\,\cr &\times\, {1
\over {2 \pi i}} \, \int_{\CC^{(0)}} \! dz \, \exp{\left[{i \over
{4 \epsilon_r}} \left({d \over P}\right) z^2\right]} \, \left(2
\sinh{\left({z \over 2}\right)}\right)^{2 - N} \cdot
\prod_{j=1}^N\, \left(2 \sinh{\left({z \over {2
\alpha_j}}\right)}\right)\,.\cr}} For instance, one can directly
check that, in the case $M = S^3$, the integral in \ZLWRZII\
reduces to our much simpler expression for $Z(\epsilon)$ in
\LOCLRM.

Similarly, the integrals for $l > 0$ arise from reducible flat
connections whose holonomies lie in a maximal torus of $SU(2)$, and
hence these connections are fixed by a $U(1)$ subgroup of the gauge
group.  As we generally saw in Section 4 when we considered higher
critical points of Yang-Mills theory, non-abelian localization at a
reducible connection leads to an integral over the Lie algebra $\Fh_0$
of the stablizer group $H_0$.  This integral over $\Fh_0$ is
represented by the contour integrals above.

In contrast, the residues in the remaining terms of \ZLWRZ\ are
identified in \LawrenceRZ\ with the local contributions from the
irreducible flat connections on $M$.  As we show later, at least
in the non-orbifold case $N=0$ and $g > 0$, the local path
integral contribution from a smooth component $\CM$ in the moduli
space of irreducible flat connections on $M$ is given by a
computation in the cohomology ring of $\CM$.  In the context of
two-dimensional Yang-Mills theory, cohomology computations on
$\CM$ are often expressed in the form of residues, and we expect
the residues in \ZLWRZ\ to arise in this fashion.

Finally, the phase of $Z(\epsilon)$ in \ZLWRZ\ is quite subtle.  As
explained in \AtiyahFM, this phase can be defined given the
choice of a 2-framing on $M$, meaning a trivialization of $TM \oplus
TM$, and for each three-manifold $M$ a canonical choice of 2-framing
exists.  The partition function can thus be presented with a canonical
phase, as originally computed in \refs{\FreedRG, \JeffreyLC} and as
given in \ZLWRZ.  The phase of $Z(\epsilon)$ which arises naturally
when we define Chern-Simons theory via localization differs from this
canonical phase, and we discuss this fact at the end of the section.

\medskip\noindent{\it Localization at the Trivial Connection}\smallskip

We now compute using localization the contribution from the trivial
connection to $Z(\epsilon)$ when $M$ is a Seifert homology sphere.
Although the results of Lawrence and Rozansky in \ZLWRZ\ hold for
gauge group $G = SU(2)$, Mari\~no has presented in \MarinoFK\ an
expression for the contribution from the trivial connection for an
arbitrary simply-laced gauge group $G$.  With our methods, the
generalization from $G = SU(2)$ to arbitrary simply-laced $G$ is
immediate, so we also consider the general case.

At the trivial connection, the moduli space $\CM$ is trivial, so
the local geometry in $\bar\CA$ is entirely described by the
normal symplectic fiber $F$ in \COTFBIII, with the appropriate
$\Fh_0$, $E_0$, and $E_1$. So we need only evaluate the canonical
symplectic integral over $F$ for this case.

We first observe that the stabilizer subgroup $H_0 \subset \CH$ for
the trivial connnection is given as in \NEWHO\ by
\eqn\NEWHOI{ H_0 \,=\, U(1)^2 \times G\,,}
where the factor $G$ arises from the constant gauge transformations on
$M$.  Because $H_0$ decomposes as a product, we decompose an arbitrary
element $\psi$ of its Lie algebra ${\Fh_0 = \BR \oplus \Fg \oplus
\BR}$ as
\eqn\NEWL{ \psi \,=\, p + \phi + a\,,}
where $p$ and $a$ generate the $U(1)$ factors of $H_0$ and $\phi$ is an
element of $\Fg$, according to the notation of Section 3.

As in \AUXHDGII, the exceptional bundles $\CE_0$ and $\CE_1$ at the
trivial connection are now given by
\eqn\BIGES{\eqalign{
\CE_0 \,&=\, \bigoplus_{t \ge 1} H^0_{\bar\partial}(\Sigma, \Fg
\otimes (\CL^t \oplus \CL^{-t}))\,,\cr
\CE_1 \,&=\, \bigoplus_{t \ge 1} H^1_{\bar\partial}(\Sigma,
\Fg \otimes (\CL^t \oplus \CL^{-t}))\,.}}
Here $\CL = \CL_0^d$ is the line $V$-bundle on $\Sigma$ which
describes $M$.

From our localization formula \ZVWHKIV\ in Section 4, the contribution
of the trivial connection to $Z(\epsilon)$ is now given formally by
the following integral over $\Fh_0$,
\eqn\TRV{ Z(\epsilon)\Big|_{\{0\}} \,=\, {{(2 \pi \epsilon)} \over
{\Vol(G)}} \, \int_{\Fh_0} \left[{{d\psi} \over {2\pi}}\right] \,
e(\psi) \, \exp{\left[-i \left(\gamma_0, \psi\right) - {{i
\epsilon} \over 2} \left(\psi, \psi\right)\right]}\,,}
where $e(\psi)$ is an infinite-dimensional determinant,
\eqn\EPSI{ e(\psi) \,=\, \det\left({{\psi} \over {2
\pi}}\Big|_{\CE_0}\right) \, \det\left({{\psi} \over {2
\pi}}\Big|_{\CE_1}\right)^{-1}\,.}
In normalizing \TRV, we have cancelled the factor $\Vol(U(1)^2)$ that
appears in the relative normalization \PREZCH\ against a corresponding
factor in $1/\Vol(H_0)$ from the localization formula \ZVWHKIV,
leaving the factor $1/\Vol(G)$.  We have also included the factor $(2
\pi \epsilon)$ from \PREZCH.

\medskip\noindent{\it Evaluating $e(\psi)$}\smallskip

We first evaluate $e(\psi)$, which turns out to be the only
non-trivial piece of our computation.  From \EPSI, we see that
$e(\psi)$ is described formally by the determinant of the operator
$\psi$ acting on the infinite-dimensional vector spaces $\CE_0$ and
$\CE_1$.  So to evaluate $e(\psi)$, we will have to decide how
to define such a determinant.

Here we employ the standard analytic technique of
zeta/eta-function regularization to define the various infinite
products that represent the determinant $e(\psi)$.  This choice is
somewhat {\it ad hoc}, and our best justification for it is the
fact that it eventually leads to agreement with the results of
Lawrence and Rozansky.  However, this method of regularization
does feature in the usual perturbative approach to Chern-Simons
gauge theory, for instance in the one-loop computation in
\WittenHF.  So, optimistically, one might be able to better
justify the use of zeta/eta-function regularization here by
comparing the localization computation with conventional
perturbation theory.  We make a few further remarks in Section
5.3.

Since the general element of $\CH$ acts on $\bar\CA$ as
\eqn\DAI{ \delta A = d_A \phi + p \, \lie_R A\,,}
we see that the determinants in $e(\psi)$ can be written concretely in
terms of $p$ and $\phi$ in \NEWL\ as
\eqn\DAII{ e(\psi) \,=\, e(p, \phi) \,=\, \det\left[{1 \over {2 \pi}}
\left(p \lie_R - \left[ \phi, \,\cdot\,
\right]\right)\Big|_{\CE_0}\right] \,
\det\left[{1 \over {2 \pi}} \left(p \lie_R - \left[ \phi, \,\cdot\,
\right]\right)\Big|_{\CE_1}\right]^{-1}\,.}
In particular, $e(p, \phi)$ does not depend on $a$ in $\Fh_0$, since
this generator acts trivially.  This fact is important later.

As $\lie_R$ acts on sections of $\CL^t$ with eigenvalue $-2 \pi i t$,
we rewrite $e(p, \phi)$ as a product over the non-zero eigenvalues of
$\lie_R$ as
\eqn\DAIII{ e(p, \phi) \,=\, \prod_{t \neq 0} \, \det\left[\left(-i t p  -
{{\left[ \phi, \,\cdot\, \right]} \over {2
\pi}}\right)\Bigg|_{\Fg}\right]^{\chi(\CL^t)}\,.}
Here $\chi(\CL^t)$ is the Euler character of $\CL^t$, so that we
incorporate the cancellation between the action of $\psi$ on elements
of $\CE_0$ and $\CE_1$, and the determinant in \DAIII\ indicates the
determinant with respect to the action on $\Fg$.

We now evaluate this finite-dimensional determinant on $\Fg$.  This
determinant is invariant under the adjoint action on $\Fg$, and
without loss we assume that $\phi$ lies in the Lie algebra $\Ft$ of a
maximal torus $T$ of $G$.  In this case, if $\beta$ denotes a root of
$\Fg$ and $g_\beta$ the corresponding generator of $\Fg$, then the
adjoint action of $\phi$ on $g_\beta$ is given by ${[\phi, g_\beta]
\,=\, i \, \langle\beta,\phi\rangle \, g_\beta}$.  Thus diagonalizing
the adjoint action of $\phi$, we see that
\eqn\DETG{\eqalign{
\det{\left( -i t p - {{\left[ \phi, \,\cdot\, \right]} \over {2
\pi}}\right)}\Bigg|_{\Fg} \,&=\, \left(-i t p\right)^{\Delta_G} \,
\prod_{\beta} \left(1 + {{\langle\beta, \phi\rangle} \over {2 \pi t
p}}\right)\,,\cr
&=\, \left(-i t p\right)^{\Delta_G} \, \prod_{\beta > 0} \left(1 -
\left( {{\langle\beta, \phi\rangle} \over {2 \pi t
p}}\right)^2\right)\,.\cr}}
Here $\Delta_G$ denotes the dimension of $G$.  In the first line of
\DETG, the product runs over all the roots $\beta$ of $\Fg$, whereas
in the second line of \DETG, we have grouped together the two terms
arising from the roots $\pm \beta$ and rewritten the product over a
set of positive roots.

Now from \DAIII\ and \DETG, we rewrite $e(p, \phi)$ as
\eqn\DAIV{ e(p, \phi) \,=\, \exp{\left(-{{i \pi} \over 2}
\eta\right)} \cdot \prod_{t \ge 1} \, \left|\left(t p\right)^{\Delta_G} \,
\prod_{\beta > 0} \left(1 - \left({{\langle\beta, \phi\rangle} \over
{2 \pi t p}}\right)^2\right)\right|^{\chi(\CL^t) + \chi(\CL^{-t})}\,.}
Here $\exp{\left(-{{i \pi} \over 2} \eta\right)}$ represents the phase
of $e(p, \phi)$, which involves an infinite product of factors $\pm
i$, and the product written explicity in \DAIV\ represents the norm.
We first evaluate this norm, as the quantity $\eta$ is much more
delicate to determine.

To start, we evaluate the exponent that appears in \DAIV.  By the
Riemann-Roch theorem in \RR,
\eqn\RRVG{ \chi(\CL^t) + \chi(\CL^{-t}) \,=\, \deg(\CL^t) +
\deg(\CL^{-t}) + 2\,.}
In general, the degree of a line $V$-bundle is {\it not}
multiplicative, so that $\deg(\CL^t) \neq t \deg(\CL)$, and the first
two terms on the right of \RRVG\ do not necessarily cancel as they do
for ordinary line bundles.

So we must work a little bit to simplify \RRVG.  As we now show, this
exponent can be simplified as
\eqn\TWOCH{  \chi(\CL^t) + \chi(\CL^{-t}) \,=\,  2 - N +
\sum_{j=1}^N \varphi_{\alpha_j}(t)\,,}
where $\varphi_{\alpha_j}(t)$ is an arithmetic function which takes
the value $1$ if $\alpha_j$ divides $t$ and is $0$ otherwise,
\eqn\NEWAF{\eqalign{
\varphi_{\alpha_j}(t) \,&=\, 1 \quad \hbox{ if } \quad \alpha_j \, |
\, t\,,\cr
&=\, 0 \quad \hbox{ otherwise}\,.\cr}}

To deduce \TWOCH, we suppose that the line $V$-bundle $\CL^t$ is
characterized on $\Sigma$ by isotropy invariants $\gamma_j$, where
\eqn\GMAS{ \gamma_j \equiv t \, \beta_j \,\, \mod \, \alpha_j\,,\quad
0 \le \gamma_j < \alpha_j\,,}
and, as before, the isotropy invariants $\beta_j$ characterize the
line $V$-bundle $\CL$ itself.  From \CHRNCL, the degree of
$\CL^t$ is given in terms of the first Chern class, which is
multiplicative, and $\gamma_j$ as
\eqn\DEGL{ \deg(\CL^t) \,=\, t \, c_1(\CL) - \sum_{j=1}^N
{{\gamma_j} \over {\alpha_j}}\,.}

On the other hand, the isotropy invariants $\bar\gamma_j$ for the
inverse line $V$-bundle $\CL^{-t}$ are given by
\eqn\BARGAZ{  \bar\gamma_j \equiv -t \, \beta_j \,\, \mod \,
\alpha_j\,,\quad 0 \le \bar\gamma_j < \alpha_j\,,}
so that in terms of $\gamma_j$,
\eqn\BARGA{\eqalign{
\bar\gamma_j \,&=\, \alpha_j - \gamma_j \quad \hbox{ if }
\gamma_j \neq 0\,,\cr
&=\, \gamma_j \,=\, 0 \quad \hbox{ otherwise}\,.\cr}}

We note from \GMAS\ that $\gamma_j$ vanishes whenever ${t \beta_j
\equiv 0 \,\, \mod \, \alpha_j}$.  Because $\beta_j$ is relatively prime
to $\alpha_j$ by assumption, the
vanishing of $\gamma_j$ is then equivalent to the condition that
$\alpha_j$ divide $t$, so that
\eqn\GMASII{ \gamma_j = 0 \quad\Longleftrightarrow\quad \alpha_j \,|\,
t\,.}

Thus, using the arithemetic function $\varphi_{\alpha_j}(t)$ defined
in \NEWAF\ in conjunction with \BARGA\ and \GMASII, we see that the
degree of $\CL^{-t}$ can be written as
\eqn\DEGLII{\eqalign{
\deg(\CL^{-t}) \,&=\, -t \, c_1(\CL) - \sum_{j=1}^N
{{\bar\gamma_j} \over {\alpha_j}}\,,\cr
&=\, -t \, c_1(\CL) - \sum_{j=1}^N \left(1 - {{\gamma_j} \over
{\alpha_j}} - \varphi_{\alpha_j}(t)\right)\,.\cr}}
From \RRVG, \DEGL, and \DEGLII, we immediately deduce \TWOCH.

Consequently, $e(p, \phi)$ now becomes
\eqn\DAV{\eqalign{
e(p, \phi) \,&=\, \exp{\left(-{{i \pi} \over 2}
\eta\right)} \cdot \prod_{t \ge 1} \, \left|\left(t
p\right)^{\Delta_G} \, \prod_{\beta > 0} \left(1 - \left(
{{\langle\beta, \phi\rangle} \over {2 \pi t
p}}\right)^2\right)\right|^{2 - N + \sum_{j=1}^N
\varphi_{\alpha_j}(t)}\,,\cr
&=\, \exp{\left(-{{i \pi} \over 2} \eta\right)} \cdot f_0(p, \phi)^2
\cdot \prod_{j=1}^N \, \left|{{f_{\alpha_j}(p, \phi)} \over {f_0(p,
\phi)}}\right|\,,\cr}}
where
\eqn\IDAV{
f_0(p, \phi) \,=\, \prod_{t \ge 1} \, \left[\left(t p\right)^{\Delta_G}
\, \prod_{\beta > 0} \left(1 - \left({{\langle\beta, \phi\rangle}
\over {2 \pi t p}}\right)^2\right)\right]\,,}
and $f_{\alpha_j}$ is related to $f_0$ by
\eqn\ARIDII{ f_{\alpha_j}(p, \phi) \,=\, f_0(\alpha_j \cdot p,
\phi)\,.}
In deducing \DAV\ from \IDAV\ and \ARIDII, we apply the following
arithmetic identity, which holds for an arbitrary function $f(t)$,
\eqn\ARID{ \prod_{t \ge 1} \, f(t)^{\varphi_{\alpha_j}(t)} \,=\, \prod_{t
\ge 1} \, f(\alpha_j \cdot t)\,.}

We finally evaluate the infinite product which defines $f_0(p,
\phi)$.  We use the well known identity below,
\eqn\PRODS{ {{\sin(x)} \over x} \,=\, \prod_{t \ge 1} \, \left(1 - {{x^2}
\over {\pi^2 t^2}}\right)\,,}
and we use the Riemann zeta-function $\zeta$ to define trivial, but
infinite, products
\eqn\PRODSII{\eqalign{
&\prod_{t \ge 1} \, p^{\Delta_G} \,=\, \exp{\left(\Delta_G \ln{p}
\cdot \zeta\left(0\right)\right)} \,=\, p^{- \Delta_G/2}\,,\cr
&\prod_{t \ge 1} \, t^{\Delta G} \,=\, \exp{\left(- \Delta_G \cdot
\zeta'(0)\right)} \,=\, (2 \pi)^{\Delta_G/2}\,.\cr}}
So from \PRODS\ and \PRODSII, we evaluate $f_0(p, \phi)$ to be
\eqn\ARIDIII{\eqalign{
f_0(p, \phi) \,&=\, \left({p \over {2
\pi}}\right)^{-\Delta_G/2} \, \prod_{\beta > 0} \, \left[ {{2 p} \over
{\langle\beta, \phi\rangle}} \, \sin\left({{\langle\beta, \phi\rangle}
\over {2 p}}\right)\right]\,,\cr
&=\, \left(2 \pi\right)^{\Delta_G/2} \, p^{-\Delta_T/2} \,
\prod_{\beta > 0} \, \left[ {2 \over {\langle\beta, \phi\rangle}} \,
\sin\left({{\langle\beta, \phi\rangle} \over {2 p}}\right)\right]\,.\cr}}
Here $\Delta_T$ denotes the dimension of the maximal torus $T$ of $G$
(hence the rank of $G$), and in passing to the second line of
\ARIDIII\ we just pull the factors of $p$ outside the product over
the positive roots of $G$.

From \DAV, \ARIDII, and \ARIDIII, we finally evaluate $e(p, \phi)$ to
be
\eqn\DAVI{\eqalign{
e(p, \phi) \,&=\, \exp{\left(-{{i \pi} \over 2} \eta\right)} \cdot
{{\left(2 \pi\right)^{\Delta_G}} \over {(p \, \sqrt{P})^{\Delta_T}}}
\,\times\,\cr
&\times\, \prod_{\beta > 0} \, \langle\beta, \phi\rangle^{-2} \,
\left|2 \sin\left({{\langle\beta, \phi\rangle} \over {2
p}}\right)\right|^{2 - N} \, \prod_{j=1}^N \, \left|2
\sin\left({{\langle\beta, \phi\rangle} \over {2 \alpha_j
p}}\right)\right|\,,\cr}}
where $P$ is defined in \QRZE\ as the product of all the $\alpha_j$.

\medskip\noindent{\it Evaluating $\eta$ and the Quantum Shift in the
Chern-Simons Level}\smallskip

We now evaluate the phase factor $\exp{\left(-{{i \pi} \over 2}
\eta\right)}$, from which we will find the famous quantum shift in
the Chern-Simons level $k \to k + \check{c}_\Fg$, where
$\check{c}_\Fg$ is the dual Coxeter number of $\Fg$.  For instance, we
recall that in the case $G=SU(r+1)$, $\check{c}_\Fg = r+1$.

To start, we consider the operator \eqn\BIGOP{ {i \over {2\pi}}
\left(p \lie_R - \left[ \phi, \,\cdot\, \right]\right)\,,} acting
on the vector spaces $\CE_0$ and $\CE_1$ in \BIGES.  The spectrum
of this operator is real, so at least formally, we see from the
definition of $e(p, \phi)$ in \DAII\ that the phase $\eta$ is
given by \eqn\BIGOPI{ \eta \, \approx \, \sum_{\lambda_{(0)} \neq
0} \, \sgn(\lambda_{(0)}) - \sum_{\lambda_{(1)} \neq 0} \,
\sgn(\lambda_{(1)})\,,} where $\lambda_{(0)}$ and $\lambda_{(1)}$
range, respectively, over the eigenvalues of the operator in
\BIGOP\ acting on $\CE_0$ and $\CE_1$.

We have not written \BIGOPI\ with an equality because the sums on
the right of \BIGOPI\ are ill-defined without a regulator.  To
regulate these sums, we follow the philosophy of \APS\ and
introduce the eta-function \eqn\BIGETA{ \eta_{(p, \phi)}(s) \,=\,
\sum_{\lambda_{(0)} \neq 0} \, \sgn(\lambda_{(0)}) \,
|\lambda_{(0)}|^{-s} - \sum_{\lambda_{(1)} \neq 0} \,
\sgn(\lambda_{(1)}) \, |\lambda_{(1)}|^{-s}\,.} Here $s$ is a
complex variable.  When the real part of $s$ is sufficiently
large, the sums in \BIGETA\ are absolutely convergent so that
$\eta_{(p, \phi)}(s)$ is defined in this case.  Otherwise,
$\eta_{(p, \phi)}(s)$ is defined by analytic continuation in the
$s$-plane. Assuming that the limit $s \to 0$ exists, we then set
\eqn\DETA{ \eta = \eta_{(p, \phi)}(0)\,.} Thus, $\eta$ is
basically the classic eta-invariant of \APS\ which is here
associated to the operator in \BIGOP\ acting on the virtual vector
space $\CE_0 \ominus \CE_1$, where the ``$\ominus$'' simply
indicates the relative sign in \BIGETA.

In our problem, because we explicitly know the spectrum of the
operator in \BIGOP, we can directly evaluate $\eta_{(p, \phi)}(0)$
without too much work.  One advantage of this direct approach is that
it very concretely displays the origin of the finite shift in the
Chern-Simons level $k$, a very subtle quantum effect to understand
otherwise.

Ultimately this shift in $k$ arises because, despite what might be
one's naive expectation from \BIGOPI, $\eta$ depends nontrivially on
$p$ and $\phi$.  To isolate this interesting functional dependence of
$\eta_{(p, \phi)}(0)$ on $p$ and $\phi$, we observe that, for $s=0$, the
sum in \BIGETA\ is invariant under an overall scaling of the
eigenvalues $\lambda_{(0)}$ and $\lambda_{(1)}$, so that $\eta_{(p,
\phi)}(0)$ is invariant under an overall scaling of the operator itself in
\BIGOP.  In particular, so long as $p > 0$ (as holds when we later set
${p = 1 / \epsilon}$), we are free to rescale the operator in \BIGOP\
by $1/p$ without changing $\eta$.

As a technical convenience, we thus introduce another eta-function
$\eta'_{(p, \phi)}(s)$ which is defined as in \BIGETA\ but is
associated to the rescaled operator
\eqn\BIGOPII{ {i \over {2\pi}} \left(\lie_R - \left[{\phi \over
p}, \,\cdot\, \right]\right)\,.}
Because ${\eta = \eta_{(p, \phi)}(0) = \eta'_{(p, \phi)}(0)}$, we see from
\BIGOPII\ that $\eta$ can only depend on $p$ and $\phi$ in the
combination $\phi / p$.

We also introduce the eta-function $\eta_0(s)$ which is associated
to the constant operator $i \lie_R / 2 \pi$, and to isolate the
functional dependence of $\eta$ on $p$ and $\phi$ we define
\eqn\DLETA{ \delta\eta(p, \phi) \,=\, \eta'_{(p, \phi)}(0) -
\eta_0(0)\,.} As we now compute directly, \eqn\DLETAII{
\delta\eta(p, \phi) \,=\, - {{\check{c}_\Fg} \over {2 (\pi p)^2}}
\left({d \over P}\right) \Tr(\phi^2)\quad{\rm mod}~2.}  The role of
the mod 2 terms is to remove the absolute value bars $|\cdot|$
that appear in \DAVI, so that $e(p, \phi)$ depends analytically on
$p$ and $\phi$ as its definition suggests.

Of course, $\eta$ itself is given by ${\eta = \delta\eta(p, \phi) +
\eta_0(0)}$.  We also discuss $\eta_0(0)$, though this constant
is much less interesting than $\delta\eta(p, \phi)$.

\medskip\noindent{\it A Warmup Computation on $S^1$}\smallskip

Before we directly evaluate $\delta\eta$, $\eta_0(0)$, and $\eta$ for
the case at hand, we find it useful to warm up with a simpler example,
originally presented in \refs{\APS, II}.  Thus, we consider the
eta-function $\eta_\nu(s)$ which is associated to the operator
$D_\nu$ acting on functions on $S^1$,
\eqn\LITETA{ D_\nu \,=\, {i \over {2 \pi}} {d \over {d x}} \,+\, \nu\,.}
Here $\nu$ is a real parameter in the interval $0 < \nu < 1$,
and $x$ is a coordinate on $S^1$ with period $2 \pi$.  If we wish, we can
equivalently consider $D_\nu$ as the covariant derivative acting on
sections of a  flat $U(1)$ bundle over $S^1$ whose connection has holonomy
parametrized by $\nu$.

Clearly the eigenvalues $\lambda$ of $D_\nu$ are given by ${\lambda
= t + \nu}$ as $t$ runs over all integers.  So we compute
\eqn\LITETAI{\eqalign{
\eta_\nu(s) \,&=\, \sum_{\lambda} \, \sgn(\lambda) \,
|\lambda|^{-s}\,,\cr
&=\, \sum_{t \ge 0} \, {1 \over {(t + \nu)^s}} - \sum_{t \ge 1} \, {1
\over {(t - \nu)^s}}\,,\cr
&=\, {1 \over {\nu^s}} - \sum_{t \ge 1} \, {{2 \nu s} \over {t^{s+1}}} +
\sum_{t \ge 1} \, s \cdot \CO\left({1 \over {t^{s+2}}}\right)\,.\cr}}
In passing from the second to the third lines of \LITETAI, we apply
the binomial expansion, and we collect into $\CO(1 / t^{s+2})$ the
terms in this expansion for which the sum over $t$ is absolutely
convergent near $s = 0$.  Thus, when we evaluate $\eta_\nu(s)$ at
$s=0$, the last term of \LITETAI\ vanishes.

On the other hand, for the term involving the sum over $1/t^{s+1}$, we
have
\eqn\RESZ{ \sum_{t \ge 1} \, {{2 \nu s} \over {t^{s+1}}} \,=\, 2
\nu s \, \zeta(1 + s)\,.}
Because $\zeta(1+s)$ has a simple pole with residue $1$ at $s=0$, we
see that \RESZ\ makes a non-zero contribution to $\eta_\nu(0)$, and
\eqn\LITETAII{ \eta_\nu(0) \,=\, 1 - 2 \nu\,.}
Physically the term involving $\nu$ arises as a finite
renormalization effect, due to the divergence in the sum over
eigenvalues in \RESZ.

\medskip\noindent{\it The Computation of $\eta$ on $M$}\smallskip

Given the formal similarity of the operators in \BIGOPII\ and \LITETA,
we now evaluate $\eta_{(p, \phi)}(0)$ just as in our warmup
computation on $S^1$.   In the case at hand, we must consider the
eigenvalue multiplicities which are associated to the dimensions of the
Dolbeault cohomology groups $H^0_{\bar\partial}(\Sigma,\CL^t)$ and
$H^1_{\bar\partial}(\Sigma,\CL^t)$, and as in our earlier computation
we must also consider the eigenvalues of the adjoint action of $\phi$ on
$\Fg$.  Taking these considerations into account, we find the following
compact expression for $\eta'_{(p, \phi)}(s)$,
\eqn\ETAPR{\eqalign{
\eta'_{(p, \phi)}(s) \,&=\, \sum_{t = -\infty}^{+\infty}
\sum_{\beta} \; \chi(\CL^t) \,
\sgn\left(\lambda\left(t,\beta\right)\right) \,
|\lambda(t,\beta)|^{-s}\,,\cr
\lambda(t,\beta) \,&=\, t + {{\langle\beta, \phi\rangle} \over{2 \pi
p}}\,.}}
Here the sum over $\beta$ is again a sum over the roots of $\Fg$,
including the roots $\beta = 0$ from the Cartan subalgebra.  We note
that the appearance of the Euler character $\chi(\CL^t)$ in \ETAPR\
accounts both for the multiplicities and the relative signs of the
eigenvalue contributions from $\CE_0$ and $\CE_1$ in \BIGETA.

We can give a similar, simpler expression for $\eta_0(s)$,
\eqn\ETAZ{\eqalign{
\eta_0(s) \,&=\, \sum_{t \neq 0} \sum_\beta \, \chi(\CL^t)
\, \sgn(t) \, |t|^{-s}\,,\cr
&=\, \sum_{t \ge 1} \sum_{\beta} \, {{\chi(\CL^t) - \chi(\CL^{-t})} \over
{t^s}}\,.\cr}}
In the general orbifold case, the index difference $\chi(\CL^t) -
\chi(\CL^{-t})$ that arises in \ETAZ\ appears to be a somewhat
complicated arithmetic function of $t$, in contrast to our simple
expression for the index sum in \TWOCH, and we will not evaluate
$\eta_0(0)$ in complete generality here.

However, if we consider the special case of a degree $d$ line-bundle
$\CL$ over a smooth Riemann surface $\Sigma$, then the
Riemann-Roch theorem immediately implies that
\eqn\INDXD{ \chi(\CL^t) - \chi(\CL^{-t}) \,=\, 2 d t\,,}
independent of the genus of $\Sigma$.  So in this special case, we
have from \ETAZ\ that
\eqn\ETAZII{\eqalign{
\eta_0(s) \, &=\, \Delta_G \, \sum_{t \ge 1} \, {{2 d t} \over
{t^s}}\,,\cr &=\, 2 d \Delta_G \, \zeta(s-1)\,.\cr}}
Thus,
\eqn\ETAZII{
\eta_0(0) \,=\, 2 d \Delta_G \, \zeta(-1) \,=\, -{{d \Delta_G} \over
6}\,.}

Having discussed $\eta_0(0)$, we now compute the more interesting
quantity $\delta\eta(p, \phi)$ in \DLETA.  Upon expressing \ETAPR\ as
in \ETAZ\ and collecting terms, we find that
\eqn\ETAPRII{\eqalign{
\eta'_{(p, \phi)}(s) - \eta_0(s) \,&=\, \sum_{t \ge
0} \sum_{\beta > 0} \, \left(\chi(\CL^t) - \chi(\CL^{-t})\right) \cdot
\left[ {1 \over {\left(t + {{\langle\beta, \phi\rangle} \over {2 \pi
p}}\right)^s}} - {1 \over {t^s}} \right] \,+\,\cr
&+\, \sum_{t \ge 1} \sum_{\beta > 0} \, \left(\chi(\CL^t) -
\chi(\CL^{-t})\right) \cdot \left[ {1 \over
{\left(t - {{\langle\beta, \phi\rangle} \over {2 \pi
p}}\right)^s}} - {1 \over {t^s}} \right]\,.\cr}}
In writing this expression, we assume without loss that the condition
below holds for each positive root $\beta$,
\eqn\SPFL{ 0 < {{\langle\beta, \phi\rangle} \over {2 \pi p}} < 1\,.}
Otherwise, when the quantity in \SPFL\ undergoes an integral shift,
then the overall phase $\exp{(-i \pi \eta / 2)}$ of $e(p, \phi)$
simply picks up a sign so as to effectively remove the absolute value
bars $|\cdot|$ appearing in \DAVI.  Hence $e(p, \phi)$ depends
analytically on $p$ and $\phi$.

We now observe from our general expressions \DEGL\ and \DEGLII\
for $\deg(\CL^t)$ and $\deg(\CL^{-t})$ that the index difference
in \ETAPRII\ depends generally on $t$ as \eqn\INDXDII{ \chi(\CL^t)
- \chi(\CL^{-t}) \,=\, 2 t \left({d \over P}\right) + \CO(t^0).}
We have used the fact that $c_1(\CL) = d / P$, since
$\CL=\CL_0^d$, and $c_1(\CL_0)=\prod_j{1/\alpha_j}= 1/P$.

If we now consider the binomial expansion of the denominators in
\ETAPRII, we see immediately that no contribution at $s=0$ can arise
from the terms of order $t^0$ in \INDXDII.  The leading terms in the
expansion which arise from these $\CO(t^0)$ terms are proportional to
${\pm \langle\beta,\phi\rangle/(2 \pi p) \cdot t^{-(s+1)}}$, and such
terms linear in $\phi$ cancel between the two sums in \ETAPRII.  The same
cancellation occurs between the leading expansion terms which arise
from the term linear in $t$ in \INDXDII, and fundamentally these
cancellations reflect the fact that no invariant linear function of
$\phi$ exists.

Thus, expanding the denominators in \ETAPRII\ to second order, we find
\eqn\ETAPRIII{ \eta'_{(p, \phi)}(s) - \eta_0(s) \,=\, 2 \left({d \over
P}\right) \, \sum_{t \ge 1} \sum_{\beta > 0} \, \left({{\langle\beta,
\phi\rangle} \over {2 \pi p}}\right)^2 \cdot {{s (s+1)} \over
{t^{s+1}}} \,+\, \sum_{t \ge 1} \sum_{\beta > 0} \, s \cdot \CO\left({1
\over {t^{s+2}}}\right)\,.}
We evaluate \ETAPRIII\ at $s=0$ to determine $\delta\eta(p,
\phi)$, which is thus given by
\eqn\DLETA{ \delta\eta(p, \phi) \,=\, 2 \left({d \over P}\right) \,
\sum_{\beta > 0} \, \left({{\langle\beta, \phi\rangle} \over {2 \pi
p}}\right)^2\,.}

To simplify the sum over roots on the right side of \DLETA, we note
that this sum defines an invariant quadratic polynomial of $\phi$
and hence must be proportional to $\Tr(\phi^2)$.  When $\Fg$ is
simply-laced, we have the following identity, as shown for instance
in \refs{\Bourbaki, VI},
\eqn\BBAKI{ \sum_{\beta > 0} \, \langle\beta, \phi\rangle^2 \,=\, -
\check{c}_\Fg \, \Tr(\phi^2)\,.}
Together, \DLETA\ and \BBAKI\ imply the main result in \DLETAII.

Thus the full determinant $e(p, \phi)$ is now given by
\eqn\DLETAIV{\eqalign{
&e(p, \phi) \,=\, \exp{\left(-{{i \pi} \over 2} \, \eta_0(0)\right)}
\cdot {{\left(2 \pi\right)^{\Delta_G}} \over {(p \, \sqrt{P})^{\Delta_T}}}
\,\times\,\cr
&\times\, \exp{\left[{{i \, \check{c}_\Fg} \over {4 \pi p^2}} \,
\left({d \over P}\right) \, \Tr(\phi^2)\right]} \, \prod_{\beta > 0}
\, \langle\beta, \phi\rangle^{-2} \, \left[2 \sin\left({{\langle\beta,
\phi\rangle} \over {2 p}}\right)\right]^{2 - N} \, \prod_{j=1}^N \,
\left[2 \sin\left({{\langle\beta, \phi\rangle} \over {2 \alpha_j
p}}\right)\right]\,.\cr}}
As we will see directly, the exponential term involving $\Tr(\phi^2)$
in $e(p, \phi)$ describes the quantum shift in the Chern-Simons level $k$.

\medskip\noindent{\it Evaluating the Integral over $\Fh_0$}\smallskip

We are finally left to consider the integral over $\Fh_0$ in \TRV.  We
first observe that the norm $(\psi,\psi)$ appearing in the exponent of
the integrand there is given explicity by
\eqn\NORMPS{\eqalign{
\left(\psi, \psi\right) \,&=\, - \int_M \kappa\^d\kappa \, \Tr(\phi^2)
- 2 p a\,,\cr
&=\, - \left({d \over P}\right) \Tr(\phi^2) - 2 p a\,.\cr}}
In passing to the second line of \NORMPS, we use the fact that $\phi$
is constant so that the integral over $M$ simply evaluates to
$c_1(\CL) = d / P$.  Second, we recall from Section 3 that the moment
map at the trivial connection satisfies
\eqn\GAMNO{ \langle\mu,\psi\rangle \,=\, (\gamma_0, \psi) \,=\, a\,.}
Hence the integral over $\Fh_0$ takes the explicit form
\eqn\TRVII{
Z(\epsilon)\Big|_{\{0\}} \,=\, {{(2 \pi \epsilon)} \over {\Vol(G)}}
\, \int_{\Fh_0} \left[{{dp} \over {2\pi}}\right]
\left[{{da} \over {2\pi}}\right] \left[{{d\phi} \over {2\pi}}\right]
\, e(p, \phi) \, \exp{\left[-i a + i \epsilon p a + {{i \epsilon} \over
2} \left({d \over P}\right) \Tr(\phi^2)\right]}\,.}

We now evaluate the integral over $a$, which is easy since $a$ only
appears in the exponent of the integrand in \TRVII.  From a previous
identity \DLTAF, this integral produces the delta function $2
\pi \, \delta(1 - \epsilon p)$.

In turn, we use the delta function to perform the integral over $p$,
setting $p = 1/\epsilon$.  In the process, we cancel the explicit
factor of $2 \pi \epsilon$ which appears in the normalization of
\TRVII, and the integral over $\Fh_0$ simplifies to an integral over
$\Fg$,
\eqn\TRVIII{
Z(\epsilon)\Big|_{\{0\}} \,=\, {1 \over {\Vol(G)}} \, \int_\Fg
\left[{{d\phi} \over {2\pi}}\right]
\, e(\epsilon^{-1}, \phi) \, \exp{\left[{{i \epsilon} \over
2} \left({d \over P}\right)  \Tr(\phi^2)\right]}\,.}

Because the integrand of \TRVIII\ is invariant under the adjoint
action on $\Fg$, we can apply the classical Weyl integral formula to
reduce the integral over $\Fg$ to an integral over the Cartan
subalgebra $\Ft$, in which form we make contact with the results in
\refs{\LawrenceRZ, \MarinoFK}.  In its infinitesimal version, the Weyl
integral formula states that, if $f$ is a function on $\Fg$ invariant
under the adjoint action, then
\eqn\WEYLIF{ \int_{\Fg} \left[ d\phi \right] \, f(\phi) \,=\, {1 \over
{|W|}} \, {{\Vol(G)} \over {\Vol(T)}} \, \int_{\Ft} \left[ d \phi
\right] \, \prod_{\beta > 0} \langle\beta,\phi\rangle^2 \, f(\phi)\,.}
Here $|W|$ is the order of the Weyl group of $G$, and the product over
the positive roots $\beta$ of $G$ appearing on the right of \WEYLIF\ is a
Jacobian factor.

Applying \WEYLIF\ and recalling the form of $E$ in \DLETAIV, we
rewrite \TRVIII\ explicitly as
\eqn\TRVIV{\eqalign{
Z(\epsilon)\Big|_{\{0\}} &= \e{\left(-{{i \pi} \over 2} \,
\eta_0(0)\right)} \, {1 \over {|W|}} \, {1 \over {\Vol(T)}} \,
\left({\epsilon \over {\sqrt{P}}}\right)^{\Delta_T} \, \int_\Ft
\left[d\phi\right] \exp{\left[{{i \epsilon} \over 2} \left({d \over
P}\right) \left(1 + {{\epsilon \, \check{c}_\Fg} \over {2
\pi}}\right) \Tr(\phi^2)\right]} \,\times\cr
&\times\, \prod_{\beta > 0}
\, \left[2 \sin\left({{\epsilon \, \langle\beta, \phi\rangle} \over
2}\right)\right]^{2 - N} \, \prod_{j=1}^N \, \left[2
\sin\left({{\epsilon \, \langle\beta, \phi\rangle} \over {2
\alpha_j}}\right)\right]\,.\cr}}
We finally make the change of variables ${\phi \rightarrow \epsilon
\phi}$ to remove some of the extraneous factors of $\epsilon$ in front
of \TRVIV, so that
\eqn\TRVV{\eqalign{
&Z(\epsilon)\Big|_{\{0\}} \,=\, \exp{\left(-{{i \pi} \over 2} \,
\eta_0(0)\right)} \, {1 \over {|W|}} \,
{1 \over {\Vol(T)}} \, \left({1 \over {\sqrt{P}}}\right)^{\Delta_T}
\,\times\,\cr
&\times \, \int_\Ft \left[d\phi\right] \,  \exp{\left[{i \over {2
\epsilon_r}} \left({d \over P}\right)  \Tr(\phi^2)\right]} \,
\prod_{\beta > 0} \, \left[2 \sin\left({{\langle\beta, \phi\rangle}
\over 2}\right)\right]^{2 - N} \, \prod_{j=1}^N \, \left[2
\sin\left({{\langle\beta, \phi\rangle} \over {2
\alpha_j}}\right)\right]\,.\cr}}
Here we introduce the usual renormalized coupling $\epsilon_r$,
\eqn\EPSR{ \epsilon_r \,=\, {{2 \pi} \over {k + \check{c}_\Fg}}\,,}
to absorb the explicit shift in the coefficient of $\Tr(\phi^2)$ that
arises from the phase $\delta\eta$ and that appears in \TRVIV.

As it stands, the integral over $\Ft$ in \TRVV\ has oscillatory, as
opposed to exponentially damped, behavior at infinity due to purely
imaginary Gaussian factor involving $\Tr(\phi^2)$.  Such oscillatory
Gaussian integrals typically arise in quantum field theory.  For instance,
we saw an earlier example in our path integral manipulations at the
end of Section 3.1, when we integrated out the auxiliary scalar field
$\Phi$ that appeared there.

Exactly as in Section 3.1, the standard analytic prescription to
define such an oscillatory integral is to shift the integration
contour slightly off the real axis.  That is, in the context of \TRVV\
we consider the complexification $\Ft \otimes \BC$ of the real Lie
algebra $\Ft$, and we define \TRVV\ by integrating over $\Ft \times (1 - i
\varepsilon)$ for a small real parameter $\varepsilon$.  This $i
\varepsilon$ prescription has the added virtue that the new contour
avoids any poles of the integrand on the real axis that generally
occur for $N > 2$.

Once we define \TRVV\ with the $i \varepsilon$ prescription, we are
free to analytically continue the contour to lie along the diagonal
$\Ft \times \e{-i \pi / 4}$, so that the Gaussian factor
in \TRVV\ becomes purely real and negative-definite\foot{We recall
that $\Tr$ is a negative-definite form.}.  To make contact with the
result of Lawrence and Rozansky in \ZLWRZ, we finally make another
change of variables ${\phi \,\rightarrow\, i \phi}$, so that
\eqn\TRVI{\eqalign{
&Z(\epsilon)\Big|_{\{0\}} \,=\, \exp{\left(-{{i \pi} \over 2} \,
\eta_0(0)\right)} \, {1 \over {|W|}} \,
{{(-1)^{(\Delta_G - \Delta_T)/2}}\over {\Vol(T)}} \, \left({1 \over
{i \, \sqrt{P}}}\right)^{\Delta_T} \,\times\,\cr
&\times \, \int_{\CC \times \Ft} \left[d\phi\right] \,  \exp{\left[-{i
\over {2 \epsilon_r}} \left({d \over P}\right)  \Tr(\phi^2)\right]} \,
\prod_{\beta > 0} \, \left[2 \sinh\left({{\langle\beta, \phi\rangle}
\over 2}\right)\right]^{2 - N} \, \prod_{j=1}^N \, \left[2
\sinh\left({{\langle\beta, \phi\rangle} \over {2
\alpha_j}}\right)\right]\,,\cr}}
where $\CC$ is the diagonal contour $\BR \times \e{{i \pi} \over 4}$,
as in \CCRC.

We immediately see that \TRVI\ has the same form as our earlier
expression in \ZLWRZII\ for the contribution from the trivial connection
in the case $G = SU(2)$, and with a suitable choice of generator for
$\Ft$ one can see that \TRVI\ agrees, up to the overall phase, with
the result of Lawrence and Rozansky.  For general $G$, our expression
takes the same form as that found by Mari\~no in \MarinoFK.

\medskip\noindent{\it The Phase of $Z(\epsilon)$}\smallskip

We now discuss the phase of our result \TRVI\ for the contribution
of the trivial connection to the Chern-Simons path integral.  In
the simplest case that $M$ is described by a smooth line-bundle of
degree $d=n$ over $\BC\BP^1$, we have computed this phase
explicitly, as determined by the constant \eqn\ETAZIII{ \eta_0(0)
\,=\, -{{d \Delta_G} \over 6}\,.} Since we have not performed a
careful analysis of the path integral phases that arise from the
$\eta$ invariant when $M$ is an orbifold, we restrict attention to
the smooth  case in the following.

If we compare our result to the result \ZLWRZII\ of Lawrence and
Rozansky for gauge group $SU(2)$, we see that the overall phase of
$Z(\epsilon)$ which arises naturally from localization does not agree
with the canonical phase.  To be more precise, the result of Mari\~no
\MarinoFK\ in the case of a general gauge group $G$ shows that the
ratio $\exp{(i \, \delta\Psi)}$ between the canonical phase of
$Z(\epsilon)$ and the phase we determine via \ETAZIII\ is given by
\eqn\PHSDIF{\eqalign{
\exp{(i \, \delta\Psi)} \,&=\, \exp{\left({{i \pi \Delta_G} \over 4} -
{{i \pi \Delta_G \check{c}_\Fg} \over {12 (k + \check{c}_\Fg)}} \,
\theta_0 + {{i \pi} \over 2} \eta_0(0)\right)}\,,\cr
&=\,\exp{\left({{i \pi \Delta_G} \over 12} (3 - d) - {{i \pi \Delta_G
\check{c}_\Fg} \over {12 (k + \check{c}_\Fg)}} \, \theta_0\right)}\,.\cr}}
Here $k$ is the Chern-Simons level.  The quantity $\theta_0$ is
defined in general in \QRZE, and in the smooth case we see that
$\theta_0$ is given by
\eqn\THETAN{ \theta_0 \,=\, 3 - d\,.}
Hence the expression in \PHSDIF\ simplifies greatly to
\eqn\PHSDIFII{ \exp{(i \, \delta\Psi)} \,=\, \exp{\left({{i
\pi k \Delta_G} \over {12 (k + \check{c}_\Fg)}} \, (3 - d)\right)}\,.}

As we now explain, the phase discrepancy in \PHSDIFII\ is not
really a discrepancy at all, and it merely reflects the fact that
our path integral computation is effectively performed in a
framing of $M$ which differs from the canonical two-framing of
Atiyah \AtiyahFM, which has been used by Lawrence and Rozansky.
We first recall from \WittenHF\ that the partition function of
Chern-Simons theory generally transforms under a change in the
framing of $M$ by \eqn\FRAMCS{ Z \,\longrightarrow\,
\exp{\left({{i \pi c} \over 12} \, s\right)} \, Z\,,\qquad c \,=\,
{{k \Delta_G} \over {k + \check{c}_\Fg}}\,,\qquad s \in \BZ\,.}
Here $c$ arises as the central charge of the two-dimensional WZW
model associated to the group $G$, and $s$ is an integer that
labels the shift in the frame.  As a result, we see immediately
from \FRAMCS\ that the phase discrepancy \PHSDIFII\ can be
eliminated by a shift in $s = (3 - d)$ units from the canonical
framing of $M$.

Of course, in evaluating the Chern-Simons path integral by
localization, we did not explicitly specify any framing of $M$.
Given the framing ambiguity \FRAMCS\ in $Z$, one might naturally
wonder how we managed to obtain a definite answer for the phase of
$Z$ in the first place.

To answer this question, we observe generally that if $M$ is an
integral homology sphere, then the choice of a locally-free $U(1)$
action on $M$ implies a canonical choice, up to homotopy, of a
framing of $M$.  Concretely, a framing of $M$ amounts to the
choice of three linearly independent, non-vanishing vector fields
on $M$, and the $U(1)$ action on $M$ immediately supplies us with
one such vector field, the generating vector field $R$ of $U(1)$.
We decompose the tangent bundle to $M$ as $TM=L\oplus W$, where
$L$ is a one-dimensional bundle generated by $R$ and $W$ is the
complement.  We are left to make a choice for the other two vector
fields, which must span the rank two sub-bundle $W$ of $TM$ which
lies in the kernel of the contact form $\kappa$. The choice of
these two vector fields amounts to a trivialization of $W$, so if
the Euler class of $W$ is non-zero, $W$ is non-trivial and our
construction fails. However, since the Euler class of $W$ lies in
the cohomology group $H^2(M, \BZ)$, which vanishes for an integral
homology sphere, $W$ is automatically trivial in this case.
Finally, because $W$ has rank two, possible changes of
trivialization of $W$ are classified by homotopy classes of maps
of $M$ to $SO(2)$.  But for a homology sphere $M$ (or even a
rational homology sphere), the space of maps to $SO(2)$ is
connected.\foot{Let $w=du$ be an angular form on $SO(2)\cong S^1$
and let $v:M\to SO(2)$ be any map.   As $M$ is a rational homology
sphere, $v^*(w)$ vanishes in de Rham cohomology, so $v^*(w)=df$
where $f:M\to \Bbb{R}$ is some real-valued function. Because
$\Bbb{R}$ is contractible, we can define a homotopy from $f$ to a
constant map from $M$ to $\Bbb{R}$ by simply setting $f_t=tf$,
$0\leq t\leq 1$. Now let $\pi:\Bbb{R}\to S^1\cong \Bbb{R}/2\pi$ be
the projection.  Then setting $v_t=\pi\circ f_t$, we get the
desired homotopy from $v$ to a constant map from $M$ to $S^1$.}
So, given the choice of the original $U(1)$ action, we produce a
unique framing of $M$ up to homotopy.

More generally, if $M$ is not assumed to be a homology sphere,
then $W$ might be nontrivial. To define the Chern-Simons invariant
of a three-manifold $M$, however, it is not quite necessary to
have a framing of $TM$.  It is enough to have a ``two-framing,'' a
trivialization of $TM\oplus TM$.  We claim that every Seifert
fibration $\pi:M\to \Sigma$ determines a natural two-framing on
$M$ (which might depend on the choice of $\pi$, as a given $M$ may
admit more than one Seifert fibration).  As $TM\oplus TM= L\oplus
L\oplus W\oplus W$, it suffices to trivialize $W\oplus W$. First
of all, $W\oplus W$ has a natural spin structure, the spin bundle
being the sum of exterior powers of $W$. A trivial bundle, which
is a product $M\times V$ for some fixed vector space $V$, also has
a natural spin bundle, namely $M\times C(V)$, where $C(V)$ is a
Clifford module for $V$, which is unique up to isomorphism. Any
trivialization of $W$ determines a spin structure, since a
trivialization of $W$ identifies it with a trivial bundle, which
as we just noted has a natural spin structure.  One condition we
want to put on a trivialization of $W\oplus W$ is that the spin
structure of $W\oplus W$ that it determines should coincide with
the natural one.  The second condition we want is that the
trivialization of $W\oplus W$ should be invariant under the $U(1)$
action on the Seifert manifold.  $W$ is a pullback from some
$SO(2)$ bundle $W_0$ over $\Sigma$, so $W\oplus W$ is the pullback
of $U=W_0\oplus W_0$. The rank four real bundle $U$ has vanishing
$w_1$ and $w_2$ (they are killed by taking two copies of $W_0$),
so it is trivial.  Compatibility with a given spin structure of a
rank $k$ real bundle $U$ -- in our application $k=4$ -- means that
changes of trivialization really come from maps to $Spin(k)$
rather than $SO(k)$.  As $\pi_i(Spin(k))=0$ for $i\leq 2$, $k\geq
3$, a trivial $SO(k)$ bundle $U$ over $\Sigma$ of rank $k\geq 3$
has up to homotopy only one trivialization compatible with a given
spin structure. So finally the Seifert fibration $\pi:M\to \Sigma$
endows $M$ with a natural two-framing (which may differ from its
canonical two-framing \AtiyahFM, which is determined by a different
construction).

In sum, then, a Seifert fibration of a homology
sphere $M$ determines a natural trivialization of the tangent
bundle $TM$, which we will call the Seifert framing, and any
Seifert fibration $\pi:M\to \Sigma$ (even if $M$ is not a homology
sphere) determines a natural trivialization of $TM\oplus TM$,
which we will call the Seifert two-framing.  If $M$ is a Seifert
homology sphere, the Seifert two-framing just arises by applying
the Seifert framing to each copy of $TM$.

Now we  consider in detail the illustrative example $M = S^3$.
$S^3$ has no one  natural framing.  However, if we identify it
with the Lie group $SU(2)$, then it has two equally natural
framings, one which is left-invariant and one which is
right-invariant.  They are exchanged by an orientation-reversing
reflection of $S^3$, so neither one is preferred. In regarding
$S^3$ as a Seifert fibration over $\Bbb{CP}^1$, we write
$\Bbb{CP}^1=S^3/U(1)$, where $U(1)$ is either part of the left
action of $SU(2)$ on itself or part of the right action.  For
either choice of $U(1)$, our construction produces a framing that
is canonically determined by the choice of $U(1)$ generator and so
is invariant under any symmetry that commutes with $U(1)$.  If the
$U(1)$ is part of the left $SU(2)$, then it commutes with the
right $SU(2)$ and so we get the right-invariant framing; and
likewise if the $U(1)$ is part of the right $SU(2)$, we get the
left-invariant framing.

We naturally expect that  the phase of $Z$ in our computation of
the Chern-Simons path integral is based on the Seifert framing. In
view of our direct computation of the phase of $Z$, the Seifert
two-framing of $M$  must differ from the canonical two-framing of
\AtiyahFM\ by $s = (3 - d)$ units. We now give a simple proof of
this fact in the case $M = S^3$ and $d = 1$ (though we will not be
careful about the sign of the shift).

When $M=S^3$, the canonical two-framing  of \AtiyahFM\ can be
described as follows.  It is the trivialization of $TM\oplus TM$
that comes from the left-invariant framing on, say, the first copy
of $TM$ and the right-invariant framing on the second.  (This is
the unique reflection-invariant two-framing of $S^3$, so it must
be the canonical two-framing.)  On the other hand, the Seifert
framing of $M$ is (for a suitable choice of fibration $\pi:S^3\to
\Bbb{CP}^1$) the left-invariant framing of $TM$, so the Seifert
two-framing comes by applying the left-invariant framing to each
of the two copies of $TM$.  Hence the comparison between the
Seifert two-framing and the canonical one is the same as the
comparison between the left-invariant two-framing and the
right-invariant two-framing for a single copy of $TM$.

The right-invariant framing of $S^3$ is determined by the basis of
right-invariant one-forms $\theta = dg \,g^{-1}$, while the
left-invariant framing is determined by the basis of
left-invariant one-forms $\hat\theta=g^{-1}dg$.  We are supposed
to compare them by writing $\theta= T \hat\theta T^{-1}$, where
$T$ is a map from $M$ to $SO(3)$.  Such a map has a ``degree,'' an
integer which measures by how many units the two framings differ.
Clearly, in this case, $T=g$, so $T$ is the ``identity'' map from
$S^3\cong SU(2)$ to itself.  This map is of degree 1 as a map to
$SU(2)$, but as a map from $S^3$ to $SO(3)=SU(2)/\Bbb{Z}_2$, it is
of degree 2. This shows, as expected, that the Seifert two-framing
 of $S^3$ differs from the canonical two-framing by $3-d=2$ units.

The degrees are appropriately counted for maps to $SO(3)$, rather
than $SU(2)$, because this is the structure group of the tangent
bundle of $M$.  To illustrate the role of $SO(3)$, let us consider
one more simple example, which is $M=SO(3)=S^3/\Bbb{Z}_2$.  This
is the case $d=2$ of the lens space considered above, so we expect
the Seifert two-framing and the canonical two-framing to differ by
$3-d=1$ unit.  The comparison again reduces to comparing the
right-invariant framing of $TM$ with the left-invariant one. So
again we have to compare $\theta= dg\,g^{-1}$ with
$\hat\theta=g^{-1}dg$.  We have again $\theta = g\hat\theta
g^{-1}$, where now $g$ is the identity map from $SO(3)$ to itself,
which is of degree 1, showing that the two two-framings differ by
one unit.

For any $d$, the general analysis of framings by Freed and Gompf
in \FreedRG\ can be used to check that the canonical two-framing
and the Seifert two-framing on $M$ differ by ${s= (3-d)}$ units.

\subsec{Localization on a Smooth Component of the Moduli Space of
Irreducible Flat Connections}

We now extend our work in the previous section to describe the local
contribution to the Chern-Simons path integral from a smooth component
$\CM$ of the moduli space of irreducible flat connections on a Seifert
manifold $M$.  We assume here for simplicity that $M$ is described by
a line bundle $\CL$ of degree $n$ over a smooth Riemann surface
$\Sigma$ of genus $g \ge 1$.  The orbifold case is also discussed by
Rozansky in \RozanskyWV\ but is somewhat more involved.

As we recall from Section 5.1, $\CM$ is literally the moduli space
of flat connections on the trivial $G$-bundle over $M$ such that
the holonomy $\rho(h)$ around the $S^1$ fiber of $M$ is a fixed
element of the center $\Gamma$ of $G$.  This moduli space is not
smooth for arbitrary $\rho(h)$ in $\Gamma$, but it is smooth in
certain cases. The main such case, and the case we consider here,
arises when the gauge group $G$ is $SU(r+1)$, $\rho(h)$ is a
generator of $\Gamma = \BZ_{r+1}$, and $n$ and $r+1$ are
relatively prime. Under these conditions, $\rho(h)^n$ also
generates $\Gamma$, and $\CM$ is smooth and can be identified with
an unramified $(r+1)^{2g}$-fold cover of the moduli space $\CM_0$
of flat Yang-Mills connections on an associated principal bundle
$P$ over $\Sigma$ with structure group $\bar G=G/\Gamma$. ($\bar G$
enters because when we project to $\bar G$, $\rho(h)$ projects to
1 and the representation $\rho$ becomes a pullback from $\Sigma$.
 But as the three-dimensional gauge group is
really $G$, the holonomies of $\rho$ around one-cycles in $\Sigma$
are defined as elements of $G$, not $\bar G$; this leads to the
unramified cover.)

Our general discussion of non-abelian localization in Section 4
implies that the path integral contribution from $\CM$ can be
expressed entirely in terms of the cohomology ring of $\CM$, or
equivalently $\CM_0$.  One of the reasons that localization on $\CM$
is interesting is that we find in Chern-Simons theory a natural
generalization of the cohomological formula \ZV\ for the path integral
contribution from $\CM_0$ in two-dimensional Yang-Mills theory.

We recall from our discussion in Section 5.1 that a local symplectic
neighborhood $N$ near $\CM$ in $\bar\CA$ is described by an
equivariant bundle
\eqn\FIBRI{F\longrightarrow N\buildrel pr\over\longrightarrow {\cal
M}\,,}
where the normal fiber $F$ takes the (by now familiar) form ${F \,=\,
\CH \times_{H_0} \left(\Fh \ominus \Fh_0 \ominus \CE_0 \oplus
\CE_1\right)}$.

By assumption, the only gauge transformations which fix the irreducible
flat connections associated to points in $\CM$ are constant gauge
transformations by elements in the center $\Gamma$ of $G$, since the
center of $G$ always acts trivially in the adjoint representation.  So the
stabilizer subgroup $H_0$ in $\CH$ is now given by
\eqn\NEWHOII{ H_0 \,=\, U(1)^2 \times \Gamma\,,}
where we recall that the torus $U(1)^2$ arises from the two extra
generators in $\CH$ relative to $\CG$.

Also, we recall that the vector spaces $\CE_0$ and $\CE_1$ are now
given over a point of $\CM$ by
\eqn\AUXHDGIII{\eqalign{ \CE_0 \,&=\, \bigoplus_{t \neq 0}
H^0_{\bar\partial}(\Sigma, \ad(P) \otimes \CL^t)\,=\, \bigoplus_{t
\ge 1} H^0_{\bar\partial}(\Sigma, \ad(P) \otimes (\CL^t \oplus
\CL^{-t}))\,,\cr \CE_1 \,&=\, \bigoplus_{t \neq 0}
H^1_{\bar\partial}(\Sigma, \ad(P) \otimes \CL^t)\,=\, \bigoplus_{t
\ge 1} H^1_{\bar\partial}(\Sigma, \ad(P) \otimes (\CL^t \oplus
\CL^{-t}))\,.\cr}}

\medskip\noindent{\it The Canonical Symplectic Integral Over
$N$}\smallskip

Having described the local geometry near $\CM$ in $\bar\CA$, we next
consider the canonical symplectic integral over $N$.  This
integral takes the form
\eqn\MI{ Z(\epsilon)\Big|_\CM \,=\, {{2 \pi \epsilon \cdot \Vol(U(1)^2)}
\over {\Vol(\CH)}} \, \int_{\Fh \times N} \left[{{d\phi} \over
{2\pi}}\right] \exp{\left[\Omega - i \, \langle\mu,\phi\rangle -
{{i \epsilon} \over 2} (\phi,\phi) + t D\lambda\right]}\,,}
where we include in the normalization of \MI\ the prefactor from
\PREZCH.  To define the integral over the non-compact directions in
$N$, we also include in \MI\ the localization form $t D \lambda$.

Our goal now is to reduce the integral over $\Fh \times N$ in \MI\ to
an integral over the moduli space $\CM$ itself.  We have already
discussed a problem of this sort in Section 4.2, when we considered
the path integral contribution from irreducible flat connections in
two-dimensional Yang-Mills theory.  As we briefly recall, in the case
of Yang-Mills theory the fiber $F$ in \FIBRI\ is modelled on the
cotangent bundle $T^* H$ (with $H$ being the group of gauge
transformations in that case), so that $N$ retracts equivariantly onto a
principal $H$-bundle $P_H$ over the the moduli space $\CM_0$.  Because
$H$ acts freely on $P_H$, the $H$-equivariant cohomology of the total
space $P_H$ can be identified with the ordinary cohomology of the
quotient $P_H/H = \CM_0$, so $H^*_H(P_H) \cong H^*(\CM_0)$.  In
particular, the $H$-equivariant cohomology classes of $[\Omega - i \,
\langle\mu,\phi\rangle]$ and $[-\ha(\phi,\phi)]$ on $P_H$ pull back
from ordinary cohomology classes $\Omega$ and $\Theta$ of degrees two
and four on $\CM_0$, and we apply this fundamental fact to reduce the
symplectic integral in Yang-Mills theory to an integral over $\CM_0$.

In the case of Chern-Simons theory, the group $H \equiv \CH$ no longer
acts freely on $N$, but we can still apply much the same logic as for the
case of Yang-Mills theory.  Here a subgroup $H_0$ of $H$ acts with
fixed points on $N$, so $N$ equivariantly retracts onto a
bundle with fiber $H / H_0$ over $\CM$.  We denote the total space
of this bundle by $N_0$, so that ${H / H_0 \longrightarrow N_0
\longrightarrow \CM}$.

Because $N_0$ is an equivariant retraction from $N$, the
$H$-equivariant cohomology ring of $N$ is the same as that of $N_0$.
As we explain in Appendix C, the formal properties of equivariant
cohomology further imply that the $H$-equivariant cohomology ring of
$N_0$ is identified under pullback with the $H_0$-equivariant
cohomology ring of $\CM$ itself.  So in total, we have the relation
$H^*_H(N) \cong H^*_{H_0}(\CM)$.

As a result, in precise analogy to the case of two-dimensional
Yang-Mills theory, the $H$-equivariant cohomology classes of
$[\Omega - i \, \langle\mu,\phi\rangle]$ and $[-\ha(\phi,\phi)]$ which
appear in the symplectic integral over $N$ can be identified as the
pullbacks from $\CM$ of elements in the ring $H^*_{H_0}(\CM)$.

To identify the elements of $H^*_{H_0}(\CM)$ which pull back to these
classes appearing in the symplectic integral over $N$, we note that
$H^*_{H_0}(\CM)$ has a very simple structure.  As we also explain in
Appendix C, because $H_0$ acts trivially on $\CM$, $H^*_{H_0}(\CM)$ is
given by the tensor product of the ordinary cohomology ring $H^*(\CM)$
of $\CM$ with the $H_0$-equivariant cohomology ring $H_{H_0}^*(pt)$ of
a point.  Thus, $H^*_{H_0}(\CM) = H^*(\CM) \otimes H_{H_0}^*(pt)$.

Finally, our previous discussion of the Cartan model of equivariant
cohomology explicitly identifies the $H_0$-equivariant cohomology ring
of a point with the ring of invariant functions on the Lie algebra
$\Fh_0$.  Thus, all elements of $H^*_{H_0}(\CM)$ can be written as
sums of terms having the form $x \cdot f(\psi)$, where $x$ is an
ordinary cohomology class on $\CM$ and $f(\psi)$ is an invariant
function of $\psi$ in $\Fh_0$.

With our concrete description of $H^*_{H_0}(\CM)$, we can immediately
identify the elements of this ring which pull back to the
$H$-equivariant classes $[\Omega - i \, \langle\mu,\phi\rangle]$ and
$[-\ha(\phi,\phi)]$ on $N$.  Let us decompose the Lie algebra $\Fh$ of
$\CH$ as a sum ${\Fh = (\Fh \ominus \Fh_0) \oplus \Fh_0}$.  As a
result, we write ${\phi = \varphi + p + a}$, where $\varphi$ is an
element of $\Fh\ominus\Fh_0$, which can be identified as the Lie
algebra of $\CG$, and, in the same notation from Section 3.4, $p$ and
$a$ are elements of the Lie algebra $\Fh_0$ of $H_0$.

We then identify the $H$-equivariant classes on $N$ appearing in \MI\
with corresponding $H_0$-equivariant classes on $\CM$ via
\eqn\HHZM{\eqalign{
\Omega - i \, \langle\mu,\phi\rangle &\,\longleftrightarrow\, \Omega \,-\,
i \, a\,,\cr
-\ha(\phi,\phi) &\,\longleftrightarrow\, n \, \Theta \,+\, p a\,.\cr}}
We abuse notation slightly in the first line of \HHZM.  On the left,
$\Omega$ is the symplectic form on $\bar\CA$ restricted to $N$, and on
the right $\Omega$ is the induced symplectic form on $\CM$ (or
equivalently $\CM_0$), exactly as in our discussion of two-dimensional
Yang-Mills theory.  In identifying the dependence of this degree two
class in $H^*_{H_0}(\CM)$ on $p$ and $a$, we use the fact, evident
from the formula for $\mu$ in \MUIV, that the value of the moment map
$\langle\mu,\phi\rangle$ evaluated at a flat connection which pulls
back from $\Sigma$ is just the constant $a$ appearing on the right of
the first line in \HHZM.

Similarly, in the second line of \HHZM, the degree
four class $\Theta$ on $\CM$ is the same degree four class that
appeared in our discussion of Yang-Mills theory.  The identification
in \HHZM\ arises by writing the degree four invariant
$-\ha(\phi,\phi)$ in terms of $\varphi$, $p$, and $a$ as
\eqn\NPHI{ -\ha(\phi, \phi) \,=\, \ha \int_M \kappa\^d\kappa \,
\Tr(\varphi^2) \,+\, p a \,=\, {n \over 2} \int_\Sigma \omega \,
\Tr(\varphi^2) \,+\, p a\,,}
where we recall that $n$ is the degree of the line-bundle $\CL$ over
$\Sigma$ which defines $M$ and $\omega$ is a unit-volume
symplectic form on $\Sigma$.  As in the case of two-dimensional
Yang-Mills theory, the term quadratic in the generators $\varphi$ of the
gauge symmetry is associated by the Chern-Weil homomorphism to the
degree four class $\Theta$.

With the identifications in \HHZM, we can rewrite the symplectic
integral over $N$ as
\eqn\MII{ Z(\epsilon)\Big|_\CM \,=\, {{2 \pi \epsilon \cdot \Vol(U(1)^2)}
\over {\Vol(\CH)}} \, \int_{\Fh \times N} \left[{{d\phi} \over
{2\pi}}\right] \exp{\left[\left(pr^*\Omega\right) \,-\, i a \left(1 -
\epsilon p\right) \,+\, i \epsilon n \left(pr^*\Theta\right) \,+\, t
D\lambda\right]}\,.}
As in the case of localization at the trivial connection, the
generator $a$ acts trivially on all of $N$ and so does not appear in
the localization form $t D\lambda$.  So we can perform the integrals
over $a$ and $p$ exactly as before, and the integral over $a$
produces a delta-function that sets $p = 1 / \epsilon$.  As a result,
the symplectic integral reduces to the form
\eqn\MIII{ Z(\epsilon)\Big|_\CM \,=\, {{\Vol(U(1)^2)} \over
{\Vol(\CH)}} \, \int_{(\Fh \ominus \Fh_0) \times N} \left[{{d\phi}
\over  {2\pi}}\right] \exp{\left[\left(pr^*\Omega\right) \,+\, i
\epsilon n \left(pr^*\Theta\right) \,+\, t D\lambda\Big|_{p =
1/\epsilon}\right]}\,.}

The only term in \MIII\ which does not pull back from $\CM$ is the
localization term $t D\lambda$, so we are left to integrate $t
D\lambda$ over the fiber $F$ of $N$.  In the case of two-dimensional
Yang-Mills theory, with $F = T^* H$, this integral gave a trivial
factor of unity.  In Chern-Simons theory, the result is much more
interesting.

\medskip\noindent{\it An Equivariant Euler Class From $F$}\smallskip

To evaluate \MIII, we consider the following integral,
\eqn\MIV{I(\psi) \,=\, {1 \over {\Vol(\CH)}} \, \int_{\wt F}
\left[{{d\phi} \over  {2\pi}}\right] \exp{\left[t
D\lambda\right]}\,,\qquad \wt F \,=\, (\Fh \ominus \Fh_0) \times
F\,,\qquad \psi \in \Fh_0\,.}
Here we let $\psi = p + a$ be an arbitrary element of $\Fh_0$, though in
general the generator $a$ will not appear in \MIV\ since $a$ acts
trivally on $N$, and we set $p = 1 / \epsilon$ at the end of the
discussion, as in \MIII.

Of course, in Section 4.3 we computed this integral over the abstract
model for $F$.  There we assumed $\CM$ to be a point, and we found the
result
\eqn\ICII{  I(\psi) \,=\, {1 \over {\Vol(H_0)}}\,
\det\left({{\psi} \over {2 \pi}}\Big|_{E_0}\right) \,
\det\left({{\psi} \over {2 \pi}}\Big|_{E_1}\right)^{-1}\,,\qquad
\psi \in \Fh_0\,.}
Unfortunately, we cannot apply this result directly to the case at
hand.  When $F$ is fibered over a non-trivial moduli space
$\CM$, then $I(\psi)$ will generally involve cohomology classes on
$\CM$ which are associated to the twisting of the bundle and which our
previous computation did not detect.

To compute $I(\psi)$ in \MIV, one approach is simply to generalize the
abstract localization computation in Section 4.3 to allow for a
non-trivial moduli space $\CM$.  We perform this computation in
Appendix D.  However, we can also make an
immediate guess, on the basis of mathematical naturality, for what the
generalization of the formula \ICII\ must be when $\CM$ is
non-trivial.  This guess relies on a more intrinsic topological
interpretation of the result \ICII\ even in the case that $\CM$ is a
point.  For this reason, it turns out to be much more illuminating to
``guess'' the generalization of \ICII\ rather than simply to compute,
so we pursue this approach now.

Let us think about what our result for $I(\psi)$ really {\it means} in
the case that $\CM = pt$.  Abstractly, the data which enter the
formula \ICII\ are the group $H_0$, which acts trivially on $\CM$,
and the finite-dimensional unitary representations $E_0$ and $E_1$ of
$H_0$.  In general, to say that $E$ is a representation of $H_0$ is the
same thing as to say that $E$ is an $H_0$-equivariant bundle over a
point, so if we like, we can consider $E_0$ and $E_1$ as $H_0$-equivariant
bundles over $\CM = pt$.

This language is useful, since whenever we have a vector bundle (even
a vector bundle over a point!) an extremely natural set of topological
invariants to consider are the characteristic classes of the bundle.
In our context, we naturally consider the $H_0$-equivariant
characteristic classes\foot{Although we will not require the
generalization here, we refer the reader to Chapter 8.5 of
\GuilleminSII\ for a general discussion of equivariant characteristic
classes.} of $E_0$ and $E_1$ as $H_0$-equivariant bundles over ${\CM =
pt}$.  These characteristic classes are valued in the $H_0$-equivariant
cohomology ring of $\CM$ --- since $\CM$ is a point, this ring
is the ring of invariant functions on the Lie algebra $\Fh_0$ of
$H_0$.

If $E$ is a unitary representation of $H_0$ and we consider $E$ as an
$H_0$-equivariant bundle over a point, then the $H_0$-equivariant
characteristic classes of $E$ have a simple description.  We let
$U(E)$ be the unitary group acting on $E$.  Since $H_0$ acts in a
unitary fashion on $E$, the relevant characteristic classes of $E$ to
consider are the equivariant Chern classes.  As is well known, the
ordinary Chern classes of a vector bundle are associated via the
Chern-Weil homomorphism to the generators $c_i$ of the ring of invariant
polynomials on the Lie algebra of the unitary group.  To describe the
corresponding $H_0$-equivariant Chern classes of $E$, we observe that,
since $E$ is a unitary representation of $H_0$, we have an induced map
${H_0 \longrightarrow U(E)}$.  Consequently, any invariant polynomial
on the Lie algebra of $U(E)$ pulls back to an invariant polynomial on
the Lie algebra $\Fh_0$ of $H_0$.  The pullbacks of the generators
$c_i$ to invariant polynomials on $\Fh_0$ are then the
$H_0$-equivariant Chern classes of $E$.  In particular, if the action
of $H_0$ on $E$ is non-trivial, then the equivariant Chern classes of
$E$ can also be non-trivial, despite the fact that $E$ is a bundle
over only a point.

The invariant polynomials appearing in $I(\psi)$, namely
\eqn\DETPSIS{ e_{H_0}(pt, E_0) \,\equiv\, \det\left({{\psi} \over {2
\pi}}\Big|_{E_0}\right)\,,\qquad e_{H_0}(pt, E_1) \,\equiv\,
\det\left({{\psi} \over {2 \pi}}\Big|_{E_1}\right)\,,}
arise from determinants.  The Chern-Weil homomorphism associates the
determinant to the top Chern class, so by our discussion above the
invariant polynomials in \DETPSIS\ can be characterized intrinsically
as the $H_0$-equivariant top Chern classes, or equivalently Euler
classes, of $E_0$ and $E_1$ as equivariant bundles over a point.
Thus, when $\CM$ is a point, we write $I(\psi)$ in \ICII\
intrinsically as
\eqn\ICIII{ I(\psi) \,=\,  {1 \over {\Vol(H_0)}} \, {{e_{H_0}(pt,
E_0)} \over {e_{H_0}(pt, E_1)}}\,.}

More generally, if $E$ is an $H_0$-equivariant vector bundle over a
complex manifold $\CM$, then we can still consider the $H_0$-equivariant
Euler class $e_{H_0}(\CM, E)$ of $E$, which takes values in the
$H_0$-equivariant cohomology ring of $\CM$.  If $H_0$ acts trivially on
$\CM$ (but not necessarily trivially on $E$), we have already
identified this cohomology ring as a product $H_{H_0}^*(\CM) \cong
H^*(\CM) \otimes H^*_{H_0}(pt)$.  We describe $e_{H_0}(\CM, E)$ in
this case explicitly below.

In our application to Chern-Simons theory, the infinite-dimensional
vector spaces $\CE_0$ and $\CE_1$ in \AUXHDGIII\ determine associated
$H_0$-equivariant bundles over the moduli space $\CM$, on which $H_0$
in \NEWHOII\ acts trivially.  Given our intrinsic interpretation of
$I(\psi)$ when $\CM$ is a point, we certainly expect that the integral
over $F$ in \MIV\ produces the natural generalization of \ICIII,
involving the $H_0$-equivariant Euler classes of the bundles
associated to $\CE_0$ and $\CE_1$ over $\CM$.  That is,
\eqn\MV{ I(\psi) \,=\, {1 \over {\Vol(H_0)}} \, {{e_{H_0}(\CM,
\CE_0)} \over {e_{H_0}(\CM, \CE_1)}}\,.}
As our direct computation in Appendix D shows, this formula is
correct.

We remark that the appearance of the equivariant Euler class of the
bundle $\CE_1$ in the denominator of \MV\ is quite standard.  This
class appears in precisely the same way in the classic
Duistermaat-Heckman formula \Duistermaat\ for abelian localization,
as was explained in \AtiyahRB.  The essentially new feature of the formula
\MV\ is the appearance of a corresponding Euler class from $\CE_0$ in the
numerator.

We set
\eqn\LE{ e(\psi) \,=\, {{e_{H_0}(\CM, \CE_0)} \over {e_{H_0}(\CM,
\CE_1)}}\,.}
Then from \MIII, \MIV, and \MV, the local contribution from $\CM$ in
Chern-Simons theory is given abstractly by
\eqn\MVI{ Z(\epsilon)\Big|_\CM \,=\, {1 \over {|\Gamma|}} \, \int_\CM
e(p)\Big|_{p = 1/\epsilon} \, \exp{\left(\Omega \,+\, i
\epsilon n \Theta\right)}\,.}
In arriving at \MVI, we note that the prefactor $\Vol(U(1)^2)$ in
\MIII\ cancels against a corresponding factor in $\Vol(H_0)$ from
$I(\psi)$.  This cancellation leaves the factor $1 / |\Gamma|$ in
\MIV, where $|\Gamma|$ is the order of the center $\Gamma$ of $G$.

As we recall in writing \MVI, since the generator $a$ in $\Fh_0$ acts
trivially on $N$, ${e(\psi) \equiv e(p)}$ depends only on $p$ in
$\Fh_0$.  Once we set $p = 1 / \epsilon$ in \MVI, $e(\epsilon^{-1})$
will become an ordinary cohomology class on $\CM$.  As in the case of
localization at the trivial connection, our computation now reduces to
determining explicitly this class.

\medskip\noindent{\it More About the Equivariant Euler Class}\smallskip

Before we evaluate the equivariant Euler classes of the
infinite-dimensional bundles corresponding to $\CE_0$ and $\CE_1$, we
first give a more explicit description of the equivariant Euler class
in a simpler, finite-dimensional situation.  To make contact with
Chern-Simons theory, we assume abstractly that $H_0$ is a torus which acts
trivially on a complex manifold $\CM$, and we assume that $E$
is a complex representation of $H_0$ which is fibered over $\CM$ to
determine an associated $H_0$-equivariant bundle.  Our goal is now to
give a concrete topological formula for $e_{H_0}(\CM, E)$, which we
will then apply to evaluate $e(\psi)$ in \LE\ for Chern-Simons theory.

In general, $e_{H_0}(\CM, E)$ incorporates both the algebraic data
associated to the action of $H_0$ on $E$ as well as the topological
data that describes the twisting of $E$ over $\CM$.  To encode the
data related to the action of $H_0$ on $E$, we decompose $E$ under
the action of $H_0$ into a sum of one-dimensional complex eigenspaces
\eqn\DECE{ E \,=\, \bigoplus_{j=1}^{\dim E} \, E_{\beta_j}\,,}
where each $\beta_j$ is a weight in $\Fh_0^*$ which describes the
action of $H_0$ on the eigenspace $E_{\beta_j}$.

To encode the topological data associated to the vector bundle
determined by $E$ over $\CM$, we apply the splitting principle in
topology, as explained for instance in Chapter 21 of \BottT.  By this
principle, we can assume that the vector bundle determined by $E$ over
$\CM$ splits equivariantly into a sum of line-bundles associated to
each of the eigenspaces $E_{\beta_j}$ for the action of $H_0$.  Under
this assumption, we let $x_j = c_1(E_{\beta_j})$ be the first Chern class
of the corresponding line-bundle.  These virtual Chern roots $x_j$
determine the total Chern class of $E$ as
\eqn\CHRRTS{ c(E) \,=\, \prod_{j=1}^{\dim E} \, (1 + x_j)\,.}
In particular, the ordinary Euler class of $E$ over $\CM$ is then
given by
\eqn\CHRRTSII{ e(\CM, E) \,=\, \prod_{j=1}^{\dim E} x_j\,.}

The equivariant Euler class $e_{H_0}(\CM, E)$ is now determined in
terms of the  weights $\beta_j$ and the Chern roots $x_j$.  We
recall that $e_{H_0}(\CM, E)$ is defined as an element of
${H^*_{H_0}(\CM, E) = H^*(\CM) \otimes H_{H_0}^*(pt)}$ since $H_0$
acts trivially on $\CM$. Thus, $e_{H_0}(\CM, E)$ will be a
function of $\psi\in \Fh_0$ with values in the cohomology of
$\CM$.  Explicitly, the $H_0$-equivariant Euler class of $E$ over
$\CM$ is given by \eqn\HEQE{ e_{H_0}(\CM, E) \,=\,
\prod_{j=1}^{\dim E} \left( {{i \, \langle\beta_j, \psi\rangle}
\over {2 \pi}} + x_j\right)\,.} We see that this expression is a
natural generalization of the ordinary Euler class \CHRRTSII\ of
$E$.  Also, when $\CM$ is only a point, the Chern roots $x_j$ do
not appear in \HEQE\ for dimensional reasons, and the product over
the weights $\beta_j$ in \HEQE\ reproduces the determinant of
$\psi$ acting on $E$ as in \DETPSIS.

\medskip\noindent{\it Evaluating $e(p)$}\smallskip

We now evaluate $e(p)$ for Chern-Simons theory\foot{We set $p =
1/\epsilon$ only at the very end of the computation.}.  First we
recall that the complex vector spaces $\CE_0$ and $\CE_1$ appearing in
\LE\ arise from the Dolbeault cohomology groups of the bundles $\ad(P)
\otimes \CL^t$ over $\Sigma$, with
\eqn\AUXHDGIV{\eqalign{ \CE_0 \,&=\, \bigoplus_{t \neq 0}
H^0_{\bar\partial}(\Sigma, \ad(P) \otimes \CL^t)\,=\, \bigoplus_{t
\ge 1} H^0_{\bar\partial}(\Sigma, \ad(P) \otimes (\CL^t \oplus
\CL^{-t}))\,,\cr \CE_1 \,&=\, \bigoplus_{t \neq 0}
H^1_{\bar\partial}(\Sigma, \ad(P) \otimes \CL^t)\,=\, \bigoplus_{t
\ge 1} H^1_{\bar\partial}(\Sigma, \ad(P) \otimes (\CL^t \oplus
\CL^{-t}))\,.\cr}}
We also recall that the action of $H_0$ on $\CE_0$ and $\CE_1$ is
determined by the operator $p \lie_R$, whose action in turn only
depends on the grading $t$ in \AUXHDGIV.  We naturally decompose
$\CE_0$ and $\CE_1$ under the action of $H_0$, and we consider the
finite-dimensional eigenspaces
\eqn\ETS{ \CE_0^{(t)} \,=\, H^0_{\bar\partial}(\Sigma, \ad(P) \otimes
\CL^t)\,,\qquad \CE_1^{(t)} \,=\, H^1_{\bar\partial}(\Sigma, \ad(P)
\otimes \CL^t)\,.}
The abelian group $H_0$ acts canonically on both $\CE_0^{(t)}$ and
$\CE_1^{(t)}$ with eigenvalue $-2 \pi i t$.

In terms of this decomposition, the quantity $e(p)$ is given by the
following infinite product,
\eqn\EPI{ e(p) \,=\, \prod_{t \neq 0} \, \left[{{e_{H_0}\bigl(\CM,
\CE_0^{(t)}\bigr)} \over {e_{H_0}\bigl(\CM, \CE_1^{(t)}\bigr)}}\right]
\,=\, \prod_{t \ge 1} \, \left[{{e_{H_0}\bigl(\CM,
\CE_0^{(t)}\bigr) \cdot {e_{H_0}\bigl(\CM,
\CE_0^{(-t)}\bigr)}} \over {e_{H_0}\bigl(\CM, \CE_1^{(t)}\bigr) \cdot
e_{H_0}\bigl(\CM, \CE_1^{(-t)}\bigr)}}\right] \,.}
Here $e_{H_0}\bigl(\CM, \CE_0^{(t)}\bigr)$ and $e_{H_0}\bigl(\CM,
\CE_1^{(t)}\bigr)$ denote the $H_0$-equivariant Euler classes of the
finite-dimensional bundles determined by $\CE_0^{(t)}$ and
$\CE_1^{(t)}$ over $\CM$.

Our basic strategy to evaluate the product in \EPI\ is to deduce a
recursive relation between the equivariant Euler classes of
$\CE_0^{(t)}$, $\CE_0^{(t-1)}$, $\CE_1^{(t)}$, and
$\CE_1^{(t-1)}$. So far, we have only specified the line-bundle
$\CL$ topologically, by specifying its degree $n$.  The
holomorphic structure of $\CL$ really was not important.  Now we
want to pick a convenient holomorphic structure on $\CL$ to
simplify our  computation.  We pick $n$ arbitrary points
$\sigma_1,\dots,\sigma_n$ on $\Sigma$ and we take $\CL$ to be
$\CO(\sigma_1+\dots+\sigma_n)$.

With this choice of ${\cal L}$, we have the following short exact
sequence of coherent sheaves on $\Sigma$, \eqn\SES{ 0
\longrightarrow \ad_\BC(P) \otimes \CL^{t-1} \longrightarrow
\ad_\BC(P) \otimes \CL^t \longrightarrow \bigoplus_{i=1}^n
\ad_\BC(P)\big|_{\sigma_i} \longrightarrow 0\,.} Here $t$ is any
integer, and $\ad_\BC(P)\big|_{\sigma_i}$ denotes the skyscraper
sheaf associated to the fiber of $\ad_\BC(P)$ over the point
$\sigma_i$. The appearance of this skyscraper sheaf explains our
need to work a bit more generally with coherent sheaves, as
opposed to more innocuous bundles.

Associated to this short exact sequence we have the usual long
exact sequence in sheaf cohomology, \eqn\LES{\eqalign{ 0
&\longrightarrow H^0\bigl(\Sigma, \ad_\BC(P) \otimes
\CL^{t-1}\bigr) \longrightarrow H^0\bigl(\Sigma, \ad_\BC(P)
\otimes \CL^t\bigr) \longrightarrow \bigoplus_{i=1}^n
H^0\bigl(\Sigma, \ad_\BC(P)\big|_{\sigma_i}\bigr) \longrightarrow
\cr &\longrightarrow H^1\bigl(\Sigma, \ad_\BC(P) \otimes
\CL^{t-1}\bigr) \longrightarrow H^1\bigl(\Sigma, \ad_\BC(P)
\otimes \CL^t\bigr) \longrightarrow 0\,.\cr}} Since a skyscraper
sheaf has no higher cohomology, we observe that ${H^1\bigl(\Sigma,
\ad_\BC(P)\big|_{\sigma_i}\bigr) = 0}$ for the last term of \LES.

Each cohomology group appearing in \LES\ can be considered as the
fiber of an associated holomorphic bundle over the moduli space
$\CM$, and the exactness of the sequence \LES\ implies the
exactness of the corresponding sequence of bundles on $\CM$.
Except for the single term involving the skyscraper sheaf, we see
that the bundles which appear in \LES\ are those associated to
$\CE_0^{(t-1)}$, $\CE_0^{(t)}$, $\CE_1^{(t-1)}$, and
$\CE_1^{(t)}$.  In analogy to \ETS, we set \eqn\BIGV{ \CV_{(i)}
\,=\, H^0\bigl(\Sigma, \ad_\BC(P)\big|_{\sigma_i}\bigr)\,.} Over
$\CM$, $\CV_{(i)}$ also fibers as a holomorphic bundle. Although
the holomorphic structure of $\CV_{(i)}$ depends on $\sigma_i$,
its topology, which is all we will care about, does not (as is
clear from the fact that the points $\sigma_i$ can be moved
continuously), so we just write $\CV$ for any of the $\CV_{(i)}$.
Thus, the exact sequence in \LES\ implies the following exact
sequence of associated bundles on $\CM$, \eqn\LESIII{0
\longrightarrow \CE_0^{(t-1)} \longrightarrow \CE_0^{(t)}
\longrightarrow \CV^{\oplus n} \longrightarrow \CE_1^{(t-1)}
\longrightarrow \CE_1^{(t)} \longrightarrow 0\,.}

This sequence is an exact sequence of bundles on $\CM$, but we need an
exact sequence of $H_0$-equivariant bundles on $\CM$, such that the
maps in the sequence are compatible with the action of $H_0$.  Because
$H_0$ acts with different eigenvalues on the equivariant
bundles $\CE_0^{(t-1)}$ and $\CE_0^{(t)}$, and similarly on
$\CE_1^{(t-1)}$ and $\CE_1^{(t)}$, the canonical action of $H_0$ is
not compatible with the maps in \LESIII.

To fix this problem, we note that we are free to consider actions of
$H_0$ on $\CE_0^{(t)}$ and $\CE_1^{(t)}$ other than the canonical
action.  That is, we consider $H_0$-equivariant bundles over
$\CM$ whose fibers are still given by the cohomology groups
${H^0_{\bar\partial}(\Sigma, \ad(P) \otimes \CL^t)}$ and
${H^1_{\bar\partial}(\Sigma, \ad(P) \otimes \CL^t)}$ but where the
action of $H_0$ is not the canonical action fixed by $t$.  In fact, so
long as $H_0$ acts uniformly on the fiber, we can take $H_0$ to act
with any eigenvalue.

Thus we let $\CE_{0,m}^{(t)}$ and $\CE_{1,m}^{(t)}$ denote the
$H_0$-equivariant bundles over $\CM$ whose fibers are determined by
$t$ as before but where $H_0$ acts with eigenvalue $-2 \pi i m$ for
some integer $m$.   In this notation, the bundles $\CE_0^{(t)}$ and
$\CE_1^{(t)}$ with the canonical action of $H_0$ are $\CE_{0,t}^{(t)}$
and $\CE_{1,t}^{(t)}$.  We similarly denote by $\CV_m$ the
$H_0$-equivariant bundle associated to $\CV$ for which $H_0$ acts
uniformly on the fiber with eigenvalue $-2 \pi i m$.

The exact sequence in \LESIII\ on $\CM$ now determines a corresponding
exact sequence of $H_0$-equivariant bundles,
\eqn\LESIV{ 0 \longrightarrow \CE_{0,m}^{(t-1)} \longrightarrow
\CE_{0,m}^{(t)} \longrightarrow \left(\CV_m\right)^{\oplus n}
\longrightarrow \CE_{1,m}^{(t-1)} \longrightarrow \CE_{1,m}^{(t)}
\longrightarrow 0\,.}
Since the action of $H_0$ is the same on every term in this sequence,
the maps are trivially compatible with the group action.

We now recall that a fundamental property of the equivariant Euler
class is that it behaves multiplicatively with respect to an exact
sequence of equivariant  bundles, just like the ordinary Euler class.
Thus, if $E_1$, $E_2$, and $E_3$ are $H_0$-equivariant bundles on
$\CM$ which fit into an exact sequence whose maps respect the action
of $H_0$,
\eqn\LESII{ 0 \longrightarrow E_1 \longrightarrow E_2 \longrightarrow
E_3 \longrightarrow 0\,,}
then the $H_0$-equivariant Euler classes of these bundles satisfy the
relation
\eqn\EEQQ{ e_{H_0}(\CM, E_2) \,=\, e_{H_0}(\CM, E_1) \cdot
e_{H_0}(\CM, E_3)\,.}
More generally, given an exact sequence of arbitrary length,
\eqn\LESIII{ 0 \longrightarrow E_1 \longrightarrow E_2 \longrightarrow
\cdots \longrightarrow E_{2N} \longrightarrow E_{2N+1} \longrightarrow
0\,,}
the relation \EEQQ\ generalizes in the natural way, with
\eqn\EEQQII{ e_{H_0}(\CM, E_2) \,\cdots\,
e_{H_0}(\CM, E_{2N}) \,=\,e_{H_0}(\CM, E_1) \,\cdots\,
e_{H_0}(\CM, E_{2N+1})\,.}

We apply this multiplicative property of the equivariant Euler class
to the exact sequence in \LESIV.  For the following, it is very
natural to introduce the ratio of equivariant Euler classes,
\eqn\RAT{ \CQ^{(t)}_m \,\equiv\, \left[{{e_{H_0}\bigl(\CM,
\CE_{0,m}^{(t)}\bigr)} \over {e_{H_0}\bigl(\CM,
\CE_{1,m}^{(t)}\bigr)}}\right]\,,}
so that $e(p)$ is given by
\eqn\EPII{ e(p) \,=\, \prod_{t \neq 0} \, \CQ^{(t)}_t\,.}
In terms of $\CQ^{(t)}_m$, the multiplicative relation \EEQQ\ applied
to \LESIV\ implies that
\eqn\RATII{ \CQ^{(t)}_m \,=\, \CQ^{(t-1)}_m \cdot
\left[e_{H_0}\bigl(\CM, \CV_m\bigr)\right]^n\,.}
Expanding the recursive relation \RATII, we find
\eqn\RATIII{
\CQ^{(t)}_m \,=\, \CQ^{(0)}_m \cdot \left[e_{H_0}\bigl(\CM,
\CV_m\bigr)\right]^{n t}\,.}

What has this work gained us?  As we now explain, we can give a very
concrete expression for the quantity on the right of \RATIII.  By
definition, the bundles over $\CM$ which determine the ratios
$\CQ^{(0)}_{\pm t}$ have fibers
\eqn\ETSII{ \CE_0^{(0)} \,=\, H^0_{\bar\partial}(\Sigma,
\ad_\BC(P))\,,\qquad \CE_1^{(0)} \,=\, H^1_{\bar\partial}(\Sigma,
\ad_\BC(P))\,.}
By our assumption that all points in the moduli space $\CM$ correspond
to irreducible connections, $\CE_0^{(0)} = 0$.  Further, as we
mentioned in Section 4.3, $\CE_1^{(0)}$ is naturally identified with the
holomorphic tangent bundle $T \CM$ of the moduli space itself, so
${\CE_1^{(0)} = T\CM}$.  We introduce the convenient notation
$\CE_{1,t}^{(0)} \equiv T\CM_t$ to indicate the $H_0$-equivariant
version of $T\CM$.  Because of this observation, we can apply the
relations \EPII\ and \RATIII\ to rewrite $e(p)$ entirely in terms of
the equivariant bundles $T\CM_t$ and $\CV_t$,
\eqn\EPVI{ e(p) \,=\, \prod_{t\neq 0} \, {1 \over
{e_{H_0}\bigl(\CM,T\CM_t\bigr)}} \cdot \left[e_{H_0}\bigl(\CM,
\CV_t\bigr)\right]^{n t}\,.}

Let us make the factors appearing on the right in \EPVI\ more
explicit.  To this end, we introduce the Chern roots
$\varpi_j$ of $T\CM$, where ${j=1\,,\ldots,\dim \CM}$, and the Chern
roots $\nu_l$ of $\CV$, where ${l=1\,,\ldots,\rk \CV}$.  Since $\CV$
arises from the fiber of the adjoint bundle $\ad_\BC(P)$, the rank of
$\CV$ is simply ${\rk \CV = \dim G \equiv \Delta_G}$.  As in our
general discussion of the equivariant Euler class, the Chern roots
$\varpi_j$ and $\nu_l$ are ``virtual'' degree two
classes in $H^*(\CM)$ which are defined in terms of the total Chern
classes of $T\CM$ and $\CV$ as
\eqn\CHRNM{ c(T\CM) \,=\, \prod_{j=1}^{\dim \CM} \, (1 +
\varpi_j)\,,\qquad c(\CV) \,=\, \prod_{l=1}^{\Delta_G} \, (1 + \nu_l)\,.}

In terms our these Chern roots, our general description of the
equivariant Euler class in \HEQE\ implies that
\eqn\RATV{ e_{H_0}\bigl(\CM,T\CM_t\bigr) \,=\,
\prod_{j=1}^{\dim \CM} \, \left(-i t p \,+\, \varpi_j\right)\,,\qquad
e_{H_0}\bigl(\CM,\CV_t\bigr) \,=\, \prod_{l=1}^{\Delta_G} \, \left(-i
t p \,+\, \nu_l\right)\,.}
The terms in \RATV\ which involve $p$ arise via the infinitesimal
action of $H_0$ on the fibers of $T\CM_t$ and $\CV_t$.  We recall
that $H_0$ acts infinitesimally as ${p \lie_R = -2 \pi i t p}$.

Together, \EPVI\ and \RATV\ imply the following formal expression for
$e(p)$,
\eqn\EPVII{ e(p) \,=\, \prod_{t\neq 0} \, \Bigg[\prod_{j=1}^{\dim \CM}
\, {1 \over {(-i t p\,+\, \varpi_j)}}\Bigg] \,
\Bigg[\prod_{l=1}^{\Delta_G} \, \left(-i t p \,+\,
\nu_l\right)^{n t}\Bigg]\,.}
This infinite product represents the determinant of a first-order operator
$\CD$ acting on $\CE_0 \ominus \CE_1$, where
\eqn\ETAOP{ \CD \,=\, {1 \over {2\pi}} \left(p \lie_R \,+\, i
\CR\right)\,.}
Here $\CR$ is the curvature operator acting on $\CE_0$ and $\CE_1$ as
bundles over $\CM$, as appears in the computation in Appendix D, and
``$\ominus$'' indicates that we actually take the inverse of the
determinant of $\CD$ acting on $\CE_1$.

The determinant in \EPVII\ is only a formal expression, and to define
it we must choose some regularization procedure.  For instance, we
considered the determinant of a similar operator $\CD_0$ in our
computation at the trivial connection in Section 5.2,
\eqn\NBIGOPII{ \CD_0 \,=\, {1 \over {2\pi}} \left(p \lie_R -
\left[\phi, \,\cdot\, \right]\right)\,.}
In that case, we defined the determinant of $\CD_0$ analytically,
using the zeta-function to define its absolute value and the
eta-function to define its phase.

We follow a similar strategy to define the determinant of $\CD$, or
more explicitly the infinite product in \EPVII.  To start, we find it
useful to rewrite the product in \EPVII\ by pulling out an overall
factor of $p$,
\eqn\EPVIII{ e(p) \,=\, p^{\dim \CM} \, \prod_{t\neq 0} \,
\Bigg[\prod_{j=1}^{\dim \CM} \, \left(-i t
\,+\, \left({\varpi_j \over p}\right)\right)^{-1}\Bigg] \,
\Bigg[\prod_{l=1}^{\Delta_G} \, \left(-i t \,+\,
\left({\nu_l \over p}\right)\right)^{n t}\Bigg]\,.}
In passing from \EPVII\ to \EPVIII, we use as in Section 5.2 the
classical Riemann zeta-function to define the trivial, but infinite,
product over $p$ which arises from \EPVII,
\eqn\PRODSIII{ \prod_{t \ge 1} \, p^{-2 \dim \CM} \,=\, \exp{\left(-2
\dim \CM \cdot \ln{p} \cdot \zeta\left(0\right)\right)} \,=\, p^{\dim
\CM}\,.}
(There is no contribution from the factors in \EPVII\ which are
associated to $\CV$ due to a cancellation between the terms for $\pm
t$.)  Thus, we are left to consider the determinant of the rescaled
operator $\CD'$,
\eqn\ETAOPII{ \CD' \,=\, {1 \over {2\pi}} \left(\lie_R \,+\, i {\CR
\over p}\right)\,,}
which represents the infinite product appearing in \EPVIII\ and which
depends on $p$ and the Chern roots only in the combinations $\varpi_j
/ p$ and $\nu_l / p$.

One interesting distinction between the operator $\CD$, or
equivalently $\CD'$, and the operator $\CD_0$ which appeared
previously is that whereas $\CD_0$ is an anti-hermitian operator, with
a purely imaginary spectrum, the operator $\CD$ has no particular
hermiticity properties and its spectrum has no particular phase.  This is
manifest in the product \EPVIII, since $-i t$ is imaginary but both the
Chern roots and $p$ are real.  In terms of \ETAOPII, both $\lie_R$ and
$\CR$ are anti-hermitian operators, but we have an explicit factor of
`$i$' in front of $\CR$.  Because $\CD'$ is neither hermitian nor
anti-hermitian, we will have to generalize the zeta/eta-function
regularization technique which we applied to define the determinant of
$\CD_0$ in Section 5.2.

Before we supply a definition for the determinant of $\CD'$, or
equivalently for the products in \EPVIII, let us consider what general
properties our definition should possess.  To start, we factorize the
product in \EPVIII\ into the two infinite products below,
\eqn\FSMVZ{\eqalign{
f_\CM(z) \,&=\, \prod_{t\neq 0} \, \prod_{j=1}^{\dim \CM} \, \left(-i
t \,+\, z \varpi_j\right)^{-1}\,,\cr
f_\CV(z) \,&=\, \prod_{t\neq 0} \, \prod_{l=1}^{\Delta_G} \, \left(-i
t \,+\, z \nu_l\right)^{n t}\,,\cr}}
where $z = 1/p$ is now a formal parameter.

The expressions in \FSMVZ\ are ill-defined as they stand.  However, if
we formally differentiate $\log{f_\CM(z)}$ and $\log{f_\CV(z)}$ with
respect to $z$ a sufficient number of times, we eventually obtain
well-defined, absolutely convergent sums.  For instance, in the case
of $f_\CM(z)$, we see that
\eqn\FSMVZII{ {{d^2} \over {dz^2}} \log{f_\CM(z)} \,=\,
\sum_{j=1}^{\dim \CM} \, \sum_{t \neq 0} \, {{\varpi_j^2} \over
{\left(-i t \,+\, z \varpi_j\right)^2}}\,=\, \sum_{j=1}^{\dim \CM} \,
{{d^2} \over {dz^2}} \log{\left[{(\pi z \varpi_j)} \over
{\sinh{(\pi z \varpi_j)}}\right]}\,.}
The second equality in \FSMVZII\ follows from the same product identity
\PRODS\ for $\sin(x)/x$ as we applied in Section 5.2.

So any reasonable definition for $f_\CM(z)$ in \FSMVZ\ must be
compatible with the relation \FSMVZII.  In particular, upon
integrating \FSMVZII, we see that $\log f_\CM(z)$ is determined up to
a linear function of $z$, and hence $f_\CM(z)$ is determined up to two
arbitrary real constants $a_0$ and $a_1$,
\eqn\FSMVZIII{ f_\CM(z) \,=\, \exp{\left[a_0 \,+\, a_1 z \,
c_1(T\CM)\right]} \, \prod_{j=1}^{\dim \CM} \, {{(\pi z \varpi_j)}
\over {\sinh{(\pi z \varpi_j)}}}\,.}
Here $c_1(T\CM) = \sum_j \varpi_j$ is the first Chern class of $\CM$.
In deducing the form \FSMVZIII, we have applied the fact, manifest
from \FSMVZ, that $f_\CM(z)$ can only depend on $z$ and the Chern
roots $\varpi_j$ in the combinations $z \varpi_j$, and we have also
used the fact that only symmetric combinations of the Chern roots have
any real meaning --- hence each Chern root $\varpi_j$ must appear with
the same coefficient $a_1$ in the exponential factor of \FSMVZIII.
Comparing to the product \FSMVZ, we also note that $f_\CM(z)$ is
formally real (for real $z$), so $a_0$ and $a_1$ must be real.

We can also apply this general analysis to $f_\CV(z)$ in \FSMVZ.  Here
we observe that $\log f_\CV(z)$ should satisfy
\eqn\FSVVZ{\eqalign{
{{d^3} \over {dz^3}} \log{f_\CV(z)} \,&=\, \sum_{t \neq 0} \,
\sum_{l=1}^{\Delta_G} \, {{2 n t \nu_l^3} \over {\left(-i t \,+\, z
\nu_l\right)^3}}\,,\cr
&=\, \sum_{t \ge 1} \, \sum_{l=1}^{\Delta_G} \, \left[  {{2 n t
\nu_l^3} \over {\left(-i t \,+\, z \nu_l\right)^3}} \,+\, {{2 n t
\nu_l^3} \over {\left(-i t \,-\, z \nu_l\right)^3}}\right]\,,\cr
&=\, 0\,.}}
In contrast to the case of $f_\CM(z)$, we must take three derivatives
of $\log f_\CV(z)$ to get a convergent sum, due to the exponent $n t$
appearing in \FSMVZ.  In passing to the second equality of \FSVVZ, we
have simply paired terms for $\pm t$.  However, to deduce the
cancellation in the third line of \FSVVZ, we must use some topological
facts about the bundle $\CV$.

We recall that $\CV$ is the bundle over $\CM$ whose fibers are given
by $H^0\bigl(\Sigma, \ad_\BC(P)\big|_\sigma\bigr)$ for some point
$\sigma$ on $\Sigma$.  This  bundle is naturally the complexification
of a real bundle over $\CM$, namely the bundle whose fibers are
$H^0\bigl(\Sigma, \ad(P)\big|_\sigma\bigr)$.  Consequently, the
non-zero Chern roots of $\CV$ are paired such that for each root
$\nu$ there is a corresponding root $\nu'$ with $\nu' = -
\nu$.  This fact implies that any odd, symmetric function of the
Chern roots vanishes.  In particular, all odd Chern classes of $\CV$
vanish.

We now consider a series expansion of the denominators in the second
line of \FSVVZ\ in terms of the nilpotent quantities $z \nu_l$.
Because of the relative signs in these denominators, and because of
the explicit cubic factor $\nu_l^3$ in the numerators, all terms of even
degree in the Chern roots $\nu_l$ automatically cancel.  However, by
our observation about $\CV$ above, the remaining terms of odd degree
in the $\nu_l$ cancel when we sum over roots.

From \FSVVZ, we see that $\log f_\CV(z)$ is determined up to a
quadratic function of $z$.  Hence $f_\CV(z)$ is determined up to
two real constants $b_0$ and $b_2$, \eqn\FSVVZII{ f_\CV(z) \,=\,
\exp{\left[i b_0 \,+\, i b_2 z^2 \, \Theta\right]}\,.} A term
linear in $z$ would necessarily appear with the first Chern class
$c_1(\CV)$, which vanishes by our observation above.  Since
$c_1(\CV) = 0$, the only degree two class that can appear in
\FSVVZII\ is the characteristic class $\Theta$.  We also observe
from the product \FSMVZ\ that $f_\CV(z)$ must be simply a phase
(for real $z$), since under complex conjugation $f_\CV(z)$ goes to
$f^{-1}_\CV(z)$.  This observation fixes the factors of `$i$' in
\FSVVZII.

Having fixed the general forms \FSMVZIII\ and \FSVVZII\ of $f_\CM$ and
$f_\CV$, we now compute the undetermined constants.  To do this, we
must still decide how to define the determinant of the operator
${\CD' = (1 / 2\pi) \, [\lie_R + i (\CR/p)]}$.  Motivated by our work
in Section 5.2, we proceed as follows.  First, although $p$ is a
positive, real variable in our problem, we will define the determinant
of $\CD'$ more generally for complex $p$.  Second, once we allow $p$
to be complex, we impose the requirement that the determinant of
$\CD'$ depend analytically on $p$.  In particular, if we evaluate the
determinant for purely {\it imaginary} $p$, of the form $p = i/y$ for
real $y > 0$ (the fact that we use $1/y$ is just for notational
convenience later), then the determinant is defined for real $p > 0$
by analytic continuation.  Finally, when $p = i/y$, we see that ${\CD'
= (1 / 2\pi) \, [\lie_R + y \CR]}$ is an anti-hermitian operator
exactly like $\CD_0$, and we can use zeta/eta-function regularization
to define the determinant of $\CD'$ for these values of $p$ as we did
in Section 5.2.

In terms of $f_\CM$ and $f_\CV$ in \FSMVZ, this definition of the
determinant of $\CD'$ amounts to the prescription to use
zeta/eta-function regularization to define the products
\eqn\FS{\eqalign{
f_\CM(z=-iy) \,&=\, \prod_{t\neq 0} \, \prod_{j=1}^{\dim \CM} \,  {i
\over {\left(t \,+\, y \varpi_j\right)}}\,,\cr
f_\CV(z=-iy) \,&=\, \prod_{t\neq 0} \, \prod_{l=1}^{\Delta_G} \, (-i)^{n
t} \left(t \,+\, y \nu_l\right)^{n t}\,.\cr}}
We first ignore the factors of `$i$' in \FS\ and we compute the
absolute values of $f_\CM$ and $f_\CV$.

For instance,
\eqn\FSII{
|f_\CM(-i y)| \,=\, \prod_{t \ge 1} \, \prod_{j=1}^{\dim
\CM}\, \left[t^2 - \left(y \varpi_j\right)^2\right]^{-1} \,=\,
\left({1 \over {2 \pi}}\right)^{\dim \CM} \cdot \prod_{j=1}^{\dim \CM}
\, {{\left(\pi y \varpi_j\right)}
\over {\sin(\pi y \varpi_j)}}\,.}
Since the Chern roots $\varpi_j$ are nilpotent, the terms in the first
product in \FSII\ are manifestly positive.  In passing to the second
equality, we apply the same identities \PRODS\ and \PRODSII\ from
Section 5.2.  This form of $|f_\CM(-i y)|$ is clearly compatible with
our general expression \FSMVZIII.

On the other hand, one can easily check that zeta-function
regularization defines the absolute value of $f_\CV$ to be trivial,
for the same topological reason that we explained following \FSVVZ, so
\eqn\FSIII{ |f_\CV(-i y)| \,=\, \prod_{t \ge 1} \,
\prod_{l=1}^{\Delta_G} \, \left[{{t \,+\, y \nu_l} \over {t \,-\, y
\nu_l}}\right]^{n t} \,=\, 1\,.}

We are left to compute the phases of $f_\CM(-i y)$ and $f_\CV(-i y)$.
We define these using the eta-function, as in Section 5.2.  More
precisely, we write
\eqn\PHFS{ f_\CM(-i y) \,=\, \exp{\left(-{{i \pi} \over 2} \,
\eta_\CM\right)} \cdot |f_\CM|\,,\qquad f_\CV(-i y) \,=\, \exp{\left(-{{i
\pi} \over 2} \, \eta_\CV\right)}\,.}
Here $\eta_\CM$ and $\eta_\CV$ denote the eta-invariants which arise
as the values at $s=0$ of the eta-functions $\eta_\CM(s)$ and
$\eta_\CV(s)$ abstractly associated to the hermitian operator $i \CD'$
as it acts on $\CE_0 \ominus \CE_1$,
\eqn\ETAOPIII{ i \CD' \,=\,  {i \over {2\pi}} \left(\lie_R \,+\, y
\CR\right)\,.}
This operator should be compared to the corresponding operator which
we considered when computing the phase of $e(p,\phi)$ at the trivial
connection,
\eqn\NBIGOPII{ {i \over {2\pi}} \left(\lie_R - \left[{\phi \over
p}, \,\cdot\, \right]\right)\,.}
We recall from Section 5.2 that the eta-invariant associated to the
operator in \NBIGOPII\ acquires an anomalous dependence on $(\phi/p)$
which produces the finite shift in the Chern-Simons level.  In the case at
hand, a similar anomalous dependence of $\eta_\CM$ and $\eta_\CV$ on
$y \CR$ gives rise to the same shift in the level.

Concretely, the eta-functions $\eta_\CM(s)$ and $\eta_\CV(s)$ are
given by the following regularized sums over the factors which appear
in $f_\CM(-iy)$ and $f_\CV(-iy)$ in \FS\ and which represent the eigenvalues
$\lambda$ of $i \CD'$,
\eqn\ETAEM{\eqalign{
\eta_\CM(s) \,&=\, \sum_{t \neq 0} \sum_{j=1}^{\dim \CM} \,
-\sgn\bigl(\lambda(t, \varpi_j)\bigr) \cdot \left|\lambda(t,
 \varpi_j)\right|^{-s}\,,\quad \lambda(t, \varpi_j) \,=\, t + y
 \varpi_j\,,\cr
\eta_\CV(s) \,&=\, \sum_{t \neq 0} \sum_{l=1}^{\Delta_G} \, n t \cdot
 \sgn\bigl(\lambda(t, \nu_l)\bigr) \cdot \left|\lambda(t,
 \nu_l)\right|^{-s}\,,\quad \lambda(t, \nu_l) \,=\, t + y \nu_l\,.\cr}}
The various constants appearing in \ETAEM\ are perhaps most clear if
we compare to the formal expressions for $f_\CM(-i y)$ and $f_\CV(-i y)$
in \FS.  Thus, the overall minus sign in $\eta_\CM(s)$ arises because
 $i$ as opposed to $-i$ appears in $f_\CM(-i y)$, which is in turn  
 associated to the fact that we consider $\CE_0 \ominus \CE_1$ as 
 opposed to $\CE_0 \oplus \CE_1$.  Similarly, the multiplicity factor
 $n t$ appears in $\eta_\CV(s)$ because of the factor $(-i)^{n t}$ in
$f_\CV(-i y)$.

Because the Chern roots are nilpotent, we note that
${\sgn(\lambda(t,x)) = \sgn(t)}$, where ${x = \varpi_j}$ or ${x=\nu_l}$ as
the case may be.  Thus, we write the regularized sums in \ETAEM\
explicitly as
\eqn\ETAEMII{\eqalign{
\eta_\CM(s) \,&=\, \sum_{t \ge 1} \sum_{j=1}^{\dim \CM} \, {{-1}
\over {\left(t \,+\, y \varpi_j\right)^s}} \,+\, \sum_{t
\ge 1} \sum_{j=1}^{\dim \CM} \, {1 \over {\left(t \,-\, y
\varpi_j\right)^s}}\,,\cr
\eta_\CV(s) \,&=\, \sum_{t \ge 1} \sum_{l=1}^{\Delta_G} \, {{n t} \over
{\left(t \,+\, y \nu_l\right)^s}} \,+\, \sum_{t \ge 1}
\sum_{l=1}^{\Delta_G} \,  {{n t} \over {\left(t \,-\, y
\nu_l\right)^s}}\,.}}
As in Section 5.2, we are left to evaluate these sums at $s=0$.

In fact, we have already done all of the required computation.  The
sum which defines $\eta_\CM(s)$ is the same as the sum \LITETAI\ which
we evaluated in the warmup computation on $S^1$ in Section 5.2.  Thus
we find
\eqn\ETAM{ \eta_\CM(0) \,=\, 2 y \sum_{j=1}^{\dim \CM}
\, \varpi_j \,=\,  2 y \, c_1(T\CM)\,.}
In deducing the second equality, we note that the trace over all Chern
roots of $T\CM$ is the first Chern class of $T\CM$.

To evaluate $\eta_\CV(0)$, we perform a computation precisely
isomorphic to our computation of $e(p,\phi)$ in Section 5.2.  Applying
our earlier results, we find
\eqn\ETAVII{ \eta_\CV(0) \,=\,  \eta_0 \,+\, n  y^2 \,
\sum_{l=1}^{\Delta_G} \nu_l^2\,,\qquad \eta_0 \,=\, - {{n \Delta_G}
\over 6}\,.}
Here $\eta_0$ is the same constant that appeared in our localization
computation at the trivial connection.  As for the term
quadratic in $\nu_l$, this term arises in the same way as the
term quadratic in $\phi$ in \DLETA.

We now recall from Section 5.2 that we applied a Lie algebra identity
\BBAKI\ involving $\check{c}_\Fg$ to rewrite the term quadratic in
$\phi$ in \DLETA\ in terms of the natural quadratic invariant $\ha
\Tr(\phi^2)$.  Under the Chern-Weil homomorphism, by
which we identify the Chern roots $\nu_l$ with the eigenvalues
of the curvature operator $i \CR / 2 \pi$, we can apply the same  Lie
algebra identity to rewrite the degree four class ${\sum_l
\nu_l^2}$ in terms of the class $\Theta$ that already appears in
the integral over $\CM$.  We find from the identity \BBAKI\ that
\eqn\CHTWO{ \sum_{l=1}^{\Delta_G} \nu_l^2 \,=\, {{\check{c}_\Fg
\Theta} \over {\pi^2}}\,,}
and $\eta_\CV(0)$ becomes
\eqn\ETAV{ \eta_\CV(0) \,=\, \eta_0 \,+\, {{n \check{c}_\Fg} \over
{\pi^2}} \, y^2 \Theta\,.}

With these results \ETAM\ and \ETAV, we evaluate $f_\CM(-i y)$
and $f_\CV(-i y)$ to be
\eqn\PHFSII{\eqalign{
f_\CM(-i y) \,&=\, \exp{\left(-i \pi y \, c_1(T\CM)\right)} \cdot
\left({1 \over {2 \pi}}\right)^{\dim \CM}
\cdot \prod_{j=1}^{\dim \CM} \, {{\left(\pi y \varpi_j\right)}
\over {\sin(\pi y \varpi_j)}}\,,\cr
f_\CV(-i y) \,&=\, \exp{\left(-{{i \pi} \over 2} \eta_0  \,-\, {{i n
\check{c}_\Fg} \over {2 \pi}} y^2 \Theta\right)}\,.\cr}}
Upon setting $z = -i y$, these expressions assume the same form as the
general expressions in \FSMVZIII\ and \FSVVZII.

We recall that $p$ is related to $y$ via $p = i/y$.  So $e(p)$, as
determined by the analytic continuation of \PHFSII, is finally
given by \eqn\ERPF{\eqalign{ e(p) \,&=\, p^{\dim \CM} \cdot
f_\CM(p) \cdot f_\CV(p)\,,\cr &=\, \exp{\left(-{{i \pi} \over 2}
\eta_0 \,+\, {\pi \over p} \, c_1(T\CM) \,+\, {{i n \check{c}_\Fg}
\over {2 \pi p^2}} \Theta\right)} \, \left({p \over {2
\pi}}\right)^{\dim \CM} \, \prod_{j=1}^{\dim \CM} \, {{\left({{\pi
\varpi_j} / p}\right)} \over {\sinh({{\pi \varpi_j} /
p})}}\,.\cr}} As we will see, this formula incorporates the famous
shift in the Chern-Simons level $k$, and leads to agreement with
the results of Rozansky.

\medskip\noindent{\it Some Further Remarks}\smallskip

Our use of zeta/eta-function regularization to define $e(p)$, and
especially the analytic continuation we performed in $p$, is
somewhat {\it ad hoc}.  The need for analytic continuation would
have been avoided if at the very beginning of this paper, we had
introduced the Cartan model of equivariant cohomology with the
differential  $D=d+ \iota_V$ rather than the choice we actually
made, $D=d+i \iota_V$.  This would have resulted in the basic
symplectic volume integral on a symplectic manifold ${\cal M}$
being $\int_{\CM}\exp(i\Omega)$ rather than the more standard
$\int_{\CM}\exp(\Omega)$; it also would clash with some conventions
of physicists about reality conditions for fermions.  However, it
would clarify our discussion of the determinants, since if all
factors of $i$ are omitted from the localization form $\lambda$,
then the operator $i\CD'$  would come out to be hermitian.  Hence,
the zeta/eta-function definition of determinants could be
implemented with no need for artificial analytic continuation.

That definition is really most natural for oscillatory bosonic
integrals such as appear in Chern-Simons theory. If a bosonic
integral \eqn\nonn{Z=\int D\Phi\exp(i(\Phi,M\Phi)),} for some
indefinite real symmetric operator $M$, is regularized by $M\to
M+i\varepsilon$, for small positive $\varepsilon$, then the phase of $Z$
is naturally $\exp(i\pi\eta(M)/2)$.  This is really why, in
Chern-Simons theory, eta-invariants appear in the one-loop
corrections.  If we take $D=d+\iota_V$, and take the localization
form $\lambda$ to be purely imaginary rather than purely real,
then all integrals in Appendix D are oscillatory Gaussian
integrals rather than real Gaussians.  This gives a natural
framework for zeta/eta-function regularization of the determinants
in our localization computation.

Our general analysis of $d^2\log f_{\CM}(z)/dz^2$ and $d^3\log
f_{\CV}(z)/dz^3$ showed that any reasonable definition of these
determinants would differ from the zeta/eta-function approach by
adding a constant to $\eta_0$ and changing the coefficients of
$c_1(T\CM)$ and $\Theta$ in \ERPF.  We will see shortly that the
coefficients as written in \ERPF\ do agree with Chern-Simons
theory; in fact, they show up in Chern-Simons theory at the
one-loop level.  Ultimately, to justify the coefficients in \ERPF\
on an {\it a priori} basis requires a more rigorous comparison
between the localization procedure and Chern-Simons theory.

\medskip\noindent{\it The Contribution From $\CM$ in Chern-Simons
Theory}\smallskip

Having evaluated $e(p)$, we now set $p=1/\epsilon$ and substitute
\ERPF\ into our expression \MVI\ for the contribution from $\CM$ to
the Chern-Simons path integral.  Thus,
\eqn\MII{\eqalign{
Z(\epsilon)\Big|_\CM \,&=\, {1 \over {|\Gamma|}} \, \exp{\left(-{{i
\pi} \over 2} \eta_0 \right)}\, \left({1 \over {2 \pi
\epsilon}}\right)^{\dim \CM} \,\times\cr
&\times\, \int_\CM \exp{\left[\Omega + \pi \epsilon \, c_1(T\CM) + i
\epsilon n \left(1 + {{\epsilon \check{c}_\Fg} \over {2
\pi}}\right)\Theta\right]}\,\prod_{j=1}^{\dim \CM} \, \left[{{\pi
\epsilon \varpi_j} \over {\sinh\left(\pi \epsilon
\varpi_j\right)}}\right]\,.\cr}}

Since we are dealing with an integral, by making changes of variables
we can rewrite the integrand of \MII\ in different ways which illuminate
different features of this result.  In the form at hand,
we note that one can define a non-trivial scaling limit of \MII\ such
that the Chern-Simons coupling $\epsilon$ goes to zero (so that the
level $k$ goes to $\infty$) and the degree $n$ of $\CL$ goes to
$\infty$ with $\epsilon n$ held fixed.  In this limit, which
physically decouples all the higher Kaluza-Klein modes of the gauge
field, we see directly that the contribution from $\CM$ in
Chern-Simons theory has the same form as the simple expression \ZV\
for the corresponding contribution from $\CM_0$ in two-dimensional
Yang-Mills theory.

To express \MII\ more compactly, we now rescale all elements of the
cohomology ring of $\CM$ by a factor $(2 \pi \epsilon)^{q/2}$, where
$q$ is the degree of the given class.  So for instance, the degree two
Chern roots $\varpi_j$ scale as $\varpi_j \to 2 \pi \epsilon \, \varpi_j$.
This trivial change of variables cancels the prefactor involving
$\epsilon$ in \MII\ and reduces the product over Chern roots in \MII\
to a well-known characteristic class, the $\hat A$-genus of $\CM$.

We recall that the $\hat A$-genus of $\CM$ is given in terms of the Chern
roots of $T\CM$ as
\eqn\AROOFMII{ \hat A(\CM) \,=\, \prod_{j=1}^{\dim \CM} \, {{\varpi_j / 2}
\over {\sinh(\varpi_j / 2)}}\,.}
In a sense, the appearance of the $\hat A$-genus in our problem is not
so surprising, since it appears in roughly the same way as in the
standard path integral derivations of the index theorem.  See
\AtiyahCS\ for a derivation of the index theorem that applies abelian
localization to a sigma model path integral;  at least formally, that
computation shares many features of our computation here.

In terms of the $\hat A$-genus, our expression in \MII\ simplifies to
\eqn\MIII{ Z(\epsilon)\Big|_\CM \,=\, {1 \over{|\Gamma|}} \,
\exp{\left(-{{i \pi} \over 2} \eta_0\right)} \,
\int_\CM \hat A(\CM) \, \exp{\left[{1 \over {2 \pi
\epsilon}} \, \Omega + \ha c_1(T\CM) + {{i n} \over {4 \pi^2
\epsilon_r}} \, \Theta\right]}\,.}
Here we have absorbed the contribution from $\eta_\CV(0)$ into a
renormalization of the coupling ${\epsilon_r = 2 \pi / (k +
\check{c}_\Fg)}$ that appears in front of $\Theta$.

Of course, we would like to write \MII\ entirely in terms of the
renormalized coupling $\epsilon_r$.  To do so, we apply a theorem of
\Drezet\ which relates the first Chern class $c_1(T\CM)$ to the
symplectic form $\Omega$ in the case of gauge group ${G = SU(r+1)}$.  In
this case,
\eqn\DRZTHM{ c_1(T\CM) \,=\, 2 (r+1) \, \Omega'\,,}
where $\Omega' = \Omega / (2 \pi)^2$ is the standard, integral symplectic
form on $\CM$.  Happily, the dual Coxeter number $\check{c}_\Fg$ of $G
= SU(r+1)$ is also given by ${\check{c}_\Fg = r+1}$, so we see that
\MIII\ can be expressed very simply using $\epsilon_r$,
\eqn\MIIV{ Z(\epsilon)\Big|_\CM \,=\, {1 \over{|\Gamma|}} \,
\exp{\left(-{{i \pi} \over 2} \eta_0\right)} \, \int_\CM \hat
A(\CM) \, \exp{\left[{1 \over {2 \pi \epsilon_r}} \left( \Omega + {{i
n} \over {2 \pi}} \, \Theta\right)\right]}\,.}
This expression is of the same form as the corresponding result of
Rozansky in \RozanskyWV.

We close with the following amusing observation.  On general grounds,
the $\hat A$-genus of $\CM$ is related to the Todd class $\Td(\CM)$
of $\CM$ by
\eqn\ARTD{ \Td(\CM) \,=\, \exp{\left(\ha c_1(T\CM)\right)} \, \hat
A(\CM)\,.}
So from \MIII, we see that an alternative expression for the path
integral contribution from $\CM$ is
\eqn\MIIVI{ Z(\epsilon)\Big|_\CM \,=\, {1\over{|\Gamma|}} \,
\exp{\left(-{{i \pi} \over 2} \eta_0\right)} \, \int_\CM \Td(\CM)
\, \exp{\left[k \, \Omega' + {{i n} \over {4 \pi^2
\epsilon_r}} \, \Theta\right]}\,.}
Although our derivation of \MIIVI\ is not valid for the trivial case ${M =
S^1 \times \Sigma}$, we see that, upon setting $n=0$, our result
\MIIVI\ takes the same form as the index formula \INDX\ for
$Z(\epsilon)$ in the trivial case.  It is satisfying to see that both the
index formula \INDX\ and the two-dimensional Yang-Mills formula \ZV\
are reproduced as special limits of our general result.

\bigbreak\bigskip\bigskip\centerline{{\bf Acknowledgements}}\nobreak

CB would like to thank John McGreevy, Sameer Murthy, Peter Ouyang,
Ronen Plesser, Peter Svr\v cek, and Martijn Wijnholt for stimulating
discussions.  CB would also like to thank the organizers and
participants of the 2004 Simons Workshop, where some of this work was
performed.

The work of CB is supported in part under National Science
Foundation Grants No. PHY-0243680 and PHY-0140311 and by a
Charlotte Elizabeth Procter Fellowship. The work of EW is
supported in part under National Science Foundation Grant No.
PHY-0070928.  Any opinions, findings, and conclusions or
recommendations expressed in this material are those of the
authors and do not necessarily reflect the views of the National
Science Foundation.

\vfill
\eject

\appendix{A}{Brief Analysis to Justify the Localization Computation in
Section 4}

In this appendix, we show that the quantity $Q \cdot
Z'(\epsilon)$ computed using $\lambda'$ in \NEWZINTII\ of Section 4.3
agrees with the same quantity defined using $\lambda$, so that
$Z'(\epsilon)$ as defined by integrating \NEWZINTII\ agrees with
$Z(\epsilon)$.  Thus we consider the following one-parameter
family of invariant forms, interpolating from $\lambda$ to $\lambda'$
on $F$,
\eqn\INTERP{ \Lambda(s) \,=\, s \, \lambda + (1-s) \,
\lambda'\,,\qquad s \in [0,1]\,,}
and to start we consider the corresponding family $Z(\epsilon,s)$ of
integrals over $F$,
\eqn\INTERPZ{ Z(\epsilon,s) \,=\, {1 \over {\Vol(H)}} \,
\int_{\Fh \times F} \left[{{d\phi} \over {2\pi}}\right] \,
\exp{\left[\Omega - i \, \langle\mu,\phi\rangle - {\epsilon \over 2}
(\phi,\phi) + t \, D\Lambda(s) \right]}\,.}
If this integral is convergent for all $s$ and also continuous as a
function of $s$, then $Z(\epsilon,s)$ is independent of $s$, so that
$Z(\epsilon) = Z(\epsilon, 1) = Z(\epsilon, 0) = Z'(\epsilon)$.  This
fact follows by differentiating the integrand of \INTERPZ\ with
respect to $s$, which produces a total derivative on $F$.

We thus need to consider the basic convergence and continuity of
$Z(\epsilon,s)$.  Very broadly, divergences in the integral over $F$
in \INTERPZ\ can only arise from integration over the non-compact
fibers $\Fh^\perp$ and $E_1$ which sit over the compact orbit $H/H_0$.
However, the first, degree one term of $\lambda'$ in \LOCLAM\ is
precisely of the canonical form to define localization on the fiber
$\Fh^\perp$, exactly as in our computation on $T^* H$.  Thus, no
divergence arises from the integral over $\Fh^\perp$, and we need only
analyze the integral over the complex vector space $E_1$.  As we have
already seen, precisely this integral over $E_1$ leads to the
dangerous, possibly singular factor in $I(\psi)$ in \ZVWHKIII.
Furthermore, in our application to Yang-Mills theory, the
corresponding vector space $\CE_1$ describes the set of
gauge-equivalence classes of unstable modes of the Yang-Mills action,
and we expect the integral over these modes to be the most delicate.

We now analyze directly the symplectic integral over $E_1$ that arises
from \INTERPZ.  To set up notation, we recall that $E_1$ is a complex
vector space, ${\dim_\BC E_1 = d_1}$, with an invariant, hermitian
metric $(\cdot,\cdot)$ and an invariant symplectic form $\wt\Omega$.
In terms of holomorphic and anti-holomorphic coordinates $v^n$ and
$\bar v{}^{\bar n}$ on $E_1$, $\wt\Omega$ is given by
\eqn\EONEOM{ \wt\Omega \,=\, -{i \over 2} \left(dv, dv\right) \,=\,
-{i \over 2} \, d \bar v_n \^ dv^n\,.}
If $\psi$ is an element of $\Fh_0$, then the corresponding vector
field $V(\psi)$ on $E_1$ is described by
\eqn\VPSI{ \delta v \,=\, \psi \cdot v\,,}
or in coordinates, $\delta v^n = \psi^n_m v^m$, and similarly for the
conjugate components of $V(\psi)$.

From \EONEOM\ and \VPSI, we see that the moment map $\wt\mu$ for the
action of $H_0$ on $E_1$ is explicitly given by
\eqn\EONEMOM{ \langle\wt\mu,\psi\rangle \,=\, {i \over 2} \left(v,\psi
\cdot v\right)\,.}
By our assumption that $(\cdot,\cdot)$ is invariant under \VPSI,
$\psi$ is anti-hermitian and the expression in \EONEMOM\ is real.

Of course, the complex structure $J$ acts on $E_1$ as ${J(dv) = -i
\, dv}$ and ${J(d\bar v) = +i \, d\bar v}$.  Thus, since
\eqn\EONES{ S = \ha (\wt\mu,\wt\mu) = {1 \over 8} (v,v)^2\,,}
we see that the canonical one-form $\lambda = J \, dS$ is given by
\eqn\EONELAM{ \lambda \,=\, -{i \over 4} \left(v,v\right)
\left(\left(v,dv\right) - \left(dv,v\right)\right)\,.}
On the other hand, from \LOCLAM\ we see that $\lambda'$ on $E_1$
reduces to
\eqn\EONELAMII{ \lambda' \,=\, i \left(\psi \cdot v, dv\right)\,.}

Thus, if we restrict the integral in \INTERPZ\ to $E_1$ and keep only
the terms relevant in the limit of large $t$ (after which we set
$t=1$), we just consider the reduced integral
\eqn\ZEONE{ Z_{red}(\epsilon,s) \,=\,  \int_{\Fh_0 \times E_1}
\left[{{d\psi} \over {2\pi}}\right] \, \exp{\left[ -i
\left(\gamma_0,\psi\right) - {\epsilon \over 2} \left(\psi,\psi\right)
+ s \, D \lambda + \left(1 - s\right) \, D \lambda'\right]}\,.}
Of the original integral over the full Lie algebra $\Fh$ of $H$, only
the integral over the subalgebra $\Fh_0$ is relevant to the integral
over $E_1$.

We first perform integral over $\psi$ in $\Fh_0$.  To illustrate the
essential behavior of the integral over $E_1$, we assume as before
that $\Fh_0 = \BR$ has dimension one.  Explicitly, $D\lambda$ and
$D\lambda'$ depend on $\psi$ as
\eqn\DEONELAM{ D\lambda \,=\, d \lambda + \ha \left(v, v\right)
\left(v, \psi \cdot v\right),}
and
\eqn\DEONELAMII{ D\lambda' \,=\, i \left(\psi \cdot dv, dv\right) -
\left(\psi \cdot v, \psi \cdot v\right),}
so the integral over $\psi$ is purely Gaussian.  Upon performing this
integral over $\psi$, we find that $Z_{red}$ is formally given by
\eqn\ZEONEII{ Z_{red}(\epsilon,s) \,=\, \int_{E_1} \, \left(4 \pi
A\right)^{-\ha} \, \exp{\left[ s \, d\lambda + {1 \over
4} \left(J, A^{-1} \, J\right)\right]}\,,}
where $A$ is defined in terms of the normalized generator $T_0$ of
$\Fh_0$ by
\eqn\BIGA{ A \,=\, {\epsilon \over 2} \,+\, \left(1-s\right) \left(T_0
\cdot v, T_0 \cdot v\right)\,,}
and $J$ in $\Fh_0$ is defined by
\eqn\BIGJ{ J \,=\, -i \, \gamma_0 +
{s \over 2} \left(v, v\right) \left(v, T_0 \cdot v\right) \, T_0 + i
\left(1 - s\right) \left(T_0 \cdot dv, dv\right) \, T_0\,.}

We are now interested in the behavior of the integral in \ZEONEII\ for
large $|v|$, where the non-compactness of $E_1$ is essential.  So long
as $s \neq 0$, then the integrand of \ZEONEII\ falls off at least as
fast as $\exp{[-(v,v)^3]}$ for large $v$, due to the term quartic in
$v$ in \BIGJ\ that arises from $\lambda$ and the term quadratic in $v$ in
\BIGA\ that arises from $\lambda'$.  Thus, the integral over $E_1$ is
strongly convergent for $s \neq 0$ and depends smoothly on $s$ away
from $0$.  Of course, this integral is also non-Gaussian and cannot be
simply expressed using elementary functions.

However, when $s=0$, the integrand of \ZEONEII\ is no longer
suppressed exponentially and decays only as a power law at infinity.
This behavior arises because the bosonic term of $D\lambda'$ is
quadratic in $\psi$, whereas the bosonic term of $D\lambda$ is linear
in $\psi$.  Because the integrand of \ZEONEII\ decays only as a power
law for $s=0$, the integral over $E_1$ does not generally converge.  The
prefactor proportional to $A^{-1/2}$ decays like $1 / |v|$, and for
$s=0$ the measure arising from the quadratic term $(J, A^{-1} \, J)$
in the exponential of \ZEONEII\ is of the form $1/|v|^{d_1} \, d^{2
d_1} v$.  Consequently, the integral over $E_1$ behaves as ${\int d^{2
d_1} v \,\, 1/|v|^{(d_1+1)}}$ for large $|v|$ and diverges.

However, we now consider the same analysis as applied to $Q \cdot
Z(\epsilon,s)$.  By our analysis above, we are only concerned with the
potentially dangerous behavior near $s=0$ and for large $|v|$, for
which we must consider the following integral over $E_1$,
\eqn\ZEONEIII{ \left(-2 {\partial \over {\partial
\epsilon}}\right)^{\ha d_1} \cdot Z_{red}(\epsilon,s) \,=\,  \int_{E_1}
\left(-2 {\partial \over {\partial \epsilon}}\right)^{\ha d_1}
\left(\left(4 \pi A\right)^{-\ha} \, \exp{\left[ s \, d\lambda + {1 \over
4} \left(J, A^{-1} \, J\right)\right]}\right)\,.}

To analyze \ZEONEIII, we first note that $\epsilon$ only appears in
the quantity $A$ in \BIGA, and $A$ satisfies
\eqn\BIGAII{ \left( -2 {\partial \over {\partial \epsilon}} + {1 \over
{(1-s)}} {{\partial^2} \over {\partial \bar v_i \, \partial
v^i}}\right) A \,=\, 0\,.}
Thus, we can rewrite \ZEONEIII\ as
\eqn\ZEONEIV{\eqalign{
\left(-2 {\partial \over {\partial \epsilon}}\right)^{\ha d_1} \cdot
Z_{red}(\epsilon,s) \,=\, \int_{E_1} &\left(- {1 \over {(1-s)}}
{{\partial^2} \over {\partial \bar v_i \, \partial v^i}}\right)^{\ha
d_1} \, \times\cr
&\times \, \left(\left(4 \pi A\right)^{-\ha} \, \exp{\left[ s \,
d\lambda + {1 \over 4} \left(J, A^{-1} \, J\right)\right]}\right)\,.\cr}}

We now apply simple scaling arguments to \ZEONEIV\ to show that this
integral is convergent at $s=0$ and behaves continuously as $s \rightarrow
0$.  First, at $s=0$, we immediately see that this integral behaves
for large $|v|$ as ${\int d^{2 d_1} v \,\, 1/|v|^{(2 d_1+1)}}$ and
hence is convergent, though just barely.

To discuss the limit $s \rightarrow 0$, we assume $s$ is fixed at a small,
non-zero value.  All terms involving $s$ which we previously dropped
for $s=0$ now appear in the argument of the exponential in \ZEONEIV.
For large $|v|$, this argument behaves schematically as
\eqn\ARGEXP{ s \, |v|^2 \left(dv, dv\right) +
{{\left(\gamma_0,\gamma_0\right)} \over {|v|^2}} + s \, |v|^2
\left(\gamma_0, T_0\right) + {{\left(dv, dv\right)} \over {|v|^2}}
\left(\gamma_0, T_0\right) + s^2 \, |v|^6 + {{\left(dv, dv\right)^2}
\over {|v|^2}}\,.}
Since our argument is only a scaling argument, we ignore all signs and
constants in writing \ARGEXP, though we do recall that the dominant
term $s^2 \, |v|^6$ leads to an exponential decay of the integrand at
large $v$.

We see three terms in \ARGEXP\ which vanish in the limit $s
\rightarrow 0$.  Of these terms, we can ignore the quadratic term $s
\, |v|^2 (\gamma_0, T_0)$, since it is subleading compared to $s^2 \,
|v|^6$ for fixed $s$ and large $|v|$.

However, we need to consider the effect of the measure $s^2 \, |v|^4
\left(dv, dv\right)^2$, which dominates the measure $(dv, dv)^2 /
|v|^2$ at $s=0$ by a relative factor of $s^2 \, |v|^6$.  We also need
to consider the terms which arise when the derivative $\partial^2 /
\partial \bar v_i \, \partial v^i$ in \ZEONEIV\ acts on $\exp{(-s^2 \,
|v|^6)}$ to bring down the term $s^2 \, |v|^4$, which dominates $1 /
 |v|^2$ by the same relative factor $s^2 |v|^6$.

These terms lead to contributions depending on $s$ in
\ZEONEIV\ which behave for large $|v|$ as
\eqn\SINTS{ \int_{E_1} d^{2 d_1} v \, {1 \over {|v|^{2 d_1 +
1}}} \, s^{2n} \, |v|^{6 n} \, \exp{(- s^2 |v|^6)}\,,\qquad
n=1,\,\ldots\,, d_1\,.}
Since these integrals only converge for $s \neq 0$, when
the integrand is exponentially damped, one might have worried
that these terms could cause the limit $s \rightarrow 0$ to be
singular.  However, we see by scaling that the expression in \SINTS\
behaves as $s^{+1/3}$ for all $n$ and hence the asymptotic
contributions to \ZEONEIV\ from these terms still go continuously to
zero as $s \rightarrow 0$.

Finally, apart from the terms in \SINTS\ with $n \ge 1$, the integrand
of \ZEONEIV\ is a smooth function $F(v, s)$ of $v$ and $s$ which behaves
asymptotically for large $|v|$ as
\eqn\ASYMP{ F(v, s) \,\sim\, {1 \over {|v|^{2 d_1 + 1}}} \, \exp{(-s^2
\, |v|^6)}\,.}
Thus, $F(v, s)$ decays exponentially for $s \neq 0$, is
integrable for all $s$, and is dominated by $F(v, 0)$, which has a pure
power law decay at infinity.  On general grounds, the integral of
$F(v, s)$ over $E_1$ then depends continously on $s$, and, for the
purpose of computing $Q \cdot Z(\epsilon)$, we can validly interpolate
from $\lambda$ to $\lambda'$ on $F$.

\appendix{B}{More About Localization at Higher Critical Points: Higher
Casimirs}

In this appendix, we continue from Section 4.3 our general discussion
of non-abelian localization at higher critical points.  We recall that
we obtained a formal expression for the canonical symplectic integral
over $F$ in terms of an integral over the Lie algebra $\Fh_0$ of the
stabilizer group $H_0$,
\eqn\BI{ Z(\epsilon) \,=\, {1 \over {\Vol(H_0)}} \, \int_{\Fh_0}
\left[{{d\psi} \over {2\pi}}\right] \, \det\left({{\psi} \over {2
\pi}}\Big|_{E_0}\right) \, \det\left({{\psi} \over {2
\pi}}\Big|_{E_1}\right)^{-1}\, \, \exp{\left[-i \left(\gamma_0,
\psi\right) - {\epsilon \over 2} \left(\psi, \psi\right)\right]}\,.}

As we discussed in Section 4.3, this integral generally fails to
converge when the ratio of determinants in the integrand has
singularities in $\Fh_0$.  In the special case $H_0 = U(1)$,
relevant for higher critical points of $SU(2)$ Yang-Mills theory, we
deal with this problem by computing not $Z(\epsilon)$ itself but a
higher derivative $Q \cdot Z(\epsilon)$, where $Q \equiv
Q(\partial/\partial \epsilon)$ is a differential operator which we
choose so that the action of $Q$ on the integrand of \BI\ brings down
sufficient powers of $(\psi,\psi)$ to cancel any poles that would
otherwise appear.

However, if we consider higher critical points of Yang-Mills theory
with general gauge group $G$, then the rank of $H_0$ can be arbitrary, and
the determinants in \BI\ cannot generally be expressed as a
functions of only  the quadratic invariant $(\psi,\psi)$.
Consequently, we cannot simply differentiate $Z(\epsilon)$ with
respect to $\epsilon$ to cancel the poles in \BI.

Nevertheless, by applying some simple ideas about the localization
construction, we can generalize our discussion in Section 4.3 to
the case that $H_0$ has higher rank. As in Section 4.1, we recall
the form of the localization integral:
\eqn\BII{ Z(\epsilon) \,=\,
{1 \over {\Vol(H)}} \, \int_{\Fh \times X} \left[{{d\phi} \over
{2\pi}}\right] \exp{\left[\Omega - i \, \langle\mu,\phi\rangle -
{\epsilon \over 2} (\phi,\phi)\right]}\,.} In the case of
Yang-Mills theory, $H = \CG(P)$ and $X = \CA(P)$ in the notation
of Section 2.

Let us consider what natural generalizations of \BII\ exist.  Of
the terms appearing in \BII, the quantity ${\Omega - i \,
\langle\mu,\phi\rangle}$ is distinguished as an element of the
equivariant cohomology ring of $X$, since it represents the
equivariant extension of the symplectic form on $X$.  However,
nothing really distinguishes the quadratic function $-\ha
(\phi,\phi)$ among all invariant polynomials of $\phi$, and we are
free to consider a general symplectic integral over $\Fh \times X$
of the form \eqn\BIII{ Z[V] \,=\, {1 \over {\Vol(H)}} \, \int_{\Fh
\times X} \left[{{d\phi} \over {2\pi}}\right] \exp{\left[\Omega -
i \, \langle\mu,\phi\rangle - V(\phi)\right]}\,.} Here $V(\phi)$
is any invariant polynomial on $\Fh$ such that the integral over
$\Fh$ remains convergent at large $\phi$.  We can take
\eqn\tako{V(\phi)=\sum_j \epsilon_j \, C_j(\phi),} where $C_j$ are the
Casimirs of $H$ -- the homogeneous generators of the ring of
invariant polynomials on $\Fh$ -- and $\epsilon_j$ are parameters.
The standard localization technique can be applied to evaluate this
integral.  The fact that $V$ is not quadratic in $\phi$ leads to no
special complications.

In the case of Yang-Mills theory on a Riemann surface $\Sigma$
with symplectic form $\omega$, we would write \eqn\VP{ V(\phi)
\,=\, \sum_{j=1}^r \, \epsilon_j \int_\Sigma \omega \cdot
C_j(\phi)\,.}   We assume that the gauge group $G$ has rank $r$,
and now $C_j(\phi)$ are the Casimirs of $G$.  We associate to each
generator a corresponding coupling $\epsilon_j$.  If we want to
compare to standard methods of studying two-dimensional Yang-Mills
theory by cut and paste methods, we should integrate over $\phi$
to express the theory in terms of the gauge field (and
noninteracting fermions) alone. Of course, if $V(\phi)$ is not
quadratic, we can no longer perform the integral over $\phi$ in
\BIII\ as a Gaussian integral. Instead, if we abstractly introduce
the Fourier transform
\eqn\FTV{ {\exp{\left[-\hat V(\phi^*)\right]}}\,\equiv\,
\int_{\Fh} \left[{{d\phi} \over {2\pi}}\right] \exp{\left[ - i \,
\langle\phi^*,\phi\rangle - V(\phi)\right]}\,,}
which is an invariant function of $\phi^*$ in the dual algebra
$\Fh^*$, then the generalized symplectic integral over $X$ takes the form
\eqn\BIV{  Z[V] \,=\, {1 \over {\Vol(H)}} \, \int_X
\exp{\left(\Omega - \hat V(\mu)\right)}\,.}

In the case of Yang-Mills theory, we recall that $\mu = F_A$.  So
in that case, \BIV\ corresponds to a generalization of Yang-Mills
theory in which the action is not the usual $\Tr\, f^2$ (with
$f=\star F$) but $\Tr\,\hat V(f)$, for some more general function
$\hat V$.  The partition function of this generalized Yang-Mills
theory can be computed by the usual cut and paste methods
\MigdalZG.  If $G$ is simply-connected and we apply the same
normalization conventions as we used in \TDYMZ\ for the case
$G=SU(2)$, the generalized partition function is
\eqn\MYT{ Z[V] \,=\, (\Vol(G))^{2g-2} \,
\sum_\CR {1\over \dim(\CR)^{2g-2}}\exp(- V'(\CR)),} where
$V'(\CR)$ is the energy of the representation $\CR$.  (We are
taking the area of $\Sigma$ to be 1; for a general area $\alpha$,
the exponential factor would be $\exp(- \alpha V'(\CR))$.)  To
compute the energy $V'(\CR)$, we start with the action $\hat V(f)$
and compute the canonical momentum $\Pi=\partial\hat V/\partial
f$. The Hamiltonian, whose eigenvalue is the energy, is then $H=f
\Pi-\hat V(f)$, which must be extremized with respect to $f$ and
regarded as a function of $\Pi$.  Thus, $H(\Pi)$ is the Legendre
transform of $\hat V(f)$. After computing $H(\Pi)$, $\Pi$ is
interpreted as the generator of the group $G$ and taken to act on
the representation $\CR$ to get the energy $V'(\CR)$. Since the
Legendre transform is a semiclassical approximation to the Fourier
transform, the Legendre transform approximately undoes the Fourier
transform, and hence $H(\Pi)=V(\Pi)+$lower order terms.   As
discussed in \WittenXU, if the representation $\CR$ has highest
weight $h$, the precise formula needed to match with the
localization computation is $V'(\CR)=V(h+\delta)$, where the
constant $\delta$ is half the sum of positive roots of the Lie
algebra of $G$.  This formula  incorporates the difference between
the Legendre transform and the Fourier transform and other
possible quantum corrections.

To generalize what we said in Section 4.3, we want to find a
polynomial $F(C_j)$ of the Casimirs of $H$ which when restricted
to $\Fh_0$ is divisible by the troublesome factor in the
denominator, namely ${w(\psi) = \det\left({{\psi} / {2
\pi}}\big|_{E_1}\right)}$. Then
$Q=F(-\partial/\partial\epsilon_j)$ is a differential operator
that when acting on $\exp(-V)$ will produce the factor $F$ and
cancel the denominator.  Thus, $Q$ generalizes the operator
$\partial^{g-1}/\partial\epsilon^{g-1}$ that we used in Section
4.3 for two-dimensional $SU(2)$ gauge theory in genus $g$.

The troublesome factor $w$ is an invariant polynomial on the Lie
algebra of $\Fh_0$ or equivalently, a polynomial on the maximal
torus of $H_0$ that is invariant under the Weyl group of $H_0$.
This polynomial can be extended, though not canonically, to a
polynomial $w'$ on the maximal torus of $H$.  We can pick the
extension to be invariant under the Weyl group of $H_0$ but not
necessarily under the Weyl group of $H$.  However, by multiplying
$w'$ by all its conjugates under the Weyl group of $H$, we make a
polynomial $\wt w$ on the maximal torus of $H$ that is invariant
under the Weyl group of $H$, and whose restriction to $H_0$ is
divisible by $w$.  The Weyl-invariant polynomial $\wt w$
corresponds to the polynomial $F(C_j)$ of the Casimirs that was
used in the last paragraph.

Finally, let us make this more explicit for Yang-Mills theory. The
denominator factor in \MYT\ that we need to cancel is
$\dim(\CR)^{2g-2}$, so it suffices to know that $\dim(\CR)^2$ is a
 polynomial of the Casimirs.  This can be proved using the Weyl
character formula, discussed in \refs{\Zelobenko, \S 123}, which
provides a general formula for $\dim(\CR)$. Parametrizing the
representation $\CR_h$ by a highest weight $h$, \eqn\WEYL{
\dim(\CR_h) \,=\, \prod_{\beta > 0} \, {{(\beta, h + \delta)}
\over {(\beta, \delta)}}\,.} The product in \WEYL\ runs over the
positive roots $\beta$, and we recall that $\delta$ is a constant,
equal to half the sum of the positive roots.  We regard this as a
function of $h'=h+\delta$.

The formula \WEYL\ exhibits a polynomial function $d$ on the
Cartan subalgebra of the Lie algebra $\Fg$ of $G$ such that
$\dim(\CR_h) = d(h')$.  The polynomial $d$ is not strictly
invariant under the action of the Weyl group on $h'$, but is
invariant up to sign, so $d^2$ is Weyl invariant. As such, $d^2$
extends to an invariant polynomial on all of $\Fg$, and thus a
polyomial in the Casimirs.  Finally, we observe that the shift $h
\to h'=h + \delta$ is the same renormalization that we introduced
for the potential $ V'(\CR_h) \,=\,  V(h + \delta)$, so that by
differentiating with respect to the couplings of each Casimir in $
V'$ we can cancel the denominator $\dim(\CR)^{2g-2}$.

\appendix{C}{A Few Additional Generalities About Equivariant Cohomology}

Following the discussion in Section 5.3, we discuss in this appendix
the identification of the $H$-equivariant cohomology of $N_0$ with the
$H_0$-equivariant cohomology of $\CM$, a fact which fundamentally leads to
the correspondence \HHZM.

To start, we find it useful to employ another topological model of
equivariant cohomology, explained for instance in Chapter 1 of
\GuilleminSII.  In this model, if $X$ is any topological space on
which a group $H$ acts, the $H$-equivariant cohomology ring of $X$ is
defined as the ordinary cohomology ring of the fiber product ${X_H = X
\times_H EH}$, where $EH$ is any contractible space on which $H$ acts
freely.  Such an $EH$ always exists, and the choice of $EH$ does not
matter, since $EH$ is unique up to $H$-equivariant homotopies.  Thus,
${H_H^*(X) = H^*(X_H)}$.

As a simple example, if $H$ acts freely on $X$, implying that $X$ is a
principal $H$-bundle over $X/H$, then $X_H$ is equivalent to a
product ${X_H = (X/H) \times EH}$.  Since $EH$ is contractible, we see
that ${H_H^*(X) = H^*(X/H)}$, a fact we applied in our discussion of
two-dimensional Yang-Mills theory.

At the opposite extreme, if $H$ acts trivially on $X$, then $X_H$ is
also a product ${X_H = X \times BH}$, where ${BH = EH/H}$ is the
classifying space associated to the group $H$.  In this case,
${H^*_H(X) = H^*(X) \otimes H^*(BH)}$.  However, by the definition of
equivariant cohomology above, the ordinary cohomology of $BH$ is the
$H$-equivariant cohomology of a point, so that ${H^*_H(X) = H^*(X)
\otimes H_H^*(pt)}$.  For the latter factor, our description of the
Cartan model in Section 4.1 clearly identifies $H_H^*(pt)$ with
the ring of invariant functions on the Lie algebra $\Fh$ of $H$.

We want the case in which $X$ is a fiber bundle over $\CM$ with
fiber $H/H_0$ for some $H$.  $H$ acts on the fibers, with fixed
subgroup $H_0$.  Now suppose that there exists a principal bundle
$Y\to \CM$, with fibers $H$, and the following properties.  We
suppose that $H\times H_0$ acts on $Y$, with $H$ acting on the
fibers on the left and $H_0$ on the right.   We also suppose that
$Y/H_0=X$.

In this situation, $H$ and $H_0$ both act freely on $Y$, the
quotient $Y/H$ being $\CM$ and the quotient $Y/H_0$ being $X$.
Moreover, $H_0$ acts trivially on $X$.

We can now argue as follows. First, $H^*_{H\times H_0}(Y) =
H^*_H(X)$, as $H_0$ acts freely on $Y$ with quotient $X$.  On the
other hand $H^*_{H\times H_0}(Y)=H^*_{H_0}(\CM)$ because $H$ acts
freely on $Y$ with quotient $\CM$.  Finally, as $H_0$ acts
trivially on $\CM$, $H^*_{H_0}(\CM)=H^*(\CM)\otimes
H^*_{H_0}(pt)$.  Putting these facts together, we have our desired
result that $H^*_H(X)= H^*(\CM)\otimes H^*_{H_0}(pt)$.

In general such a $Y$ only exists rationally (which is good enough
for de Rham cohomology), but for our problem with Chern-Simons
theory on a Seifert manifold, a natural $Y$ can be constructed as
follows.

First of all, over any symplectic manifold $\bar{\cal A}$, a
``prequantum line bundle'' ${\cal L}$ is a unitary line bundle
with connnection whose curvature is the symplectic form.  For
Chern-Simons theory, ${\cal L}$ exists and is unique up to
isomorphism as $\bar{\cal A}$ is an affine space. We let ${\cal
L}_0$ be the bundle of unit vectors in ${\cal L}$.  So ${\cal
L}_0$ is a circle bundle over $\bar{\cal A}$.

In general, any connected Lie group of symplectomorphisms of a
symplectic manifold that has an invariant moment map lifts to an
action on the prequantum line bundle. For Chern-Simons theory on a
Seifert manifold, the group ${\cal G}$ of gauge transformations
does not have a moment map, but its central extension $\wt \CG$
does. We recall that $\wt\CG$ is an extension of $\CG$ by a
subgroup $U(1)_Z$ that acts trivially on $\bar{\cal A}$ but has
moment map 1. Having a moment map, $\wt\CG$ acts on ${\cal L}$,
and hence on ${\cal L}_0$. $U(1)_Z$  acts by rotating the fibers
of the fibration ${\cal L}_0\to \bar{\cal A}$.  This action is
free.

Finally, the Hamiltonian group ${\cal H}$ that we really use for
our quantization is a semidirect product of $\wt\CG$ with
a $U(1)_R$ that rotates the fibers of the Seifert fibration.
$U(1)_R$ acts on ${\cal L}$ and ${\cal L}_0$, but not freely.  To
get the desired space $Y$ on which $U(1)_R$ acts freely, we simply
take $Y=U(1)\times {\cal L}_0$, where $U(1)_R$ acts by rotation on
$U(1)$ together with its natural action on ${\cal L}_0$.  So in
fact $H_0=U(1)_R\times U(1)_Z$ acts freely on $Y$.

We now want to restrict this construction from $\bar{\cal A}$, the
space of all connections, to $N_0$, the space of flat connections,
whose quotient $N_0/H$ is ${\cal M}$, the moduli space of
gauge-equivalence classes of flat connections.  We let $Y_0$ be the
restriction to $N_0$ of the fibration $Y\to \bar {\cal A}$.   So
$H\times H_0$ acts on $Y_0$; $H_0$ acts freely on $Y_0$ with quotient
$N_0$, and $H$ acts freely on $Y_0$ with quotient ${\cal M}$.
Finally, $H_0$ acts trivially on ${\cal M}$.  With these facts at
hand, the general argument presented above shows that
$H^*_H(N_0)=H^*(\CM)\otimes H^*_{H_0}(pt)$.

\appendix{D}{More About Localization at Higher Critical Points:
Localization Over a Nontrivial Moduli Space}

In this appendix, we consider the general case that our abstract model
for $F$ is fibered over a non-trivial moduli space $\CM$.  Our goal is
to compute the equivariant cohomology class on $\CM$ which is produced
by the canonical symplectic integral over $F$,
\eqn\DI{ I(\psi) \,=\, {1 \over {\Vol(H)}} \, \int_{\wt F}
\left[{{d\phi} \over  {2\pi}}\right] \exp{\left[t
D\lambda\right]}\,,\qquad \wt F \,=\, (\Fh \ominus \Fh_0) \times
F\,,\qquad \psi \in \Fh_0\,.}

We begin with some geometric preliminaries.  Very briefly, we recall
that we model $F$ as a vector bundle with fiber $\Fh^\perp \oplus E_1$
over a homogeneous base $H/H_0$.   Here ${\Fh^\perp = \Fh \ominus
\Fh_0 \ominus E_0}$, and explicitly,
\eqn\BIGF{ F \,=\, H \times_{H_0} (\Fh^\perp \oplus E_1)\,.}

To describe the total space $N$ of the fiber bundle ${F \longrightarrow
N \longrightarrow \CM}$, we introduce a principal $H$-bundle $P_H$
over $\CM$.  Besides the given action of $H$ on $P_H$, we assume
that $P_H$ also admits a free action of $H_0$ which commutes with the
action of $H$.  As a result, we can describe the bundle $N$ concretely
in terms of $P_H$ as
\eqn\BIGN{ N \,=\, P_H \times_{H_0} (\Fh^\perp \oplus E_1)\,.}
Upon setting $P_H = H$, where $H$ acts on the right and $H_0$ acts on the
left, this model for $N$ reduces to the model for $F$ itself, with
$\CM$ being a point.

Of course, the key ingredient in our localization computation is to
choose a good representative of the canonical localization form
$\lambda$ on $N$.  As in Section 4.3, we introduce another
localization form $\lambda'$ which (under the same caveats as in
Section 4.3 and Appendix A) is homotopic to $\lambda$ on $N$ and takes
the form
\eqn\LOCLAMK{ \lambda' \,=\, \lambda'_\perp \,+\, \lambda'_{E_0} \,+\,
\lambda'_{E_1}\,,}
with
\eqn\LOCLAMKII{\eqalign{
&\lambda'_\perp \,=\, \left(\gamma\,, \theta\right)\,,\cr
&\lambda'_{E_0} \,=\,  -i \left(\theta_{E_0}, g \phi g^{-1} + i
\CR(\theta)\right)\,,\qquad \CR(\theta) \,=\, d\theta -
\ha[\theta,\theta]\,,\cr
&\lambda'_{E_1} \,=\, i \left(\left(g \phi g^{-1}\right)_{\Fh_0} \cdot
v\,, dv - \theta_{\Fh_0} \cdot v\right)\,.\cr}}
In these expressions, we recall that $\gamma$ is an element of
$\Fh^\perp$, $g$ is an element of $H$, $\phi$ is an element of $\Fh$,
and $v$ is an element of the vector space $E_1$.  Finally, $\theta$ is
now a connection on the principal $H$-bundle $P_H$.  In particular,
$\theta$ is a globally-defined one-form on $P_H$.  As usual, we let
$\CR(\theta)$ denote the curvature of $\theta$.

Our choice for $\lambda'$ is precisely analogous to the choice we made
in Section 4.3 in the case that $P_H = H$, and in \LOCLAMK\ we have
simply grouped the terms in $\lambda'$ in a natural way for the
localization computation.  The only term present in \LOCLAMKII\ which
was not present in Section 4.3 is the term involving the curvature
$\CR(\theta)$ in $\lambda'_{E_0}$.  The curvature of $\theta$ is a
horizontal form on $P_H$, meaning that it is annihilated by
contraction with the vector fields $V(\phi)$ which generate the action
of $H$ on $P_H$, so this curvature term could not appear when $\CM$
was only a point.  Equivalently, if the connection $\theta$ takes the
global form ${\theta = dg \, g^{-1}}$ as in Section 4.3, then
$\CR(\theta)$ vanishes identically.

In \LOCLAMK\ and \LOCLAMKII\ we have written $\lambda'$ as an
invariant form on the direct product ${P_H \times (\Fh^\perp \oplus
E_1)}$, but one can check exactly as in Section 4.3 that $\lambda'$
descends under the quotient by $H_0$ to an invariant form on $N$.

Although $\lambda'$ is globally defined on $N$, we have written
$\lambda'$ in coordinates on $P_H$ with respect to a local
trivialization of this bundle about some point $m$ on the base $\CM$.
The integral we perform will be an integral over the fiber $F_m$ above
this point $m$, and since $m$ is arbitrary, this local computation
suffices to determine the cohomology class on $\CM$ that arises after
we perform the integral over all the fibers of ${F \longrightarrow N
\longrightarrow \CM}$.  In particular, upon pulling $\theta$ back to
the fiber $F_m$, $\theta$ takes the canonical form,
\eqn\THETAM{ \theta\big|_{F_m} \,=\, dg \, g^{-1}\,.}
However, since the curvature $\CR(\theta)$ can be non-zero, in general
$d\theta \neq \ha [\theta,\theta]$ at points in the fiber over $m$.

At this point, we repeat our earlier computation of $D\lambda'$,
allowing for the presence of the curvature $\CR(\theta)$.  We find
\eqn\DLOCLAMK{\eqalign{
&D\lambda'_\perp \,=\, \left(d\gamma, \theta\right) \,-\,
i\left(\gamma, \phi
+ i \, d\theta\right)\,,\cr
&D\lambda'_{E_0} \,=\, -i\left(d\theta_{E_0}, \phi + i \,
\CR(\theta)\right) \,+\, i\left(\theta_{E_0}, \left[\theta, \phi + i
\, \CR(\theta)\right]\right) \,-\, \left(\phi_{E_0}, \phi + i \,
\CR(\theta)\right)\,,\cr
&D\lambda'_{E_1} \,=\, i\left(\phi_{\Fh_0} \cdot dv, dv\right) \,-\,
\left(\phi_{\Fh_0} \cdot v, \left(\phi + i \,
\CR(\theta)\right)_{\Fh_0} \cdot
v\right) \,+\, \CX\,,}}
with
\eqn\BIGX{ \CX \,=\, i\left(\left[\theta,\phi\right]_{\Fh_0} \cdot v,
dv\right)\,+\, i\left(\phi_{\Fh_0}\cdot v,
\ha\left[\theta,\theta\right]_{\Fh_0}\cdot v\right)\,\quad \mod \,
\theta_{\Fh_0}.}
As before, in writing these expressions we make the change of variable
from $\phi$ to $g \phi g^{-1}$ at the end of the calculation to simplify
the result.  Also, the terms appearing in $\CX$ are at least of
cubic order in the ``massive'' variables $\theta$, $v$, and $dv$ and
so are irrelevant in the limit $t \to \infty$.  Finally, we are free
to work modulo terms involving $\theta_{\Fh_0}$ since $D\lambda'$ is a
pullback from the quotient $P_H \times_{H_0} (\Fh^\perp \oplus E_1)$.

We now compute directly the integral below in the limit $t \rightarrow
\infty$,
\eqn\DII{
I(\phi_{\Fh_0}) \,=\, {1 \over {\Vol(H)}} \, \int_{\wt F_m}
\left[{{d\phi} \over  {2\pi}}\right] \exp{\left[t D\lambda'_\perp + t
D\lambda'_{E_0} + t D\lambda'_{E_1}\right]}\,,\qquad \wt F_m \,=\, (\Fh
\ominus \Fh_0) \times F_m\,.}
This integral behaves essentially the same as the integral in Section
4.3, so we will be brief.

We first consider the integral over $E_1$, which we perform as a
Gaussian integral using the terms from $t D\lambda'_{E_1}$ in the large
$t$ limit.  Explicitly, the integral over $E_1$ is given by
\eqn\DIII{ \int_{E_1} \exp{\left[i t\left(\phi_{\Fh_0} \cdot dv,
dv\right) \,-\, t\left(\phi_{\Fh_0} \cdot v, \left(\phi + i \,
\CR(\theta)\right)_{\Fh_0} \cdot v\right) \,+\,
t \CX\right]}\,.}
Since $\CX$ is of at least cubic order in the massive variables
$\theta$, $v$, and $dv$, this term can be dropped from the integrand
when $t$ is large.  Keeping the other terms quadratic in $v$ and $dv$ in
\DIII, the Gaussian integral over $E_1$ immediately produces
\eqn\GVINTC{ \det\left({1 \over {2
\pi}} \left(\phi_{\Fh_0} + i \,
\CR(\theta)_{\Fh_0}\right)\Big|_{E_1}\right)^{-1}\,.}

We now integrate over both $\gamma$ and $\phi$ in ${\Fh^\perp = \Fh
\ominus \Fh_0 \ominus E_0}$.  We see from \DLOCLAMK\ that $\gamma$
still appears only linearly in $t D\lambda'$, so the integral over
$\gamma$ produces a delta-function of $\phi_\perp$, where $\phi_\perp$
denotes the component of $\phi$ in $\Fh^\perp$.  As is evident from the
form of $t D\lambda'_\perp$, this delta-function sets $\phi_\perp = -i
d\theta_\perp$.  (As in Section 4.3, the factors of $t$ cancel between
the integral over $\gamma$ and the integral over $\phi_\perp$.)

We are left to integrate over $\phi_{E_0}$ and over the base $H/H_0$
of $F_m$.  Of course, upon Taylor expanding the exponential
$\exp{(d\gamma,\theta)}$ from $D\lambda'_\perp$ to produce the measure
for $\gamma$, we also produce the canonical measure on the tangent
directions to $H/H_0$ lying in $\Fh^\perp$.  So infinitesimally we
have only to integrate over the remaining tangent directions to
$H/H_0$ which lie in $E_0$ in addition to $\phi_{E_0}$.

So we are left to integrate over $E_0$ using the terms in $t
D\lambda'_{E_0}$.  This integral takes the form
\eqn\DIV{\eqalign{
\int_{E_0} &\exp{\left[-i t\left(\theta_{E_0},
\left[\phi_{\Fh_0} + i \, \CR(\theta)_{\Fh_0},
\theta_{E_0}\right]\right) \,+\, t\left(\CR(\theta)_{E_0},
\CR(\theta)_{E_0}\right)\right]}\,\times\cr
\times\,&\exp{\left[- 2 i t\left(\CR(\theta)_{E_0},
\phi_{E_0}\right) \,-\,
t\left(\phi_{E_0},\phi_{E_0}\right)\right]}\,.\cr}}
In deducing \DIV, we have expanded and simplified various terms in
$D\lambda'_{E_0}$ in \DLOCLAMK.  For instance, the curvature term
$\left(\CR(\theta)_{E_0}, \CR(\theta)_{E_0}\right)$ arises from the
linear combination of terms ${\left(d\theta_{E_0},\CR(\theta)\right) -
\left(\theta_{E_0},\left[\theta,\CR(\theta)\right]\right)}$ in
$D\lambda'_{E_0}$.  To see this, we rewrite this expression as
${\left(d\theta_{E_0} - [\theta,\theta_{E_0}],
\CR(\theta)_{E_0}\right) \equiv
\left(\CR(\theta)_{E_0},\CR(\theta)_{E_0}\right)}$, where ``$\equiv$''
indicates that the equality holds modulo $\theta_{\Fh_0}$ and
$\theta_\perp$, which is good enough since these forms do not
contribute to the integral over $E_0$.

In writing \DIV, we also note that when we set ${\phi_\perp = -i \,
d\theta_\perp}$ in $D\lambda'_{E_0}$, we effectively cancel similar
terms in $D\lambda'_{E_0}$ which involve the components of the curvature
$\CR(\theta)$ in $\Fh^\perp$.  So $\CR(\theta)_\perp$ does not appear
in \DIV.

We first perform the Gaussian integral over $\phi_{E_0}$ in \DIV.  The
result of this integral produces a term proportional to
$\exp{\left[-t\left(\CR(\theta)_{E_0},
\CR(\theta)_{E_0}\right)\right]}$ which precisely cancels the term
quadratic in the curvature $\CR(\theta)_{E_0}$ in the first line of \DIV.
Consequently, once we collect factors of $t$ and $2\pi$ exactly as in
Section 4.3, the term quadratic in $\theta_{E_0}$ in \DIV\ produces
another determinant,
\eqn\GVINTCII{ \det\left({1 \over {2 \pi}}\left(\phi_{\Fh_0} + i\,
\CR(\theta)_{\Fh_0}\right)\Big|_{E_0}\right)\,.}

Including the factor $\Vol(H) / \Vol(H_0)$ that arises from the
integral over $H/H_0$ and setting $\phi_{\Fh_0} \equiv \psi$ for
notational simplicity, we find our final result for the integral in \DII,
\eqn\DV{ I(\psi) \,=\, {1 \over {\Vol(H_0)}} \,  \det\left({1 \over {2
\pi}}\left(\psi + i\, \CR(\theta)_{\Fh_0}\right)\Big|_{E_0}\right) \,
\det\left({1 \over {2 \pi}} \left(\psi + i \,
\CR(\theta)_{\Fh_0}\right)\Big|_{E_1}\right)^{-1}\,.}

Since both $E_0$ and $E_1$ are representations of $H_0$, the
associated bundles ${P_H \times_{H_0} E_0}$ and ${P_H \times_{H_0} E_1}$
determine $H_0$-equivariant bundles over $\CM$ once we divide by the
action of $H$ on $P_H$.  The determinants appearing in \DV\ are then the
Chern-Weil representatives of the $H_0$-equivariant Euler classes of
these bundles.

\listrefs

\end